\documentclass[fleqn,usenatbib]{mnras}

\usepackage[T1]{fontenc}
\usepackage{ae,aecompl}
\usepackage[usenames,dvipsnames]{xcolor}  

\usepackage{graphicx}	
\usepackage{amsmath}	
\usepackage{amssymb}	
\usepackage{caption}
\usepackage{subcaption}
\usepackage{tikz}
\usetikzlibrary{arrows.meta,calc,decorations.markings,math,arrows.meta}
\usepackage[dvipsnames]{xcolor}
\usepackage{newtxtext,newtxmath}

\title[PKS 1502+106]{A ring accelerator? Unusual jet dynamics in the IceCube candidate PKS 1502+106}

\author[S. Britzen et al.]{
S. Britzen$^{1}$\thanks{E-mail: sbritzen@mpifr.de},
M. Zaja\v{c}ek$^{1, 2, 3}$,
L.\v{C}. Popovi\'{c}$^{4, 5}$,
C. Fendt$^{6}$,
A. Tramacere$^{7}$,
\newauthor
I.N. Pashchenko$^8$,
F. Jaron$^{1, 9, 10}$,
R. P\'{a}nis$^{11}$,
L. Petrov$^{12}$,
M.F. Aller$^{13}$, and
H.D. Aller$^{13}$
\\
$^{1}$Max-Planck-Institut f\"ur Radioastronomie, Auf dem H\"ugel 69, 53 121 Bonn, Germany\\
$^{2}$I. Physikalisches Institut, Universit\"at K\"oln, Z\"ulpicher Str. 77, K\"oln, Germany\\
$^{3}$Center for Theoretical Physics, Polish Academy of Sciences, Al. Lotnik\'{o}w 32/46, 02-668 Warsaw, Poland\\
$^4$Astronomical observatory Belgrade, Volgina 7, P.O.Box 74 11060, Belgrade, 11060, Serbia\\
$^5$Department of Astronomy, Faculty of Mathematics, University of Belgrade, Studentski Trg 16, 11158 Belgrade, Serbia\\
$^6$Max Planck Institute for Astronomy, K\"onigstuhl 17, 69117 Heidelberg, Germany\\
$^7$Department of Astronomy, University of Geneva, Ch. d'Ecogia 16, 1290, Versoix, Switzerland\\
$^8$Astro Space Center, Lebedev Physical Institute, Russian Academy of Sciences\\
$^9$Chalmers University of Technology, Department of Space, Earth and Environment, Onsala Space Observatory, SE-439 92 Sweden\\
$^{10}$Department of Geodesy and Geoinformation, TU Wien, Wiedner Hauptstra{\ss}e 8-10, 1040 Vienna, Austria\\
$^{11}$Research Centre for Theoretical Physics and Astrophysics, Institute of Physics, Silesian University in Opava, Bezru\v{c}ovo n\'{a}m. 13,\\ CZ-74601 Opava, Czech Republic\\ 
$^{12}$NASA, Goddard Space Flight Center, 8800 Greenbelt Rd, Greenbelt MD 20771, USA\\
$^{13}$Astronomy Department, University of Michigan, Ann Arbor, MI 48109-1107, USA\\
}

\date{Accepted XXX. Received YYY; in original form ZZZ}

\pubyear{2020}

\begin{document}
\label{firstpage}
\pagerange{\pageref{firstpage}--\pageref{lastpage}}
\maketitle

\begin{abstract}
On 2019/07/30.86853 UT, IceCube detected a high-energy astrophysical neutrino candidate. The Flat Spectrum Radio Quasar PKS 1502+106 is located within the 50 per cent uncertainty region of the event. Our analysis of 15 GHz Very Long Baseline Array (VLBA) and astrometric 8 GHz VLBA data, in a time span prior and after the IceCube event, reveals evidence for a radio ring structure which develops with time. Several arc-structures evolve perpendicular to the jet ridge line. We find evidence for precession of a curved jet based on kinematic modelling and a periodicity analysis. An outflowing broad line region (BLR) based on the C IV line emission (Sloan Digital Sky Survey, SDSS) is found. We attribute the atypical ring to an interaction of the precessing jet with the outflowing material. We discuss our findings in the context of a spine-sheath scenario where the ring reveals the sheath and its interaction with the surroundings (narrow line region, NLR, clouds). We find that the radio emission is correlated with the $\gamma$-ray emission, with radio lagging the $\gamma$-rays. Based on the $\gamma$-ray  variability timescale, we constrain the $\gamma$-ray emission zone to the BLR (30-200~$r_{g}$) and within the jet launching region. We discuss that the outflowing BLR provides the external radiation field for $\gamma$-ray production via external Compton scattering. The  neutrino is most likely  produced  by proton-proton interaction in the blazar zone (beyond the BLR), enabled by episodic encounters of the jet with dense clouds, i.e. some molecular cloud in the NLR.
\end{abstract}



\begin{figure}
\includegraphics[width=8.5cm]{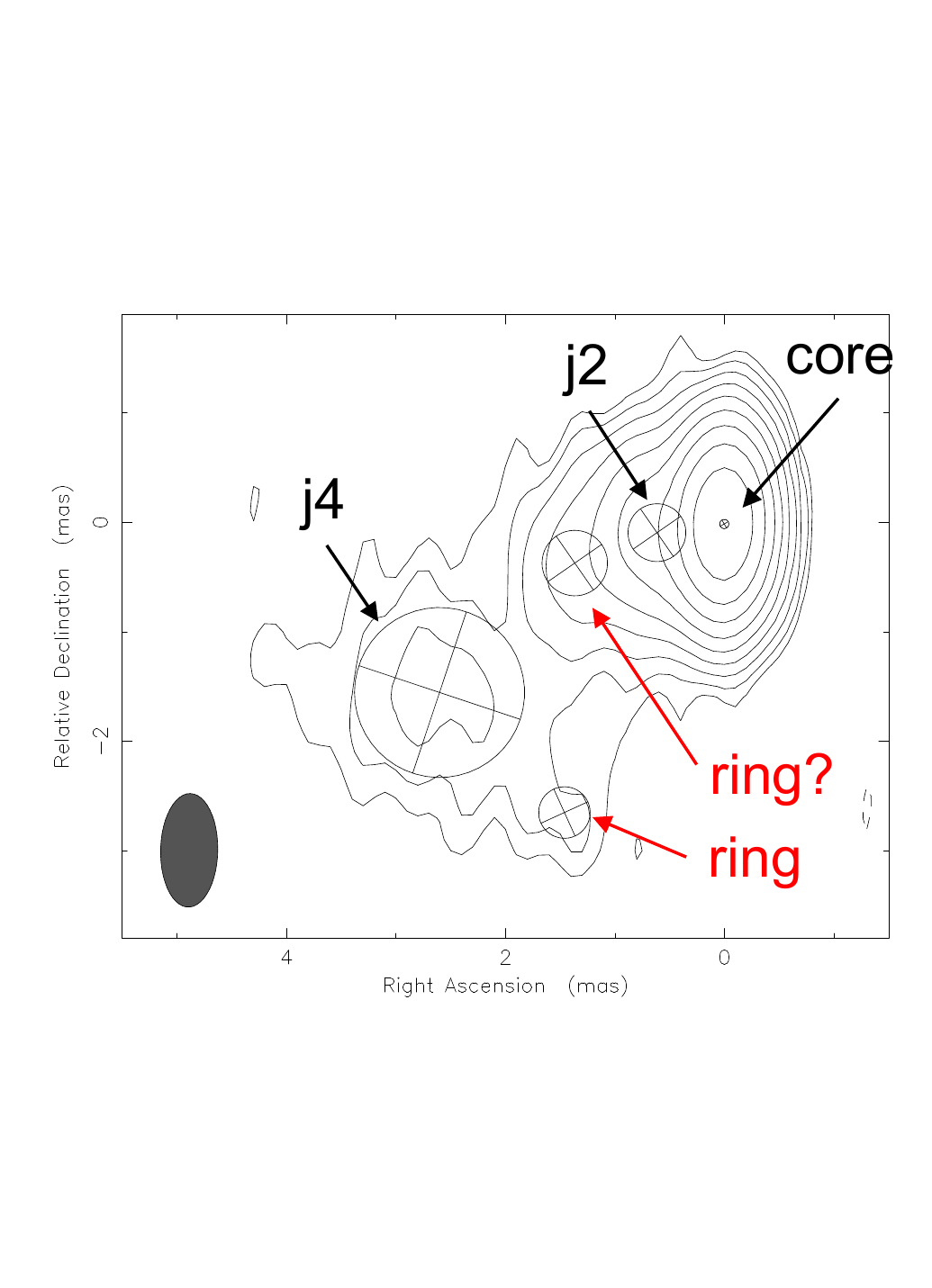}
\vspace{-2.5cm}
\caption{The pc-scale jet morphology of PKS 1502+106 as obtained in 15 GHz observations with the VLBA (2008/11/19). The contours are on a logarithmic scale between -0.2 and 51.2 per cent of the peak flux of 1.59 Jy beam$^{-1}$. The beam size is 1.04~mas $\times$ 0.523~mas at -1.11 deg. Superimposed are the modelfit components. We indicate the core region, the jet components and mark those components which might belong to the projected ring structure. The ring is also indicated in Fig.~\ref{fig_jet_ring_flux} and discussed in the text.}
\label{map0}
\end{figure}
\begin{figure}
\centering
\includegraphics[width=9.2cm]{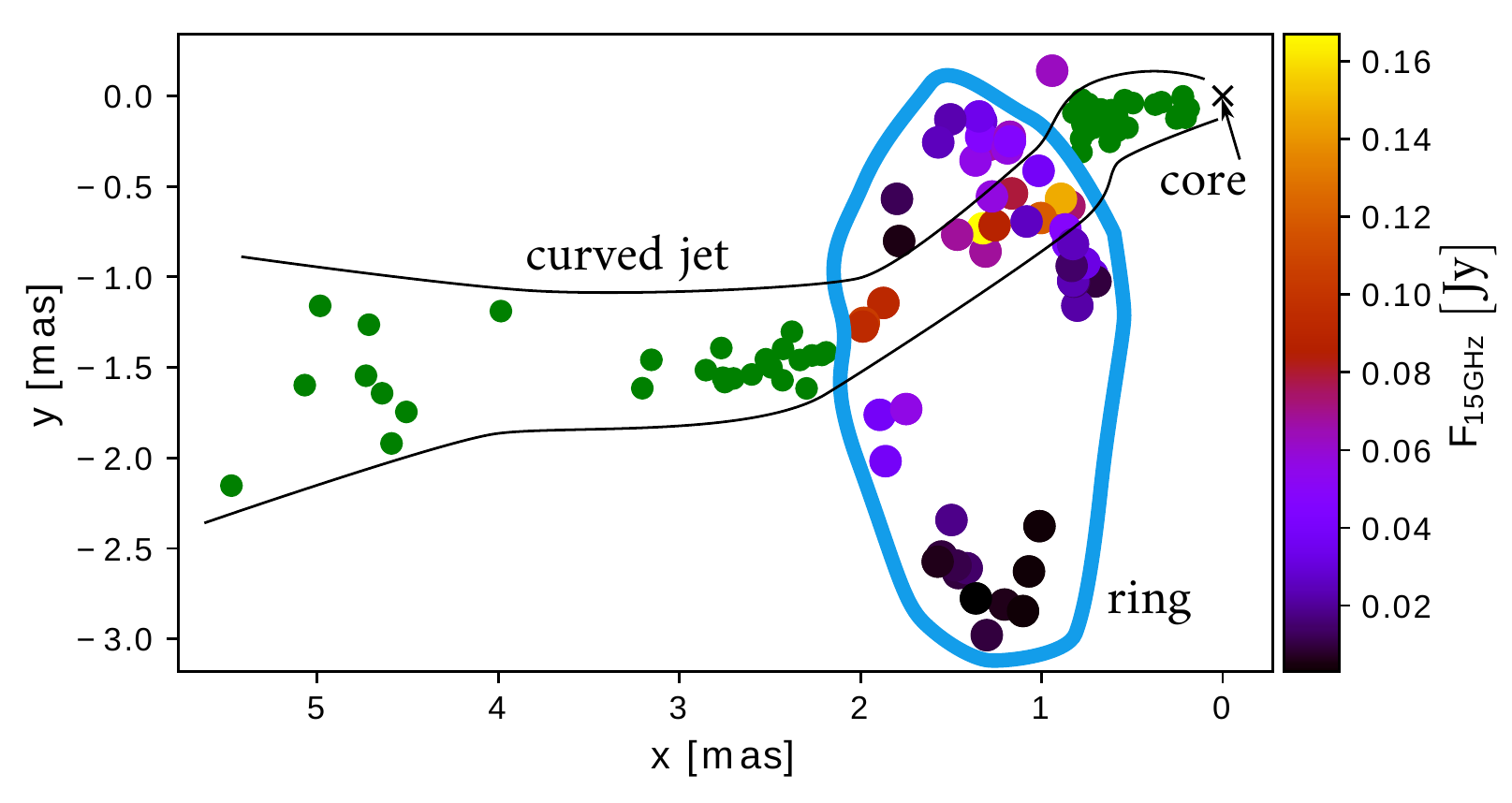}
\caption{All jet features from all epochs and marked in different colours to indicate their origin in jet or ring. According to the
colour axis to the right, the colour-coded flux density distribution in the ring region (in Janskys) is shown. In the figure, we also depict the jet path according to the identified jet components j1-j4 (all are coloured as green for clarity). The core is denoted with an x.}
\label{fig_jet_ring_flux}
\end{figure}

\section{Introduction}
TXS 0506+056 was the first blazar to be identified as a neutrino emitter by the IceCube collaboration \citep{IceCube2018a, IceCube2018b}. Modelling the Spectral Energy Distribution (SED) of both IceCube-170922A and the 2014/15 neutrino flare within the same scenario remains difficult \citep[e.g.,][]{rodrigues, keivani}. \citet{britzen_txs} presented a kinematic study of TXS 0506+056 which showed that the single neutrino and the neutrino flare could have been produced in a collision of jetted material within this BL Lac object. A very small viewing angle and a precessing inner jet provided the proper conditions.

While radio-loud Active Galactic Nuclei (AGN) and especially the blazars were among the suspected source candidates, only TXS 0506+056 could so far be identified as a neutrino emitter \citep{IceCube2018a, IceCube2018b}. Several blazars have been identified as likely neutrino emitting candidates \citep{aartsen}. It remains enigmatic, why the rest of the 2500 \textit{Fermi}-bright blazars \citep[e.g.,][]{abdo} do not seem to produce neutrinos detectable with current neutrino detectors. For a general overview of astrophysical sources of high-energy neutrinos, please see \citet{meszaros}.

On 2019/07/30.86853 UT IceCube detected another high-energy astrophysical neutrino candidate \citep{atel}. The FSRQ PKS 1502+106 is located within the 50 per cent uncertainty region of the event. According to the 
SED class and based on the SED peak location, this source is classified as an LSP (Low-spectral peaked ($<$ 10$^{14}$ Hz)) HPQ (highly polarised quasar; quasar with fractional linear optical polarisation above 3 per cent on at least one occasion) (MOJAVE\footnote{Monitoring Of Jets in Active galactic nuclei with VLBA Experiments webpage, see \url{https://www.physics.purdue.edu/MOJAVE/}}). The redshift of PKS 1502+106 is 1.838 \citep{paris}. This source is bright in the $\gamma$-ray regime, and routinely monitored with \textit{Fermi}-LAT \citep{fermi}.  PKS 1502+106 revealed a rapid (5 days duration), high-energy (E $>$ 100 MeV) $\gamma$-ray outburst in 2008 \citep{abdo}. The luminosity (at energies $>$ 100 MeV) is about 1.1$\times$10$^{49}$ erg s$^{-1}$, and the black hole mass is close to 10$^9$M$_\odot$ \citep{abdo}. 

Compared to TXS 0506+056, PKS 1502+106 is a high-$z$ source which has implications for determining the high-energy portion of the SED. No TeV detections have been reported for this target, nor are they expected from such a high redshift source with the current technology. 

PKS 1502+106 is also part of the MOJAVE sample and has been monitored in radio interferometric observations with the VLBA \citep{Lister2018}. In this paper, we perform an analysis of the kinematic evolution of the PKS 1502+106 jet (subsections \ref{jk} and \ref{jk1} similar to the analysis presented for TXS 0506+056 in \citealt{britzen_txs}), present the polarisation properties (subsection \ref{pol}), model jet precession (subsection \ref{precession}), perform a periodicity and correlation analysis of the radio and $\gamma$-ray light curves (subsection \ref{analysis}), and explore the Sloan Digital Sky Survey (SDSS) spectrum (subsection \ref{civ}) to determine the likely physical processes that make this AGN a high-energy source and a possible neutrino emitter (see discussion in section \ref{discussion}).

\subsection{PKS 1502+106 in previous observations}
PKS~1502+106 has been studied in several VLBI observations and shows a core-jet structure \citep{murphy, fey, fomalont, zensus, karamanavis}. \citet{an} discern a complex curved jet in multi-frequency observations taken with the VLBA, the European VLBI Network (EVN), and the Multi-Element Radio Linked Interferometer Network (MERLIN). Based on a radio core with a brightness temperature approaching the equipartition limit, the authors assume highly relativistic plasma beamed towards us.

\citet{anb} report evidence for extreme apparent superluminal motion of up to 37.3$\pm$9.3 $c$ and also for a jet bending at 3-4 mas, from PA $\sim$ 130~deg to PA $\sim$ 80~deg (their Fig.~2). \citet{karamanavis} report a compact core-jet morphology and fast apparent speeds ($5–-22~c$) based on VLBI observations at 15, 42, and 86 GHz.

They also find that the viewing angle differs between the inner ($\sim$ 3 deg) and outer jet ($\sim$ 1 deg) with the jet bending towards the observer beyond 1 mas. The total jet power has been determined to be log(P$_{\rm jet}$)=47.1 \citep{ghisellini09}.
\citet{ding} investigate the origin of the multi-band variability in this source and find that a fast $\gamma$-ray dominated outburst in 2015 may have been triggered through magnetic reconnection. An optical outburst in 2017 may have also been triggered by a transverse shock.

PKS 1502+106 revealed abnormally-high optical polarisation of 47$\pm$0.1 per cent with a position angle of 85.2$\pm$0.1~deg (December 9, 2017) based on observations with the 1.54m Kuiper Telescope on Mt. Bigelow, Arizona and the SPOL spectropolarimeter \citep{smith}. 
Polarisation information has also been investigated by \citet{shao}. They find that within uncertainties, the locations of the $\gamma$-ray and optical emitting regions are roughly identical, and localized within 1.2 pc from the jet base. They derive a value for the magnetic field of 0.36 G in the optical and $\gamma$-ray emitting regions. The Synchrotron self-Compton process (SSC), the external inverse Compton processes (EC), or a combination of both can explain the linear correlation between the logarithm of the $\gamma$-ray flux density and that of the V-band flux. According to \citet{shao}, changes of the viewing angle is the dominant mechanism causing the variability of the fluxes, spectral indices, and polarisation degrees for PKS 1502+106.

Throughout the paper we adopt the following parameters: a luminosity distance D$_{\rm L}$= 14366.8$\,$Mpc at the source redshift of $z=1.838$ with cosmological parameters corresponding to a $\Lambda$CDM Universe with $\Omega_{m}$=0.308, $\Omega_{\lambda}$=0.691, and $\rm{H_0 =67.8 km s^{-1}}$ Mpc$^{-1}$ \citep{planck}. Thus, a proper motion of 1\,mas\,yr$^{-1}$ corresponds to an apparent superluminal speed of 80.1$c$, while 1$\,$mas = 8.648$\,$pc.

\section{Observations \& data analysis}
\subsection{VLBA data re-analysis and uncertainty estimation}
We remodelled and reanalysed 25 VLBA observations (15$\,$GHz, MOJAVE\footnote{\url{https://www.physics.purdue.edu/MOJAVE/}}) obtained between 1997 and 2020. The MOJAVE team reported a VLBA flux scaling issue\footnote{\url{https://www.astro.purdue.edu/MOJAVE/}}. This scaling issue led to correlated flux densities that are between 10 per cent and 20 per cent
too low in the data for all the sources after early May 2019. Three epochs of MOJAVE data studied in this paper are affected. We corrected this problem by scaling the original modelfit flux densities (and the corresponding uncertainties) from core and jet components to match the OVRO data for the last three epochs (2019/08/23, 2019/08/27, and 2019/10/11) (see Fig.~\ref{recal}).

We also re-analysed six X-band VLBA observing sessions from three astrometric programs: a complete sample of 2MASS galaxies program \citep{condon}, the second epoch VLBA Calibrator Survey observations: VCS-II \citep{gordon}, and the radio follow-up on all unassociated
$\gamma$-ray sources from the third Fermi Large Area Telescope source catalog programs \citep{schinzel}. PKS 1502+106 was observed as a calibrator in these experiments. These latter data sets are publicly available\footnote{\url{http://astrogeo.org/}}. The X-band data were observed between 2012 and 2019. 
Gaussian circular components were fitted to all the data in the \textit{uv}-plane to obtain the optimum set of parameters within the
{\it difmap}-modelfit program \citep{shepherd}. 
Every epoch was fitted independently from all the other epochs. The following parameters were fitted to the data: the flux density of the component, the radial distance of the component centre from the centre of the map, the position angle of the centre of the component (measured from north to east) with respect to an imaginary line drawn vertically through the map centre, and the full width at half maximum (FWHM) axis of the circular component. The modelfit procedure was performed blindly so as  not to impose any specific outcome. 
\begin{figure*}
  \begin{minipage}{12cm}
    \includegraphics[width=\textwidth]{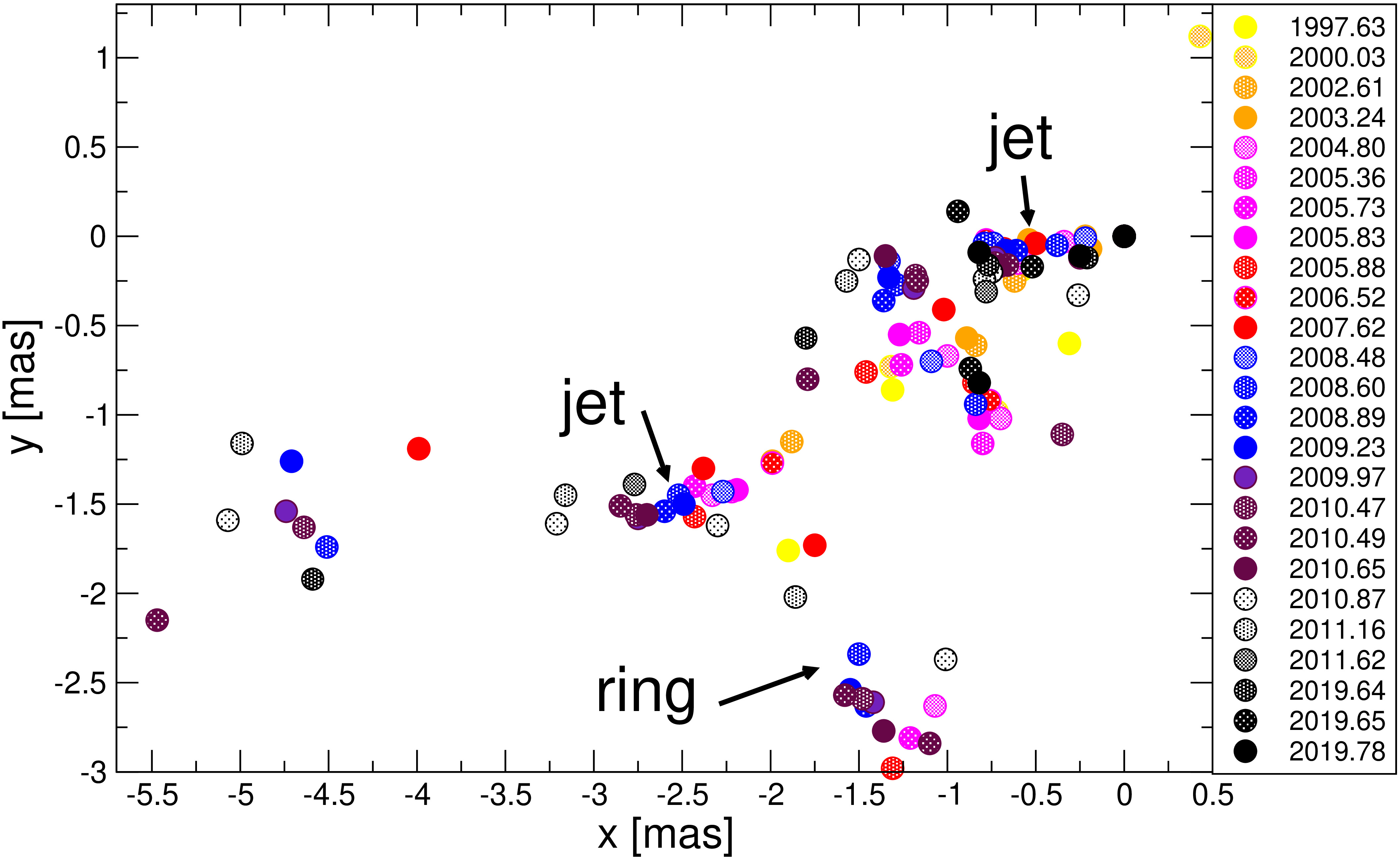}
    [a]
  \end{minipage}
  \begin{minipage}{16cm}
    \includegraphics[width=16cm]{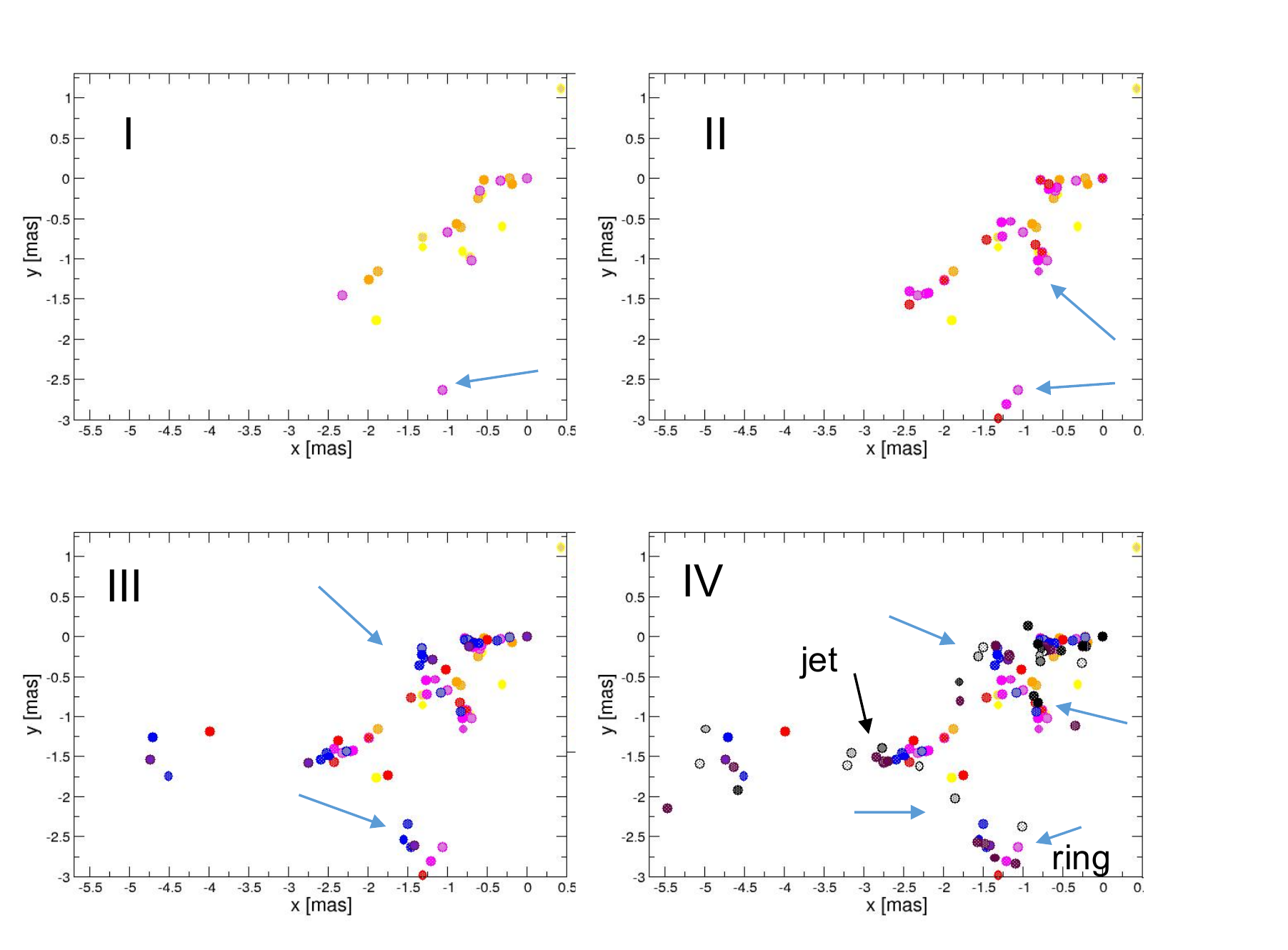}
    [b]
  \end{minipage}
\caption{[a] The pc-scale jet morphology in xy-coordinates with the individual epochs of observation marked in different colours (all 15 GHz VLBA observations). The jet and ring are indicated by arrows for clarity. [b] From I to IV: With time, more data points appear along the projected ring structure. The light blue arrows indicate the regions along the ring where the data points preferentially appear.}
\label{ring}
\end{figure*}
\begin{figure*}
\begin{minipage}{8.4cm}
  \includegraphics[width=8.4cm]{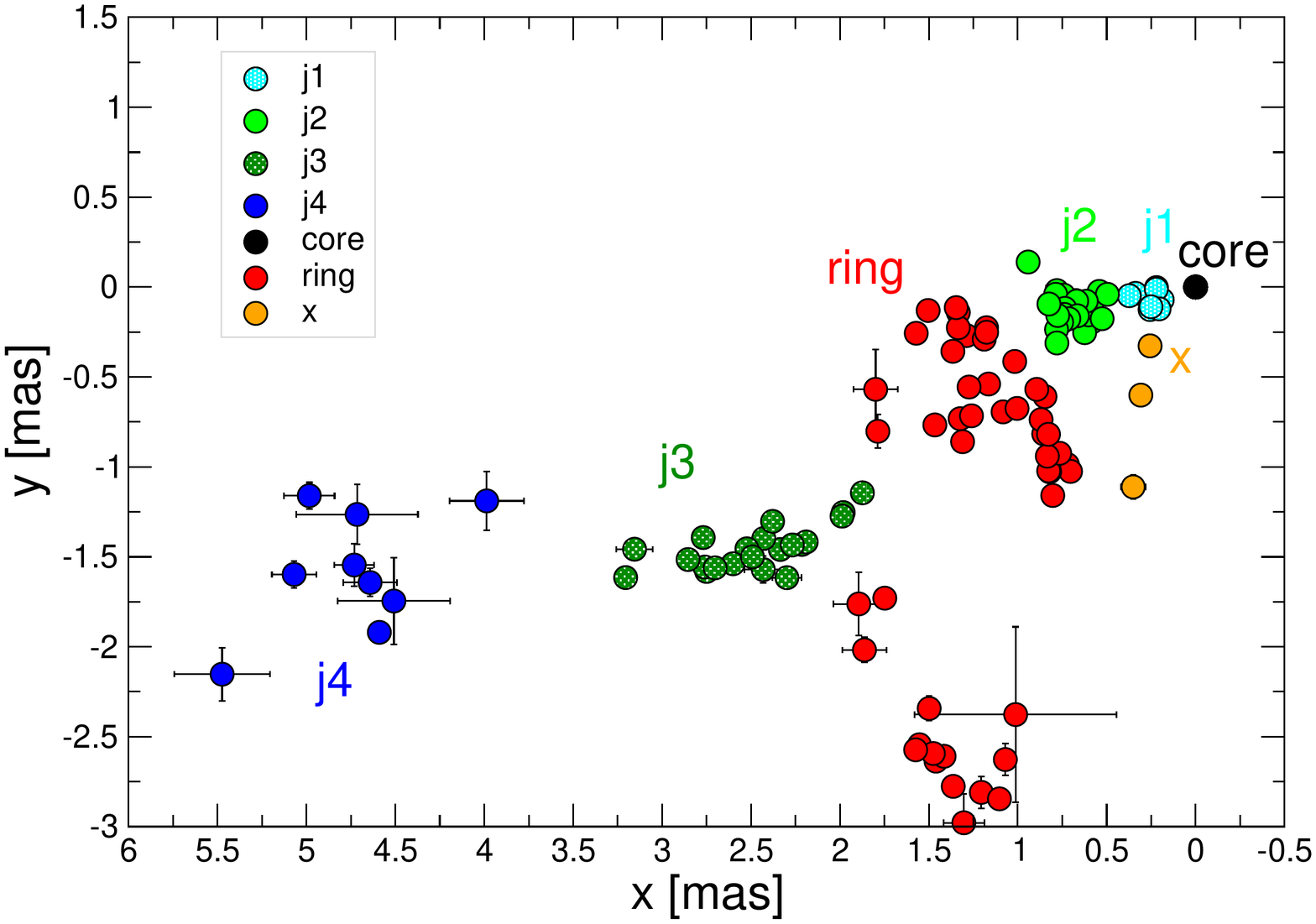}
  [a]
\end{minipage}
\begin{minipage}{8.4cm}
  \includegraphics[width=8.4cm]{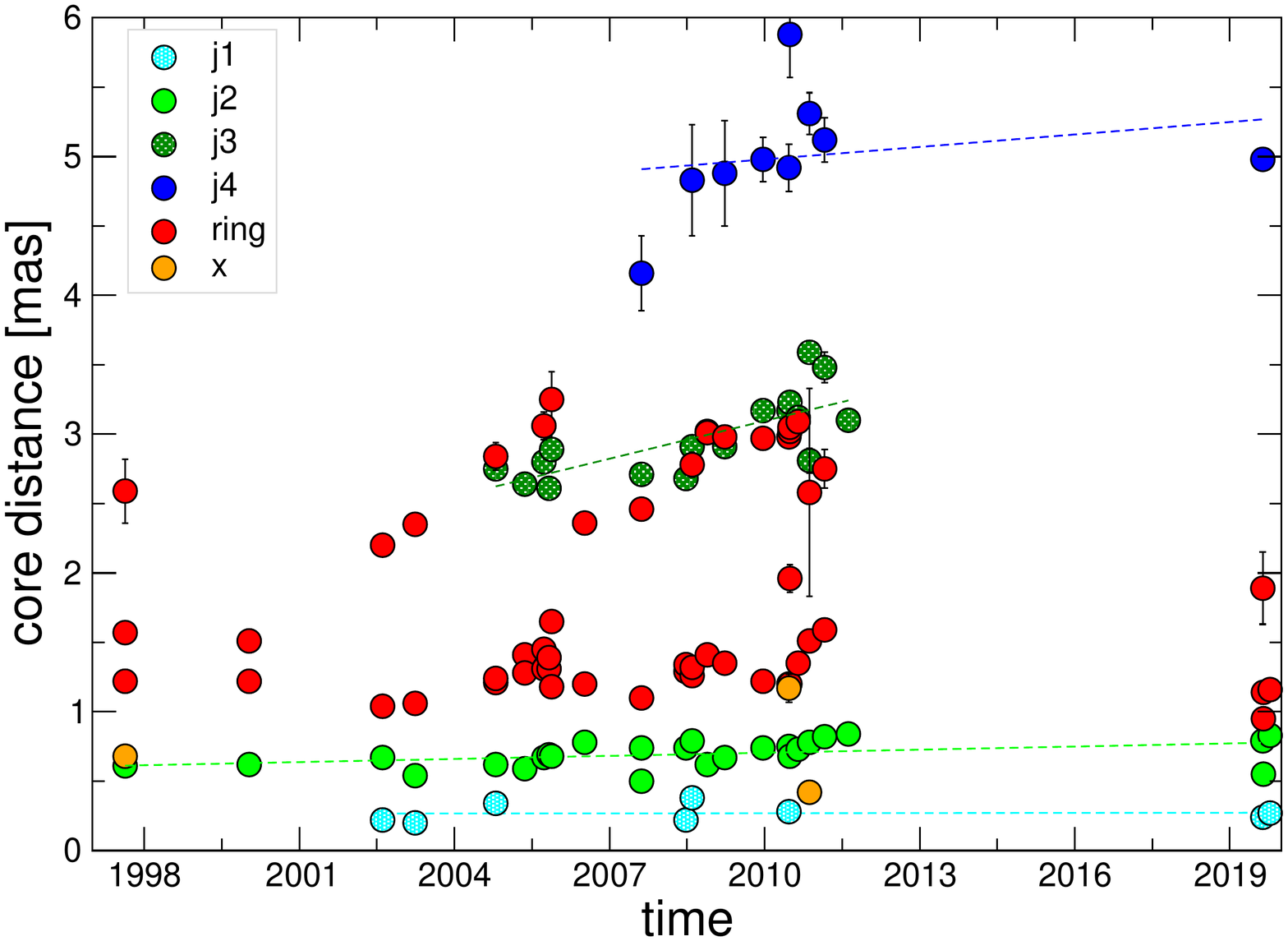}
  [b]
\end{minipage}
\begin{minipage}{8.4cm}
  \includegraphics[width=8.4cm]{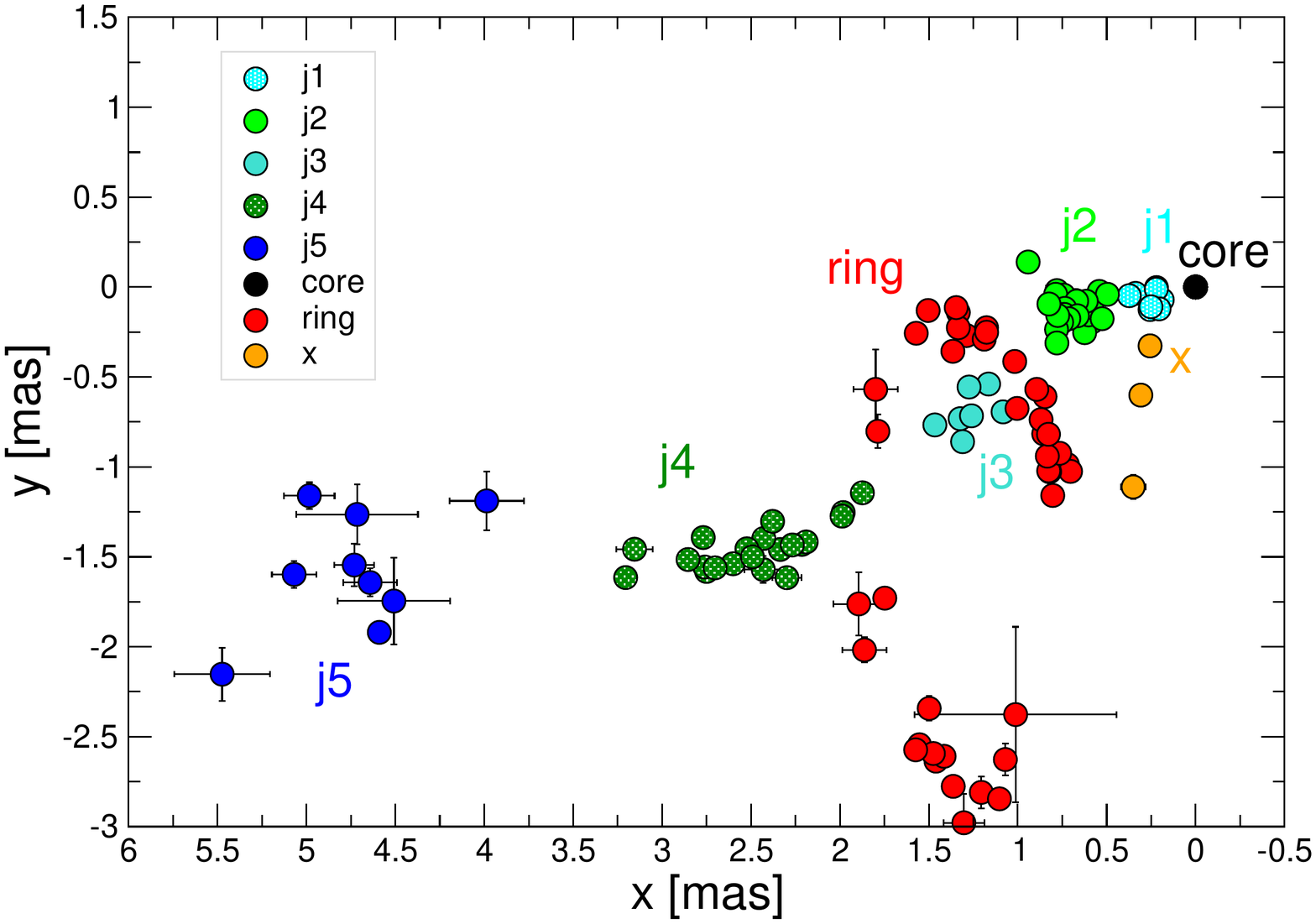}
  [c]
\end{minipage}
\begin{minipage}{8.4cm}
  \includegraphics[width=8.4cm]{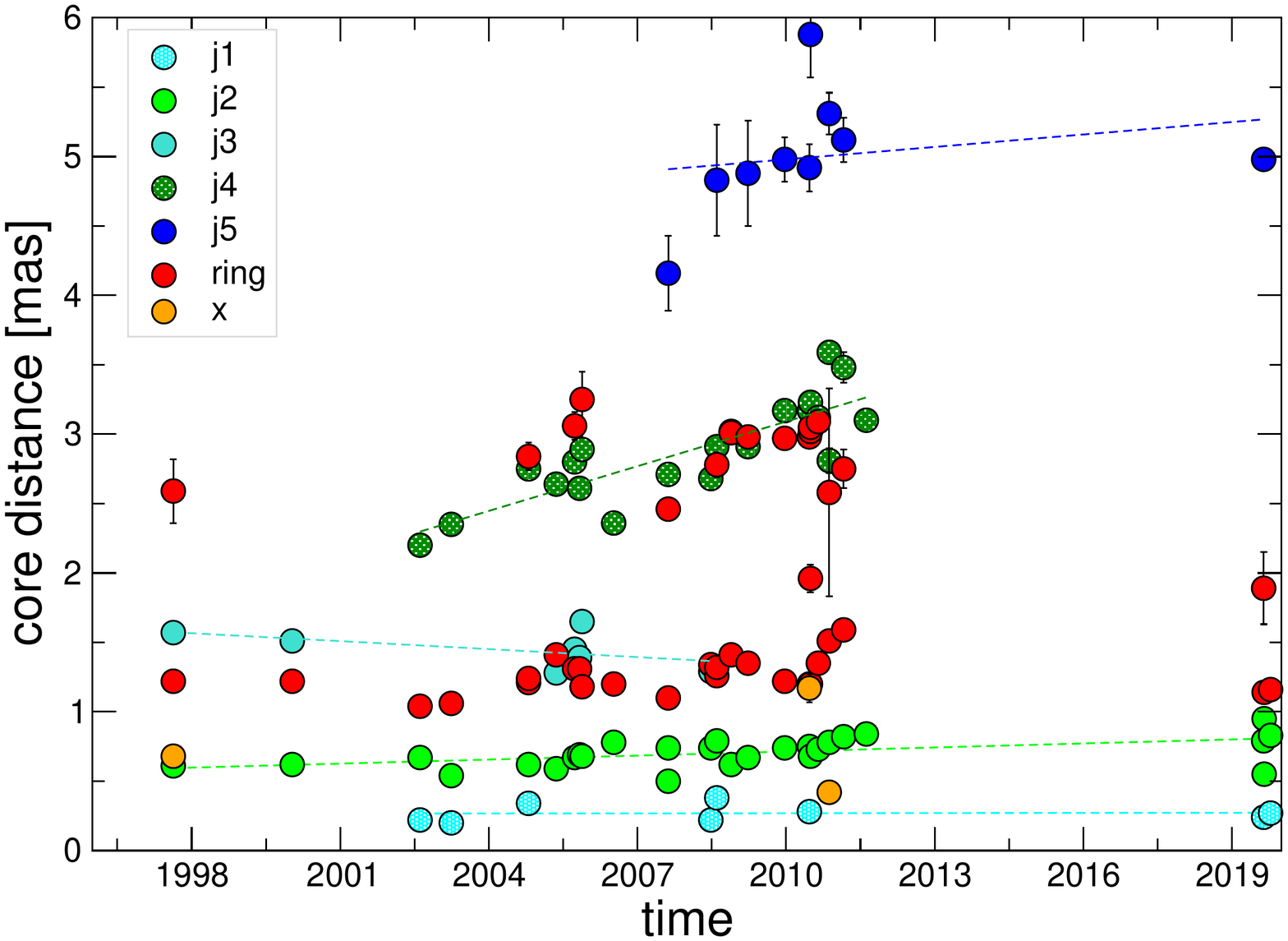}
  [d]
\end{minipage}
\caption{[a] Jet component positions are here shown with error bars and marked according to their likely physical origin (core, jet (j), ring, unidentified (x)). The main jet ridge line plus a feature resembling a ring in projection can be seen. [b] The core distance versus time for those components shown in [a]. [c] and [d] show the same relations as in [a] and [b] with a slightly different (alternative) identification of the components and additional jet component identification.}
\label{identify}
\end{figure*}
Special care was taken to identify correctly the core component in every individual data set. 
We used the brightest jet feature as a reference point for all the jet components. This implicitly assumes that the core is always the brightest feature in every epoch. We show one of these maps with the modelfit components superimposed in Fig.~\ref{map0}.

Uncertainties of the model component parameters were determined using the following procedure. We obtained both self-calibrated and raw data from the MOJAVE database\footnote{\url{https://www.cv.nrao.edu/2cmVLBA/data/1502+106/}}. Using these data we calculated amplitude and phase self-calibration corrections for each visibility. Then we created many artificial data sets using our best {\it difmap} model visibilities and added back self-calibration correction and different realizations of the noise estimated from the original data. The resulting data sets were self-calibrated and modelfitted in {\it difmap} with our best model being the initial guess. Thus, we obtained a distribution for each model parameter. Its standard deviation was used to estimate the corresponding uncertainty.
For the Radio Fundamental Catalog (rfc) data we derived the positional uncertainties using results of MCMC fit of visibilities with the same number of components as in the {\it difmap} modelfit. For sampling the posterior distribution of the parameters we employed \textit{Diffusive Nested Sampling} algorithm \citep{brewer2011} implemented in \texttt{DNest4} package \citep{dnest4}. We also compared MCMC models of the source structure with \textit{difmap} models for some of the MOJAVE 15 GHz data and found consistent results.
\begin{figure}
\includegraphics[width=7.3cm]{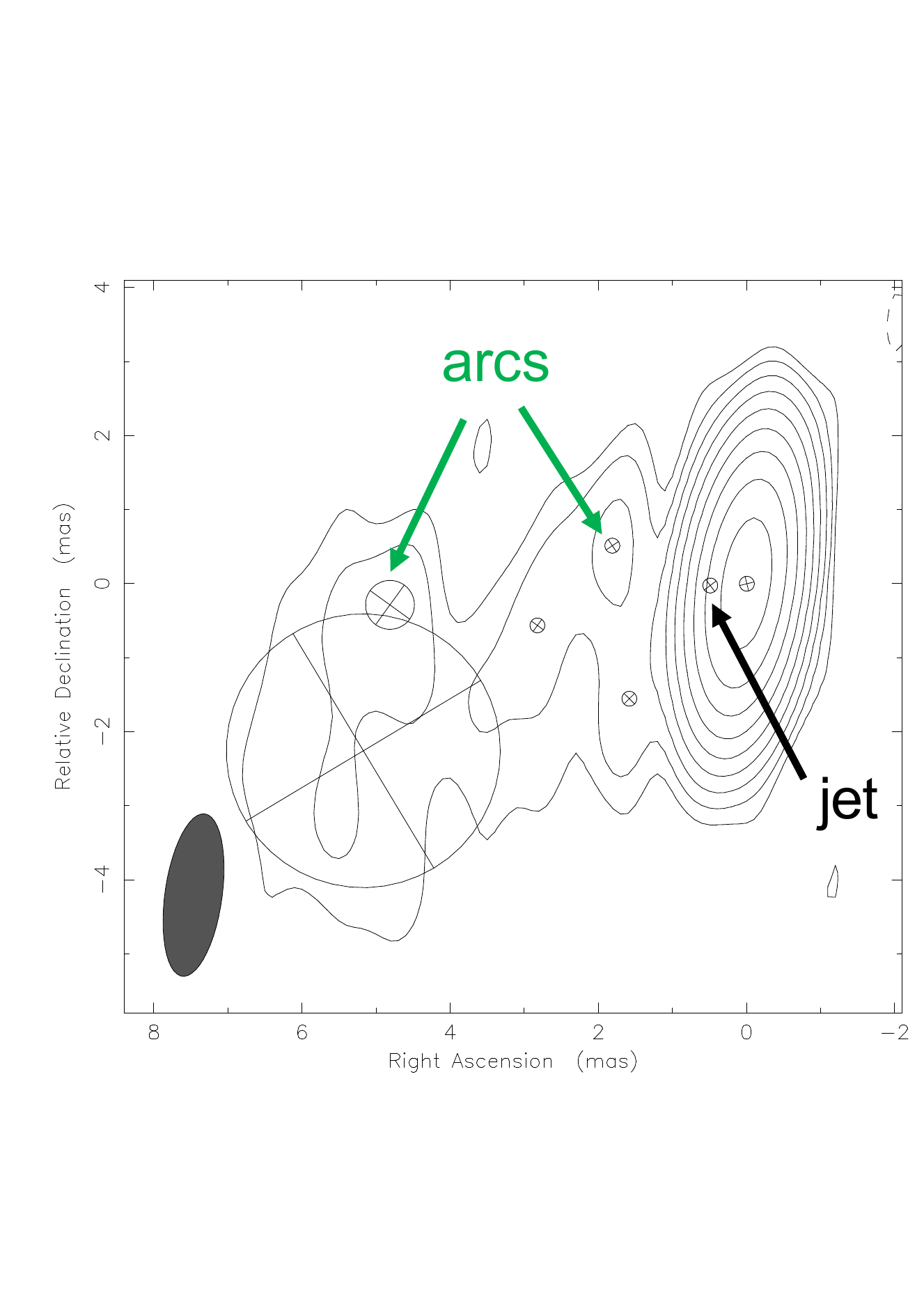}
\vspace{-2.1cm}
\caption{The pc-scale jet morphology of PKS 1502+106 as obtained in 8 GHz observations with the VLBA (2016/07/15). The contours are on a logarithmic scale between -0.25 and 64 per cent of the peak flux of 1.79 Jy beam$^{-1}$. The beam size is 2.21~mas $\times$ 0.773~mas at -7.94 deg. Superimposed are the modelfit components. We indicate the origin of the components (jet or arc).}
\label{mapx}
\end{figure}
\begin{figure*}
	{\includegraphics[width=13cm]{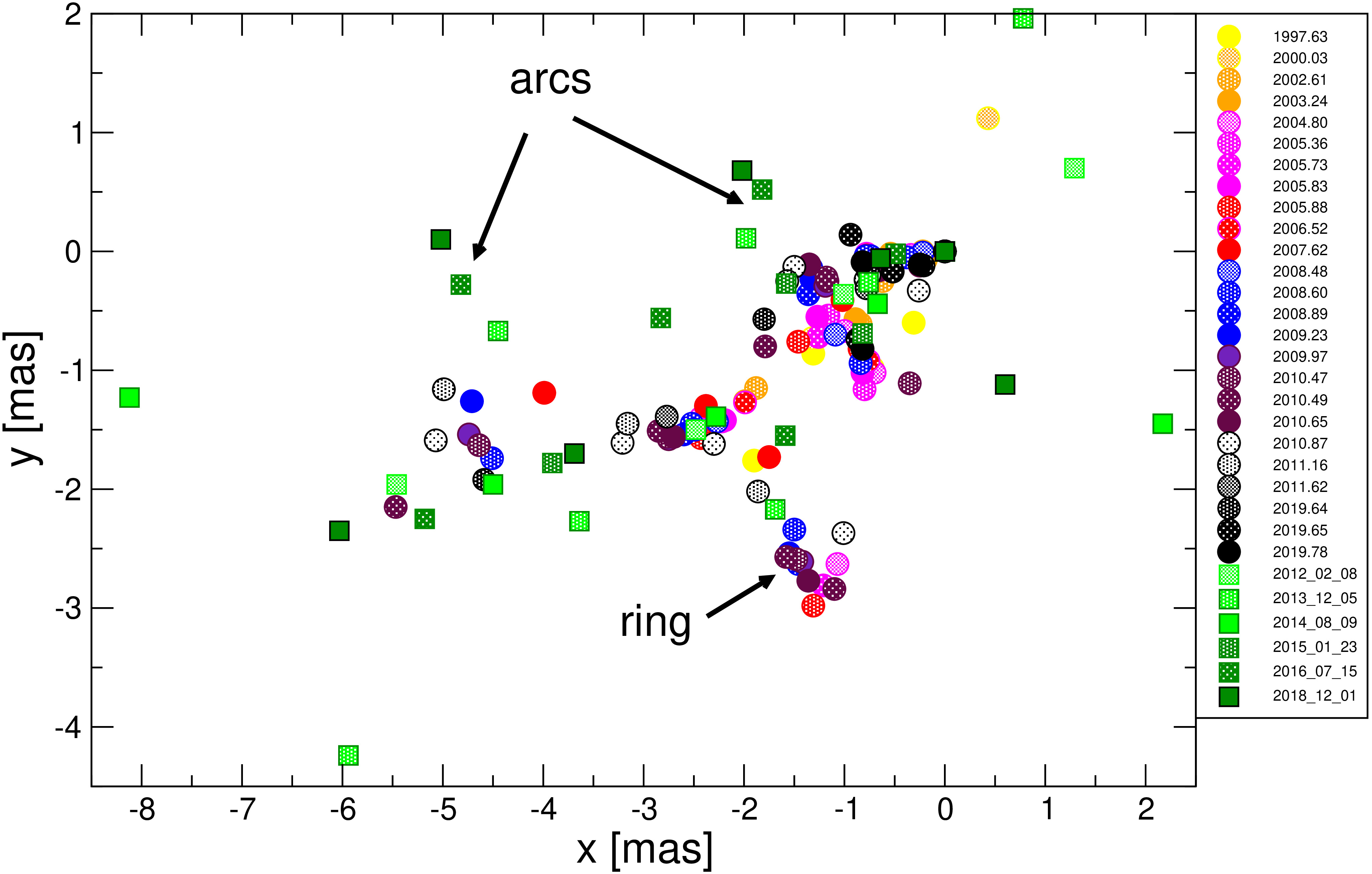}}
    \caption{Same as in Fig.~\ref{ring}, different colours mark the individual epochs of observation. In addition to Fig.~\ref{ring}, X-band observations mainly taken between 2012 and 2019 (astrometric VLBI, see text for explanation) are added (green data points). We label the ring (based on the 15 GHz data) and the arcs (based on the 8 GHz data, green).}
    \label{ring_x}
\end{figure*}
\begin{figure*}
\begin{minipage}{0.8\columnwidth}
  \includegraphics[width=\linewidth]{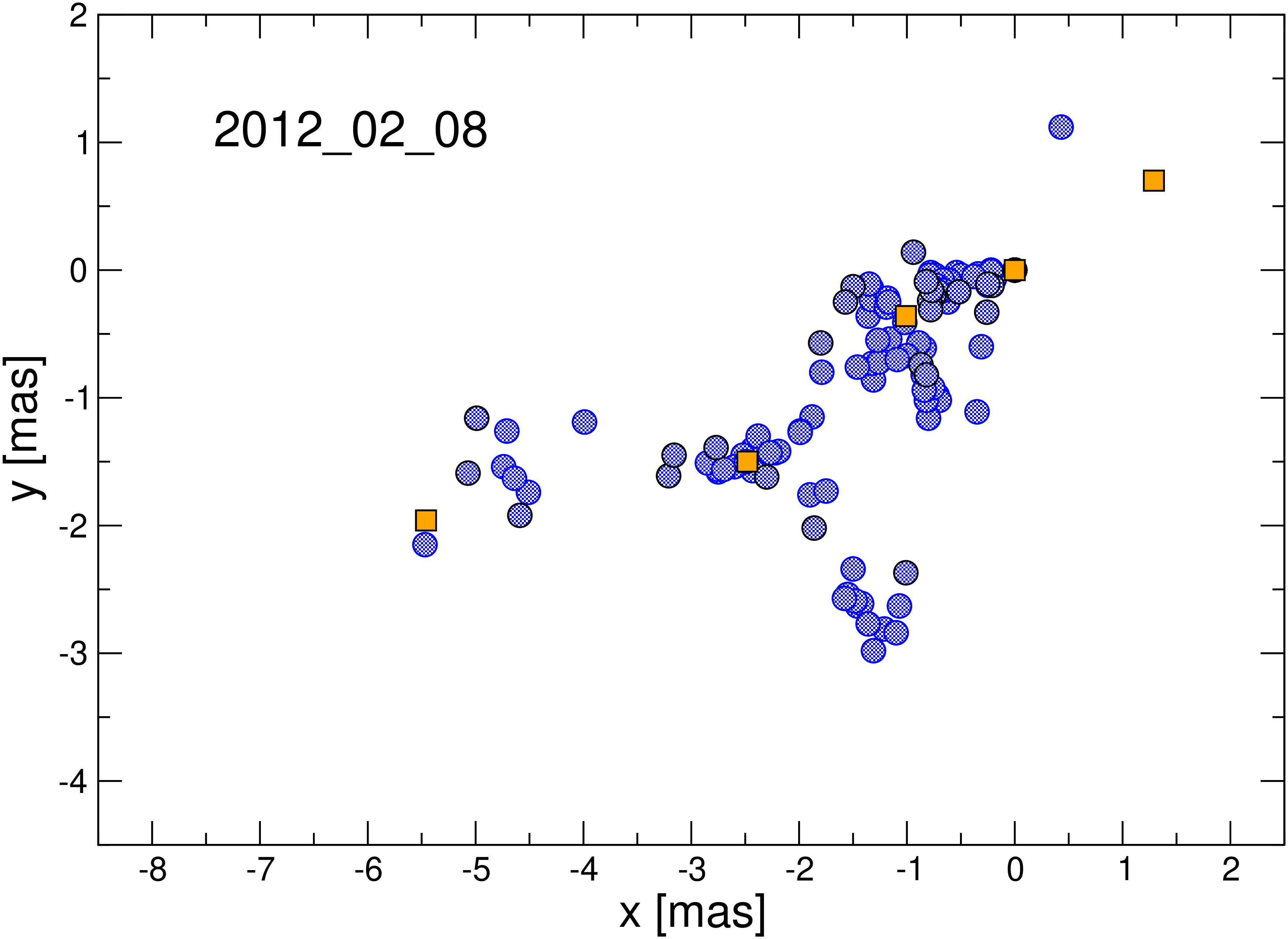}
  [a]
\end{minipage}
\begin{minipage}{0.8\columnwidth}
  \includegraphics[width=\linewidth]{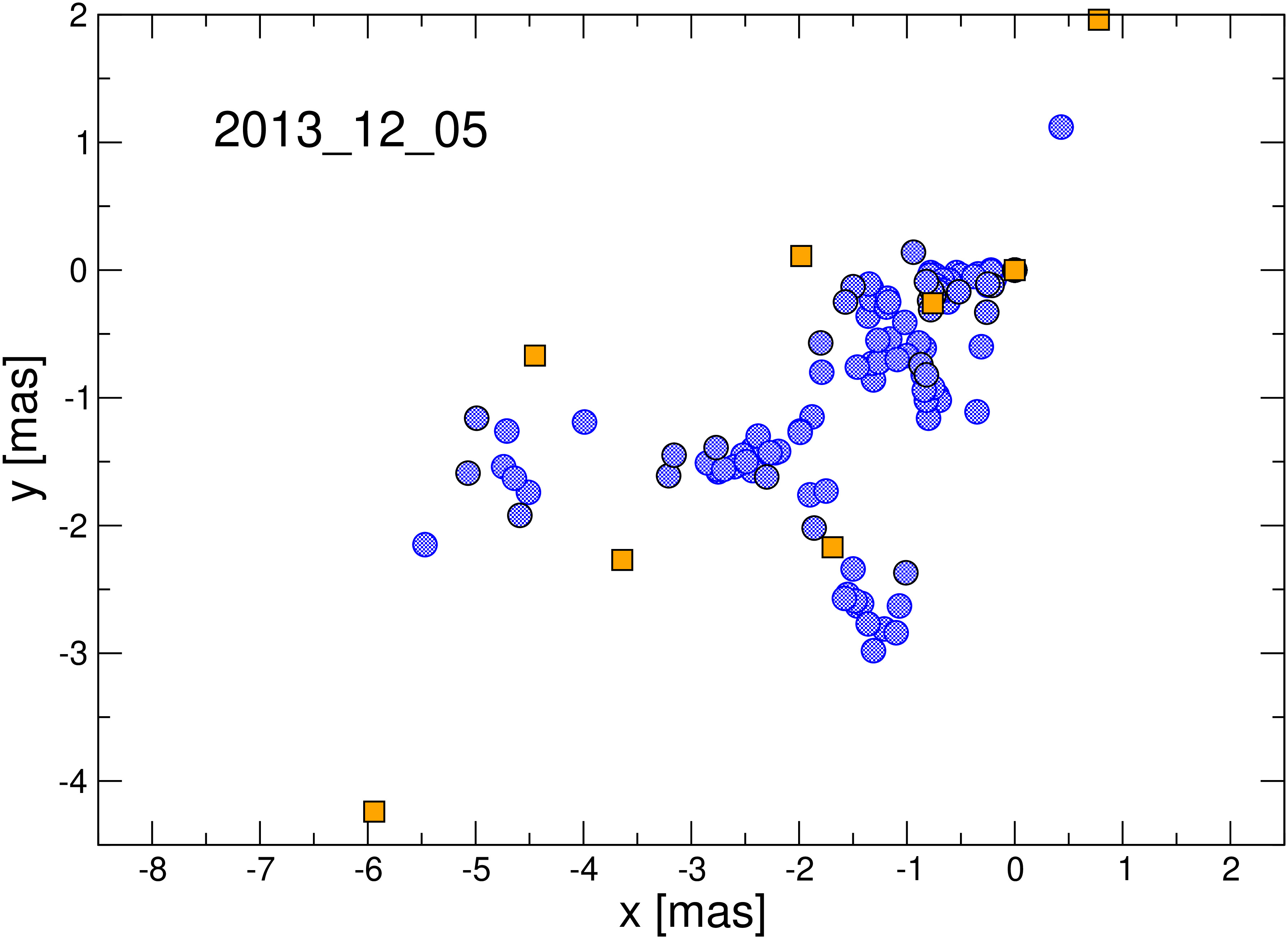}
  [b]
\end{minipage}
\begin{minipage}{0.8\columnwidth}
  \includegraphics[width=\linewidth]{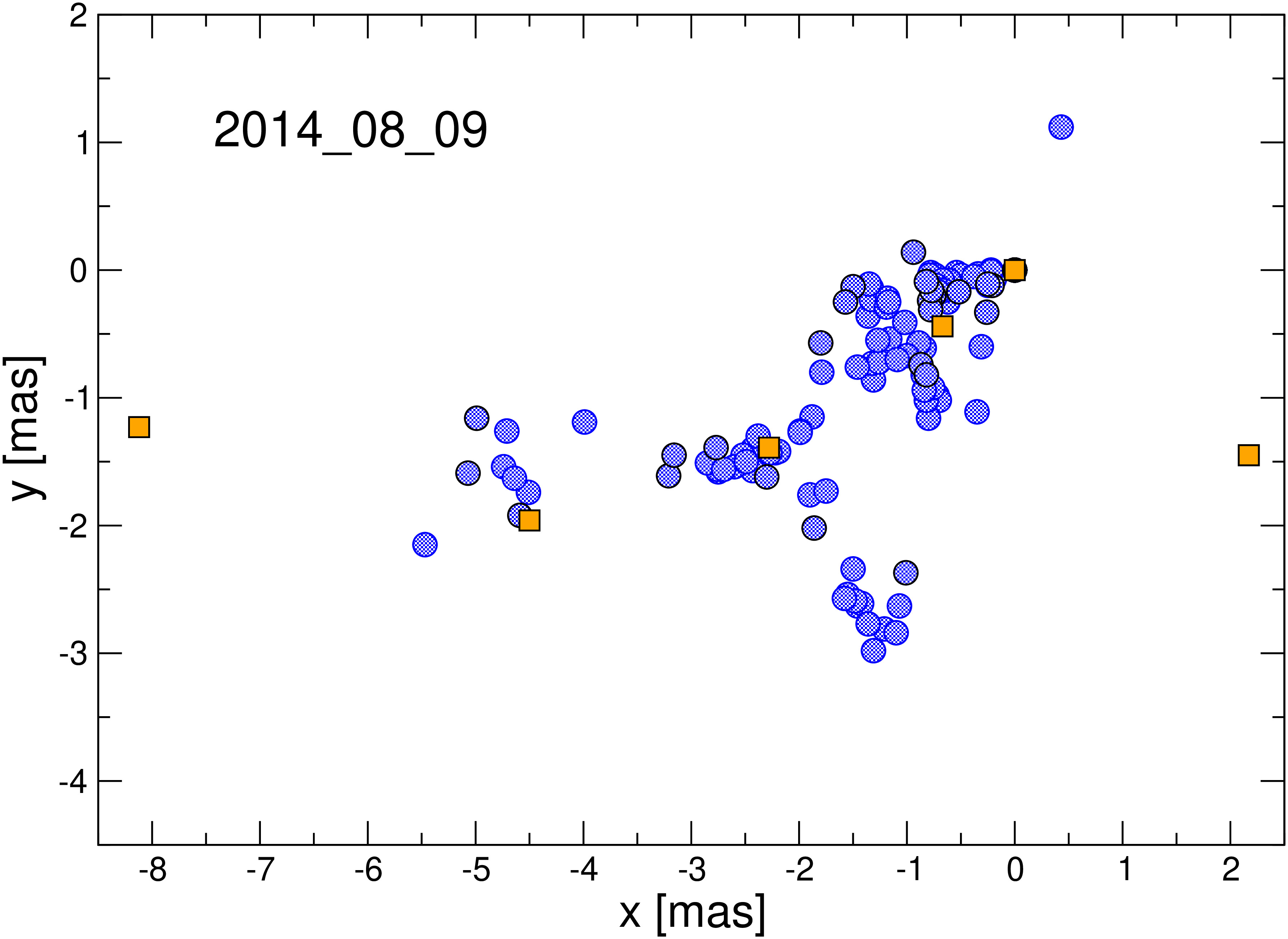}
  [c]
\end{minipage}
\begin{minipage}{0.8\columnwidth}
  \includegraphics[width=\linewidth]{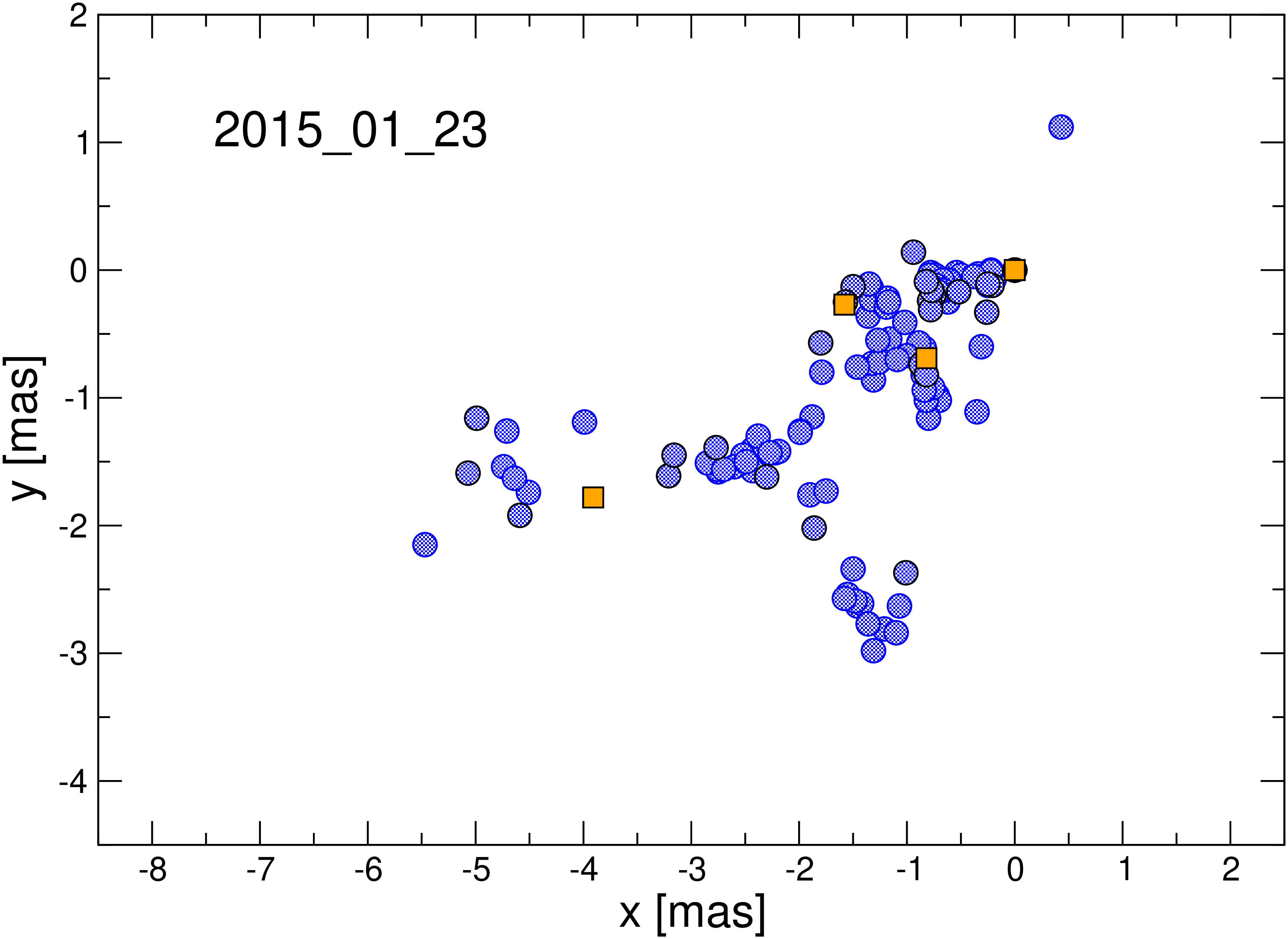}
  [d]
\end{minipage}
\begin{minipage}{0.8\columnwidth}
  \includegraphics[width=\linewidth]{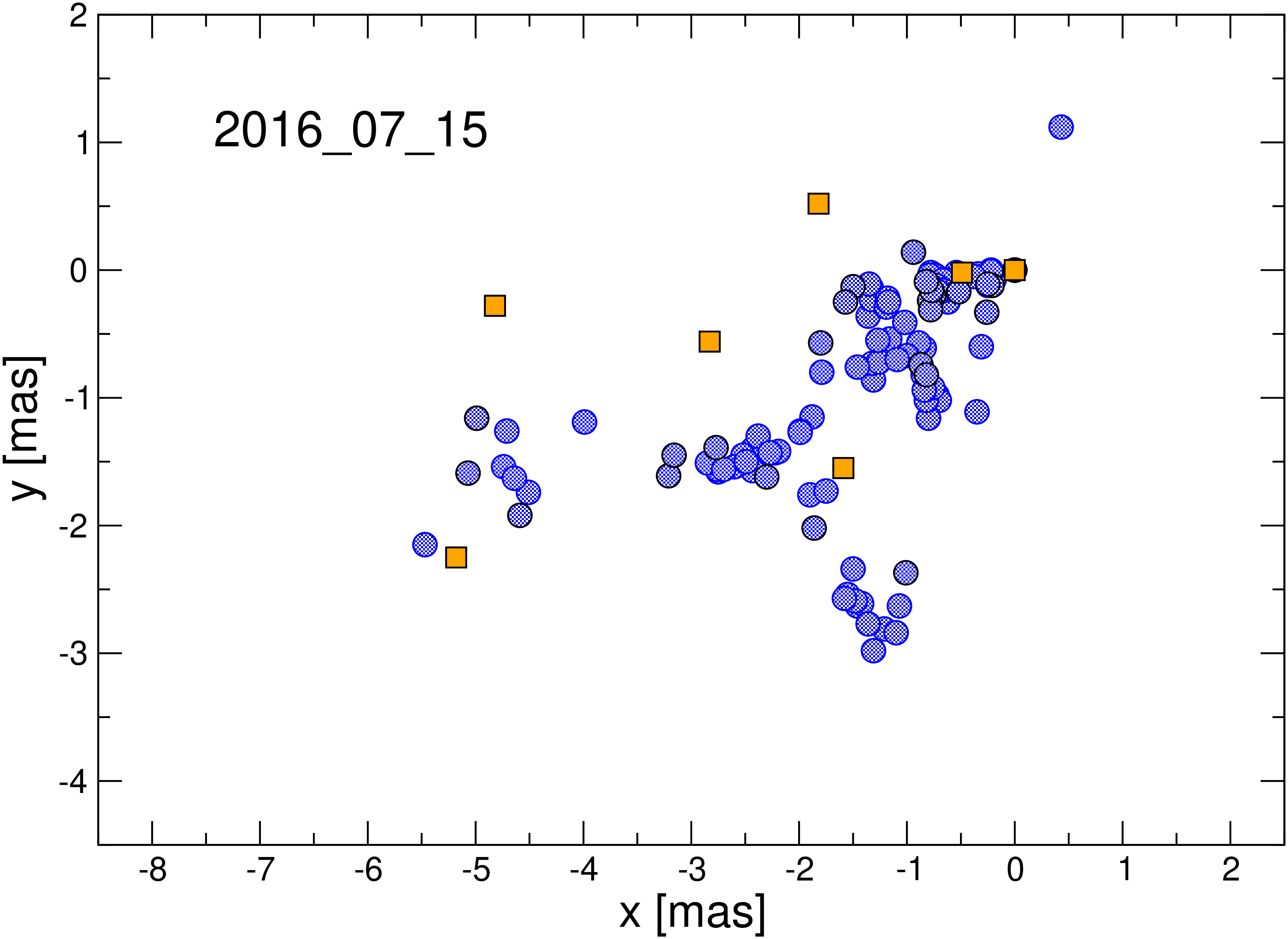}
  [e]
\end{minipage}
\begin{minipage}{0.8\columnwidth}
  \includegraphics[width=\linewidth]{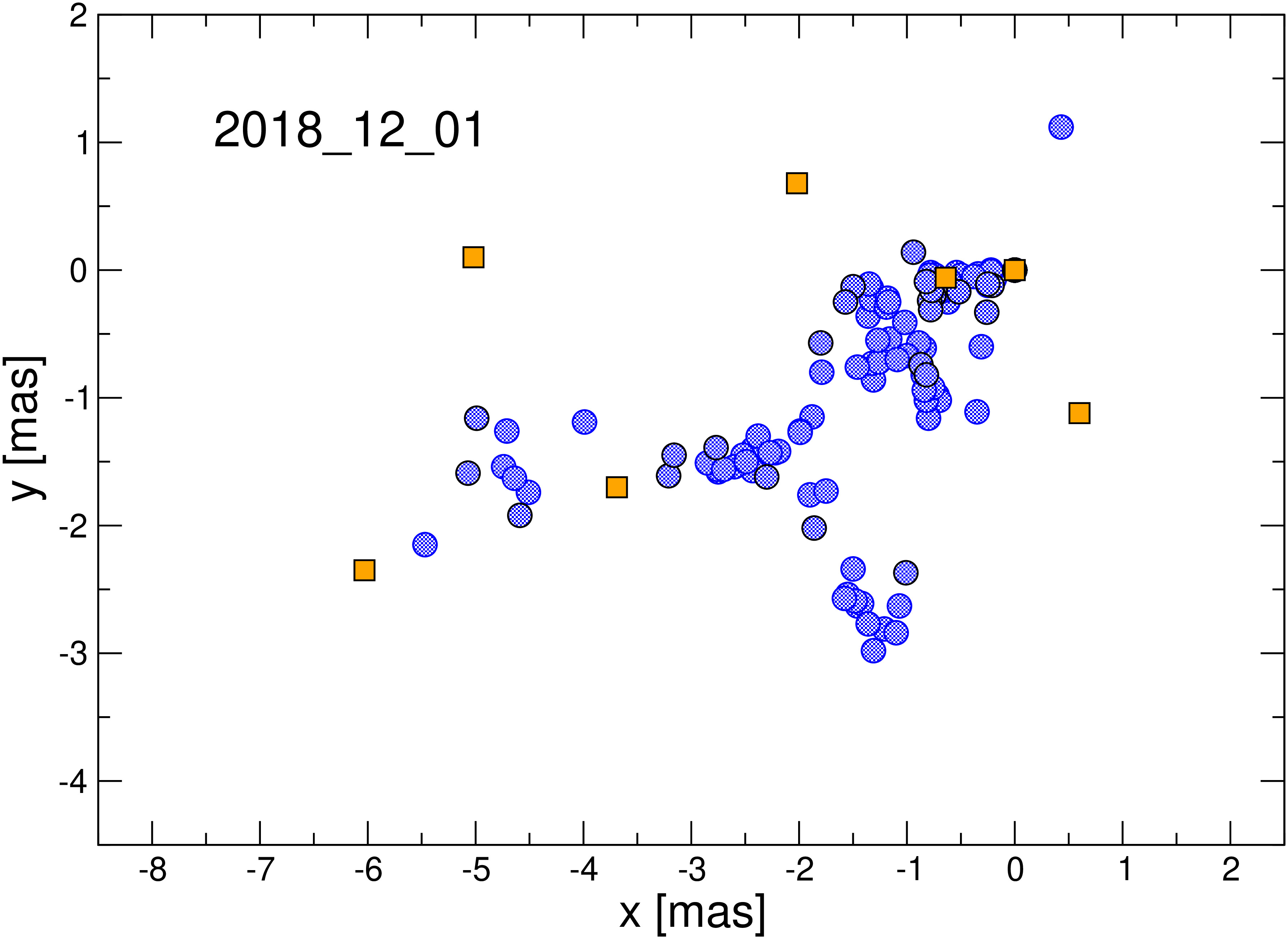}
  [f]
\end{minipage}
    \caption{[a]-[f] show the structure already displayed in Fig.~\ref{ring}. We superimpose the individual X-band epochs (in orange) to highlight the evolution of the radio morphology between 2012 and 2019 from epoch to epoch.}
    \label{xontop}
\end{figure*}

\subsection{\textit{Fermi}-LAT data analysis}
We used version 2.0.0 of the FermiTools\footnote{Available from \url{https://fermi.gsfc.nasa.gov/ssc/data/analysis/software/}} to extract a light curve from pass-8 photon data, downloaded from the \textit{Fermi} data server\footnote{\url{https://fermi.gsfc.nasa.gov/cgi-bin/ssc/LAT/LATDataQuery.cgi}}. For our analysis we used a region of interest of 14\textdegree{} around PKS 1502+106. The source of interest itself was fitted with a log-parabola of the form
\begin{equation}
    \frac{\mathrm{d}N}{\mathrm{d}E} = N_0\left(\frac{E}{E_b}\right)^{-(\alpha + \beta\log(E/E_b))},
\end{equation}
where the parameters $N_0$, $\alpha$, and $\beta$ were left free for the fit. The scale parameter $E_b$ was fixed to its catalogue value. All parameters of sources within a radius of 3\textdegree{} around our source of interest were left free for the fit as well. Sources until a radius of 25\textdegree{} were included in the fit with all parameters fixed to their catalogue values. We set the maximum zenith angle to 90\textdegree{} to exclude photons from the Earth limb. The Galactic diffuse emission was modelled with \texttt{gll\_iem\_v07.fits}, and for the isotropic emission we used the template \texttt{iso\_P8R3\_SOURCE\_V2\_v1.txt}. We divided the photon data into time bins of width 15~days and fitted this model to each of these time bins by performing an unbinned likelihood analysis. The energy range chosen for the analysis is $E = 0.1 - 300$~GeV. For the timing analysis we generated another light curve with a bin size of width 0.5~days. The source PKS~1502+106 was still significantly detected (i.e., TS~$\geq 25$) in a sufficiently large number of the time bins, as outlined in Sect.~3.7.

\begin{figure*}
    \includegraphics[width=\linewidth]{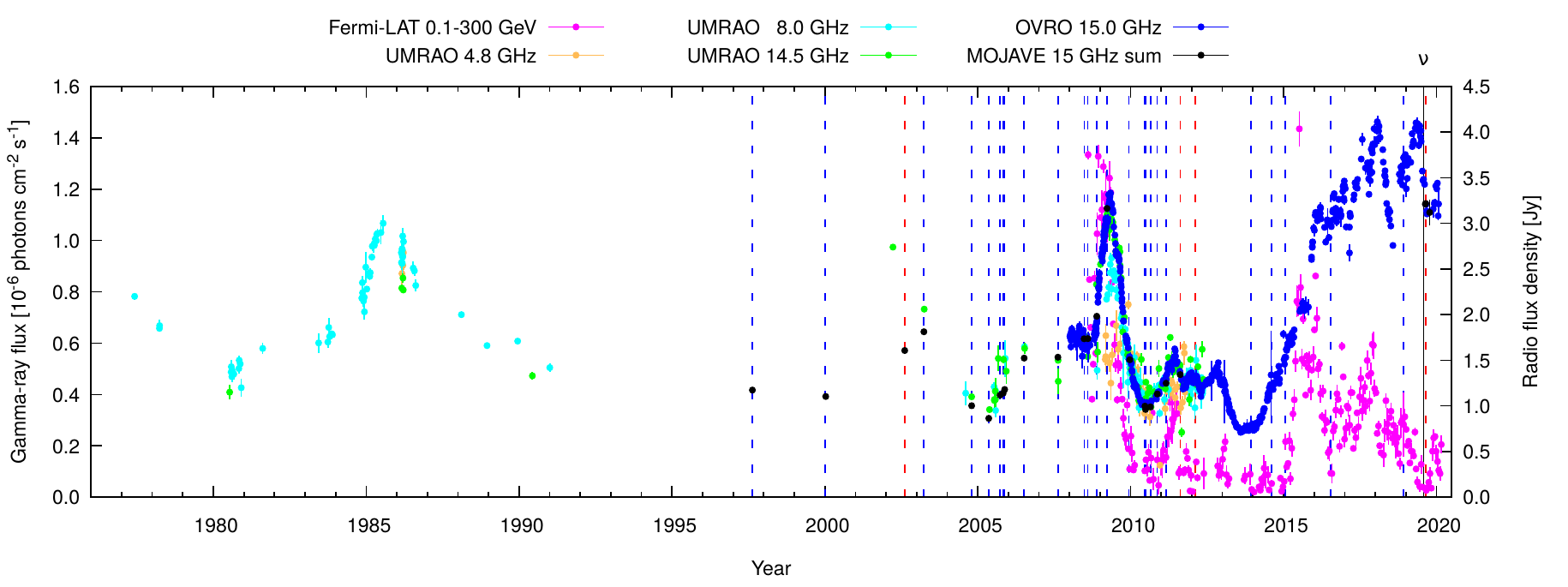}
    \caption{Superposition of the radio light-curves by OVRO and UMRAO, and the $\gamma$-ray light-curve (\textit{Fermi}-LAT, purple). The time of the neutrino detection by IceCube is marked by the black line and the symbol~$\nu$. Blue vertical dashed lines indicate epochs when VLBI observed a ring-like structure, red dashed lines refer to VLBI observations when such a structure was absent.}
    \label{radio_gamma_big}
\end{figure*}
\begin{figure}
\centering
\includegraphics[width=10.4cm]{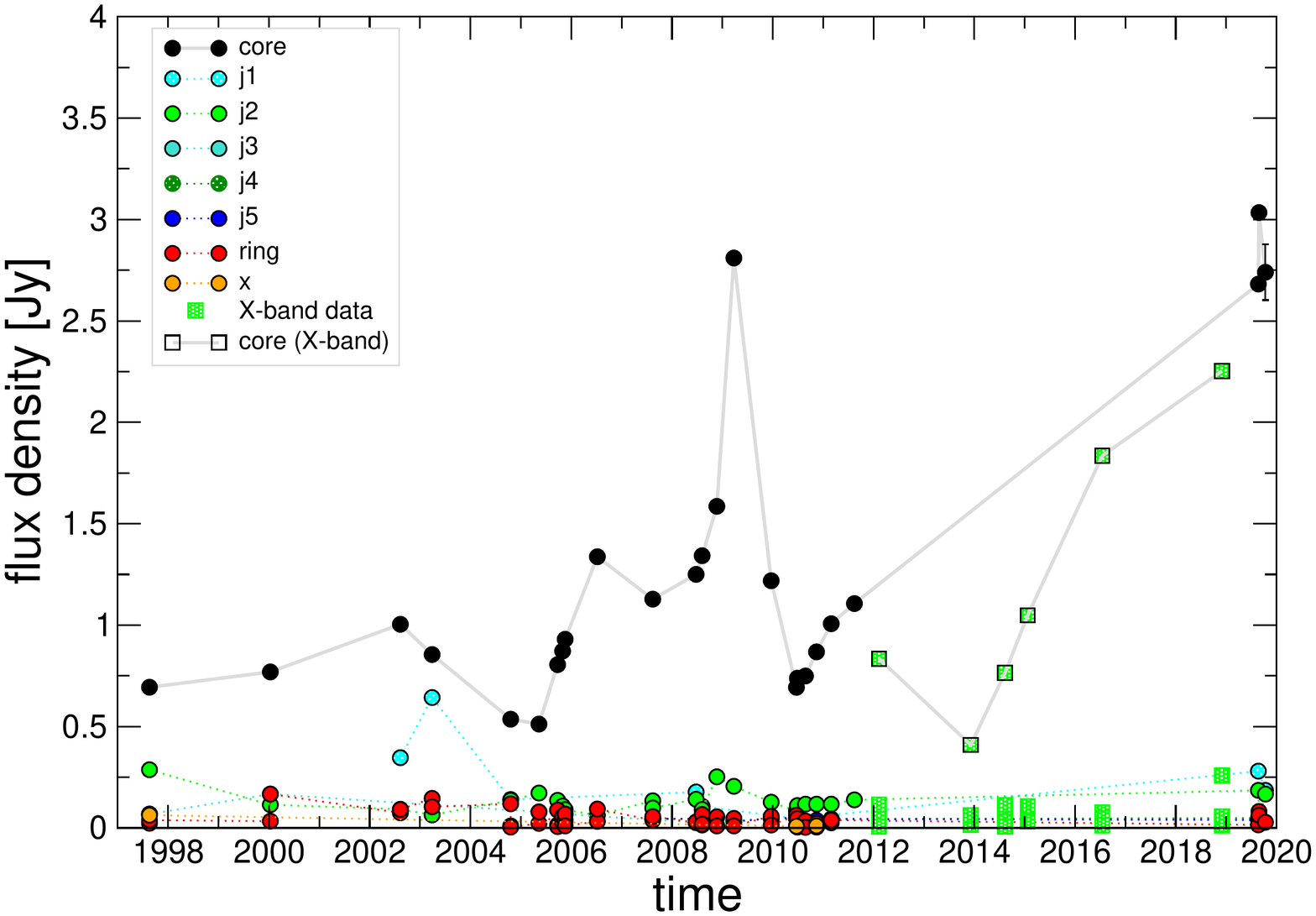}
\caption{The flux densities of the individual components derived from the VLBA data analysis (MOJAVE data) and astrometric VLBA data (X-band). The grey line indicates the core flux-density.}
    \label{radio_gamma}
\end{figure}

\section{Results}
We confirm the one-sided jet morphology (see Fig.~\ref{map0}) which has been observed before \citep[e.g., ][]{an,karamanavis}. However, the pc-scale jet of PKS 1502+106 reveals unexpected morphological structures and atypical kinematics which will be presented in the following subsections.

\subsection{Nontypical jet morphology: jet ridge line + ring structure}
\label{jk}
As an example for the performed modelfit analysis, we show in Fig.~\ref{map0} the result for one epoch only.  The modelfit components are superimposed on the map. Each of these modelfit components maps to specific xy-coordinates (mas) offset from the core (0,0) and is shown in Fig.~\ref{fig_jet_ring_flux} labeled according to the origin (jet or ring), and colour coded (by epoch) in Fig.~\ref{ring}.

Fig.~\ref{ring} shows the evolution of the pc-scale jet morphology, as seen in the VLBA data obtained at 15 GHz in the xy-coordinates of the component positions. The individual plots tracing the evolution in xy-coordinates (epoch per epoch) are shown in the Appendix in Section~\ref{xy_plots} in Figs.~\ref{xy_plot1} and~\ref{xy_plot2}. Fig.~\ref{ring}[a] shows all of the xy-data points obtained in all the epochs (at 15 GHz, VLBA) in one plot.
Clearly visible in Fig.~\ref{ring} is a jet ridge line. However, in addition to the jet ridge line at around (-1 mas, -2 mas) in x-coordinate, an additional radio structure appears with time. 

When plotting all the data from all the epochs in one figure (Fig.~\ref{ring}[a]), the additional radio emission resembles a ring-like structure in projection. We find some further evidence for additional emission in an arc-form between 0 and $-0.5$ mas (Fig.~\ref{ring}[a]).
We try to visualize how emission features appear with time in Fig.~\ref{ring}[b] where arrows indicate those regions on the apparent ring structure where emission tends to appear.
With regard to the time evolution of this additional ring structure, we find the following results: From about 2004 onward, significant emission outside of the jet ridge line towards the South becomes visible. From 2008 on, emission apart from the jet towards the North appears. 

From 2011.62 onward, only the jet is visible and the additional structure is not visible any more. Thus, significant additional radio emission is prominently visible for roughly eight years in the observer's frame. 

\subsection{Jet kinematics}
\label{jk1}
In our analysis of the jet kinematics, individual jet components are traced across the epochs. In this particular case, the jet features seem to belong to two different phenomena: the jet and an additional radio structure around and perpendicular to the jet. We call this latter radio feature the "ring". In Fig.~\ref{identify}[a] and [b] we show the same data as in Fig.~\ref{ring}[a] but in different colours marking the phenomena (jet or ring). The core distance of the individual features with time is displayed in Fig.~\ref{identify}[b]. The proper motions we derive for four jet knots dependent on this identification scheme are listed in Table~\ref{speeds}. However, an alternative component assignment is possible as well. In Fig.~\ref{identify}[c] and Fig.~\ref{identify}[d] we show this alternative identification scenario. Here the components within the ring are identified as jet components as well. The proper motions and apparent speeds for both scenarios are listed in Table~\ref{speeds}. The negative apparent speed for j3 in the second scenario (based on a small number of data points only) seems unphysical. We thus adopt the first scenario (Fig.~\ref{ring}[a], and Fig.~\ref{ring}[b]) in the following analysis.

\subsection{Polarisation information}
\label{pol}
Polarisation information traces the magnetic field structures of jets. Polarisation also serves as hydrodynamical tracer of shocks, bends, and shear (e.g., \citealt{homan}). The MOJAVE webpage provides polarisation maps for PKS 1502+106. In particular, these maps show the fractional linear polarisation as well as the electric polarisation vector angles (EVPA). 
The additional radio structure (ring) derived from the VLBA total intensity images can be seen in the polarisation maps as well. Figure~\ref{pola} shows polarisation information for two epochs ([a] in epoch 2010/06/19, [b] in epoch 2010/08/27). Both maps show -- in addition to the polarised jet -- also the polarised ring structure. The polarisation vectors of the jet seem to be perpendicular to that of the ring. From Fig.~\ref{pola}[a] to Fig.~\ref{pola}[b] a change in the polarised emission can be seen: in [b] the highly polarised outer ring becomes even more polarised (green/blue becomes black). Radio images with polarisation fraction over 15 per cent are unusual, with polarisation fraction 50 per cent are rare \citep{pushkarev2017}. Usually, cores are less polarised ($<$5 per cent) than jets ($\sim$ 10 per cent) \citep{wardle}. In the component south of the main jet we see a highly polarised structure with a polarisation degree of 50 per cent that is neither a core, nor a jet.

In Fig.~\ref{pola}[c] and [d] the polarisation information shortly before (2019/08/27) the neutrino event and after (2019/10/11) the event is shown. In the latter two images, the additional stronger polarised emission is missing and mainly the core region is polarised. The core polarisation increased compared to [a] and [b].

\subsection{More unexpected radio structures around the jet}
The data shown in Fig.~\ref{ring}[a] have been obtained at 15 GHz. Unfortunately, these data do not cover the time between 2012 and the neutrino event. To study the evolution of the pc-scale morphology in this time span as well, we analysed astrometric X-band data taken with the VLBA between 2012.11 and 2018.92. In Fig.~\ref{mapx} we show one of these maps with the modelfit components superimposed. In addition, we label some of the observed features according to their possible origin in jet or arc.

In Fig.~\ref{ring_x} we show these data superimposed on the (astronomical) 15 GHz data (see Fig.~\ref{ring}[a] for comparison). The reason why we think it is justified to superimpose VLBA images obtained at 8 and 15 GHz, stems from the following consideration. The UMRAO 8 and 14.5 GHz data generally track well in the pre 2012.5 data. This suggests that there is no significant self absorption in this source (at least during those time windows at cm band). Assuming that the spectral behaviour is not dramatically different at later times, this allows us to fill in the gaps in the MOJAVE VLBA data coverage with 8 GHz VLBA data.

All the data points resulting from the astrometric VLBA are shown in different green tonalities (indicating the different epochs). While the data until about 2012 (15 GHz) revealed only one ring structure over time, the astrometric data after 2012 show evidence for several additional structures along but perpendicular to the jet ridge line. At least two prominent "peaks" around ($-1$~mas, $-2$~mas) and around ($-4$~mas, $-5$~mas) are detected. These peaks most likely are phenomena which have a similar physical origin as the ring. However, these features extend further away from the jet ridge line compared to the ring structure which appeared earlier (and could be observed until 2012).  As they are related to the ring but do not form a full circle (in projection) we call them arcs hereafter.

The evolution with time of these later appearing radio structures can be traced in Fig.~\ref{xontop}[a-f]. For better comparison, the 15 GHz data (mainly until 2012) are shown in light-blue while the astrometric data (starting 2012.11) are superimposed and shown in orange. 

Two of the six epochs (2012/02/08 in Fig.~\ref{xontop}[a], 2014/08/09 in Fig.~\ref{xontop}[c]) show no evidence for a deviation from a straight jet ridge line, while the other four epochs clearly do. The strongest deviation with the largest amplitude is found in 2018/12/01 (Fig.~\ref{xontop}[f]). Because of this larger amplitude at the time of the higher OVRO radio flux-density, we think this phenomenon is related to the source activity. The arcs could be segments of further ring-like structures along the jet.

Weak emission on the counter-jet side is found in one epoch only at 15 GHz and four epochs at 8 GHz. This emission is faint compared to the emission on the jet side. More data and more detailed studies are required to figure out whether these counter-jet features are real or artifacts due to image noise.

\begin{table}
	\centering
	\caption{The parameters resulting from the kinematic analysis of the pc-scale jet (15 GHz MOJAVE data). Column 1 denotes for the figure where the identification scenarios are shown. Column 2 for the component identification, column 3 for the proper motion, and column 4 for the apparent speed.}
	\label{speeds}
	\begin{tabular}{llcc} 
		\hline
Figure&	Id.  & p.m. & app. speed\\
	&	& [mas/yr]& [$c$]\\
		\hline
	Fig.~\ref{identify}[b]	&	j1 & 0.0004$\pm$0.004 & 0.03$\pm$0.32 \\
	&	j2 & 0.007$\pm$0.003 &  0.56$\pm$0.24 \\
	&	j3 & 0.091$\pm$0.021 &  7.29$\pm$1.68\\
	&	j4 & 0.030$\pm$0.048 &  2.40$\pm$3.85\\
		\hline
	Fig.~\ref{identify}[d]&	j1 & 0.0004$\pm$0.004 & 0.03$\pm$0.32 \\
	&	j2 &0.010$\pm$0.003  & 0.80$\pm$0.24  \\
	&	j3 &-0.019$\pm$0.014  &-1.52$\pm$1.12   \\
	&	j4 &0.107$\pm$0.017  & 8.57$\pm$1.36 \\
	&	j5 &0.030$\pm$0.048&2.40$\pm$3.85 \\
		\hline
	\end{tabular}
\end{table}

\subsection{Jet precession: the mechanism behind the ring structure?}
\label{precession}
The filling of the ring structure with time suggests that the jet as a whole could be changing its direction. This motivated us to test the precession model using the individual components j1-j4, which are located at different offsets from the core. In a first approximation, these components j1-j4 can be treated as almost stationary, which is mostly applicable for the subluminal components j1 and j2. In comparison with previous applications of the jet-precession model to component kinematics and flux density variability studies, see in particular the studies of OJ 287 \citep{britzen_OJ287}, 3C 84 \citep{britzen_3C84}, and TXS 0506+056 \citep{britzen_txs}, we fitted the precession model individually to each component for the purpose of this study.

\begin{figure*}
    \centering
    \includegraphics[width=\textwidth]{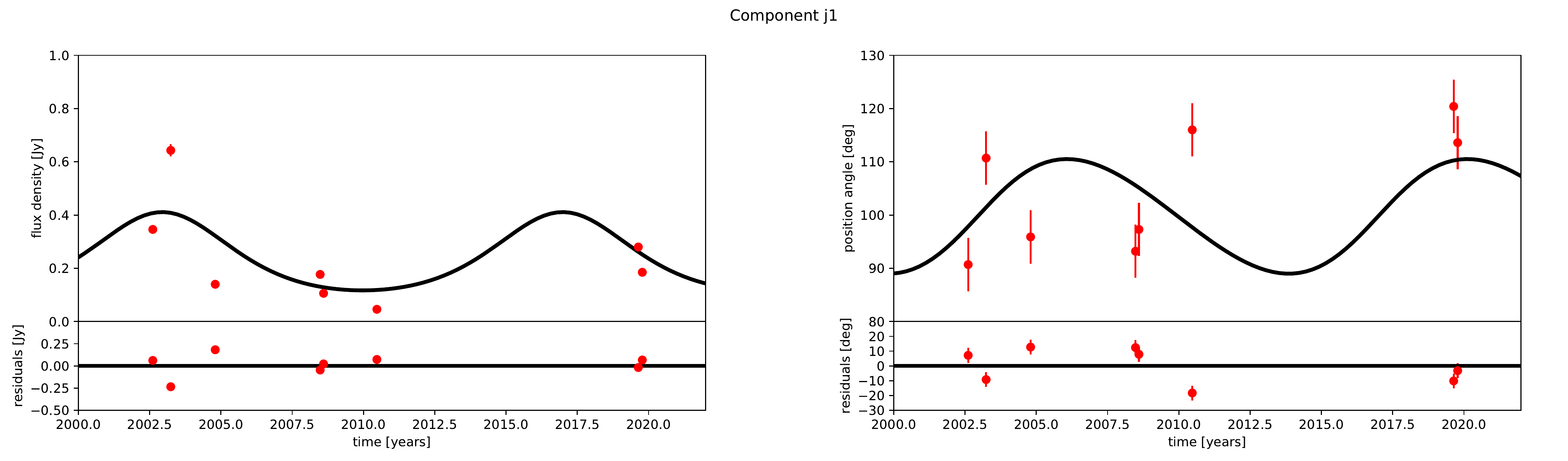}
    \includegraphics[width=\textwidth]{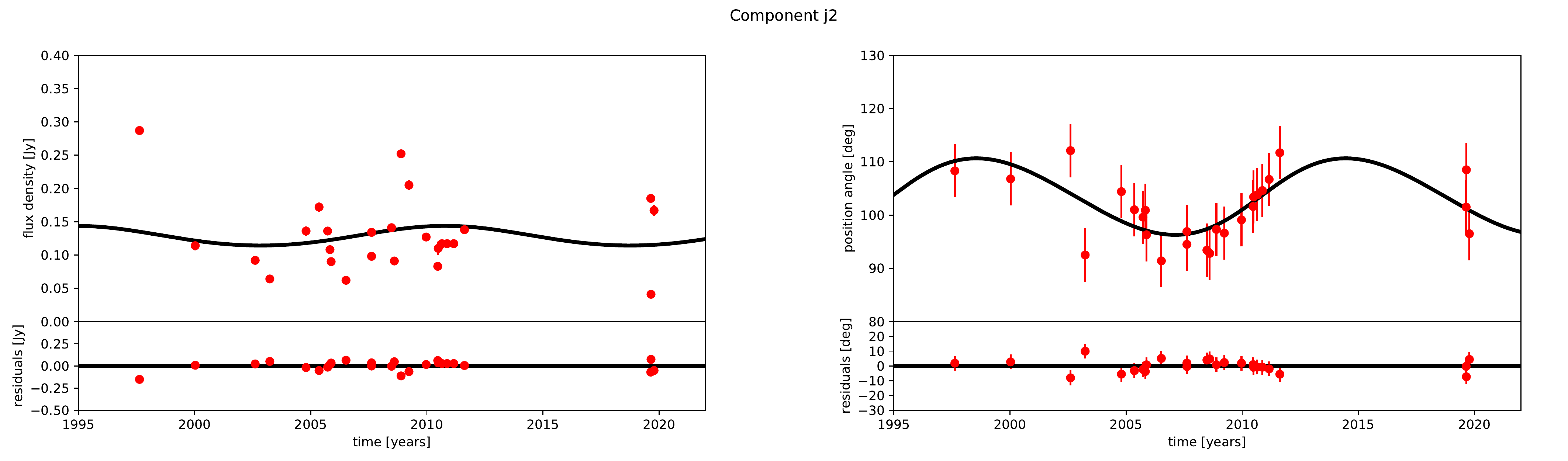}
    \includegraphics[width=\textwidth]{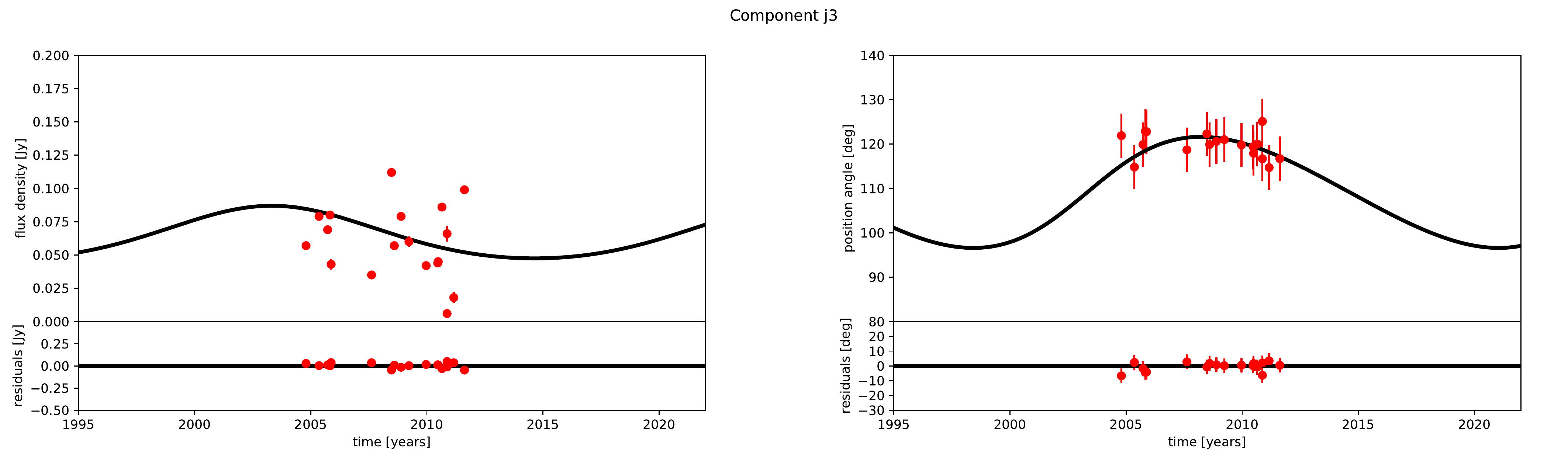}
    \includegraphics[width=\textwidth]{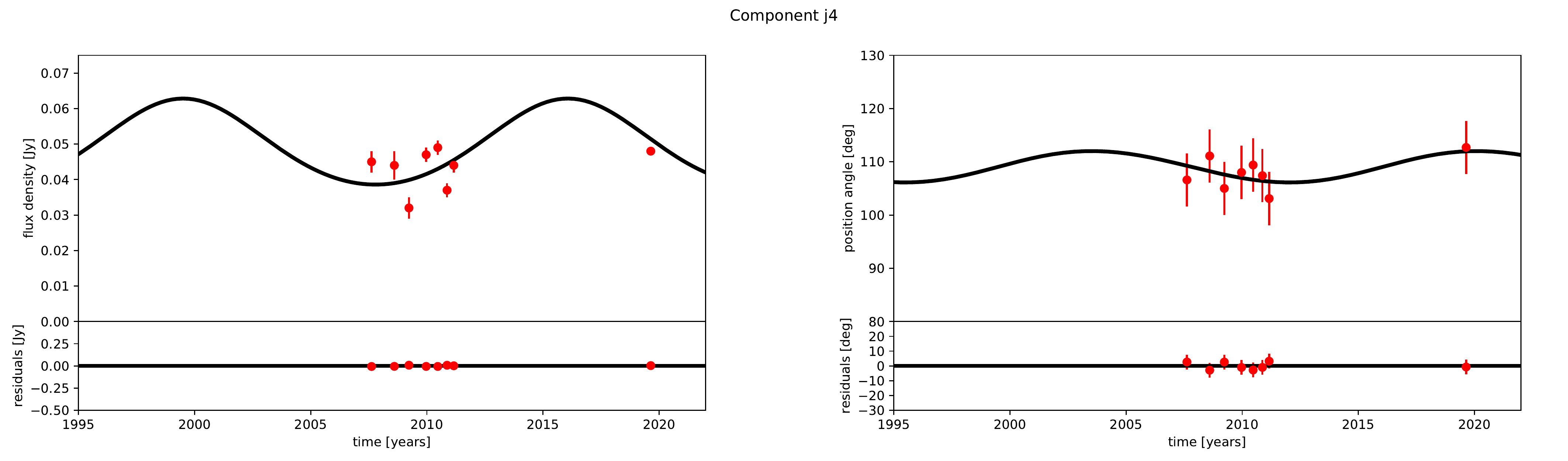}
    \caption{Bulk precession motion of the jet fitted to the flux density (left panels) and the position angle (right panels) of the jet components j1, j2, j3, j4 (from the top to the bottom panels, respectively). Residuals are displayed at the bottom of the corresponding panels. For the best-fit precession models presented here (with a larger viewing angle of the precession cone axis, $\phi_0>10^{\circ}$), the best-fit parameters are listed in Table~\ref{tab_precession_model}.}
    \label{fig_precession_model_j1-j4}
\end{figure*}
In the precession model, we assume that the whole jet is precessing - i.e. the components follow the ballistic trajectory in the jet frame and the whole jet body precesses with respect to the observer.

We use the same set of equations introduced in \citet{britzen_txs}, and references therein, which we summarize here for clarity. The jet component moving with velocity $\beta$ at the angle of $\phi$ with respect to the line of sight has the apparent velocity of,
\begin{equation}
    \beta_{\rm app}=\frac{\beta \sin{\phi}}{1-\beta \cos{\phi}}\,,
    \label{apparent velocity}
\end{equation}
where $\beta$ is related to the Lorentz factor $\gamma=(1-\beta^2)^{-1/2}$. When the jet is precessing, the component viewing angle~$\phi$ as well as its position angle on the observer's sky $\eta$ are functions of time:
\begin{align}
    \phi & =\arcsin{\sqrt{x^2(t)+y^2(t)}}\,\notag\\
    \eta & =\arctan{\left[\frac{y(t)}{x(t)}\right]}\,.
    \label{eq_phi_eta}
\end{align}
The coordinates $x(t)$ and $y(t)$ are time-dependent Cartesian coordinates of the component in the observer's frame of reference,
\begin{align}
    x & =A\cos{\eta_0}-B\sin{\eta_0}\,\notag\\
    y & =A\sin{\eta_0}+B\cos{\eta_0}\,
    \label{eq_phi_eta}
\end{align}
where $\eta_0$ is the projected position angle of the axis of the precession cone. The coefficients $A$ and $B$ depend on the half-opening angle $\Omega$ of the precession cone, the viewing angle $\phi_0$ of the axis of the precession cone, the angular frequency $\omega=2\pi/P_{\rm prec}$ in the observer's frame, and the reference epoch~$t_0$:
\begin{align}
    A & = \cos{\Omega} \sin{\phi_0}+\sin{\Omega}\cos{\phi_0}\sin{[\omega(t-t_0)]}\,\notag\\
    B & = \sin{\Omega}\cos{[\omega(t-t_0)]}\,.
    \label{eq_AB_coefs}
\end{align}
These equations are further complemented by the relation for the observed flux density $S_{\rm obs}$ of a component that is moving with velocity $\beta$ at angle $\phi$ with respect to the line of sight of the observer. This motion modulates the intrinsic component flux density $S_0$ as
\begin{equation}
    S_{\rm obs}=S_0 \delta^{\epsilon}\,,
    \label{eq_Doppler_boosting}
\end{equation}
where $\delta$ is a Doppler-boosting factor, $\delta(\gamma, \beta)=[\gamma(1-\beta \cos \phi)]^{-1}$, and $\epsilon$ is the boosting exponent, which can be expressed using the spectral slope and the geometrical factor as $\epsilon=\alpha+g$. While fitting, we fixed $\epsilon=3$, which for the geometrical factor of $g=3$ (discrete spherical components) or $g=2$ (a continuous cylindrical jet) yields $\alpha=0-1$, i.e. using the convention $S_0\propto \nu^{-\alpha}$, it corresponds to flat- or steep-spectrum synchrotron emission.

For each component, we performed the simultaneous least-squares-fitting to its flux density and its position angle. Due to a large scatter in the observed flux density values and at the same time a rather coherent evolution of the position angles, we considered two scenarios for the initial values of the fit parameters:
\begin{enumerate}
  \item scenario with the larger viewing angle close to $10$ degrees or more, which leads to the periodic variations in both flux density and position angle,
  \item scenario with the small viewing angle less than $10$ degrees, which leads to only small variations in flux density. 
\end{enumerate}

For the results of the first fitting scenario, see Fig.~\ref{fig_precession_model_j1-j4}, where we show the fits to the flux density and the position angle of components j1, j2, j3, and j4. In the bottom part of each panel, we display residuals of the fit. The best-fit parameters are listed in Table~\ref{tab_precession_model} for each component including the mean values of the precession model in the last column.

\begin{table*}
    \centering
     \caption{Precession parameters -- $t_0$, $P_{\rm prec}$, $\gamma$, $\Omega$, $\phi_0$, $\eta_0$, $S_0$, and the exponent $\epsilon$ -- listed for components j1, j2, j3, j4, and two ring positions denoted as $R_1$ (closer than 2 mas from the core) and $R_2$ (further than 2 mas from the core) for the fitting scenario (i) with the larger viewing angle of the precession axis, $\phi_0>10^{\circ}$. The last column contains the mean precession parameters for $P_{\rm prec}$, $\gamma$, $\Omega$, $\phi_0$, and $\eta_0$.}
    \begin{tabular}{c|c|c|c|c|c|c||c}
    \hline
    \hline
    Parameter & j1 & j2 & j3 & j4 & $R_1$ & $R_2$ & mean values\\
    \hline
     $t_0\, ({\rm yr})$    & $1992.41$ & $1982.96$ & $1941.13$ & $2003.65$ & $1978.52$ & $2009.80$ & -\\
    $P_{\rm prec}\,({\rm yr})$   & $14.05$ & $15.90$ & $22.62$ & $16.56$ & $23.35$ & $20.68$ & $18.9 \pm 3.5$ \\
    $\gamma$          & $2.85$ & $1.79$ & $2.24$ & $5.13$ & $2.39$ & $5.86$ & $3.4 \pm 1.5$ \\
    $\Omega\,({\rm deg})$  & $4.63$ & $1.97$ & $3.35$ & $1.25$ & $5.11$ & $2.50$ & $3.1 \pm 1.4$   \\
    $\phi_0\,({\rm deg})$  &  $25.62$ & $15.93$ & $15.67$ & $25.18$ & $16.40$ & $12.01$ & $18.5 \pm 5.1$\\
    $\eta_0\,({\rm deg})$  & $99.75$ & $103.47$ & $109.11$ & $109.07$ & $116.04$ & $138.31$ & $112.6 \pm 12.6$ \\
    $S_0\,({\rm mJy})$  & $18.91$ & $6.16$ & $1.95$ & $9.41$ & $1.56$ & $0.23$ & - \\
    $\epsilon$ (fixed) &    $3.0$ & $3.0$ & $3.0$ & $3.0$ & $3.0$ & $3.0$ & - \\  
    \hline
    \end{tabular}
    \label{tab_precession_model}
\end{table*}

The best-fit parameters for the second scenario with the small viewing angle of the precession axis are in Table~\ref{tab_precession_model_smallphi}. The precession period of $17.2 \pm 2.7$ years is comparable within uncertainties to the best-fit period in scenario (i) and so is the mean position angle of the precession axis, $\eta_0=117.0^{\circ} \pm 11.4^{\circ}$. The mean viewing angle of $\phi_0=4.5^{\circ}\pm 1.4^{\circ}$ of the precession axis is significantly smaller than in scenario (i). The half-opening angle of the precession cone $\Omega=0.8^{\circ}\pm 0.4^{\circ}$ and the mean Lorentz factor of $\gamma=1.5 \pm 0.4$ are also smaller.

\begin{figure}
    \centering
    \includegraphics[width=\columnwidth]{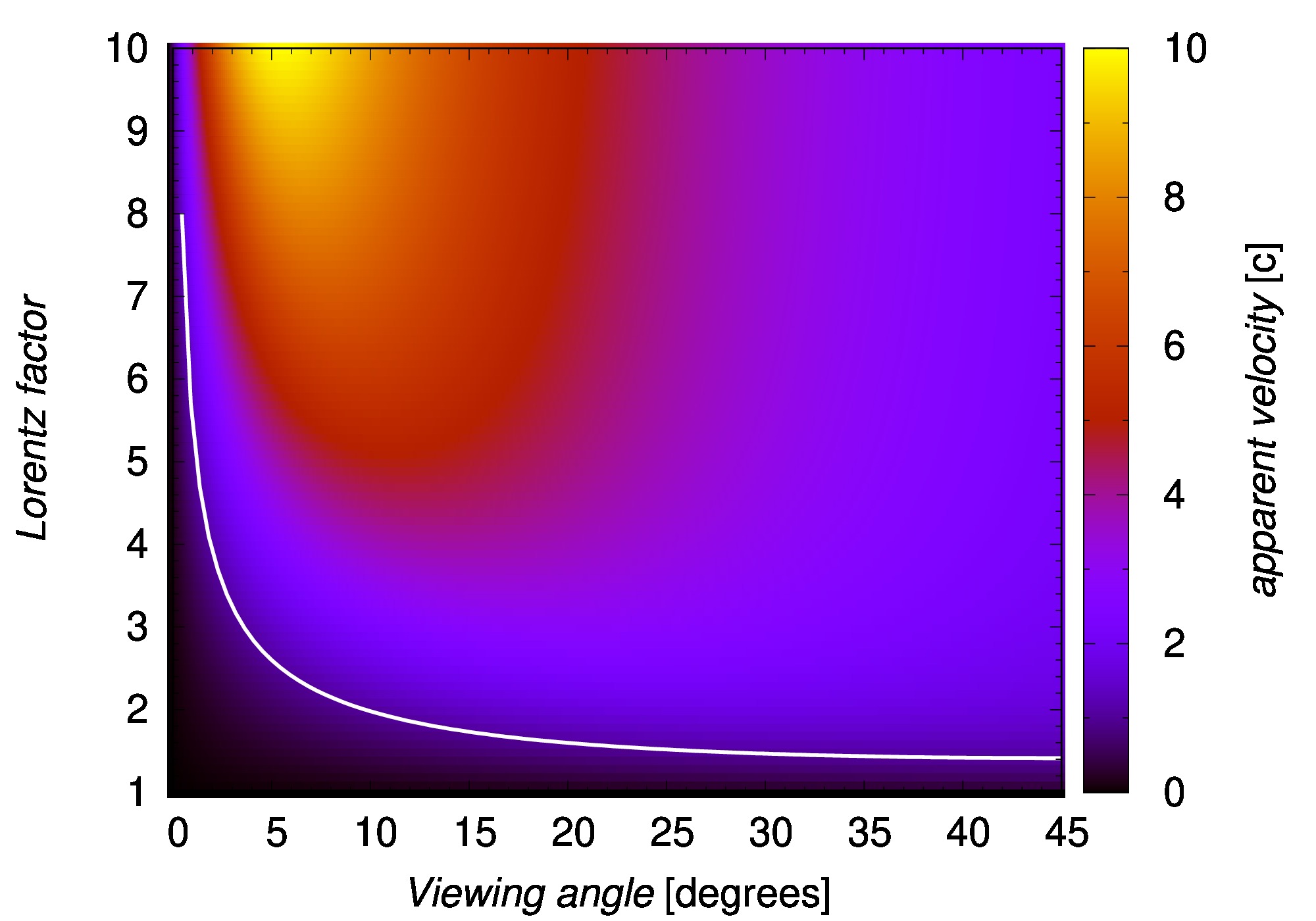}
    \caption{The jet component apparent velocity (expressed in the fraction of the light speed) as a function of the viewing angle in degrees and the Lorentz factor. The white line represents the case with $\beta_{\rm app}=1$.}
    \label{fig_bapp_gamma_phi}
\end{figure}

Scenario (ii) appears to be more consistent with the observations of the inner parts of the jet, namely for subluminal components j1 and j2, not only because of the smaller viewing angle but also other quantities are more consistent with observations of these components. We plot the temporal evolution of the component viewing angles, apparent velocities as well as Doppler-boosting factors in Fig.~\ref{fig_precession_time} for the first and the second scenarios with the dashed and solid lines, respectively. The smaller viewing angle in scenario (ii) in combination with the small Lorentz factor of $\sim 1.5$ also results in apparent subluminal velocites for all components, which is, on the other hand, inconsistent with superluminal components j3 and j4. This suggests that at larger separations than the ring, there is either a change in the viewing angle of the jetted material or the change in its Lorentz factor. In particular, if the jet is curved and its viewing angle changes with the distance from the core, such a curved jet would be precessing as a whole, which is not considered in our simple model. In Fig.~\ref{fig_bapp_gamma_phi}, we plot the apparent velocity (expressed as a fraction of the light speed) as a function of the viewing angle (in degrees) and the Lorentz factor. Detected subluminal components require sufficiently low Lorentz factors, especially for larger viewing angles. Also, according to Fig.~\ref{fig_precession_time}, scenario (ii) yields  nearly constant Doppler-boosting factors, which implies less variable observed flux density of the components due to the precession of the jet. On the other hand, the emission of the components moving close to the line of sight is enhanced by the constant factor in the observer's frame. 

\begin{table*}
    \centering
     \caption{Precession parameters -- $t_0$, $P_{\rm prec}$, $\gamma$, $\Omega$, $\phi_0$, $\eta_0$, $S_0$, and the exponent $\epsilon$ -- listed for components j1, j2, j3, j4, and two ring positions denoted as $R_1$ (closer than 2 mas from the core) and $R_2$ (further than 2 mas from the core) for the fitting scenario (ii) with the smaller viewing angle of the precession axis $\phi_0<10^{\circ}$. The last column contains the mean precession parameters for $P_{\rm prec}$, $\gamma$, $\Omega$, $\phi_0$, and $\eta_0$.}
    \begin{tabular}{c|c|c|c|c|c|c||c}
    \hline
    \hline
    Parameter & j1 & j2 & j3 & j4 & $R_1$ & $R_2$ & mean values\\
    \hline
     $t_0\, ({\rm yr})$    & $1975.03$ & $1997.16$ & $2048.71$ & $2006.08$ & $1988.59$ & $1990.76$ & -\\
    $P_{\rm prec}\,({\rm yr})$   & $20.19$ & $18.20$ & $20.53$ & $13.82$ & $13.91$ & $16.56$ & $17.2 \pm 2.7$ \\
    $\gamma$          & $1.64$ & $1.73$ & $1.06$ & $2.09$ & $1.64$ & $1.05$ & $1.5 \pm 0.4$ \\
    $\Omega\,({\rm deg})$  & $0.74$ & $0.60$ & $0.56$ & $0.18$ & $1.24$ & $1.24$ & $0.8 \pm 0.4$   \\
    $\phi_0\,({\rm deg})$  &  $1.90$ & $5.07$ & $6.24$ & $3.48$ & $4.86$ & $5.39$ & $4.5 \pm 1.4$\\
    $\eta_0\,({\rm deg})$  & $113.62$ & $103.38$ & $115.97$ & $107.76$ & $122.60$ & $138.49$ & $117.0 \pm 11.4$ \\
    $S_0\,({\rm mJy})$  & $9.51$ & $4.45$ & $21.71$ & $0.75$ & $2.13$ & $11.47$ & - \\
    $\epsilon$ (fixed) &    $3.0$ & $3.0$ & $3.0$ & $3.0$ & $3.0$ & $3.0$ & - \\  
    \hline
    \end{tabular}
    \label{tab_precession_model_smallphi}
\end{table*}

\begin{figure}
    \centering
    \includegraphics[width=\columnwidth]{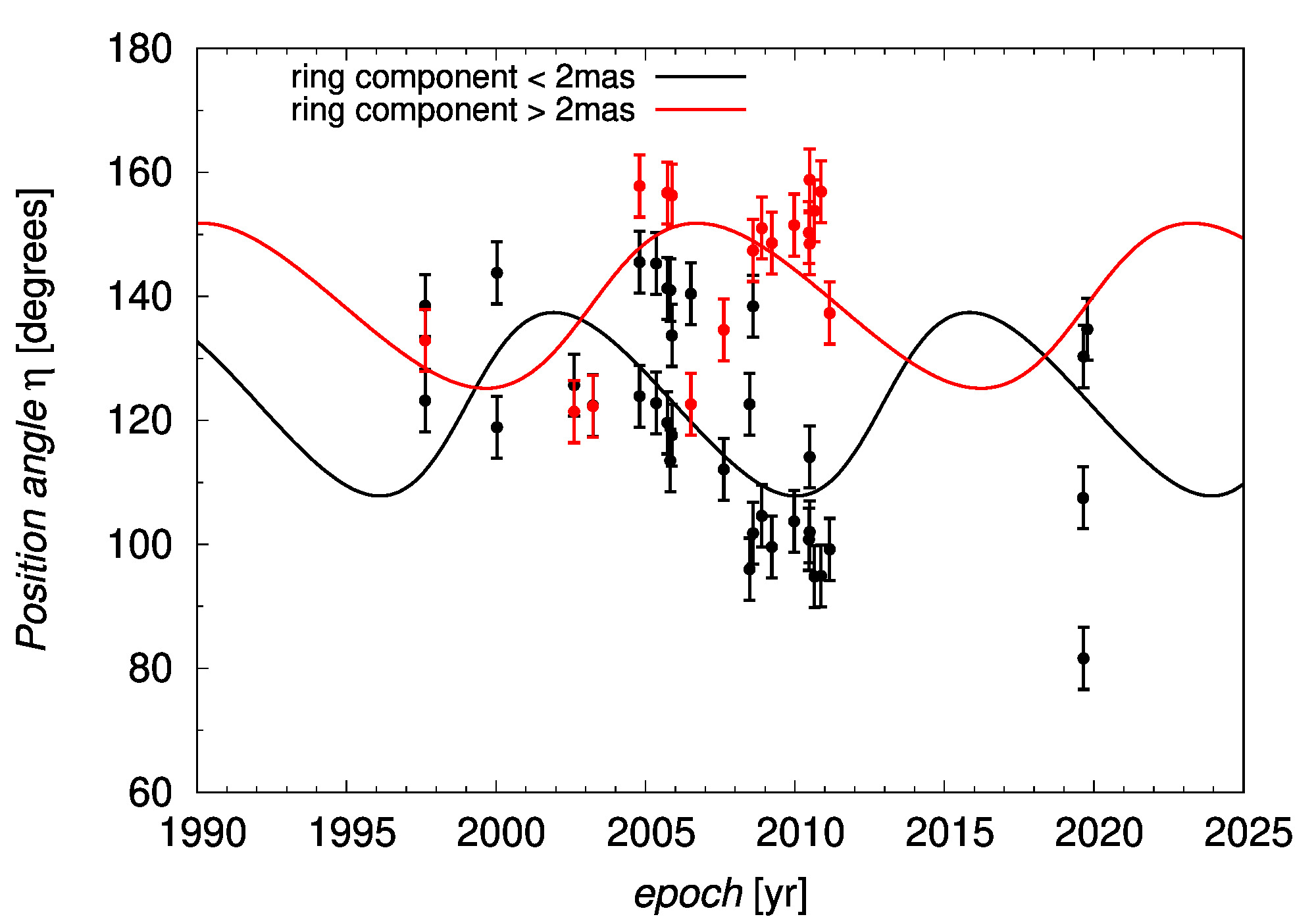}
    \caption{Temporal evolution of the ring components closer (black points) and further (red points) than 2 mas from the core. The lines depict the best-fit precession model for these two sections.}
    \label{fig_precession_ring}
\end{figure}
Although our precession model is partially degenerate with respect to the viewing angle and other parameters, we do see clear signatures of it and the model has a potential to be improved with future VLBA data. In particular, the temporal evolution of the position angle of the subluminal component j2 shows a smooth change typical of the precessing jet, see Fig.~\ref{fig_precession_model_j1-j4}, a second panel from above. This suggests that the ring structure could be a manifestation of the jet precession or of the collision of the precessing jet with the surrounding medium. We show this explicitly in Fig.~\ref{fig_precession_ring}, where we demonstrate that the components along the ring, namely their position angles, seem to precess in a similar way as the jet components j1-j4. There are two parts of the ring precessing with a different phase but a similar period close to 20 years closer than 2~mas and further than 2~mas from the core, see also Tables~\ref{tab_precession_model} and \ref{tab_precession_model_smallphi} for the best-fit parameters for these ring sections. 

The apparent ring size is also consistent with the inferred half-opening angle of the precession cone. Given the projected ring distance of $2\,{\rm mas}\sim 2\times 8.648\,{\rm pc}=17.3\,{\rm pc}$, the estimated deprojected distance is $\sim 17.3\,{\rm pc}/\tan{3^{\circ}}=330\,{\rm pc}$, where we used the viewing angle as derived by \citet{karamanavis}. In case the ring structure is formed due to the precession at a similar distance from the core, the apparent ring diameter of $3\,{\rm mas}\sim 26\,{\rm pc}$ is close to its actual diameter. Then the half-opening angle of the ring is $\Omega_{\rm ring}\approx 1/2\, (26/330)\, (180^{\circ}/\pi)=2.3^{\circ}$, which is within uncertainties consistent with our estimates based on the precession modelfits to individual components. The best-fit precession parameters in Tables~\ref{tab_precession_model} and \ref{tab_precession_model_smallphi} incorporate the corrections (to solve for the VLBA scaling problem reported by the MOJAVE team) and are within the uncertainties consistent with the fit results without the flux density correction for 2019 epochs.
\begin{figure}
    \centering
    \includegraphics[width=\columnwidth]{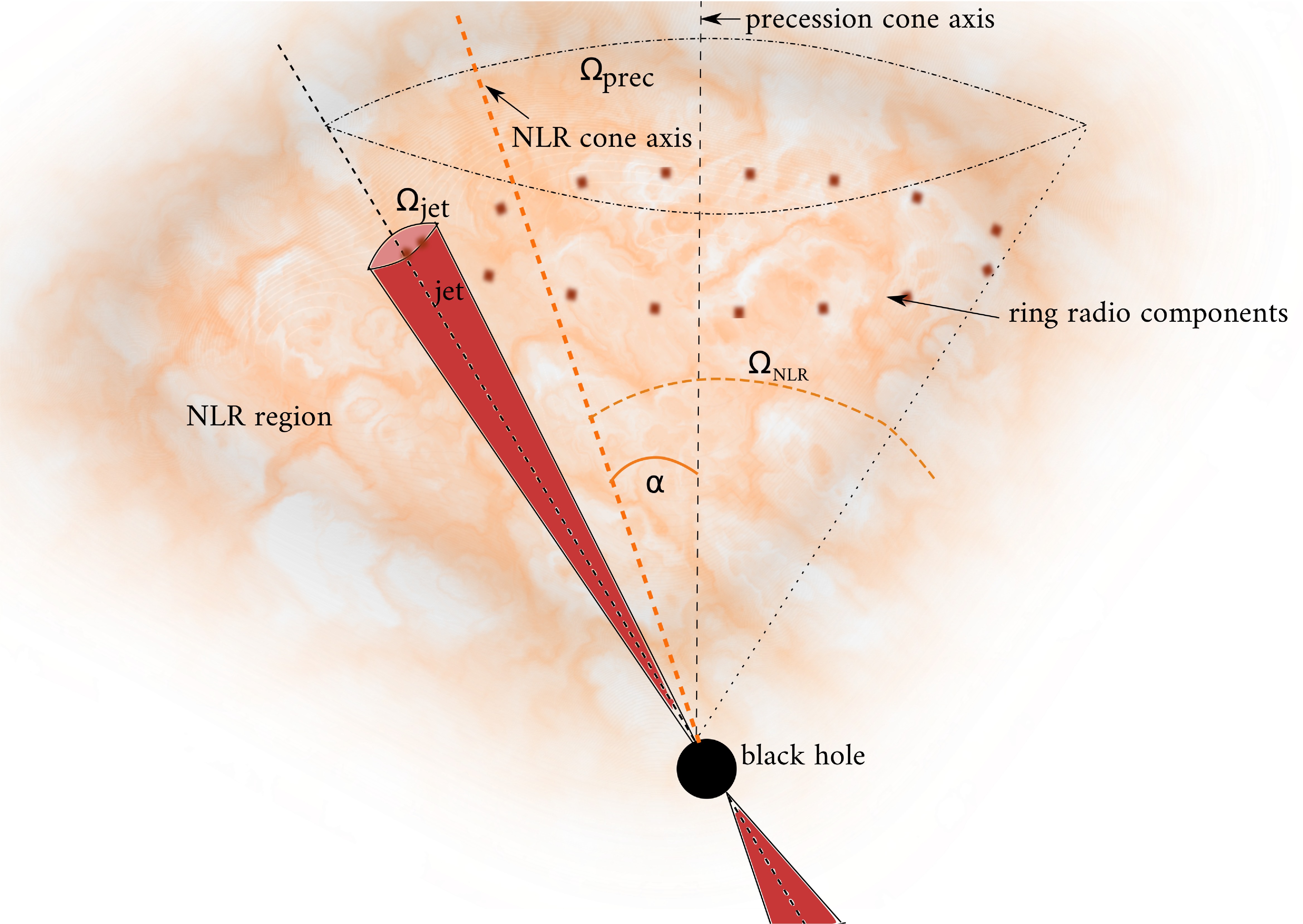}
    \caption{Illustration of the general alignment between the Narrow-Line Region ionisation cone and the precessing jet.}
    \label{fig_precession_NLR}
\end{figure}
It is expected, that PKS 1502+106 has a typical active galactic structure, where an extensive, cone-like, narrow line region (NLR) is present (see e.g. \citealt{peterson}). In order for the precessing jet to collide with the denser clumps within the NLR/ionisation cone, it requires at least a partial geometrical alignment. 
We consider the precessing cone with the half-opening angle of $\Omega_{\rm prec}$ and the jet with the half-opening angle of $\Omega_{\rm jet}$, see Fig.~\ref{fig_precession_NLR} for a comprehensive illustration. The NLR cone can be assumed to be generally inclined by the angle $\alpha$ with respect to the axis of the precession cone. Then the basic condition for the minimal interaction to occur is to require for the half-opening angle of the NLR cone $\Omega_{\rm NLR}$ the condition $\Omega_{\rm NLR}^{\rm arc}>|\Omega_{\rm prec}-\alpha|-\Omega_{\rm jet}$. This would lead to the formation of one-sided arcs until the NLR cone reaches the half-opening angle of at least $\Omega_{\rm NLR}^{\rm ring}>\Omega_{\rm prec}+\alpha-\Omega_{\rm jet}$, which leads to the formation of the full collisional ring, see Fig.~\ref{fig_precession_NLR} for the illustration of such a case. The ring half-opening angle of $2.3^{\circ}$ thus provides the observational lower limit for the NLR cone half-opening angle under the assumption that the NLR cone axis is aligned with the precession cone axis. For a general case, the additional angle $\alpha$ expressing the NLR cone misalighment needs to be added.

\subsection{Radio and $\gamma$-ray light curves}
In Fig.~\ref{radio_gamma_big} we show a light-curve comprised of the radio data (UMRAO, MOJAVE, and OVRO, at 15 GHz) and the $\gamma$-ray data obtained with \textit{Fermi}-LAT.
PKS 1502+106 showed a very strong outburst in the radio around 1985 (based on UMRAO data, Fig.~\ref{radio_gamma_big}).
Radio observations by the RATAN-600 radio telescope (2.3, 3.9, 7.7, 11.2, 21.7 GHz), and on the 32-m Zelenchuk and Badary radio telescopes (5.05, 8.63 GHz) \citep{konnikova} observed another flare around 2002 (these data are not included in Fig.~\ref{radio_gamma_big}).
Unfortunately, only few UMRAO data points are available for the time between the 1985-flare and the next major radio outburst in 2009. MOJAVE sum denotes the flux densities derived from the VLBA observations: We show the sum of the flux densities of the individual VLBA jet components and core derived per epoch. 

The major radio flare in 2009 occurred almost simultaneously with a $\gamma$-ray flare observed with the \textit{Fermi}-LAT. The next major peak in the $\gamma$-rays occurred in 2015 and was accompanied by a long-term rise in the radio. Smaller flares are superimposed on the $\gamma$-ray flare and these sub-flares seem to correlate with the sub-flares seen in the radio regime. The $\gamma$-ray flare in 2015 is not -- as before -- accompanied by a similarly sharp radio flare but the radio flare has a much broader distribution and continues for at least six years. 
The flux density at 15 GHz measured with the OVRO 40m telescope monitored this long-term outburst that started in 2014. The 15 GHz flux density reached an all-time high (since the beginning of the OVRO measurements in 2008, \citealt{atel}) of about 4 Jy twice between 2017 and 2020.

In Fig.~\ref{radio_gamma_big} we plot the flux densities derived from the VLBI observations (MOJAVE) as well as the single-dish flux densities obtained with the OVRO telescope. It seems that PKS 1502+106 reveals very sharp and high-amplitude flares in the radio regime (as traced by UMRAO and OVRO) but that since 2014 the flux density behaviour for this source changed significantly. 

To figure out, where the $\gamma$-ray emission of the 2009-flare in PKS 1502+106 originates, we plot the flux density of the individual components derived from the re-analysis of the MOJAVE data. As can be seen in Fig.~\ref{radio_gamma}, the sharp radio flare from 2009 originates from the core region as the core feature dominates the flux density contribution of the jet. The VLBA scaling issue has been corrected for the data studied in this paper and is of maximal 20 per cent for the last three data points at 15 GHz. It does not affect this dominant result. In Fig.~\ref{radio_gamma} the VLBI flux densities at X-band are plotted as well. The dominant contribution to the flux density between 2012 and 2019 at X-band as well stems from the core region. Especially the long-term rise from 2014 onward is clearly dominated by the core region.

\subsection{Periodicity and correlation analysis of radio and $\gamma$-ray light curves}
\label{analysis}
The 15 GHz OVRO and the \textit{Fermi}-LAT $\gamma$-ray light curves with 686 and 282 data points, respectively, are sufficiently dense to perform a periodicity analysis. The data points need not only to be densely sampled but to cover several cycles in the time window, which is the case. The total length of the time series is $4398.9$ and $4215.0$ days for the 15 GHz and the $\gamma$-ray light curves, respectively. 

First, we searched for the candidate periods using the periodogram that maximizes Cauchy-Schwarz Quadratic Mutual Information \citep[QMICS; ][]{huijse12}. For both the OVRO and \textit{Fermi}-LAT light curves, we went through an array of frequencies starting from $f=2\times 10^{-4}\,{\rm d^{-1}}$, which is close to the total duration of the observations, up to the Nyquist limit. For the OVRO light curve, the average sampling rate is $6.42$ days, which implies the Nyquist limit at $1/12.84\,{\rm d^{-1}}$. Since the \textit{Fermi}-LAT light curve is sampled every 15 days, the Nyquist limit is at $1/30\,{\rm d^{-1}}$. 

In Fig.~\ref{fig_QMI_periodograms}, we show periodograms for both OVRO (15 GHz) and \textit{Fermi}-LAT light curves in the left and the right panels, respectively. At high frequencies, we still see frequent peaks associated with aliases, therefore we restrict the further search to $f\gtrsim 0.1\,{\rm d^{-1}}$ and $f\gtrsim 0.01\,{\rm d^{-1}}$ for the OVRO and the \textit{Fermi}-LAT light curves, respectively. We checked different ways for the period determination, including Phase Dispersion Minimization periodogram \citep{stellingwerf1978}, Lafler-Kinman's string length \citep{clarke2002}, and orthogonal multiharmonic analysis of variance \citep[MHAOV; ][]{schwarzenberg-czerny1996}, for which we used the python package P4J, where these methods are incorporated alongside the QMI method. 
For the OVRO 15 GHz light curve, the MHAOV method proved the most efficient in suppressing the aliases and it located the best peak at $f_{\rm best}=0.000818\,{\rm d^{-1}}$, which corresponds to $1222.5$ days or $3.35$ years in the observer's frame. The MHAOV method generally increases the sensitivity and is efficient in damping alias periods. In comparison with QMICS method, the MHAOV increases the sensitivity in particular towards lower frequencies where the red noise seems to increase in the QMICS power, see Fig.~\ref{fig_QMI_periodograms} (top left panel). 

Therefore we also applied the MHAOV to the \textit{Fermi}-LAT light curve for comparison. For the $\gamma$-ray light curve, we obtained the best peak at $f_{\rm best}=0.0003275\,{\rm d^{-1}}$, which corresponds to $3053.4$ days or $8.36$ years. It is followed by the peak at $3.84\times 10^{-3}\,{\rm d^{-1}}$, which corresponds to $\sim 260.2$ days, and the peak at $2.7\times 10^{-3}\,{\rm d^{-1}}$ that corresponds to $\sim 371$ days and hence is likely an alias. However, the peak at $f=0.0008021\,{\rm d^{-1}}$ ($1246.7$ days or $3.41$ years), which is consistent with the best peak of the OVRO light curve, is also among the prominent peaks. The peak at $\sim 8\times 10^{-4}\,{\rm d^{-1}}$ is very close to the slightly higher peak at $9.126\times 10^{-4}\,{\rm d^{-1}}$, which corresponds to 1096 days or 3 years and hence it is most likely an alias. The highest peak at $8.36$ years is close to the end of the \textit{Fermi}-LAT light curve, hence it is also not considered. The OVRO and the \textit{Fermi}-LAT MHAOV periodograms can be seen in the bottom panels of Fig.~\ref{fig_QMI_periodograms}.    

Second, we apply the Lomb-Scargle (LS) periodogram \citep{lomb, scargle} for an independent analysis, which is especially suitable for unevenly-sampled data. In particular, we use the fast version of the LS periodogram as implemented in the gatspy package\footnote{\url{https://www.astroml.org/gatspy/periodic/lomb_scargle.html}}. For the 15 GHz light curve we find the peak in the LS periodogram at $\tau_{\rm radio}=1220.6\,{\rm days}=3.34\,{\rm years}$, see Fig.~\ref{fig_lomb_scargle} (left panel). An indication of a broad peak around the same period is also found in the $\gamma$-ray LS periodogram, see Fig.~\ref{fig_lomb_scargle} (right panel), for which the optimized search found the peak at $\tau_{\rm gamma}=1225.9\,{\rm days}=3.36\,{\rm years}$.

We test the statistical significance of the periodicity of the OVRO light curve using a bootstrap method. We generate 200 light curves by random resampling, for which one can assume that they do not exhibit a periodic behaviour at the best frequency we found (null hypothesis). Then we construct a histogram from the peak frequencies inferred by the MHAOV method and fit it using a Gumbel probability density function (PDF). We pick different significance levels corresponding to 1$\sigma$ up to 6$\sigma$ and from the fitted PDF we calculate the confidence intervals. 

Our best frequency inferred from the OVRO light curve lies above the 5$\sigma$ confidence interval, therefore we can reject the null hypothesis at this level and the peak can be considered statistically significant, see also Fig.~\ref{fig_histogram_confidence}. 

For the \textit{Fermi}-LAT light curve, the significance of the candidate peak at $\sim 3.4$ years is only at the $1\sigma$ level as well as for all its other peaks, see Fig.~\ref{fig_histogram_confidence} (bottom panels). Therefore we do not treat any peak in the $\gamma$-ray domain as significant and with the current \textit{Fermi}-LAT data no periodicity is thus present. 

To visually better show how the light curves evolve during one phase given by the candidate period of $\sim 3.4$ years, we fold both light curves using this common period. The folded light curves are displayed in Fig.~\ref{fig_folded_light_curves}. In the radio light curve (top panel), we see that at the phase $\phi=-0.5$, the flux density is decreasing or stays the same, which is followed by a smaller peak or a plateau at $\phi\sim 0.0$, and towards $\phi=0.5$, the flux  density is increasing or stays the same. This is even more distinct in the pattern of the $\gamma$-ray light curve in the bottom panel. This could suggest that there is a common process behind both the radio and the $\gamma$-ray emission. However, the significant periodicity is currently only present in the radio light curve. The potential connection between the $\gamma$-ray and radio emission is evaluated via the cross-correlation function below.

To investigate the periodicity at longer timescales, we studied the whole radio light curve consisting of both OVRO 15 GHz data and UMRAO data at 4.8, 8.0, and 14.5 GHz, see Fig.~\ref{radio_gamma_big}. We applied the MHAOV method as before for the shorter duration of the overlapping OVRO and the \textit{Fermi}-LAT dataset. The whole light curve consists of 1006 measurements, with the total time-span of 42.6 years and the mean sampling rate of 15.5 days. The peak at 3.35 years is still present, while at smaller frequencies even a larger peak in the periodogram is detected at $11.22$ years in the observer's frame. The significance analysis using the moving block bootstrap suggests the significance above 8$\sigma$ for the longer period, see Fig.~\ref{fig_radio_periodicity_all}. The radio light curve folded with the best-peak frequency is shown in Fig.~\ref{fig_folded_radio_all}. The longer period can capture the main peaks -- there is a prominent peak on May 2009, the period indicates the peak close to epoch 1998, where we see an increase, although not fully covered by the UMRAO monitoring, then the peak in 1986, which is well covered by the UMRAO monitoring. The predicted peak close to 2020 epoch is currently associated with the long-term increase in the radio flux density since 2014 and it is more difficult to interpret as it consists of well-defined sub-structure.

Also, we stress here that the observed light curve is a sum of the emission in the core region as well as the emission of the more extended jet. These two contributions are difficult to disentangle as the available core emission light curve is much shorter and shows only one prominent peak around 2009 and then an increase towards 2020, see Fig.~\ref{radio_gamma}. The core emission is expected to show the presence of quasi-periodic peaks due to Doppler-boosting in case the bulk jet precession is present. In addition, because of the stochastic nature of accretion, the underlying core variability is governed by the red noise \citep{timmer}. The detailed periodicity analysis should include the red noise subtraction to better reveal the periodic, non-stochastic process \citep{vaughan}, which will be the subject of our future study. The extended jet emission can exhibit emission spikes due to the interaction of the jet with the surrounding environment, i.e. denser clumps in the narrow-line region. The combination of these processes can explain the complicated nature of the increase in the radio emission after the minimum in 2014.

The two detected periods in the radio light curve -- 3.35 and 11.22 years -- correspond to the rest-frame timescales of $1.18$ and $3.95$ years, respectively. The longer timescale could in principle be associated with the jet precession. The best-fit precession periods for individual components are within uncertainties comparable in the observer's frame, see Tables~\ref{tab_precession_model} and \ref{tab_precession_model_smallphi}. The shorter period suggests a second-order jet motion, potentially due to the jet nodding motions or nutation. Previously, a similar coupling between a longer and a shorter variability periodicity in the framework of the precession-nutation model was analysed for the X-ray binary SS433 \citep{margon} as well the blazar OJ287 \citep{britzen_OJ287}. The precession and the nutation motion of the jet are natural for binary black hole systems, where the accretion disc around a primary is under the gravitational influence of a secondary black hole. In close binary systems, \citet{katz} derived a coupling relation between the nutation angular frequency on the one hand and the orbital and the precession angular frequencies on the other hand,
\begin{equation}
    \omega_{\rm n}=2(\omega_{\rm orb}-\omega_{\rm p})\,,
    \label{eq_nutation_precession}
\end{equation}
where the precession frequency should have an opposite sign with respect to the orbital frequency. Then the rest-frame binary orbital period that would induce the precession and the nutation can be estimated as,
\begin{equation}
    P_{\rm orb}=P_{\rm p}\left(\frac{P_{\rm p}}{2P_{\rm n}}-1 \right)^{-1}\,.
    \label{eq_orb_period}
\end{equation}
Since the precession period should be longer than the orbital period, from Eq.~\eqref{eq_orb_period} we obtain the minimum precession period of $P_{\rm p}=4P_{\rm n}=4.72$ years (13.4 years in the observer's frame) that could be associated with the nutation motion with the periodicity of $1.18$ years and the binary orbital period of $4.72$ years. A longer precessional period would result in the shortening of the binary period according to Eq.~\eqref{eq_orb_period}. We note that the longer precessional period derived based on Eq.~\eqref{eq_orb_period} is consistent with the precession periods inferred from the precession-model fitting, see Tables~\ref{tab_precession_model} and \ref{tab_precession_model_smallphi}. Therefore, the longer periodicity of 11.22 years (3.95 years in the rest frame) is a bit shorter than expected for a binary black hole system but this could be attributed to the uncertainties resulting from the sparse sampling of the UMRAO radio light curve. With the future monitoring of PKS 1502+106, these periodicities should be revised.

In addition and to quantify the correlation between the radio and the $\gamma$-ray light curves, we perform the auto-correlation study of each of the light curves as well as their cross-correlation. For this purpose, we use the z-transformed discrete correlation function \citep[zDCF; ][]{alexander1997}, with the implemented determination of the best peak and its uncertainty using the maximum likelihood. The zDCF is better suited for irregular light curves than DCF since it uses equal-population binning instead of regular time binning. The auto-correlation function of the radio light curve reveals clear peaks that are separated 456 days and 509 days in the observer's frame, see Fig.~\ref{fig_zDCF} (left panel). The autocorrelation function of the $\gamma$-ray light curve has a similar shape as the radio autocorrelation function when shifted by 217 days to correct for the different starting epoch of the light curves, except for the epoch of $\sim 3380$ days when the radio light curve exhibits a minimum in the autocorrelation function, while the \textit{Fermi}-LAT reveals the maximum. In addition, shortly before the neutrino emission event, both light curves exhibit an increase in the autocorrelation function.

In Fig.~\ref{fig_zDCF} (right panel), we show the cross-correlation of both light curves, which reveals a time-lag of $\tau=151^{+40}_{-95}$ days between the radio and the $\gamma$-ray light curves in the observer's frame, with the radio emission lagging behind the $\gamma$-ray emission. This time lag can be associated with the travel time of jet components and the associated emergence of $\gamma$ and radio emission as they cross the surface of unit opacity. Then the maximal length-scale between the $\gamma$-ray and the radio emission zones is $l_{\gamma,\mathrm{radio}}\lesssim  c \tau/(1+z)\sim 44.7\,{\rm milli-parsec (mpc)}$, which suggests a nearby spatial origin of the radio and $\gamma$-ray emission, most likely in the nuclear region. Since the OVRO 15 GHz light curve was obtained by a single-dish telescope, it could be contaminated by the radio emission of the larger-scale extended jet. On the other hand, a comparison of the OVRO measurements with MOJAVE 15 GHz emission (sum) suggests that the bulk of the 15 GHz emission originates from the compact core-jet system, see Fig.~\ref{radio_gamma_big}. The time-delay of $\sim 151$ days is characterized by a distinct, broad peak, which, however, has a relatively low zDCF value of $0.43$. The maximum zDCF value is at the time-delay of $3703^{+15}_{-25}$ days in the observer's frame with zDCF$=0.8$. In the rest-frame of the source, this time-delay would correspond to the distance of $\sim 1.1\,{\rm pc}$.

\begin{figure*}
    \centering
    \includegraphics[width=\columnwidth]{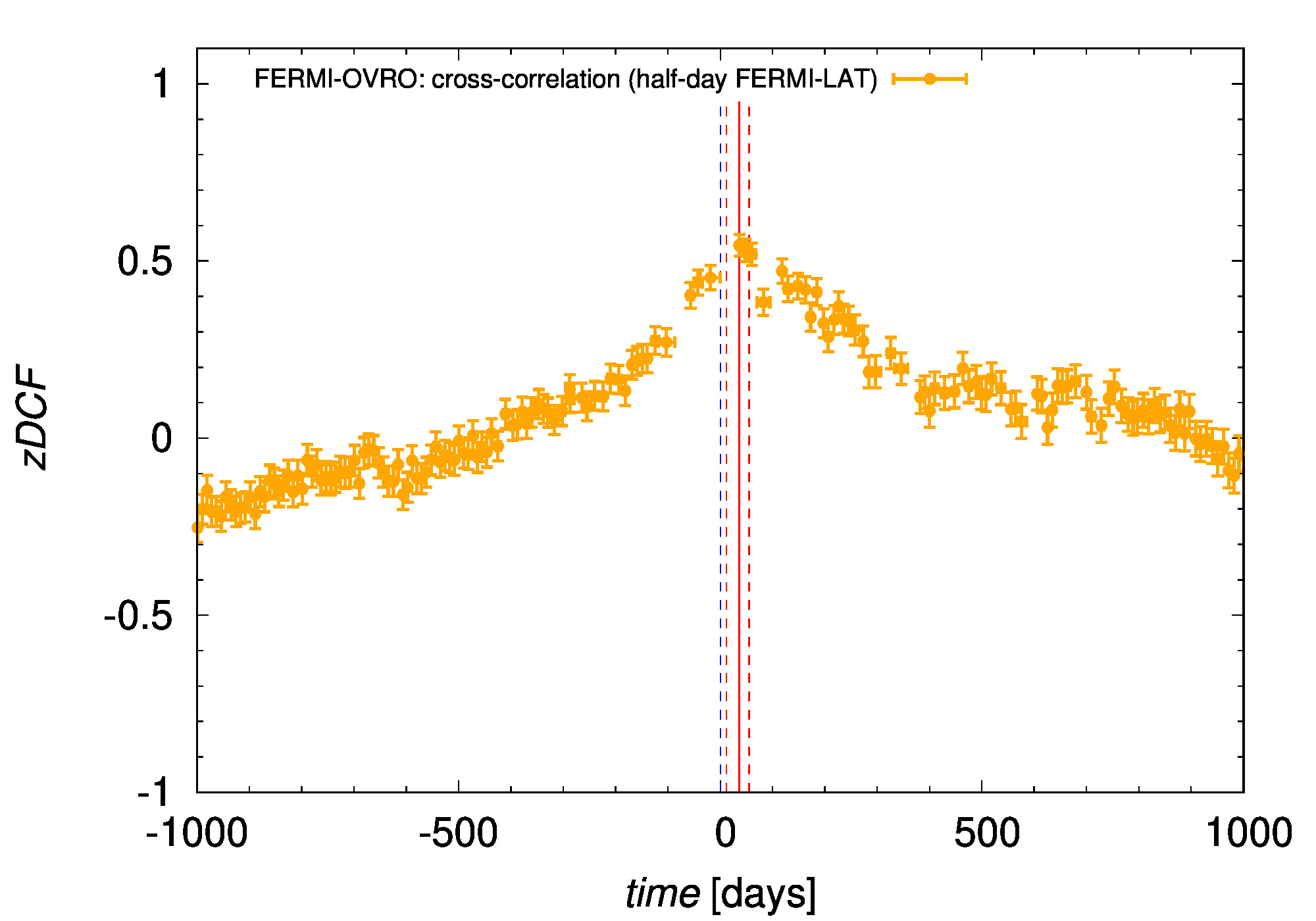}
    \includegraphics[width=\columnwidth]{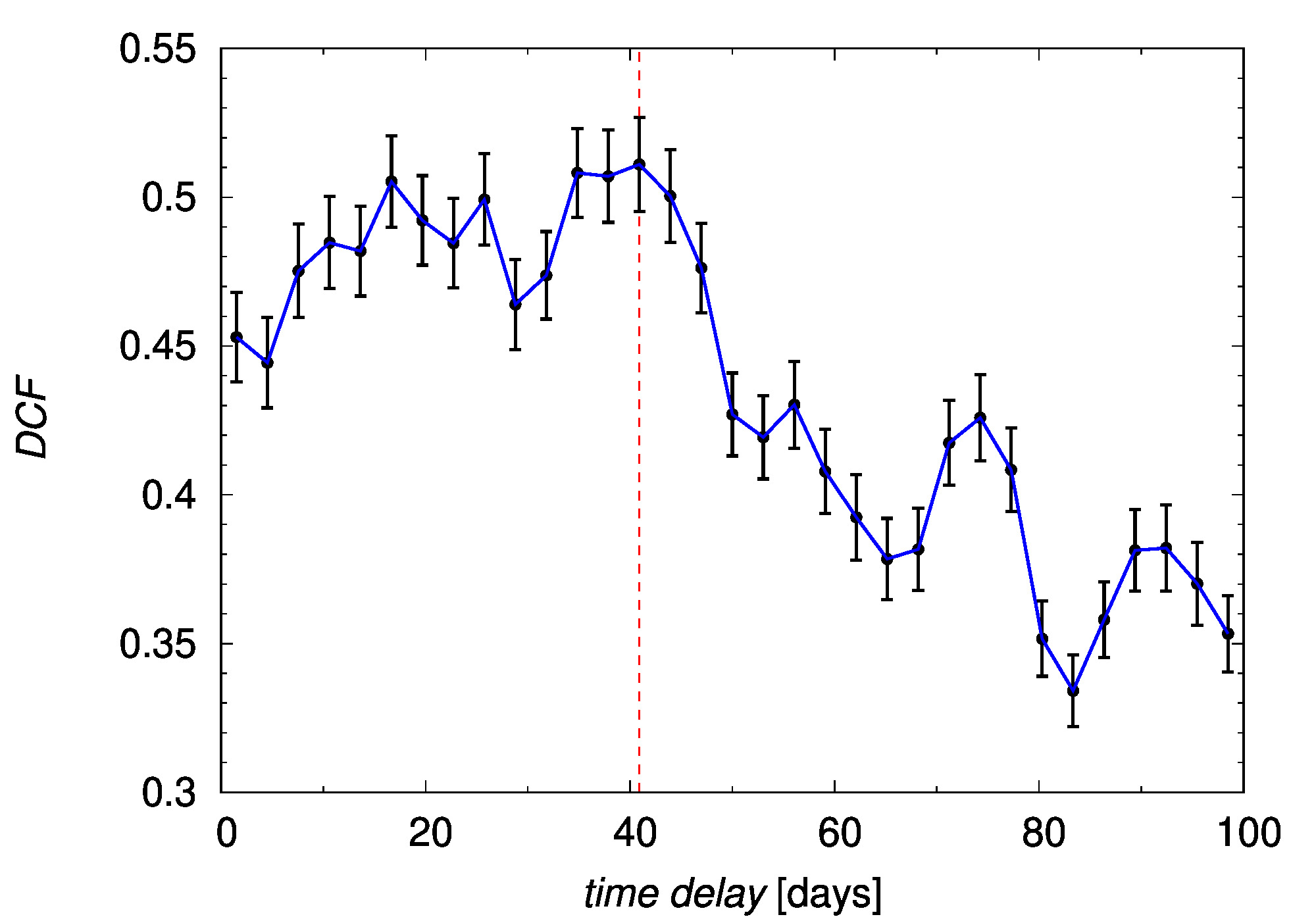}
    \caption{Cross-correlation between \textit{Fermi}-LAT (half-day sampling) and OVRO 15 GHz light curves. Left panel: $z$-transformed dicrete correlation function (zDCF) as a function of the time delay in the observer's frame, indicating a positive time-lag peak at $37^{+19}_{-24}$ days in the observer's frame, depicted by vertical red lines. The blue dashed line depicts the zero time lag. Right panel: Discrete correlation function (DCF) with the time-step of 3 days. The maximum DCF value of $0.51$ is reached for the time delay of 41 days in the observer's frame.}
    \label{fig_crosscorr_gamma_radio_1d}
\end{figure*}

When we use the \textit{Fermi}-LAT light curve with a smaller sampling interval of half day, the time-lag peak between $\gamma$-ray and radio emission shifts to a shorter lag of $\tau=37^{+19}_{-24}$ days in the observer's frame, see Fig.~\ref{fig_crosscorr_gamma_radio_1d}. This corresponds to the light-travel distance of $l_{\gamma,\mathrm{radio}}\approx c \tau/(1+z)\sim 11\,{\rm mpc}$. This value of the time lag is also consistent within uncertainties with the value of $40\pm 13$ days determined by \citet{max-moerbeck14}. In the study by \citet{max-moerbeck14}, PKS 1502+106 is one of three sources that exhibit a $\gamma$-radio cross-correlation at the significance larger than 2.25$\sigma$ level (97.54 per cent significance; the total studied sample contained 41 objects). We note here that the peak value and the uncertainty of the time lag for the zDCF method depends on the sampling of the \textit{Fermi}-LAT light curve. With the 15-day time step, we obtained 151 days, shortening to one-day time step, we got 69 days. Finally, with a half-day time step, we obtained 37 days in the observer's frame. We stress that the shortening of the time step is also beneficial for the discrete correlation function value -- for 15-day time step, zDCF$=0.43$ at the time-lag peak, for one-day step, zDCF$=0.48$, and for a half-day time step, we arrive at zDCF$=0.54$.

To verify the time-lag between the radio and the $\gamma$-ray emission, we also apply the ordinary discrete correlation function (DCF), which uses an equal time step instead of an equal population binning. We specifically focus on the interval between 0 and 100 days. For the calculation of the correlation function, we use OVRO data in combination with the \textit{Fermi} light curve with the 1-day sampling. The DCF is shown in Fig.~\ref{fig_crosscorr_gamma_radio_1d} (right panel) with the time-step of 3 days. The maximum DCF value of $0.51$ is reached for the time-lag of 41 days in the observer's frame. The Gaussian fit of the surroundings of the peak yields the mean value of $39.4 \pm 17.5$ days, which is consistent with the zDCF peak.

The measured time-lag $\tau$ between the $\gamma$-ray and the radio emission provides an estimate of the time interval between the emergence of the $\gamma$-ray and the radio emission. In the first approximation, we can imagine a component moving with the velocity of $\beta c$ along the jet. When the moving emission region crosses the surface of unit $\gamma$-ray opacity, $\gamma$ rays are observed. Likewise, when it crosses the surface of unit radio opacity, radio waves are emitted with a certain time lag with respect to the $\gamma$-ray emission. The distance travelled by the component can then be estimated as \citep{max-moerbeck14},
\begin{equation}
    d_{\rm comp}\sim \frac{\gamma \delta \beta c \tau}{1+z}\,.
    \label{eq_distance_comp}
\end{equation}
For estimating kinematic parameters, we adopted the best-fit values from scenario (ii) of the precession model with a small viewing angle, see Table~\ref{tab_precession_model_smallphi}. Specifically, we considered $\gamma\sim 1.5$, from which $\beta\sim 0.75$, and for a small viewing angle of $\phi\sim 3^{\circ}$, we get the Doppler-boosting factor $\delta\sim 2.7$. From Eq.~\eqref{eq_distance_comp}, we obtain $d_{\rm comp}\sim 33.3\,{\rm mpc}$.

The length-scales inferred from the cross-correlation analysis imply that the $\gamma$-ray and the radio emission zones are spatially very close, within subparsec distances, given the rest-frame time delay of only 13 light days. It is, however, not clear how far away the $\gamma$-ray emission zone is from the nuclear region. To estimate this, we first calculate the $\gamma$-ray variability timescale $t_{\rm var}$, following \citet{zhang99},
\begin{equation}
    t_{\rm var}=\frac{F_1+F_2}{2}\frac{t_2-t_1}{|F_2-F_1|}
    \label{eq_tvar}
\end{equation}
where $F_1$ and $F_2$ are flux densities at times $t_1$ and $t_2$, respectively. Using Eq.~\eqref{eq_tvar}, we study the \textit{Fermi}-LAT lightcurve. This time we use the light curve with half-day bins. We obtained in total 1471 detections, with a test statistic (TS) greater than or equal to 25, and 5337 upper limits (with $TS<25$). Following \citet{zhang99}, we filtered out the timescales with errors larger than 20 per cent. 

The minimum variability timescale is $0.32 \pm 0.03$ days, however, it is below the Nyquist limit of 1 day. The minimum variability timescale just above the limit is $1.01 \pm 0.17$ days, which we adopt as the representative variability time and could be considered as an upper limit of the intrinsic variability timescale. Then we estimate the distance of the $\gamma$-ray emission zone using,
\begin{equation}
    d_{\gamma}\sim \frac{2c\Gamma^2t_{\rm var}}{1+z}\,,
    \label{eq_distance_gamma}
\end{equation}
where for the bulk Lorentz factor, we take the values inferred from the precession models in Tables~\ref{tab_precession_model} and \ref{tab_precession_model_smallphi}, $\Gamma\sim 1.5-4.0$. In general, for a small viewing angle and subluminal apparent velocities, we expect a moderate Lorentz factor according to Fig.~\ref{fig_bapp_gamma_phi}. The estimate of the $\gamma$-ray emission zone distance then is $d_{\gamma}\sim 4.15\times 10^{15}\,{\rm cm}-2.95\times 10^{16}\,{\rm cm}=1.3\,{\rm mpc}-9.6\,{\rm mpc}$, which is fully consistent with the location of the $\gamma$-emitting zone inside the broad line region (BLR), assumed to be present in active galaxies (see \citealt{peterson}). The distance range derived above is consistent with the location within the unresolved VLBI core where the jet is launched and accelerated. Given the black hole mass of $M_{\bullet}\sim 10^9\,M_{\odot}$, the gravitational radius is $r_{\rm g}=GM_{\bullet}/c^2\approx 1.5\times 10^{14}\,{\rm cm}$, which implies that $d_{\gamma}\sim 28-197\,r_{\rm g}$. This spatial scale is at least two orders of magnitude smaller than reported by \citet{max-moerbeck14} for PKS 1502+106, who estimated $d_{\gamma}=22\pm 15\,{\rm pc}$ for a conical jet and $12\pm 9\,{\rm pc}$ for a collimated jet.

\subsubsection{Summary of the periodicity analysis}
To summarize, we find evidence for a period of $\tau_{\rm o}\sim 3.4$ years in the observer's frame based on radio light curve periodograms using two different, statistically robust methods. We do not detect any significant periodicity for the $\gamma$-ray light curve; however, the peak at 3.4 years is also present. A future monitoring in both the radio and the $\gamma$-ray domain is necessary to confirm the period as currently the peaks in the periodograms, associated with it, are quite broad. However, folding the OVRO and the \textit{Fermi}-LAT light curves with this period suggests that the main peaks and dips in both light curves are associated with it. 

By adding UMRAO radio data to the radio light curve, we find an indication of a longer periodicity of $\sim 11.2$ years in the observer's frame. These periodic patterns could be linked to the precession (longer timescale) and the nutation/wobbling of the jet/disc system (shorter timescale) similar to OJ 287 \citep{britzen_OJ287}. The minimum precession period coupled to the shorter nutation period of 3.4 years is 13.4 years in the observer's frame, which is consistent with the precession periods inferred from fitting precession models to individual components as well as marginally with the longer candidate period from the radio light curve. 

From the visual inspection of the long-term radio light-curve (Fig.~\ref{radio_gamma_big}) it seems likely that a longer periodicity time scale could be present in the radio data. Whether precession takes place on a longer timescale in the radio and $\gamma$-ray data cannot be inferred on a significant level from the \textit{Fermi}-LAT and the OVRO light curves due to the sparse coverage of the radio data before 2008. 

Based on the cross-correlation analysis, we find that the radio emission lags behind the $\gamma$-ray emission by $\sim 150$ days, which suggests spatially close emission domains at $\sim 45\,{\rm mpc}$ distance. Using the variability timescale of the $\gamma$-ray emission of a few days, we can constrain the location of the $\gamma$-ray emission zone in the BLR region. 

\subsection{Further evidence for deterministic processes from nonlinear analysis}
The nonlinear data analysis provides insights into the deterministic properties of the physical systems.
The advantages of the nonlinear analysis in comparison with the widely used stochastic methods are more in the direct connection to the physical properties.
Taking into account only one-dimensional input whose behaviour resembles randomness, the nonlinear methods can reveal some patterns in such data in terms of deterministic chaos \citep{Panis2020}.
The essence of the whole approach is the topological connection between the original and reconstructed phase space.
The one-dimensional time series are embedded into a higher-dimensional space and later analysed by sophisticated algorithms, see \citet{2015Chaos..25i7610B} and \citet{Kantz}. 

The interpolated light curves in $\gamma$-ray and the radio spectrum embeddings have been constructed with the estimates of time lag $\tau = $ 8 and 18 by average mutual information and embedding dimension $ m  = $ of 9 and 7 by the Cao algorithm \citep{1997PhyD..110...43C}, respectively, where the embedding dimension also estimates the number of degrees of freedom of the underlying physical system.

The Recurrence Plot (RP) is a tool which provides graphical insights into the properties of the system based on the description of some formed structures \citep{1987EL......4..973E}. The numerical description of RP is Recurrence Quantification Analysis (RQA) \citep{2008EPJST.164....3M}.
The RQA serves as a useful tool when investigating physical properties of nonlinear systems because of its straightforward interpretability within physics due to its simple algorithms.

\begin{figure*}
\centering
\includegraphics[width=\textwidth]{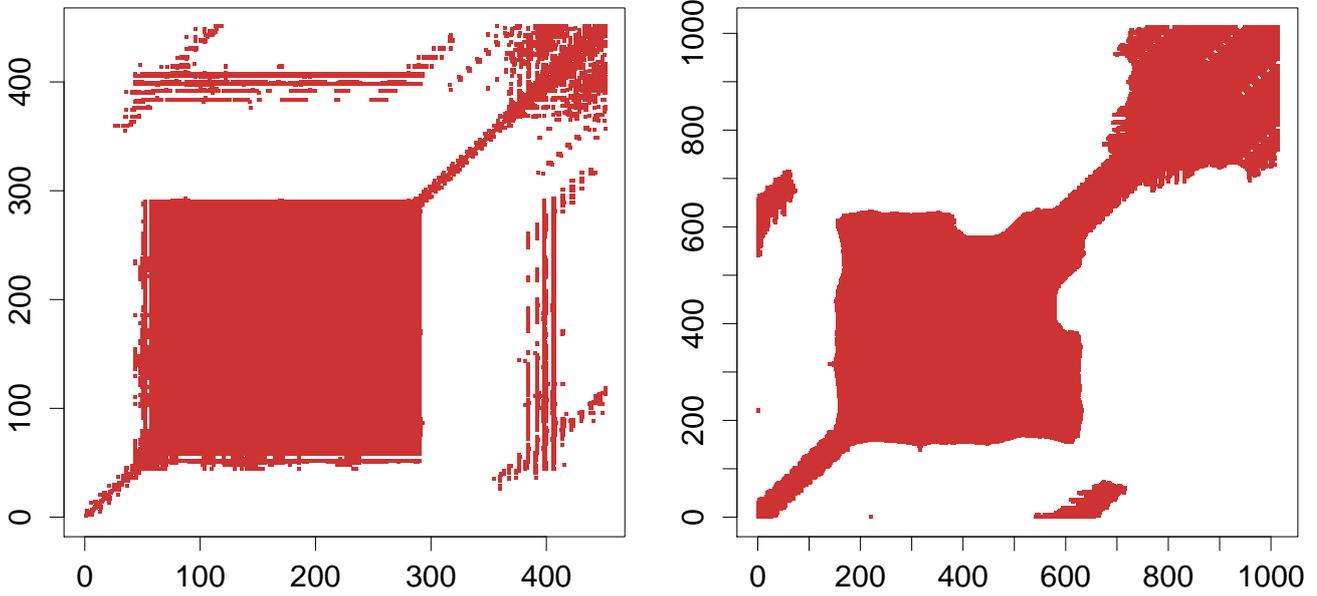}
\caption{Recurrence plot of the \textit{Fermi}-LAT $\gamma$-ray (left)  and the OVRO radio light curves (right) at the  Recurrence Rate RR = 30 per cent revealing similar structures in the $\gamma $ and the radio spectrum.}  
\label{fig1}
\end{figure*}

We performed the following calculations based on a $\gamma$-ray light curve with seven day binning. During the analysis, the connection between the radio and $\gamma$-ray light curves was found since the RPs of both light curves in Fig.~\ref{fig1} show very similar structures at the Recurrence Rate (RR) =  30 per cent as also used in \citet{2020MNRAS.497.3418P}.

\begin{figure*}
\centering
\includegraphics[width=\textwidth]{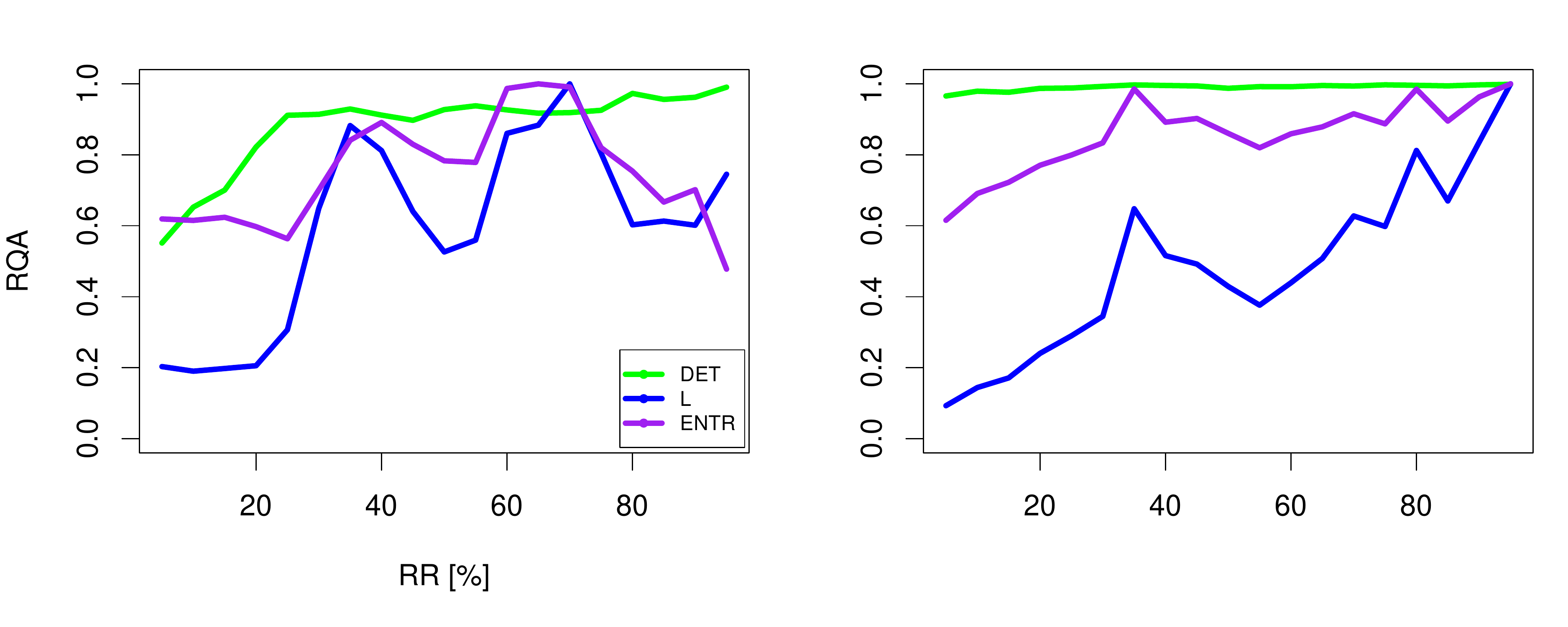}
\caption{Recurrence quantification analysis measures, namely, DET - determinism, L - average line length and ENTR - entropy showing an approximately similar behaviour across various RR $ \in [5, 10 \dots, 95]$, where especially the L and ENTR measures build peaks at similar RRs. The RQA measures have been transformed into the interval [0,1] in order to fit to one graphics.
The average (mean) of RQA measures mDET, mL, mENTR for the \textit{Fermi}-LAT $\gamma$-ray light curve 
 is (0.88, 18.2, 2.23) and for the OVRO radio light curve, the mean values are (0.99, 92.73, 4.14).}  
\label{fig2}
\end{figure*}

The connection is also visible in Fig.~\ref{fig2} using the RQA measures DET, L, and ENTR (see the legend), which follow a similar pattern across various RRs in the range RR [\%]$\, \in\, [5, 10 \dots, 95] $, 
where DET represents the deterministic properties, L the measure of predictability and ENTR the amount of information content in the time series, for a more detailed description of RQA measures, see \citet{2019EPJC...79..479P}.

In addition, we also provide the averaged RQA measures of both light curves across RR [\%]$\, \in\, [5, 10 \dots, 95] $ presented in Fig.~\ref{fig2}, which can provide more accurate information rather than comparing the data at only one RR value. However, the magnitudes of RQA measures depend on more factors and should not be taken literally. For more insights see \citet{bhatta} where the source PKS 1502+106 came out as the most deterministic among all 20 studied blazar sources with the same approach of RQA. 

\subsection{The C IV line shape - an  outflowing BLR in PKS 1502+106}
\label{civ}
\begin{figure}
\includegraphics[width=8.6cm]{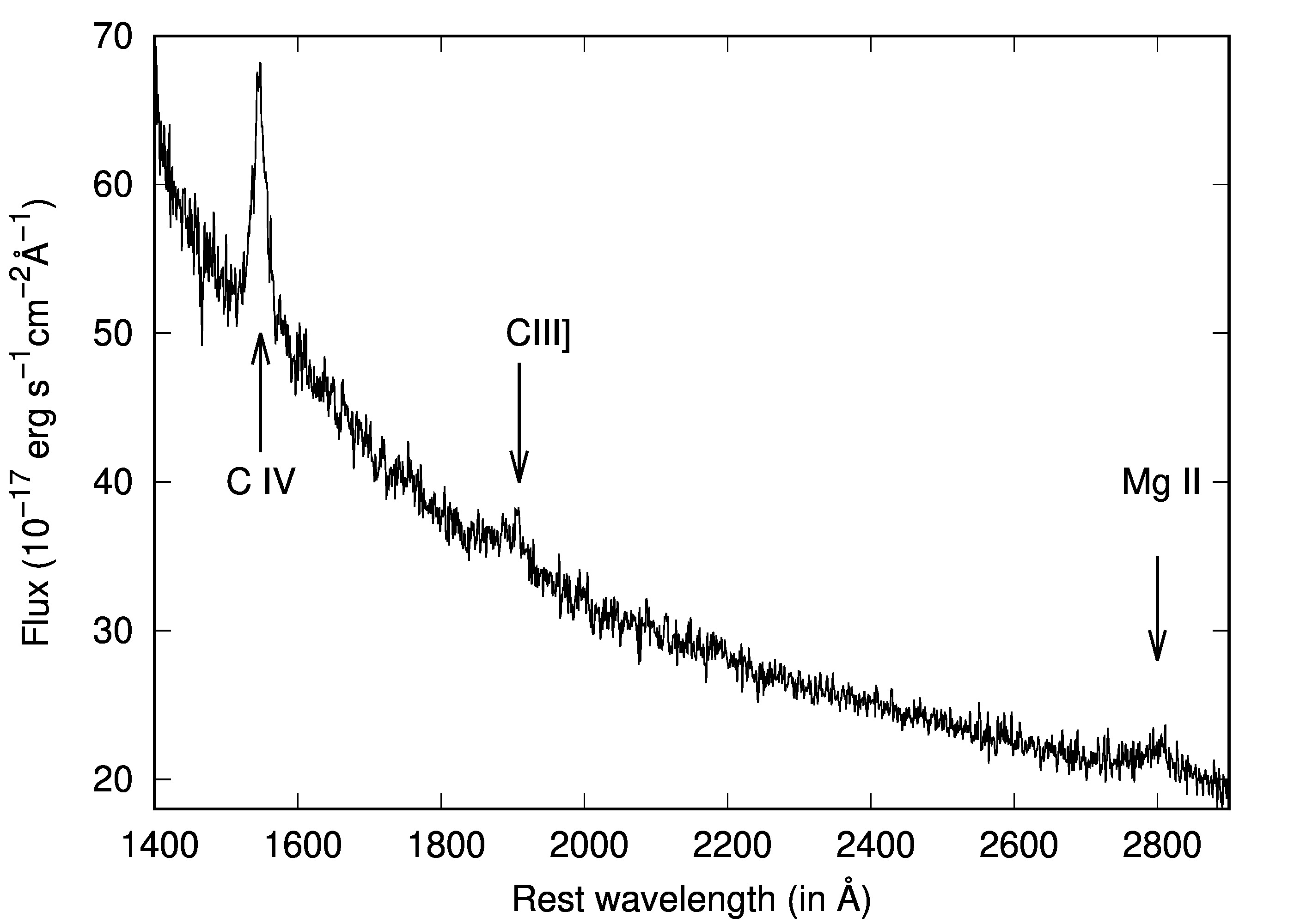}
   \caption{The UV spectrum of  PKS 1502+106. Only three broad emission lines are well seen (shown with arrows): the most intensive line C IV$\lambda$1549\AA, and also weak C III]$\lambda$1909\AA\ and Mg II$\lambda$2800\AA\ lines.}
    \label{fig-uv}
\end{figure}

We explore the PKS 1502+106 spectrum obtained within the Sloan Digital Sky Survey (SDSS) in the UV band that covers C IV $\lambda$ 1550\AA, C III] $\lambda$ 1909\AA  and Mg II $\lambda$ 2800\AA. As it can be seen in Fig.~\ref{fig-uv}, there is a very intensive C IV broad line and weak CIII] and Mg II lines. The spectral energy distribution, with the strong blue part, is typical for blazars. Using the luminosity of the C IV line (e.g. \citet{kong})
and the UV continuum (e.g. \citealt{lira}) it is possible to roughly estimate the dimension of the BLR (see the review by \citealt{popovic} and references therein). In both cases we find that the BLR is much smaller compared to the PKS 1502+106 ring structure. 
The bolometric luminosity of the BLR in PKS 1502+106 is 3.7$\times$10$^{45}$erg s$^{-1}$ \citep{Liu2017}, implying that the BLR is located at a radius of 2$\times$10$^{17}$ cm.

We find that the C IV line shape seems to be shifted to the blue around 600 km s$^{-1}$. In Fig.~\ref{civ}[a] we show the line with the continuum and in Fig.~\ref{civ}[b] we show the line after the continuum subtraction. The line appears to be blue-shifted which might indicate that the line originates from some kind of outflows, and probably is not virialised (i.e. relation given above cannot be used for estimates of the dimension of the BLR or black hole mass estimates). This might be due to the continuum, which might stem partly from the jet. Additionally we fitted the C IV line with one Gaussian (typical for kinematics of a number of emitting clouds in a BLR) and logarithm-like profile \citep[typical for an outflow-like BLR, ][]{kollatschny}, and find (see Fig.~\ref{civ}[c, d]) that a logarithmic-like profile (Fig.~\ref{civ}[d]) very well fits the PKS 1502+106 C~IV line profile (better than a Gaussian one, see Fig.~\ref{civ}[c]).

The ring-like structure we find, may be connected with the NLR, that, as it was mentioned above, is also expected to be present in blazars (or face-on oriented quasars). In the case of a face-on oriented quasar, 
an observer probably is looking  through the NLR cone, and the ring may originate in the shock-wave formed at the edge of  a cone which is detected in some AGN \citep[e.g. in the case of IC 5063, see][]{mo07}. Shock waves can produce $\gamma$-ray emission (e.g., \citet{madejski}), and probably also the radio emission. The C IV line is relatively strong compared to Mg II, that seems to be produced in some kind of shock waves \citep{fromerth}. The Mg II line is also blue-shifted around 800 km s$^{-1}$ indicating an outflowing BLR. 
The question concerning the dimensions of the outflowing Mg II BLR is still open, since reverberation can only give dimensions of a virialised BLR.  The outflowing BLR can be significantly larger than a virialised BLR. The broadening in the virialised BLR is caused by the central black hole mass, and the BLR is compact. The broadening in the outflowing material however, is caused by a velocity dispersion in the outflow. These velocities can be in a large interval -  from zero (far away from the central source) to several 1000s km s$^{-1}$ (close to the central source).
In addition, as it is shown in Fig.~\ref{civ}[c, d], the line shape of the C IV line is not Gaussian, but has some logarithmic-like profile (Fig.~\ref{civ}[d]) that is typical for outflowing material \citep{kollatschny}. 

In summary, the broad line profile  of the C IV line follows  a logarithmic-like profile (narrow peak and extensive wings) that is typical for the emission of an outflow.  Additionally,  the broad C IV and Mg II lines are shifted to the blue which strongly indicates an outflow origin of these broad lines.

The presence of an outflowing BLR  has been suggested for blazars (e.g., \citealt{paltani} for 3C 273; \citealt{finke} for 3C 454.3). In these sources, the outflowing BLR could  serve  as  an  alternative  source  of  seed  photons  for the inverse Compton scattering \citep{leon}.

Our findings are supported by recent modeling performed by \citep{rodrigues2020} who find that two hadronic models, a proton synchrotron model and a leptohadronic model, can both describe the multi-wavelength emission. In the case of the leptohadronic model, the $\gamma$-ray emission is mostly dominated by external Compton  scattering. According to the authors, this is possible due to the location of a blob (neutrino-emitting region in the jet) near the perimeter of the BLR.

\begin{figure*}
\begin{minipage}{8.1cm}
  \includegraphics[width=\linewidth]{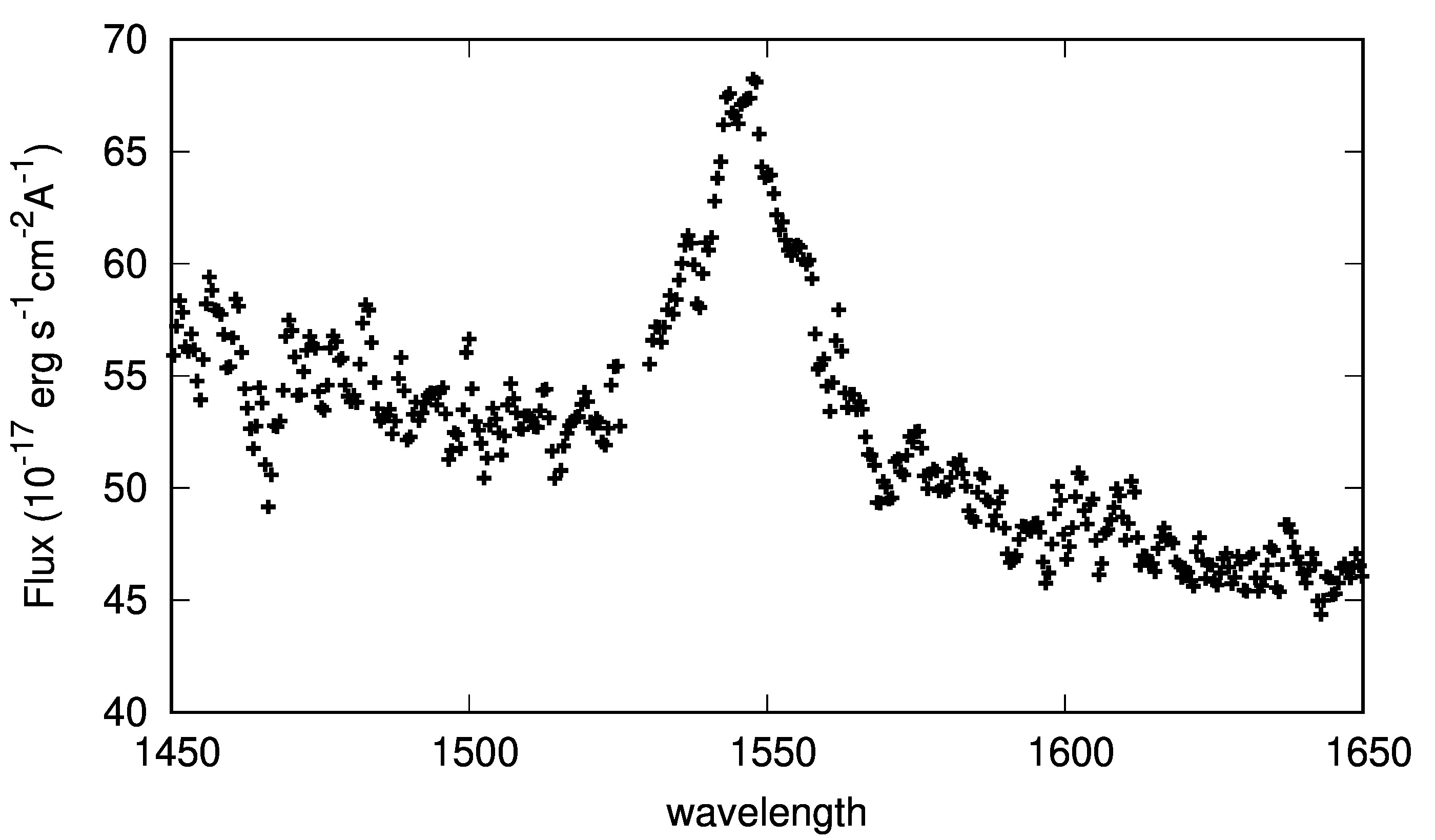}
  [a]
\end{minipage}
\begin{minipage}{8.1cm}
  \includegraphics[width=\linewidth]{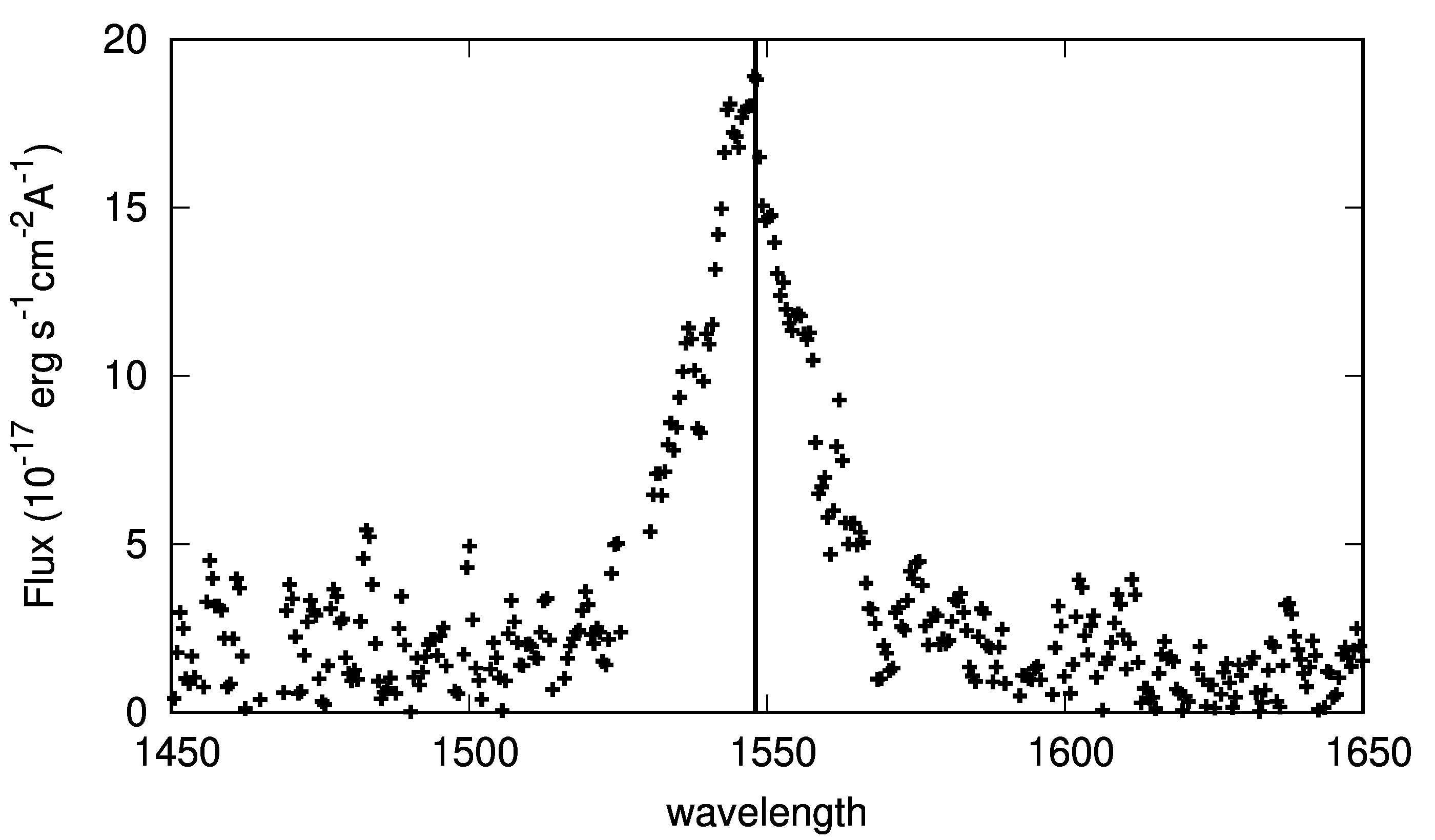}
  [b]
\end{minipage}
\begin{minipage}{8.1cm}
  \includegraphics[width=\linewidth]{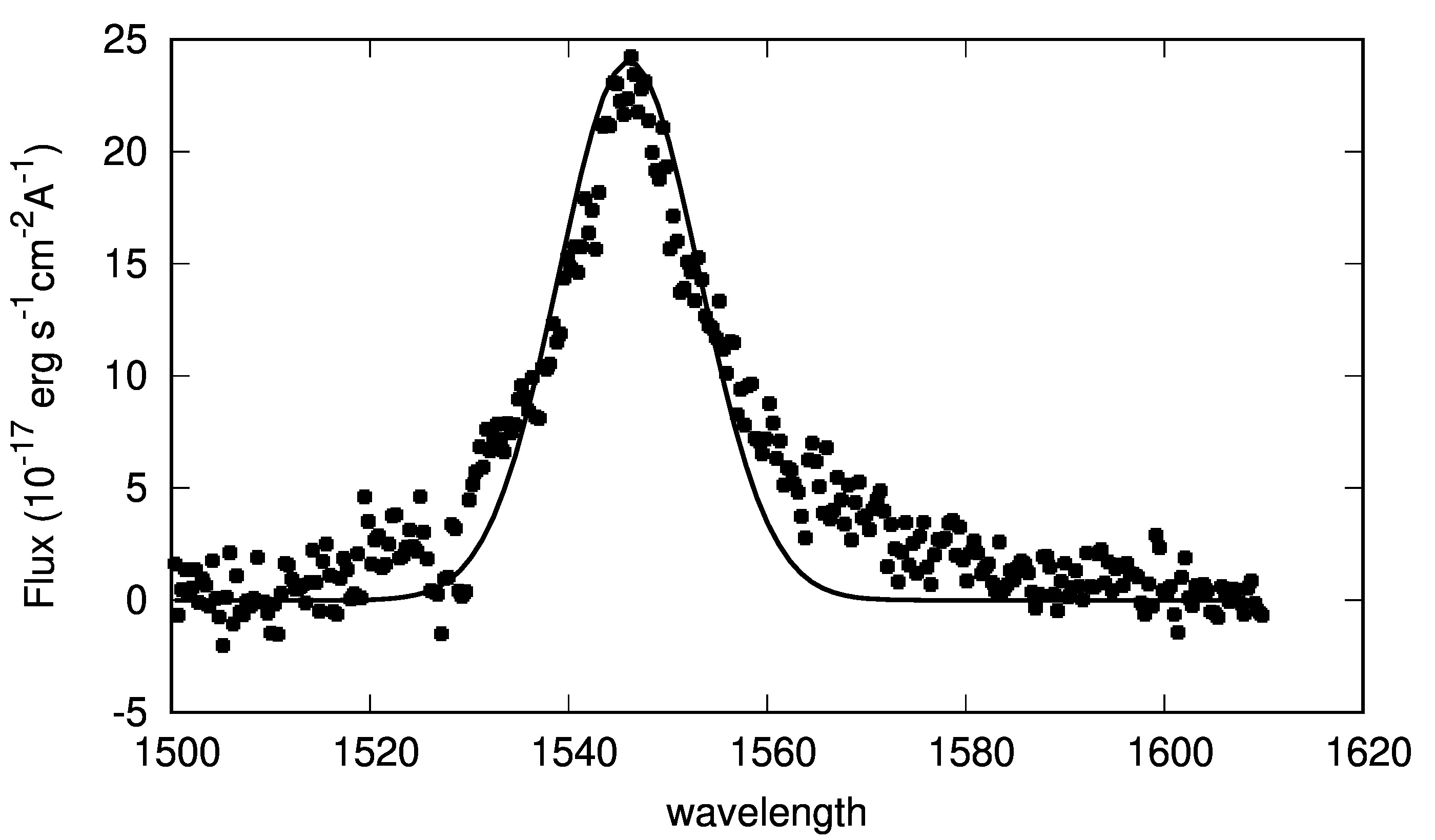}
  [c]
\end{minipage}
\begin{minipage}{8.1cm}
  \includegraphics[width=\linewidth]{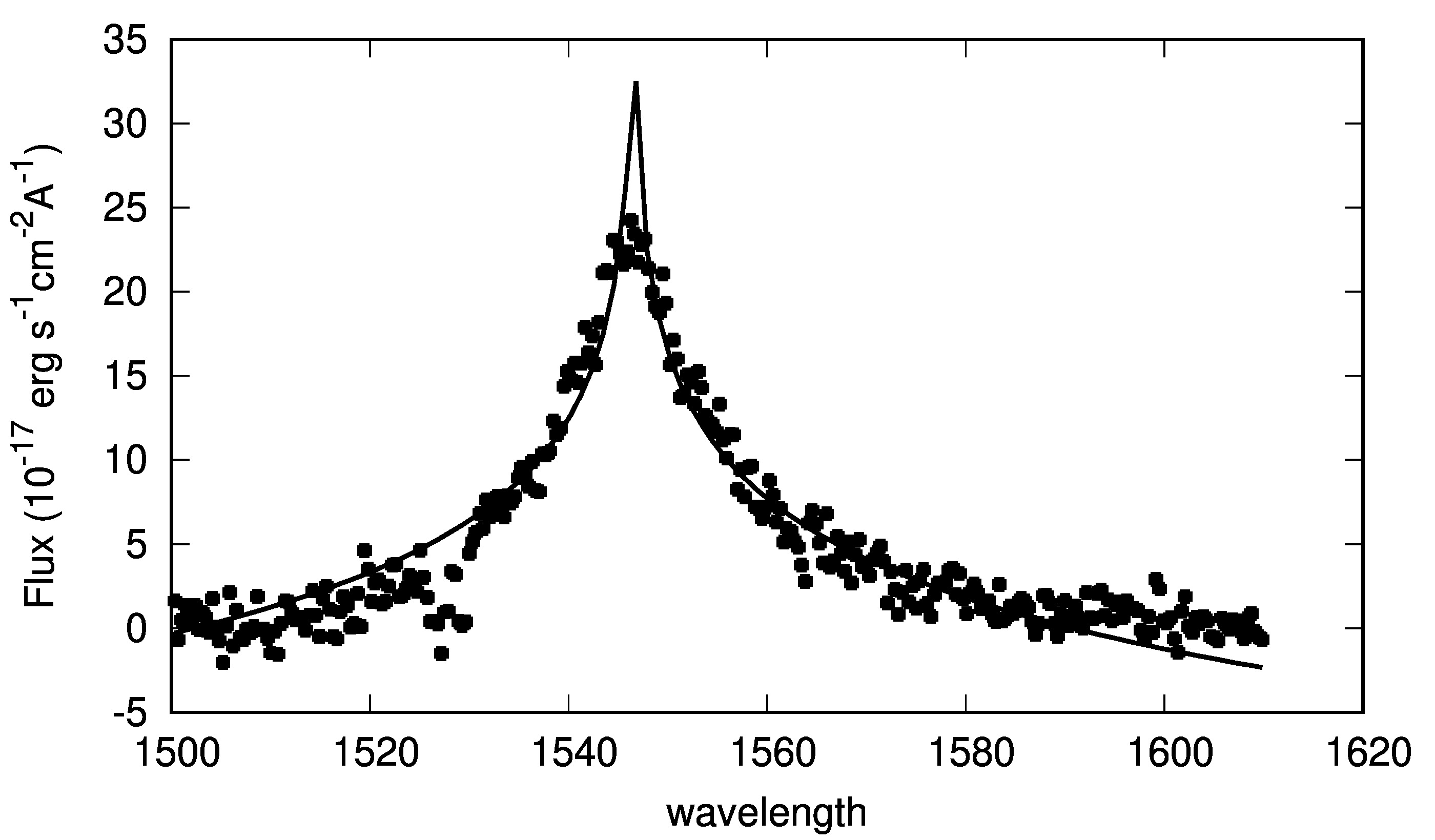}
  [d]
\end{minipage}
\caption{The C IV line extracted from the SDSS spectrum of PKS 1502+106: [a] with the continuum  and [b] after linear continuum subtraction. The vertical line on panel [b] represents the rest wavelength of C IV line. The C IV line fitted with one Gaussian is shown in panel [c] and with a logarithm-like profile in panel [d].}
    \label{civ}
\end{figure*}

\begin{figure*}
\includegraphics[width=13.9cm]{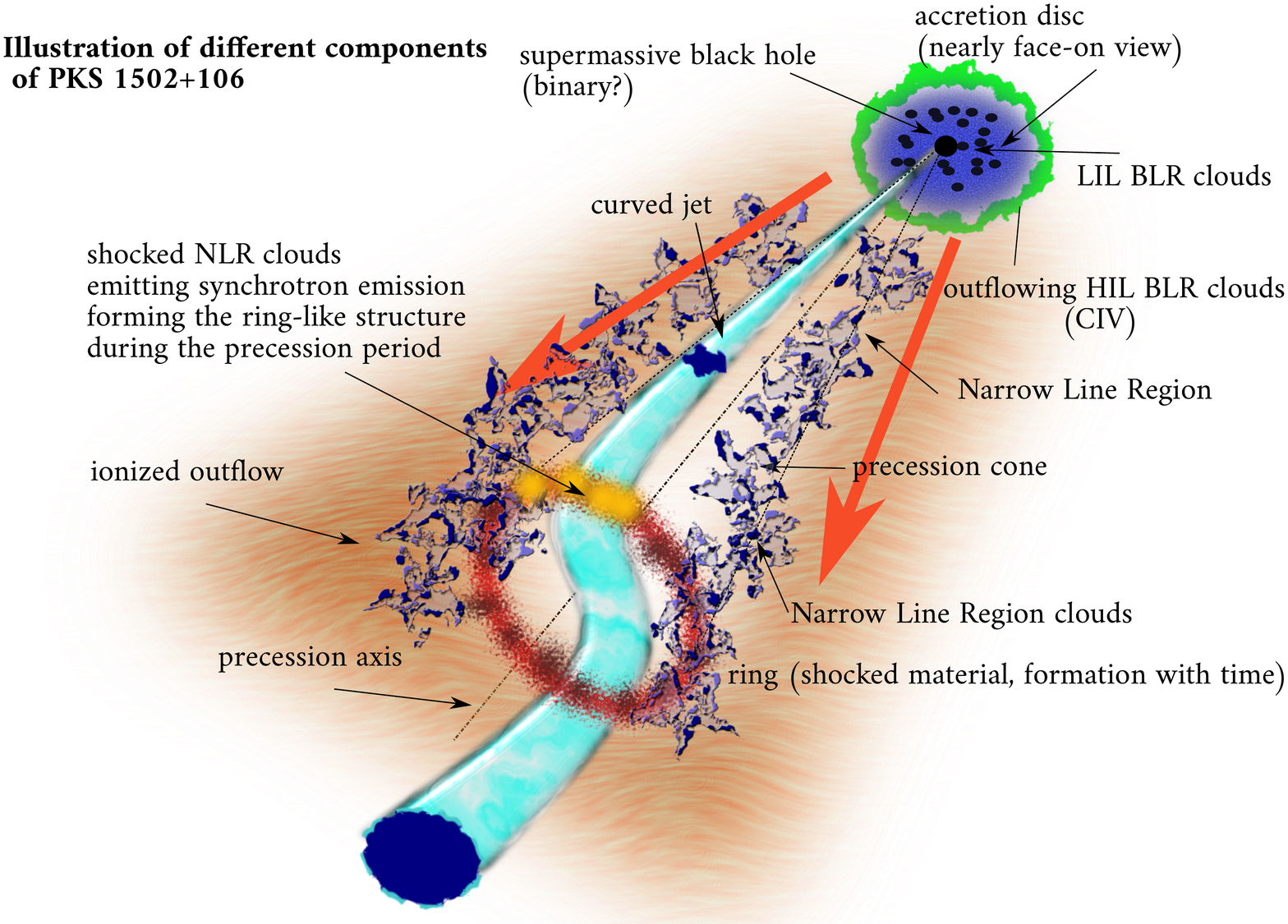}
   \caption{Illustration of different components of PKS 1502+106. In particular, we depict a curved jet whose axis is close to the line of sight. The jet as a whole is precessing around a precession axis. The interaction of the curved precessing jet with the ionised outflow, in particular the denser narrow-line region clouds, may be responsible for the formation of the ring structure with time. Closer to the supermassive black hole or potentially a binary black-hole system, we show the narrow line region and the low- and high-ionisation broad line region clouds (LIL and HIL BLR clouds). The HIL BLR material traced by C~IV broad line is blueshifted and therefore is outflowing.}
    \label{sketch}
\end{figure*}

\section{Discussion}
\label{discussion}
The main motivation for our analysis was to check whether PKS 1502+106 reveals any atypical radio morphology on pc-scales that could help to explain why and how this AGN produces high-energy emission and possibly a neutrino. 

The central findings we report in this paper are evidence for additional and atypical radio emission along and perpendicular to the main jet ridge line in the form of a ring and arcs. We find evidence for a precessing jet deduced from the properties of the jet components. Both the radio and $\gamma$-ray light curves show periodic modulation by the same process (which could be precession). In addition, we provide evidence for an outflowing BLR based on an analysis of the C IV line emission. 
In the following, we discuss our findings. Our main emphasis is to stress the implications of our results for a better understanding of the generation of the variable radio and $\gamma$-ray emission and -- potentially -- the neutrino emission.

\subsection{Jet kinematics}
\label{jet_kinematics}
Several authors have reported very fast apparent speeds (>10$c$) for jet component motion in PKS 1502+106 \citep{an,karamanavis}. 
\citet{karamanavis} obtained very large Doppler factors (between $\sim$7 up to $\sim$50).
We cannot confirm the reported fast motions. Only for one component (j3 in Fig.~\ref{identify}[b] in the first identification scenario and j4 in the second scenario in Fig.~\ref{identify}[d]) is apparent superluminal motion clearly detected in both scenarios listed in Table~\ref{speeds}. This component however, seems to be moving at a rather typical apparent speed (for quasars). The inner component j2 reveals subluminal motion in both identification approaches. For jet knots j1 and j4 the uncertainties are too large to determine a reliable value for the apparent speed. 
Our results differ from the apparent motions that have been reported in the literature. The most likely reason is that the results presented by other groups did not include the ring detection. It is possible, that by not detecting the ring and by misidentifying this structure with standard jet features, higher apparent speeds have been obtained. The misidentification or ring-components as jet components also leads to the derivation of higher Doppler factors. 

The difference between the inner slow apparent speed and the fast speed for the outer component most likely confirms a change in the viewing angle between the inner and outer jet structure proposed by \citet{karamanavis}. They report very small viewing angles: $\sim$3 deg for the inner jet and $\sim$1 deg for the outer jet. 
We confirm an internal bending of the jet and discuss our further findings in the following sections.

\subsection{Evolution of a ring and multiple arcs with time}
The most important result we derive for PKS 1502+106 is the temporary evolution of a ring and multiple arc structures which appear along but perpendicular to the main jet ridge line. Evidence for the ring is not only seen in total intensity but also in the polarisation maps. While the ring and arcs are interesting in themselves, their appearance gains more relevance in a neutrino emitting AGN. 
While it is still possible that the ring/arc structure is not physically involved in the neutrino generation process, this seems a highly unlikely coincidence. Especially since a similar structure has only been observed once in one other AGN (3C454.3, \citealt{britzen_3c454.3}). 

There is some indication that the ring structure we detect was present already earlier (see Fig.~\ref{comparison}[a], where we reproduce a figure from \citealt{an}). 
It looks like that the ring might have been present in the S-band data of 1994.52 already.
On the other hand, the results of \citet{an} suggest that the ring might not be present after 1998.11. 
In the data discussed in the present paper, the ring structure becomes prominent again around 2004 and is then visible till the end of 2011. 
Our findings concerning the position angle as a function of the core distance (see Fig.~\ref{comparison}[b]) are consistent with those presented earlier. 

In Fig.\ref{comparison} we compare the position angle versus core distance at different frequencies by \citet{an} and
those based on the analysis of a much larger data set at 15 GHz as presented in this paper. We thus find some evidence based on the literature data and the data analysed by us, that the ring is not persistent but seems to occur temporarily. In Fig.~\ref{radio_gamma_big} we indicate the times (around 2003 and 2012) when these atypical structures were {\it not} present in the data (red dashed lines).

The X-band data presented in this paper show evidence for multiple arcs which appear temporarily along but perpendicular to the jet. The arcs re-appear at similar core distances but with different diameters (perpendicular to the jet ridge line). As shown in Fig.~\ref{ring_x} and Fig.~\ref{xontop} the first major arc occurs at a similar position as the ring before. Thus, both phenomena, ring and arcs, might be of similar origin. The arcs could be segments of further ring-like structures along the jet. We find that the diameter of the arc seems to increase before the neutrino event. This increase in diameter correlates with a higher radio flux density as observed with OVRO. It seems possible, that the arc as interaction site between the jet and the outflowing medium, expands due to the excited activity. There seems to be a correlation between the appearance of the ring and the arcs and the flaring (radio, $\gamma$-rays) of the source (see Fig.~\ref{radio_gamma_big}).

We find some emission at the faint level in some epochs on the counter-jet side. More data and studies are required to analyse this in more detail and to confirm that this is counter-jet emission. As we think that the ring is the result of an interaction between the precessing jet and the outflowing NLR, 
this process can also take place on the counter-jet side.

\subsection{Location, size, and physical nature of the ring}
The ring is located at the projected distance of 1 mas (its edge closest to the radio core) to about 2 mas (its apparent centre). Considering the viewing angle of about 3 degrees of the inner jet \citep{karamanavis}, this gives us a spatial distance of $\sim$19 -- 38 mas, which corresponds to $\sim$ 164-329 pc (using 1 mas $\simeq$ $8.648\,{\rm pc}$). The ring is a macroscopic structure with the projected scales of $\sim 3\,{\rm mas} \times 1\,{\rm mas}\approx 26\,{\rm pc} \times 9\,{\rm pc}$. 
We attribute the brightness distribution of radio components in the ring-like structure due to two factors - the geometry and the collision with the ambient medium as we describe in the following two sections.

\subsubsection{Geometrical explanation of the ring: jet precession}
As shown in Section \ref{precession} and Fig.~\ref{fig_precession_time}, the flux density and position angle evolution with time of several jet components can be fitted by a precession scenario. Further support for the precession origin of the ring comes from Fig.~\ref{fig_precession_ring} where we show that the position angles of the ring-features precess in a similar way as the jet components j1-j4. The ring is composed of two parts which precess with a different phase but similar period close to 20 years. Additional support for the precession origin comes from the apparent ring size which is consistent with the inferred half-opening angle of the precession cone. Precession thus provides a convincing scenario explaining our observations.

The ring is located at the favourable location where the jet curves towards the line of sight, which according to \citet{karamanavis} occurs at $1\,{\rm mas}$ from the core, close to where the ring structure starts to be apparent. As the jet interacts with the surrounding material and in combination with the jet precession on the timescale of $\sim 20$ years, the ring structure can be formed in projection. The localisation of the ring could be explained by the occurrence of ionised denser clouds in the narrow-line region of PKS 1502-106 at the de-projected length-scale of $\sim 100\,{\rm pc}$ from the radio core.

Alternative mechanisms that can affect the global jet structure could be related to kink instabilities (e.g., \citealt{barniol, zhang, mizuno}). A mechanism which could generate a re-orientation of the jet can be given by a global kink-type instability (e.g., \citealt{barniol}). However, the jet gets disrupted, when turned into an instability. 
In this case, we would not observe a repetition of jet events and structures which we clearly observe. 
However, the jet spine (within the Blandford-Znajek model \citealt{blandford}; hereafter BZ model) might be able to stabilise the overall jet structure while the surrounding (disk) jet layer becomes kink unstable. Three-dimensional jet formation simulations revealed such a scenario \citep{ouyed, mckinney2009}.

\subsubsection{Considerations on the viewing angle}
The jet seems to pass the ring somewhat offset from the ring centre, implying that the ring structure and jet are not perfectly aligned. 
As we have to de-project the observed jet length considering the viewing angle, the same is true when we want to derive the true shape of the ring-like structure.
Interestingly, we find a possible mismatch between the jet and ring inclination, as we will explain in the following.


If we assumed the same inclination (in fact $(3-90)\degr$) for the ring structure, and further assumed that the ring structure is
oriented perpendicular to the jet, then the de-projected ring would not be circular but a highly elongated ellipse, as we need to de-project the observed minor axis of the ring according to the $3 \degr$ angle, resulting in an 8:1 axial ratio with the long axis along the jet.
On the other hand, if we assume that the ring is intrinsically circular 
(just because we observe such a nicely shaped elliptical structure over time), then we may calculate a different viewing angle of the ring structure based on this assumption.
De-projecting the minor axis length of the observed ring to the length of the observed major axis (the observed ratio is 2.36), we find a $25 \degr$ inclination of the observed ring structure, respectively we would look with $65\degr$ viewing angle along the ring axis, a value that is quite different from the jet inclination proposed by \citet{karamanavis}.

What could be the astrophysical implications of these considerations?
In the first case, the elongated ellipse may well represent the outer layers of an opening cone of some ambient medium around the jet. The jet is misaligned with the axis of this cone, potentially suggesting precession.

In the second scenario, the misalignment between the jet axis and the ring axis is hard to understand, as they differ quite a lot.
One may in principle question the $3 \degr$ inclination from the literature and suggest that the jet follows a different inclination.
However, with such a large inclination, we may not explain the superluminal motion observed in some of the knots.

At this point, we tend to follow the first scenario, with the jet precessing through and interacting with an opening cone of interstellar material (which will be discussed in the next subsection).
This is also consistent with the precession scenario (ii) resulting from a smaller line-of-sight inclination as presented in Sect.~3.5 above.

\subsubsection{Collisional cloud model of the ring -- ``Synchrotron pearls''}
 The deprojected distance of the ring at 160--330 pc suggests that it could be associated with the NLR, which is located at this distance range as well. It consists of clouds with densities in the range $10^3-10^6\,{\rm cm^{-3}}$ and temperatures ranging from $10^4$ to $2.5\times 10^4\,{\rm K}$ \citep[see][for an overview of the NLR characteristics]{peterson}. Since the jet of PKS 1502+106 precesses, it likely sweeps across the NLR material. The total power carried by the jet of PKS 1502+106 in the form of radiation, magnetic field, electrons, and protons inferred by \citet{ghisellini09} based on the SED modelling is $L_{\rm jet}\simeq 1.22 \times 10^{47}\,{\rm erg\,s^{-1}}$. The jet pressure at the deprojected distance of $z=200\,{\rm pc}$ can be estimated using the relation,
\begin{equation}
P_{\rm jet}\simeq \frac{L_{\rm jet}}{\pi R_{\rm j}^2c}=\frac{L_{\rm jet}}{\pi z^2 \tan^2{\theta}c}\,,
\label{eq_jet_pressure}
\end{equation}
where $R_{\rm j}$ is the jet radius at distance $z$ from the core, and $\theta$ is a half-opening angle, for which we take $\theta\simeq 1.90^{\circ} \pm 0.25^{\circ}$ according to \citet{karamanavis}. Finally, we obtain $P_{\rm jet} \sim 3.09\times 10^{-3}z_{200}^{-2}\,{\rm erg\,cm^{-3}}$. For the NLR, the thermal pressure is $P_{\rm therm}=n_{\rm NLR}k_{\rm B}T_{\rm NLR}\simeq 1.38\times 10^{-7}n_5\,T_4\,{\rm erg\,cm^{-3}}$, i.e. four orders of magnitude less than the jet kinetic pressure. The NLR thermal pressure may be considered as an upper limit because of its density range $n_{\rm NLR}\sim 10^3-10^5\,{\rm cm^{-3}}$. On the other hand, the ram-pressure of the orbiting NLR clouds, $P_{\rm ram}\sim \rho_{\rm NLR} v_{\rm NLR}^2$, whose orbital velocity may be estimated based on the FWHM of the narrow lines, $v_{\rm NLR}\sim {\rm FWHM}=500\,{\rm km\,s^{-1}}$, can be comparable to the jet kinetic pressure. The equilibrium distance can be estimated using the relation derived from the jet and the ram-pressure equilibrium $P_{\rm jet}\simeq P_{\rm ram}$,
\begin{align}
z & \approx \left(\frac{L_{\rm jet}}{\pi c \mu m_{\rm H} n_{\rm NLR}v_{\rm NLR}^2 \tan^2{\theta}} \right)^{1/2} \notag \\
& \approx 300\left(\frac{n_{\rm NLR}}{6.6 \times 10^5\,{\rm cm^{-3}}} \right)^{-1/2}\left(\frac{v_{\rm NLR}}{500\,{\rm km\,s^{-1}}} \right)^{-1}\,{\rm pc}.
\end{align}
Hence, the equilibrium of the jet and the ram pressure implies that the denser NLR clouds can influence the jet at the distance comparable to the distance of the ring under the assumption that the NLR clouds move perpendicular to the jet. This suggests a potential connection between the ring and the jet curvature towards the observer close to 1~mas.
    
The nature of the interaction between the jet and the NLR clumps can be approximately described using the basic velocities and involved timescales associated with this interaction. As the NLR cloud approaches the jet (we assume an approximately perpendicular motion in these considerations), its orbital speed of a few $100\,{\rm km\,s^{-1}}$ exceeds its internal sound speed of $13\,{\rm km\,s^{-1}}$ (for the temperature of $10^4\,{\rm K}$) and a shock passes through the cloud. First of all, the NLR clump will not fully enter the jet if its velocity perpedicular to the jet sheath is less than the shocked-gas sound speed, $v_{\rm NLR}<v_{\rm sc}$, where $v_{\rm sc}=c\sqrt{\Gamma_{\rm j}\rho_{\rm j}/\rho_{\rm NLR}}$ with $\Gamma_{\rm j}$ being the bulk Lorentz factor of the jet, $\rho_{\rm j}$ is the jet mass density, and $\rho_{\rm NLR}$ is the NLR mass density. Concerning the jet velocity, we consider the value $\beta_{\rm j}\sim 0.93$, which leads to the apparent velocity of $\beta_{\rm app}=0.68$ for the viewing angle of $3^{\circ}$, which is close to the inferred apparent velocity of j2 component ($0.56$ -- $0.80\,c$) that is the closest to the ring structure. The corresponding jet Lorentz factor then is, $\Gamma_{\rm j}=(1-\beta_{\rm j}^2)^{-1/2}\sim 2.72$. The jet density can be inferred from the relation for the jet luminosity $L_{\rm jet}=\pi R_{\rm j}^2(\Gamma_{\rm j}-1)\rho_{\rm j}v_{\rm j}c^2$, which yields $\rho_{\rm j}\approx 9.533\times 10^{-25}\,{\rm g\,cm^{-3}}$ at the distance of $z=300\,{\rm pc}$, or the particle density of $n_{\rm j}\approx 1.15\,{\rm cm^{-3}}$ for a fully ionised gas. Hence, the number density of NLR clouds is three to five orders of magnitude larger than the number density of the jet at the location of the ring. The sound speed in the shocked gas then is in the range of $v_{\rm sc}\approx 1680-16800\,{\rm km\,s^{-1}}$, which is larger than expected velocity of NLR clouds, $350\,{\rm km\,s^{-1}} \lesssim v_{\rm NLR}\lesssim 400\,{\rm km\,s^{-1}}$ \citep{peterson}. Therefore, the NLR clouds do not enter the jet, but get shocked at its sheath layer. In other words, the jet acts like a rigid-like obstacle for the NLR clouds. Given the jet radius at $z=300\,{\rm pc}$, $R_{\rm j}\approx z\tan{\theta}\sim 10\,{\rm pc}$, the NLR clouds with the minimum size of $l_{\rm NLR}\gtrsim 10^{18}\,n_{3}^{-1}\,{\rm cm}=0.3\,n_3^{-1}\,{\rm pc}$, with $n_3$ being the number density in units of $10^3\,{\rm cm^{-3}}$, the clouds moving perpendicular to the jet will be fully shocked on the timescale of $t_{\rm d}=l_{\rm NLR}/v_{\rm sc}\sim 19-190$ years, depending on the shock-crossing speed. The shocked gas will expand along the jet sheath, forming a flattened structure, which appears arc-like in projection.  
    
The radio synchrotron emission emerging from the shock in the NLR clouds can form an apparent ring in the radio maps with time. To calculate the peak synchrotron flux, we consider the NLR cloud dize of $l_{\rm NLR}\approx 10^{18}\,{\rm cm}$, the density of $n_{\rm NLR}\approx 10^5\,{\rm cm^{-3}}$, the temperature of $T_{\rm NLR}\approx 10^4\,{\rm K}$, and the jet diameter of $d_{\rm j}\approx 2z\tan{\theta}\approx 20\,{\rm pc}$. We set the shock-propagation speed to the  value of $v_{\rm sc}\approx 10^4\,{\rm km\,s^{-1}}$. In the shocked layer, the gas is heated up to above $10^8\,{\rm K}$, more precisely, as given by the jump condition $T_{\rm sc}\approx 3\mu m_{\rm H} v_{\rm sc}^2/16k_{\rm B}\approx 3\times 10^8\,{\rm K}$. The density in the shocked layer is increased to $n_{\rm sc}=4n_{\rm NLR}\approx 4\times 10^5\,{\rm cm^{-3}}$.  For calculating radio synchrotron flux densities originating in collisions of clouds with the jet, we also calibrate the post-shock magnetic field using $B^2=8\pi \epsilon_{\rm B}\mu m_{\rm H} n_{\rm sc} v_{\rm sc}^2$. The coefficient $\epsilon_{\rm B}$ expresses the enhancement of the magnetic field, where we consider the cosmic ray amplification like e.g. in supernova remnants, where $\epsilon_{\rm B}\approx 10^{-2}-10^{-3}$ \citep[see e.g.,][]{vink}. For $\epsilon_{\rm B}=0.01$, we obtain $B\approx 0.14\,{\rm G}$. Electrons that emit the synchrotron radiation in the given magnetic field with the peak frequency at $\nu_{\rm c}=15\,{\rm GHz}$ (MOJAVE survey) have the Lorentz factor of $\gamma\sim 160$ and the energy of $E_{\nu}= 82\,{\rm MeV}$. The corresponding cooling time is,
\begin{equation}
t_{\rm cool}=\frac{9m_{\rm e}^3 c^5}{4 e^4 B^2 \gamma} \sim 8\,{\rm yrs}\,.
\label{eq_cooling_time}
\end{equation}
Now we need to estimate the expansion timescale to determine whether we are in the adiabatic or rather the radiative regime. The timescale of expansion of the shocked NLR cloud can be estimated using the following relation,
\begin{equation}
t_{\rm exp}\sim \frac{l_{\rm NLR}}{v_{\rm sc}}\approx 63\,{\rm yrs}\,.
\label{eq_expansion_timescale}
\end{equation}
Since $t_{\rm exp}>t_{\rm cool}$, the shock is in the radiative regime, i.e. the shocked gas can radiate most of its kinetic energy before the expansion converts the thermal energy into the bulk motion.
    
For calculating the expected synchrotron flux density at 15 GHz, we first consider the denser NLR material with $n_{\rm NLR}\sim 10^5\,{\rm cm^{-3}}$, $n_{\rm sc}\sim 4n_{\rm NLR}$, and the propagating shock velocity of $v_{\rm sc}\sim 1680\,{\rm km\,s^{-1}}$. The thermal energy formed in a shocked single NLR cloud can be estimated as,
\begin{equation}
L_1\approx \mu m_{\rm H} n_{\rm sc} v_{\rm sc}^3 l_{\rm NLR}^2\sim 1.5 \times 10^{42}\,{\rm erg\,s^{-1}}\,.
\label{eq_thermal_singleNLR}
\end{equation}
The synchrotron emission requires that a fraction of the kinetic energy in Eq.~\eqref{eq_thermal_singleNLR} is deposited into relativistic electrons. We assume that the synchrotron power is a fraction of the thermal energy, $L_{\rm synch}\sim \epsilon L_1$, where $\epsilon=0.01$. Taking into account the luminosity distance of $D_{\rm L}=14366.8\,{\rm Mpc}$, the radio flux density at 15 GHz originating in the shocked NLR cloud is,
  \begin{equation}
       S_1=\frac{L_{\rm synch}}{4\pi D_{\rm L}^2 \nu}=4.05\times 10^{-29}\,{\rm erg\,s^{-1}\,cm^{-2}\,Hz^{-1}}=4.05 \times 10^{-6}\,{\rm Jy}\,,
       \label{eq_flux_15GHz_one_NLR}
   \end{equation}
which is about four orders of magnitude less than the radio flux density measured for the ring components, which is of the order of $0.01\,{\rm Jy}$. However, during the jet precession, the jet can interact with several thousands of NLR clouds at any time, which can be estimated as $N_{\rm cl}\sim f_{\rm V} V_{\rm j}/l_{\rm NLR}^3$, where $f_{\rm V}$ is the volume filling factor of the NLR that is estimated to be $f_{\rm V}\sim 0.01$. The volume associated with the jet at the length-scale of the NLR ($h_{\rm NLR}\sim 200\,{\rm pc}$) is $V_{\rm j}\approx \pi R_{\rm j}^2 h_{\rm NLR}$. The total number of NLR clouds interacting with the jet is thus of the order of $N_{\rm cl}\sim 10^4$. The overall radio flux density that arises due to the shocked NLR gas then is $S_{\rm NLR}\sim N_{\rm cl} S_1\sim 0.04\,{\rm Jy}$ at any place along the precession cone, which is comparable to the detected flux density along the ring.

\begin{figure}
\centering
\includegraphics[width=7.8cm]{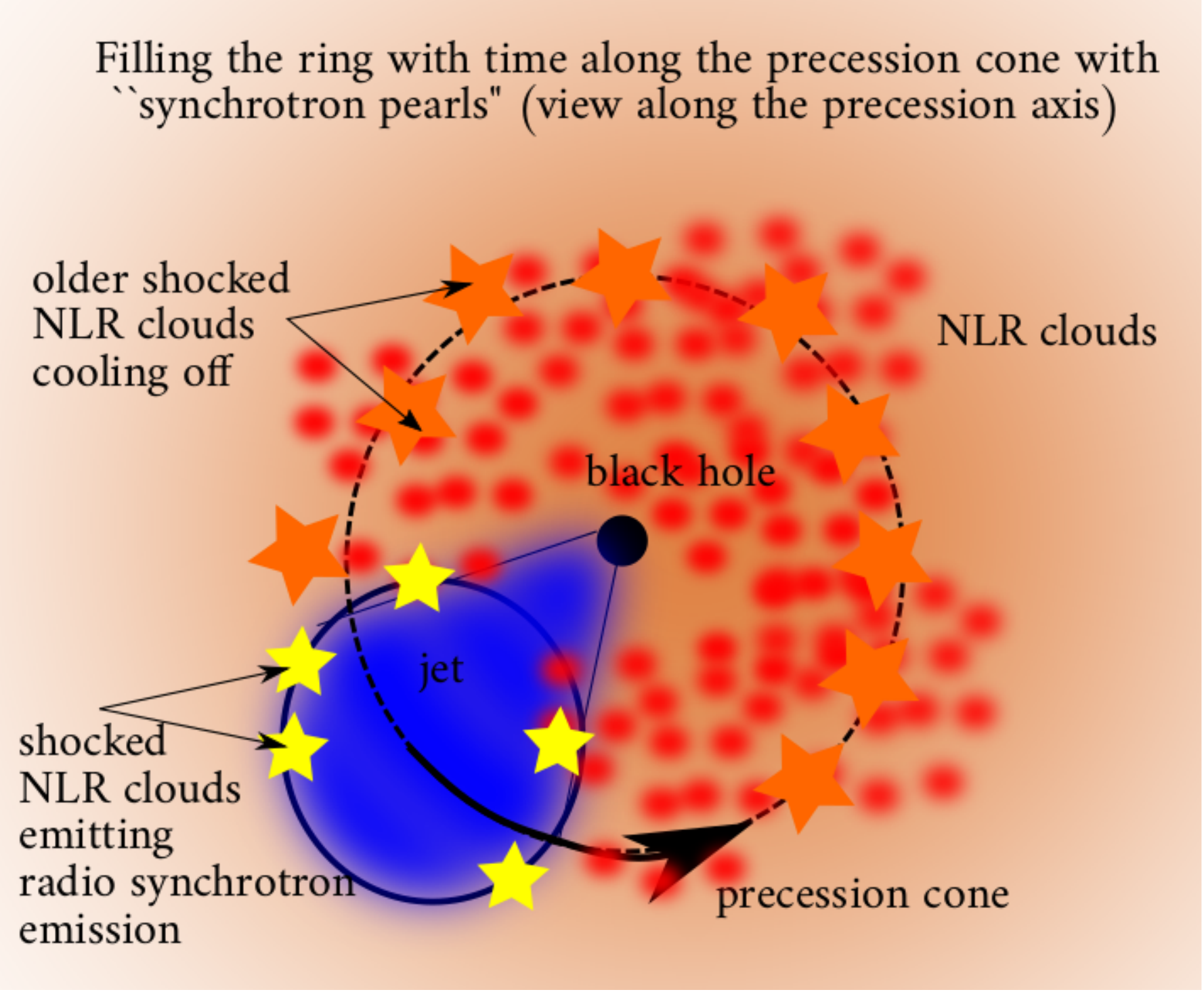}
\caption{Illustration of the model of the formation of ``synchrotron pearls'' along the precession cone. The shocked NLR clouds emit the radio synchrotron emission that can reach the levels of $0.04-0.4\,{\rm Jy}$ for a cumulative contribution of all shocked clouds at any time. Older shocked clouds also contribute to the radio emission but at lower flux density levels. Overall, we obtain a varying distribution of flux densities along the precession cone, which forms a ring-like structure close to the line of sight.}
\label{fig_model}
\end{figure}

For the more diluted NLR gas, we consider $n_{\rm NLR}\sim 10^{3}\,{\rm cm^{-3}}$, $n_{\rm sc}\sim 4n_{\rm NLR}$, and the sound speed in the shocked material $v_{\rm sc}\sim 16800\,{\rm km\,s^{-1}}$. The thermal luminosity, the radio flux density for a single shocked cloud, and a total radio flux density at any time are, respectively,
\begin{align}
      L_1 & \approx \mu m_{\rm H} n_{\rm sc} v_{\rm sc}^3 l_{\rm NLR}^2\sim 1.57 \times 10^{43}\,{\rm erg\,s^{-1}}\,,\\ 
      S_1 & =\frac{L_{\rm synch}}{4\pi D_{\rm L}^2 \nu}\notag\\
          &=4.24\times 10^{-28}\,{\rm erg\,s^{-1}\,cm^{-2}\,Hz^{-1}}=4.24 \times 10^{-5}\,{\rm Jy}\,,\\
      S_{\rm NLR}& \sim N_{\rm cl}S_1\approx 0.4\,{\rm Jy}\,.
\end{align}
Overall, the calculated flux density range $S_{\rm NLR}\sim 0.04-0.4\,{\rm Jy}$ is consistent with the flux densities inferred for radio components along the ring structure. The mean flux density of the ring components is $0.043\pm 0.037\,{\rm Jy}$. The minimum flux density of $0.003\,{\rm Jy}$ was measured in $2010.65$, while the maximum flux density of $0.167\,{\rm Jy}$ was detected in $2000.03$, which implies the timescale of $\sim 10$ years on which the ring flux density evolves. This is comparable to the cooling timescale estimated using Eq.~\eqref{eq_cooling_time}. Since the cooling time of the synchrotron emission, $t_{\rm cool}\sim 8\,{\rm yrs}$ is shorter than the precession timescale $P_{\rm prec}\sim 20\,{\rm yrs}$, the older ring components would disappear towards the completion of the precession cycle, making the ring structure variable during one precession period. 
We show a collage of all the physical phenomena we derived for PKS 1502+106 in Fig.~\ref{sketch}. We illustrate the overall mechanism of filling the precession ring with ``synchrotron pearls'' in Fig.~\ref{fig_model}. 

Hence, the observed flux density distribution along the ring, see Fig.~\ref{fig_jet_ring_flux}, can be explained by two effects:
\begin{itemize}
    \item[(i)] the brighter components are the most recent that have interacted with the jet,
    \item[(ii)] the brighter components are due to the interaction between the jet and less dense NLR clouds, while the dimmer components are due to the interaction with the denser NLR clouds.
\end{itemize}
It is quite likely that the observed flux density distribution in Fig.~\ref{fig_jet_ring_flux} arises due to both effects (i) and (ii). Geometrically, it seems that the brightest components lie in the current path of the jet or are the jet components themselves, which lie in the ring region. 
\subsubsection{Would a precession-only model be sufficient to explain the observations?}
Precessing jets seem to be a frequent phenomenon in AGN (e.g., \citealt{caproni} for BL Lac; \citealt{abraham} for 3C 273). The most convincing case is the jet of OJ 287 which precesses on a timescale of roughly 23 yr \citep{britzen_OJ287}. In this subsection we discuss, why PKS 1502+106 does not fall in this category of typical precessing jets.
In the case of OJ 287 and on the basis of VLBA observations, the precession (and nutation) can be traced as a wandering of the whole jet in the sky \citep{britzen_OJ287}. In the case of PKS 1502+106, we find similar bulk precession as for OJ287 (see Fig.~\ref{fig_precession_model_j1-j4}), but in comparison with OJ287, the jet seems more perturbed by the surroundings, which makes its kinematics more complex.
At the time, when no 15 GHz observations are available, we trace the jet component motion at 8 GHz and find that the jet again deviates from a straight ridge line. This deviation in part coincides with the former ring structure, but it also seems that this deviation propagates outward (away from the main jet ridge line). This is untypical for a pure precession as observed in other AGN.

PKS 1502+106 is a quasar, while OJ 287 is a BL Lac Object. Quasars are known for broad line emission. As shown in this paper, we find evidence for an outflowing BLR. The radio variability and periodicity of the variability in OJ 287 can be explained by precession and nutation only \citep{britzen_OJ287}. From the light curve of PKS 1502+106 shown in Fig.~\ref{radio_gamma_big} it seems that the repetitive pattern of flares is superimposed by an outburst which originates in the core region (see Fig.~\ref{radio_gamma}). The core seems to produce excess radiation from 2014 onwards. We conclude that we find evidence for an excited phase (with additional radiation from the core). Thus, while a geometric (or deterministic) precession model seems sufficient to explain the flux density evolution and the jet motion in OJ 287, the precession alone is insufficient to explain the flux density evolution in the case of PKS 1502+106. A geometric component (precession) in addition with a radiative component (excess emission from the core) is required to explain the flux density evolution in the radio.
The combination of jet precession, an additional core outburst, and the outflowing BLR together most likely produces the observational results. These are also the components to produce the observed $\gamma$-ray emission via the EC-process.

\subsubsection{An interaction of the helical magnetic field with the NLR as an alternative explanation}
Astrophysical jets are thought to be magnetized fluid. Following current models of jet formation, the jet magnetic field has a helical structure, with the pitch angle of the helix changing across the
jet. The helix is dominated by the poloidal field component along the spine (BZ), while for the outer disk the jet is toroidally dominated (a super-Alfv\'enic BP jet) \citep[e.g.][]{zakamska,broderick}. It looks like that this different orientation of the field structure becomes visible in the polarisation observations shown in this paper. Indeed, such orientation of EVPA vectors -- transverse at the jet edges and longitudinal at the jet centre -- is predicted for the aforementioned magnetic field configurations \citep{lyutikov,clausen-brown}. 

Another explanation for the observational results we present in this paper could thus be, that the VLBA observations show the interaction and interaction site of the helical magnetic field of a jet with the ambient NLR. The ring and the arcs we detect could be signatures of the helix of the magnetic field. Following this idea further, the arcs are due to loops of the helical magnetic field. The interaction would then unmask the helical magnetic field of the jet.
In this scenario, the observed features we see, which are not part of the jet or the arcs, belong to the hydrodynamic part of the jet, the flow of plasma. A changing viewing angle could be the reason for the interaction site to appear as ring or as arc. A more detailed investigation of this possibility is beyond the scope of the current paper and is planned for a future paper.

\subsubsection{Polarisation information as smoking gun of an explosive event}
At the time of the neutrino event, the polarised ring emission is not seen any more in the VLBA images (see Fig.~\ref{pola}). The strong polarised ring emission is not visible (or not as strong) for about two years. Most recent data on the MOJAVE webpage (August 30, 2020) however show, that there again is strong polarised emission at the position where we found the ring. This supports our finding that the ring structure is related to a precession phenomenon which persists. 
The core polarisation increased at the time of the neutrino event and remained higher for about two years. In the most recent data, the core polarisation decreased again. The core polarisation change can be associated with the change of the intrinsic magnetic fields. 

The core polarisation change can also be due to a varying depolarisation. The high-energy emission itself can not change the polarisation but the high-energy event most likely changed the environment. Comparable to an explosion, the distribution of the gas or magnetic field changes, and thus also the polarisation. This can be explained by an instability which develops, propagates, and subsequently compresses or otherwise changes the magnetic field topolgy. The change of the polarisation can thus be a consequence of the high-energy event. A higher density may give rise to higher depolarisation or Faraday rotation. Also, the self-absorption should be higher. The explosive event may also cause turbulence which should also have a depolarising effect. We conceive this as a consecutive issue. The high-energy radiation may not change the polarisation but the process that
produced the high energy radiation also changes the environment.

\subsection{$\gamma$-ray production in PKS 1502+106}
The polarisation information available for the pc-scale jet of PKS 1502+106 reveals a different orientation of the magnetic field of the jet
compared to those of the ring. The field orientations seem to be perpendicular to each other.
\citet{attridge} discuss a similar atypical polarisation structure of the jet in 1055+018. Both, PKS 1502+106 and 1055+018, are LSP HPQs and in both cases, the pc-scale radio emission consists of the jet emission plus additional radio emission in the direct vicinity (environment, ambient medium). 
A difference between PKS 1502+106 and 1055+018 is, that in the former case, the additional structure develops with time (and is only visible when plotting all xy-coordinates), while in the latter case, the boundary is present in one image. 
The two-component jet in the source 1055+018 consists of an inner spine with a transverse magnetic field and a distinct boundary layer with a longitudinal magnetic field according to \citet{attridge}. 
\citet{attridge} explain the different polarisation of the jet and the boundary layer in the context of the spine-sheath model. According to them, the polarisation distribution in the spine strongly supports shocked-jet models, while that in the boundary layer suggests interaction with the surrounding medium.
Further evidence for spine-sheath jet structures based on polarisation information has been presented in other sources as well, e.g., \citep{pushkarev, ros}.
While both sources reveal a two-component jet and different magnetic field orientations, 1055+018 shows a cylindrical form and PKS 1502+106 a jet plus ring. It seems possible that the different morphology can be explained by a different viewing angle.

The spine-sheath model, with a spine moving faster than the sheath, has already been invoked by several authors \citep{sol, celotti, ghisellini}. Quantitative estimates of the spine-sheath model in the context of the Blandford-Payne \citep[][hereafter BP]{payne} and BZ model have been made by e.g. \citet{xie}, while numerical simulations comparing the efficiency of BZ jets and magnetized disk winds have been published only recently \citep{qian2017, qian2018, vourellis2019}.

The existence of a magnetic field plays an essential role for jet launching \citep{blandford, payne, mckinney2009}. The typical field structure of magnetohydrodynamic jets derived by theoretical studies is that of a helical field \citep{payne, casse2002, vourellis2019}. The field geometry in these jets becomes increasingly toroidal with distance from the jet axis. However, internal and external shock may re-arrange the field geometry and also increase the field energy. Shocked gas in a compressed field may then become visible in synchrotron emission.

Also in particle simulations, the magnetic field of the jet plays a crucial role (see e.g. \citet{nishikawa2016, yao2019}). In the case of a toroidal magnetic field, both the jet and the ambient medium are seen when simulating the interaction \citep{kramer}.
In the case of the poloidal field, the jet is not seen. Only in the case of the toroidal magnetic field, both components are clearly seen (in the case of a light jet). The signature of the ambient medium seems to be perpendicular to the jet. Our results for PKS 1502+106 could provide evidence for a toroidal magnetic field.

The outflowing ionised medium and the presence of arc-like structures  can be related to the ejection of winds from the disc, and their interaction with the pre-existing jets. 
The jet-wind interactions, however, could be potentially difficult to explain because the typical opening angle of the winds observed in low-luminosity AGN is about 60 deg \citep{elvis}, too wide compared to the opening angle of a relativistic jet.
Also NLR and BLR could be related to the formation of the ring due to their interaction with the jet, but in this case the possible clumpiness of these regions could be difficult to reconcile with the observed morphology.

\citet{bruni} show the spine-sheath structure for 3C 273 in  Space-VLBI {\textit RadioAstron} observations. Especially the limb-brightening at 1.6 GHz is non-uniform. The ring we observe in PKS 1502+106 could be a manifestation of the sheath and its interaction with the surroundings (NLR clouds).

The interpretation of the ring phenomenology requires less fine tuning if framed within the spine-sheath scenario. In this scenario, a mildly-relativistic external layer of plasma surrounds a faster highly-relativistic spine.
This spine-sheath scenario has been invoked, from a theoretical point of view by \cite{ghisellini} to reconcile the low bulk Lorentz factors inferred from VLBI observation of TeV blazars, contrasting with the high values of Lorentz factors required to model their $\gamma$-ray spectral and temporal emission. In addition to this, the spine-sheath scenario provides also an external photon field, that plays a relevant 
role as a possible seed photon field for the EC scenario. In the context of FSRQs, this  external photon field can support the EC emission beyond the BLR and dusty torus scale. 

More recently, \cite{MacDonald2015} have suggested the presence of rings, 
related to the spine-sheath scenario, to explain $\gamma$-ray `orphan` flares, and have applied their model to the FSRQ PKS 1510$-–$089. In their analysis the authors also point out that, due to the interaction of the velocity shear of the jet with the ambient medium, the magnetic field in the sheath should be more aligned toward the outer edges, implying a stronger polarisation with respect to the spine. This is difficult to test with the observations presented here since we are probing larger core distances.

In addition to the implications discussed so far in terms of radiative properties, the spine-sheath scenario has a relevant consequence in terms of particle acceleration. Velocity stratification is present in relativistic jets starting from a possible jet wind interaction very close to the jet launching site, and can proceed at longer scales due to the interaction of the jets with the ambient medium. In the case of AGN, and in particular for blazars, this velocity stratification can provide an efficient mechanism to accelerate particles \citep{Rieger2019}. Due to the microphysics of this acceleration process, and in particular to the particle mean free path, the shear acceleration is more efficient for protons than for electrons \cite{Rieger2004}. Interestingly \cite{Rieger2004} suggest a phenomenological picture where non-gradual shear accelerates electrons and protons at the jet boundary, and gradual longitudinal shear acts in the inner part of the jet, with an efficiency decreasing with the radial coordinate. The same authors conclude that this scenario predicts a larger concentration of energetic particles toward the jet boundary. At least from a qualitative point of view, this could be in agreement with the observation of ring-like structures.
It is worth noting that  shear acceleration could provide a continuous  proton acceleration, as requested by the proton synchrotron model presented in \cite{rodrigues2020}, even though this would not mitigate the  fact that this model requires super-Eddington regime even during  the long quiescent states.
 
\subsection{Neutrino production in PKS 1502+106}
There seems to be no correlation between neutrino and gamma-ray activity in PKS 1502+106 (see Fig.~\ref{radio_gamma_big}), or, it seems that the neutrino has been emitted during the lowest gamma-ray state.
Regarding neutrino production, we observe that according to \cite{Liu2017}, protons of $10^{18}$ eV can be produced by shear acceleration in blazar jets, within a confinement scale $L<<10^{15}$ cm, for magnetic field of B $>1$ G.
These energies are sufficient to produce neutrinos of $\approx 10^{15}$ eV via proton-proton interaction. In any case, as discussed in \cite{murase}, the neutrino production in the blazar zone, for the {\it pp} process, can not be efficient, requiring a jet too heavy to remain relativistic. On the contrary, in the case of flaring activities, episodic encounters of the jet with dense clouds could overcome this problem.
The blazar zone for {\textit Fermi} blazars has to be located outside of the BLR, due to missing gamma-gamma absorption features \citep{costamante}.

Another possibility is that during the neutrino emission, the $\gamma$-ray emitting region is optically thick to gamma rays, hence, in this case the neutrino and gamma production 
site might be cospatial. Further analysis will be required to finally solve this point. 

The photohadronic process (e.g., \citet{dermer, atoyan, beckmann}) instead results to be more problematic because it requires a strong X-ray photon field, and for FSRQs the X-ray flux density is typically subdominant in the EC emission.
A possible solution is provided by a hybrid lepto-hadronic model, as discussed in \cite{rodrigues2020}. In this case  the  photon emission has a mainly leptonic contribution, and the hadronic components  are constrained by the X-ray data. This model seems to provide a viable solution, requiring a lower energetic budget, anyhow, the predicted  number of IceCube events, since the beginning of the real time alert system, is in mild tension with the IceCube non-detection \citep{rodrigues2020} during flares, and would require an archival search to give further constraints.

We cannot directly resolve the jet-BLR interaction or jet-blazar zone interaction at smaller scales because of the resolution and the compactness of the interaction region. However, on larger scales, we can see the disturbance of the jet sheath due to the interaction with the NLR/ionisation cone ambient medium, which is manifested by the ring formation with time. Since the wide-angle nuclear outflow with the base at the BLR scale,  as manifested by the significantly blueshifted broad CIV and MgII lines (see Subsection~\ref{civ}), likely transitions into the NLR zone at 100 pc scale, we argue that the ring and the neutrino generation are (indirectly) connected - both are the result and the manifestation of the jet-ambient medium interaction, however, at different scales.

In the case that the neutrino is generated by the interaction between the jet and the denser cloud, then this effect is enhanced by the jet precession. Jet-cloud interaction in general is less likely because of the geometry -- in particular the flattened BLR is in the plane perpendicular to the jet, also the immediate surroundings of the jet are expected to be cleared off by radiation and the ram-pressure of the jet. However, when the jet precesses, it is more likely that it hits a denser material in its surroundings. This also applies to the formation of the ring at the scales of several 100 pc. Hence, any interaction of the jet with the denser material in its surroundings is enhanced by the change in the jet direction, i.e. the precession is the most natural phenomenon to account for this.

It seems that neither the external radiation fields in FSRQs alone nor the spine-sheath structures are sufficient to explain the detected cosmic neutrinos. We here propose that an additional phenomenon -- e.g., jet precession -- might be required.

\subsection{Are there any common properties between TXS 0506+056 and PKS 1502+106?}
In the case of TXS 0506+056, the single event neutrino and the neutrino flare (from archival data) were observed in a BL Lac Object. Our VLBA data re-analysis revealed evidence for a collision of jetted material, a special viewing angle, and precession of the inner jet material. Based on these findings it seemed likely, that the neutrinos were produced within the jet but within an atypical event (a jetted collision). 

Several other models to explain the neutrino emission in TXS 0506+056 have meanwhile been discussed in the literature. \citet{ros} discuss a spine-sheath structure based on the analysis of 43 GHz data of TXS 0506+056 taken at two epochs (2 and 8 months) after the neutrino event. 
However, a spine-sheath VLBI morphology alone seems not sufficient to predict the neutrino emission since several VLBI jets show similar wide-opening-angle jet morphologies \citep{ros}.
Based on a reanalysis of multi-epoch  multi-frequency archive VLBI data \citet{li} claim a viewing angle of 20 deg and co-spatial $\gamma$-ray and neutrino emission. Similar to \citet{britzen_txs}, they find evidence for precession but with a different precessing period of 5 to 6 years. They expect that the precession could originate from instabilities.  According to \citet{li}, the neutrino  event could be related to energetic  particle  injection  into  the  jet. Based on the precession scenario, the timescale for neutrino and gravitational wave emission has been estimated by \citet{debruijn}. 

In the case of PKS 1502+106, and as discussed in the current paper, 
the neutrino emission might have been generated via proton-proton interaction. Although the neutrino production in PKS 1502+106 seems to originate in more general radiative processes, precession seems to play an essential role here as well.

Both neutrino emitting AGN, TXS 0506+056 as well as PKS 1502+106 appear to be non-standard members of the blazar class. 

\section{Summary}
In this paper we have presented evidence for unusual jet dynamics in the flat spectrum radio quasar PKS 1502+106. To our knowledge, this is the first time, such phenomena have been reported. In addition, we provide evidence for an outflowing BLR. We introduce a model for the complex dynamics that is based on a precessing jet interacting with the ambient medium. We link the observed jet dynamics with the radio and $\gamma$-ray flares. The set up supports the EC emission mechanism in PKS 1502+106. Our results are detailed in the following.
\begin{enumerate}
\item We demonstrate that a detailed study of the evolution of the $x$- and $y$-coordinates of the jet curvature with time is a powerful tool to investigate AGN physics.
\item We find clear evidence for a smoothly curved jet structure and additional radio components that are aligned in ring-like and arc-like configurations which develop with time and are not present all the time.
\item The arcs with the largest diameter become apparent shortly before the neutrino event.
\item Evidence for the ring structure is not only seen in total intensity maps, but also in the polarisation maps.
\item The overall geometry as we observe, suggests a connection or a correlation between these components -- the jet and the ring, and the jet and the arcs.
\item The jet axis is clearly offset from the ring axis.
\item We find two periods in the radio light curve -- 3.35 and 11.22 years -- corresponding  to  the  rest-frame timescales of 1.18 and 3.95 years, respectively. The longer timescale could be associated with jet precession. The shorter period suggests a second-order jet motion, potentially due to the jet nodding motion or nutation (comparable to OJ 287, \citealt{britzen_OJ287}). 
\item The ring structure can be explained by the existence of a precessing jet.
\item Based on the C IV-line emission, we find evidence for an outflowing BLR.
\item The appearance of ring-like emission can be explained within a collisional cloud model, as an interaction of the precessing jet with the NLR as the jet sweeps across the ambient NLR material. This interaction with NLR clouds takes place at deprojected scales of about 330 pc.
\item As an alternative explanation we suggest that the ring is the interaction site of the helical magnetic field of a jet with the ambient NLR. The ring and  the  arcs could  be  signatures  of  the helix of the magnetic field. In this picture, the arcs are due to loops of the helical magnetic field. The interaction would then unmask the helical magnetic field of the jet. The jet features which are not part of the jet or the arcs, could belong to the hydrodynamic part of the jet, the flow of plasma.
\item The interpretation of the ring phenomenology requires less fine tuning if framed within the spine-sheath scenario. The latter can provide a seed photon field. The high-energy $\gamma$-ray emission is thus most likely produced via the EC process.
\item A cross-correlation analysis reveals that the $\gamma$-ray and radio emission zones are spatially close - within sub-parsec distances. The variability timescale of the $\gamma$-ray emission of a few days constrains the location of the $\gamma$-ray emission zone to be in the BLR region and within the jet launching region.
\item We find evidence for deterministic processes from nonlinear analysis of the $\gamma$-ray lightcurve. This result is confirmed by \citet{bhatta}.
\item Shear acceleration can produce a larger concentration of energetic particles toward the jet boundary. This could be in agreement with the observed ring-like structure. Shear acceleration is more efficient for protons than for electrons.
\item The ring-interaction precedes the neutrino event, and the extent of the arc structure appears to be largest shortly before the neutrino event. 
\item The neutrino is most likely produced by proton-proton interaction in the blazar zone (beyond the BLR), enabled by episodic encounters of the jet with dense clouds. An encounter with some molecular cloud in the NLR seems possible. To prove this, ALMA observations might help to confirm the structure of the molecular material. In addition, modeling will be required for an improved understanding of the processes. This is planned for a future paper.
\item It seems likely that the unusual jet dynamics and neutrino generation are causally connected. The details will have to be explored in further studies.
\item PKS 1502+106 is a prototypical source which bears potential to provide crucial insight into the high energy production in AGN in general. Higher resolution observations (e.g., Event Horizon Telescope) and simultaneous multi-wavelength studies will be of importance.
\end{enumerate}

The results presented in this paper, and in particular the complexity of the jet structure and its interaction with the ambient medium -- might have implications for the modelling of jet sources, the modelling of high-energy emission, and the understanding of the production mechanisms of neutrinos. Yet, not all processes that play a role in this unusual jet source are understood. However, this object allows us to study and understand general radiative processes in AGN in much more detail than before.

\section*{Acknowledgements}
The authors thank the anonymous referee for many very valuable comments. N. MacDonald provided numerous helpful suggestions and ideas that improved the paper. We also thank X. Rodrigues, A. Fedynitch, A. Franckowiak, P. Pani, S. Garappa, L. Sidoni, R. Lico, A. Witzel, and P. Biermann for inspiring discussions. MZ acknowledges the support from the National Science Centre, Poland, grant No. 2017/26/A/ST9/00756 (Maestro 9). This research has made use of data from the University of Michigan Radio Astronomy Observatory which has been supported by the University of Michigan and by a series of grants from the National Science Foundation, most recently AST-0607523. FJ thanks Helge Rottmann for providing time at the high-performance computer cluster of the MPIfR. This work has made use of public \textit{Fermi}-LAT data obtained from the High Energy Astrophysics Science Archive Research Center (HEASARC), provided by NASA Goddard Space Flight Center. We acknowledge the NASA Fermi grants NNX09AU16G, NNX10AP16G, and NNX11AO13G. This research has made use of data from the OVRO 40-meter monitoring program \citep{Richards}, which is supported in part by NASA grants NNX08AW31G, NNX11A043G, and NNX14AQ89G and NSF grants AST-0808050 and AST-1109911. This research has made use of data from the MOJAVE database, which is maintained by the MOJAVE team \citep{Lister2018}. This work was done with datasets BZ022, BK068, BR077, BL111, BL123, BH133, BL137, BL149, BH125, BL149, BL178, BL229, BC201AC, S5272D, BG219E,
BG219H, S7104A3, SB072B6 collected with the VLBA instrument of the NRAO and available at \url{https://archive.nrao.edu/archive}. The National Radio Astronomy Observatory is a facility of the National Science Foundation operated under cooperative agreement by Associated Universities, Inc. L. \v C. P. is supported by the Ministry of Education, Science and Technological Development of R. Serbia  (the contract 451-03-68/2020-14/200002). RP acknowledges the institutional support of the Silesian University in Opava and the grant SGS/12/2019. RP was also supported by the Student Grant Foundation of the Silesian University in Opava, Grant No. $\mathrm{SGF/4/2020}$, which has been carried out within the EU OPSRE project  entitled ``Improving the quality of the internal grant scheme of the Silesian University in Opava'', reg. number: $\mathrm{CZ.02.2.69/0.0/0.0/19\_073/0016951}$.

\section{Data availability}
The data underlying this article were accessed from the MOJAVE webpage ({\url{https://www.physics.purdue.edu/MOJAVE/sourcepages/1502+106.shtml}}), the OVRO 40m telescope webpage ({\url{https://www.astro.caltech.edu/ovroblazars/}}), the \textit{Fermi} data server ({\url{https://fermi.gsfc.nasa.gov/cgi-bin/ssc/LAT/LATDataQuery.cgi}}), the SDSS webpage ({\url{https://www.sdss.org/}}), and the Astrogeo Center ({\url{http://astrogeo.org/}}). The data generated in this research will be shared on reasonable request to the corresponding author.





\appendix
\section{Correcting the VLBA flux density scaling issue for three MOJAVE data sets}
\begin{figure}
\centering
\includegraphics[width=9.6cm]{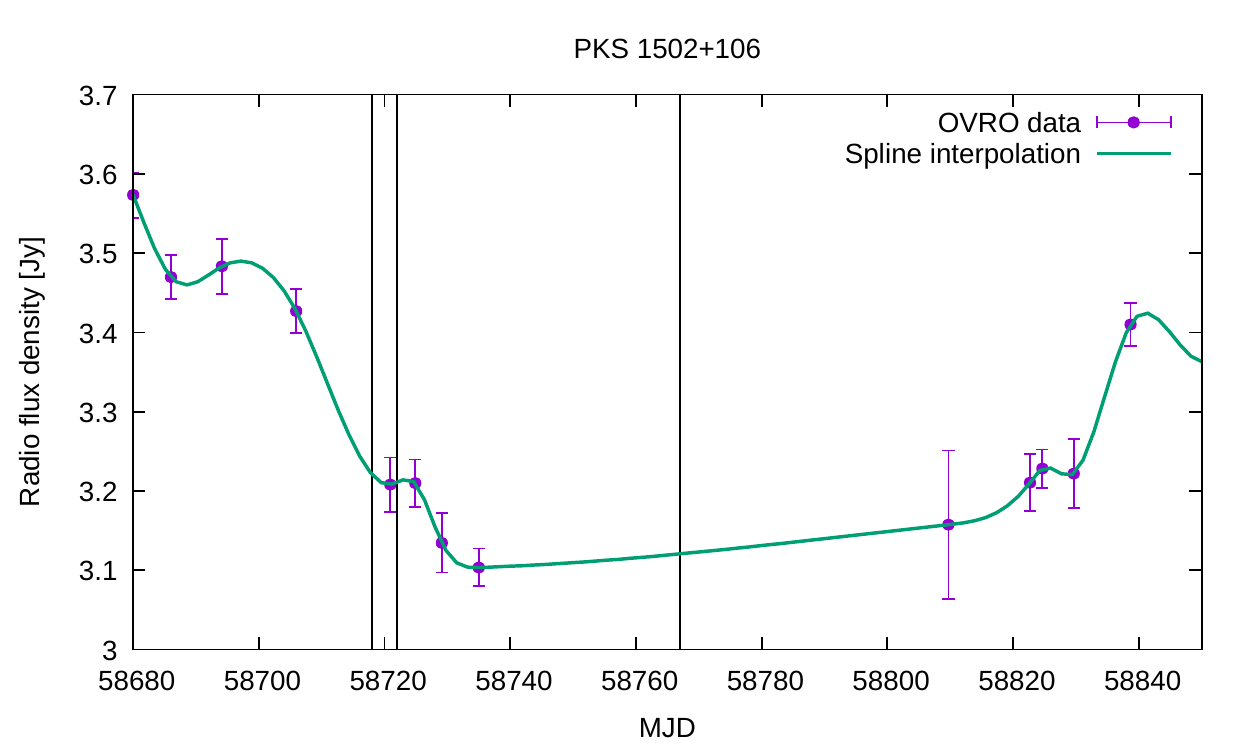}
\caption{To correct for the VLBA flux density scaling issue, we scaled the original modelfit flux-densities (and the uncertainties) from core and jet components to match the OVRO data for the last three epochs studied in this paper (2019/08/23, 2019/08/27, 2019/10/11). The black vertical lines indicate the dates of the MOJAVE observations.}
\label{recal}
\end{figure}
\section{Evolution of the ring with time}
\label{xy_plots}

\begin{figure*}
\centering
\includegraphics[width=0.32\textwidth]{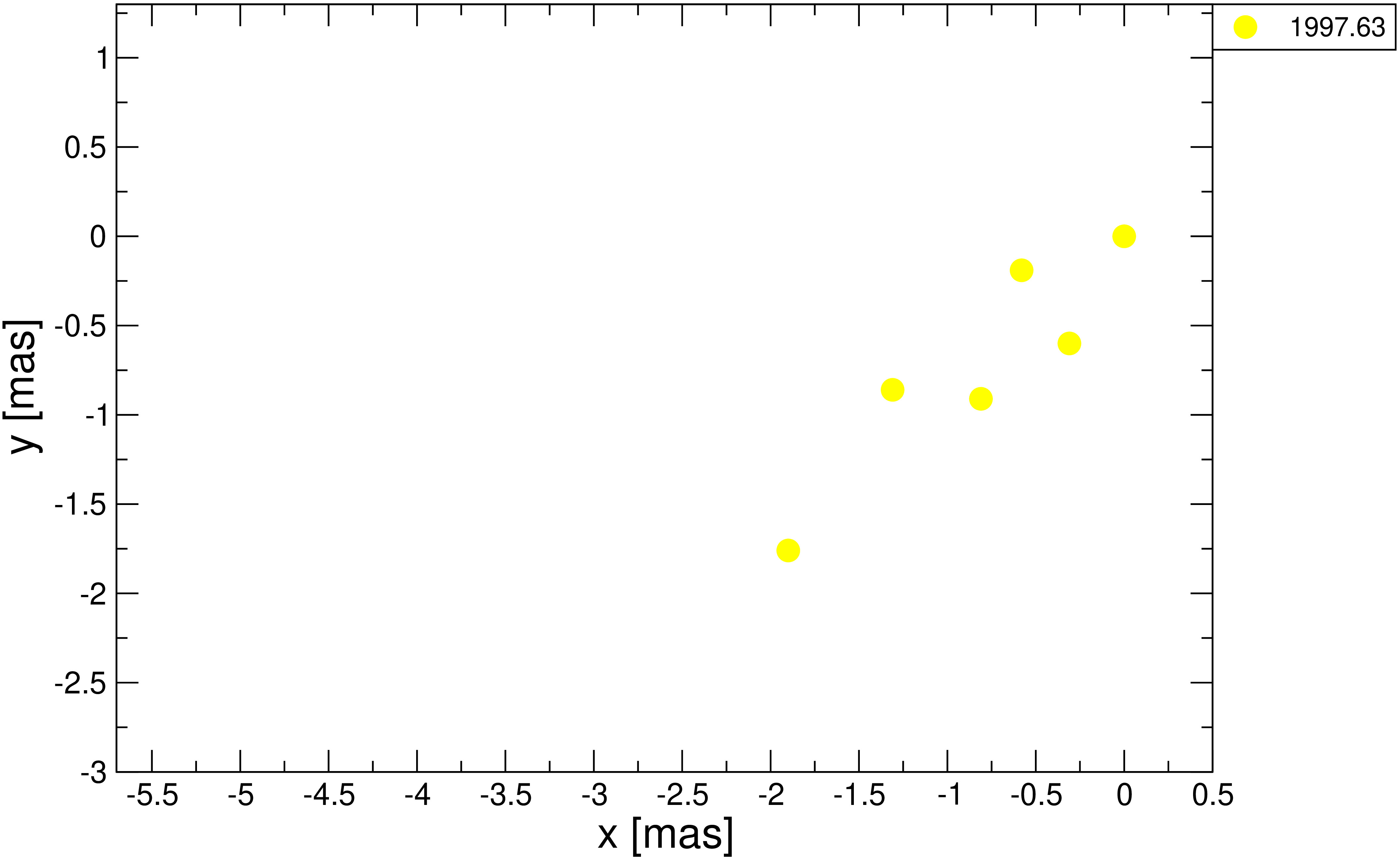}
\includegraphics[width=0.32\textwidth]{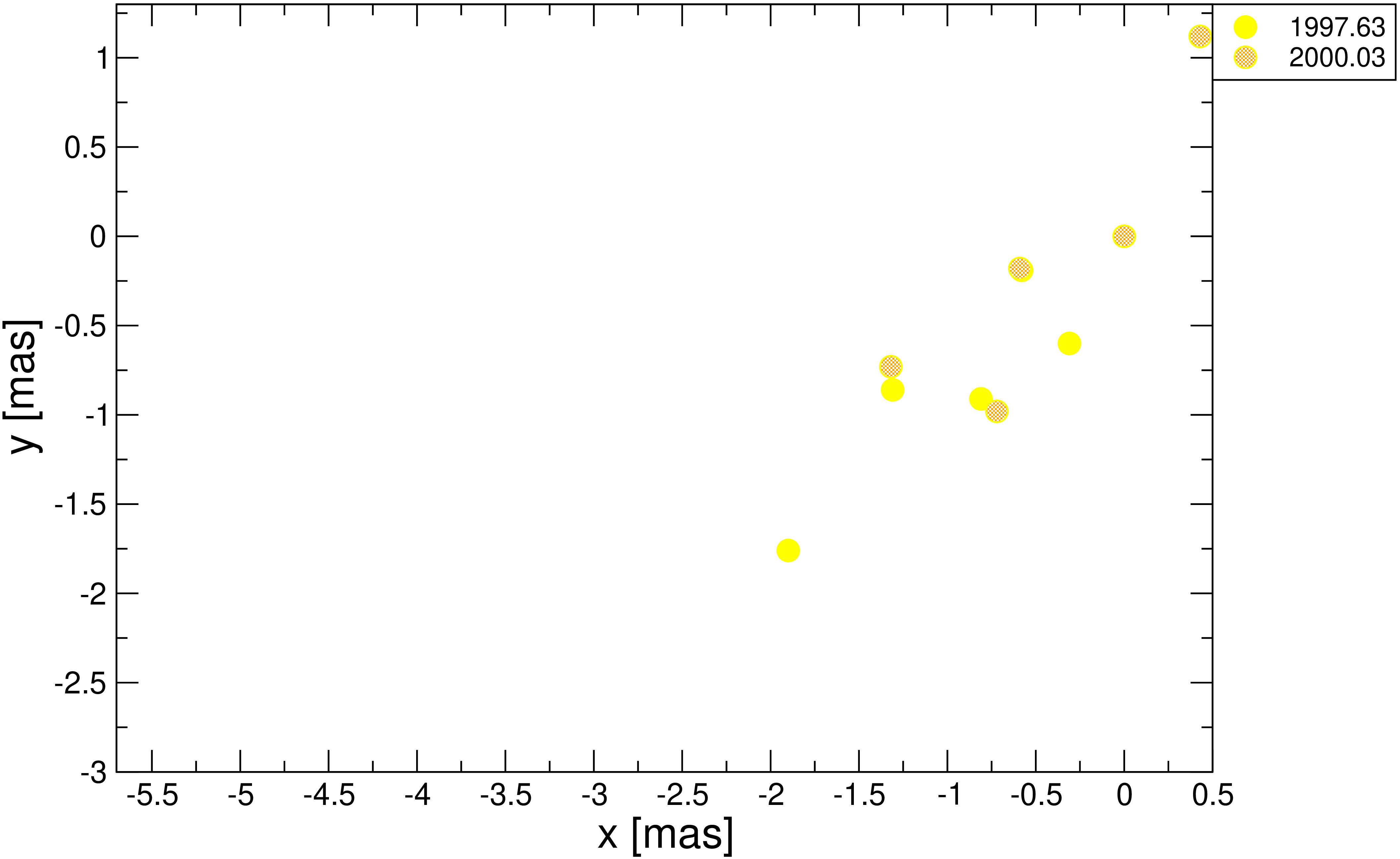}
\includegraphics[width=0.32\textwidth]{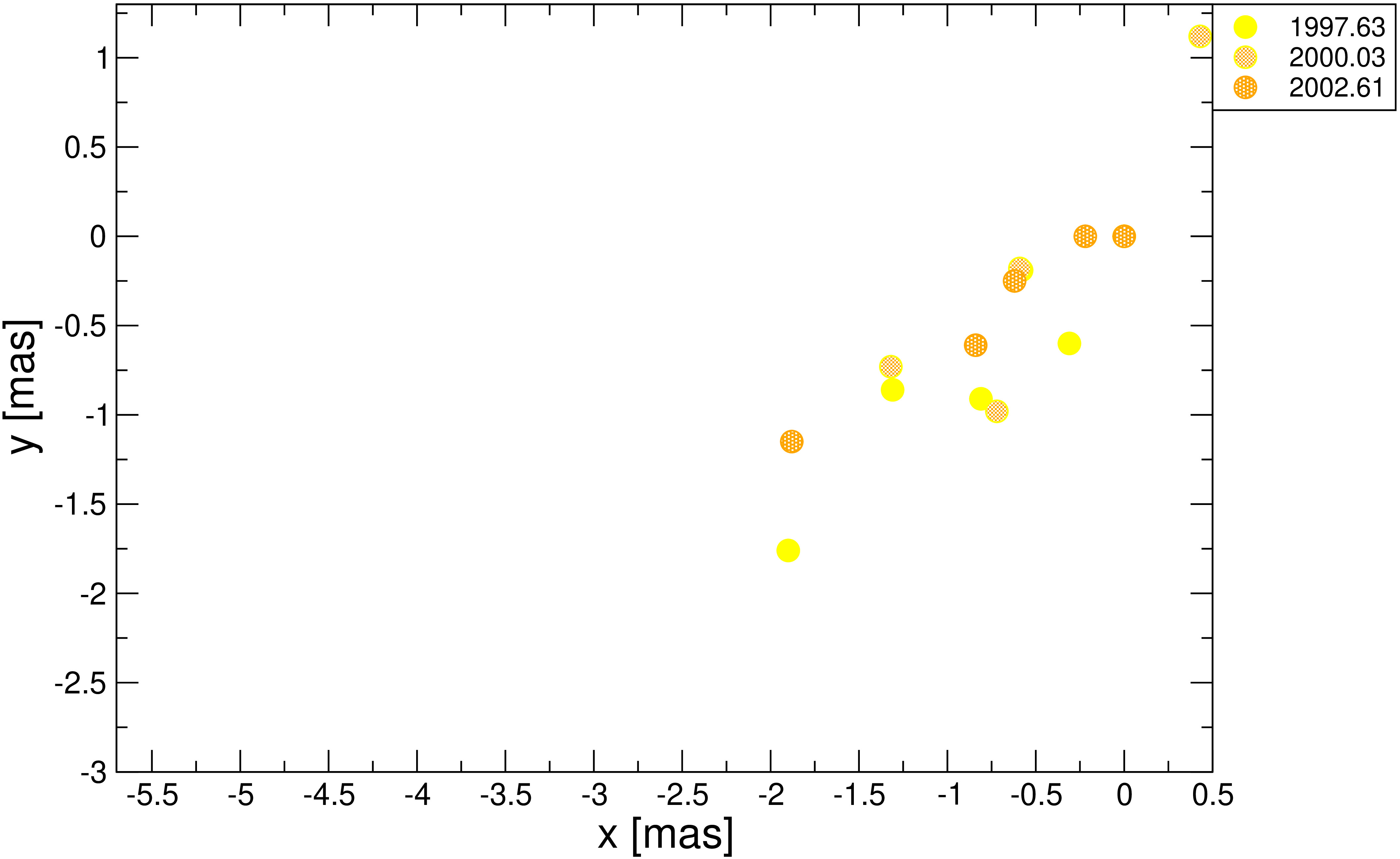}
\includegraphics[width=0.32\textwidth]{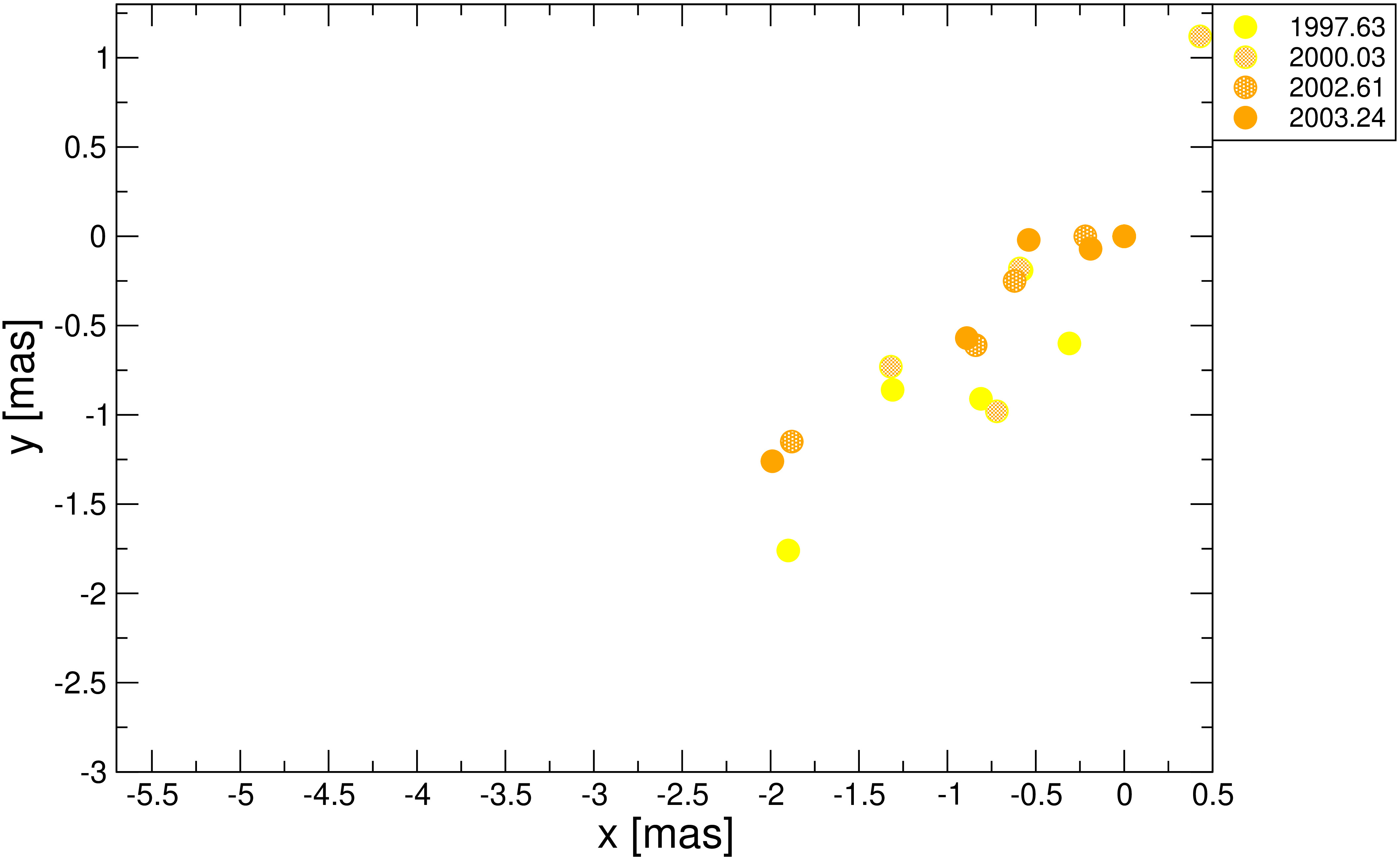}
\includegraphics[width=0.32\textwidth]{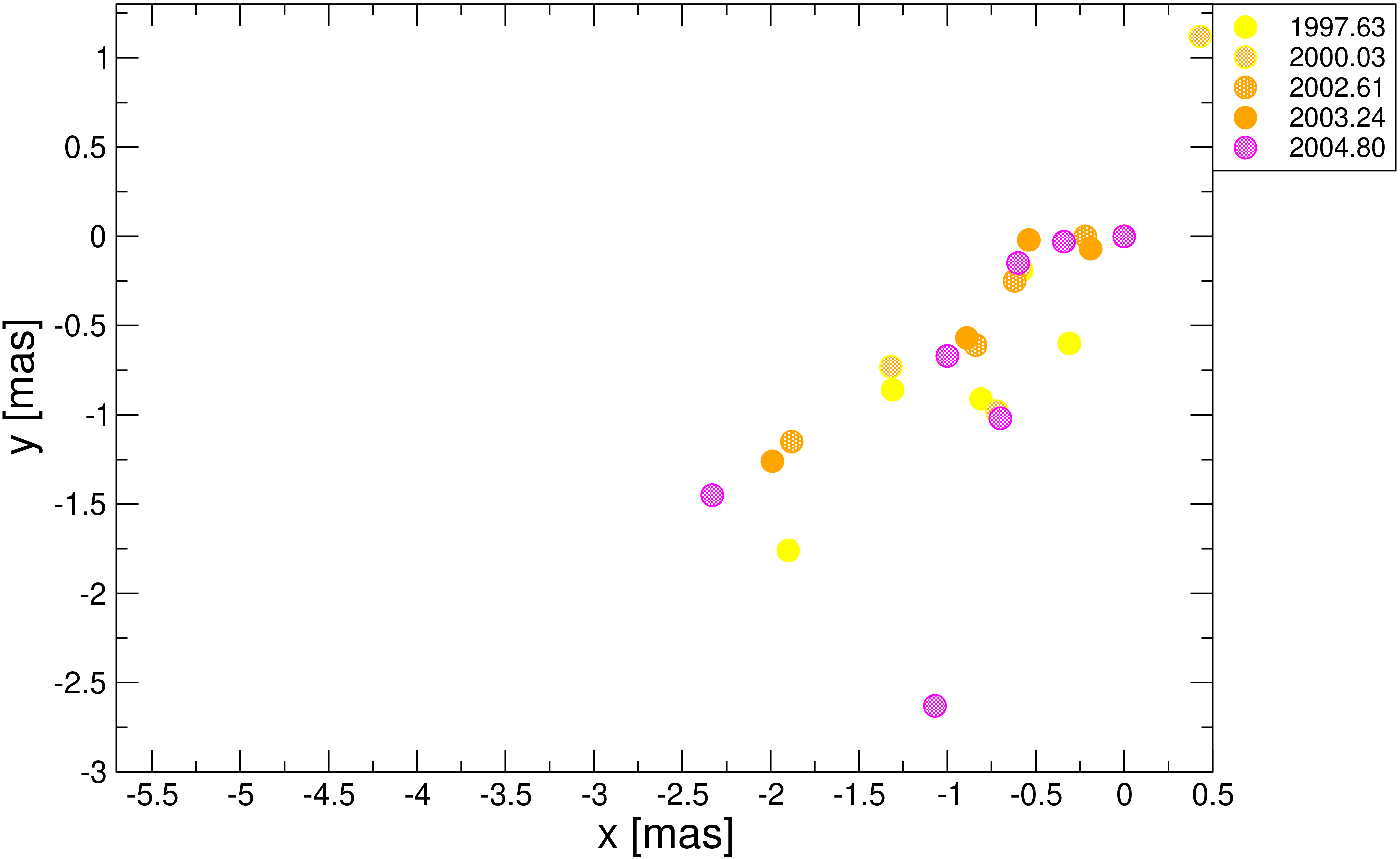}
\includegraphics[width=0.32\textwidth]{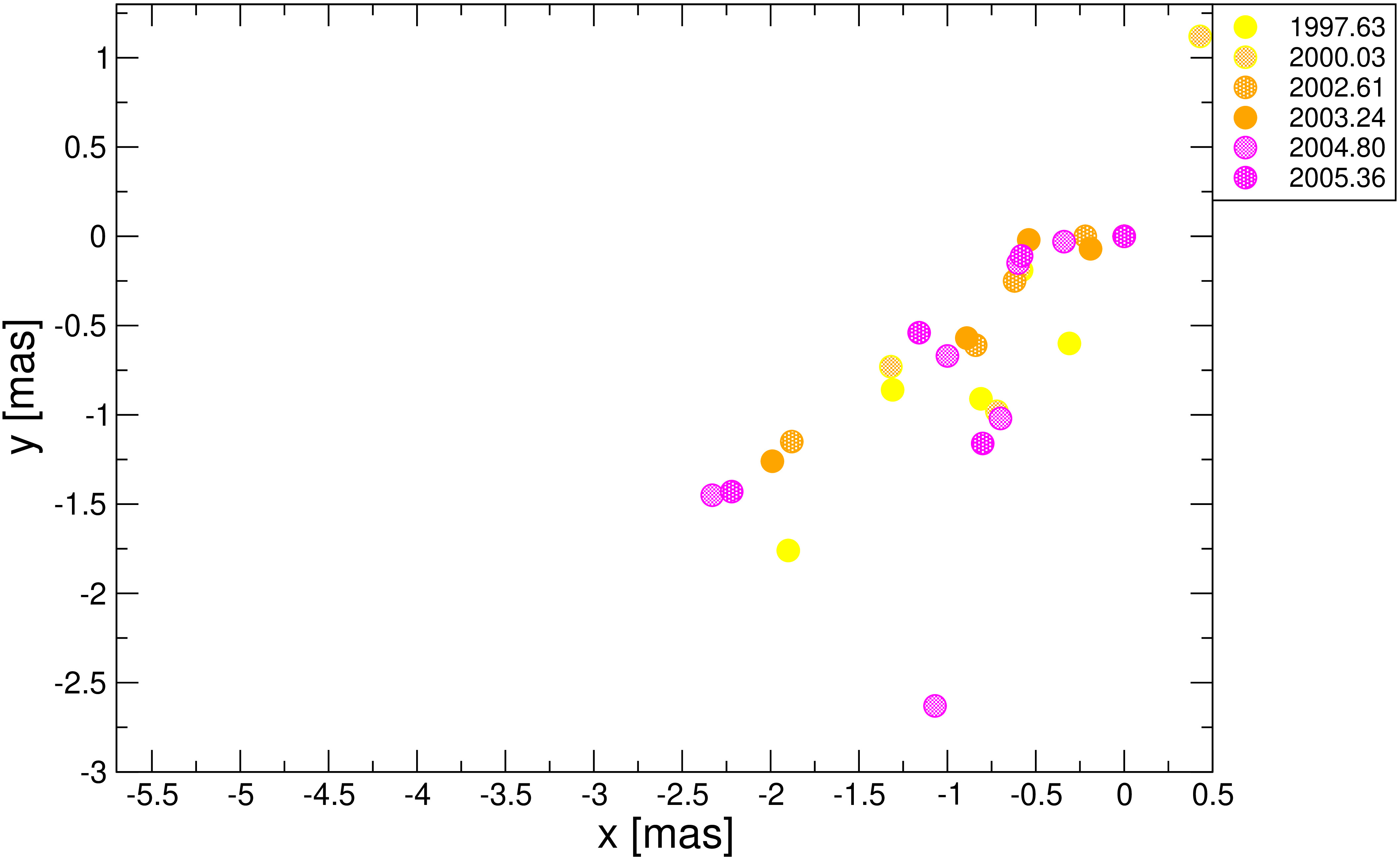}
\includegraphics[width=0.32\textwidth]{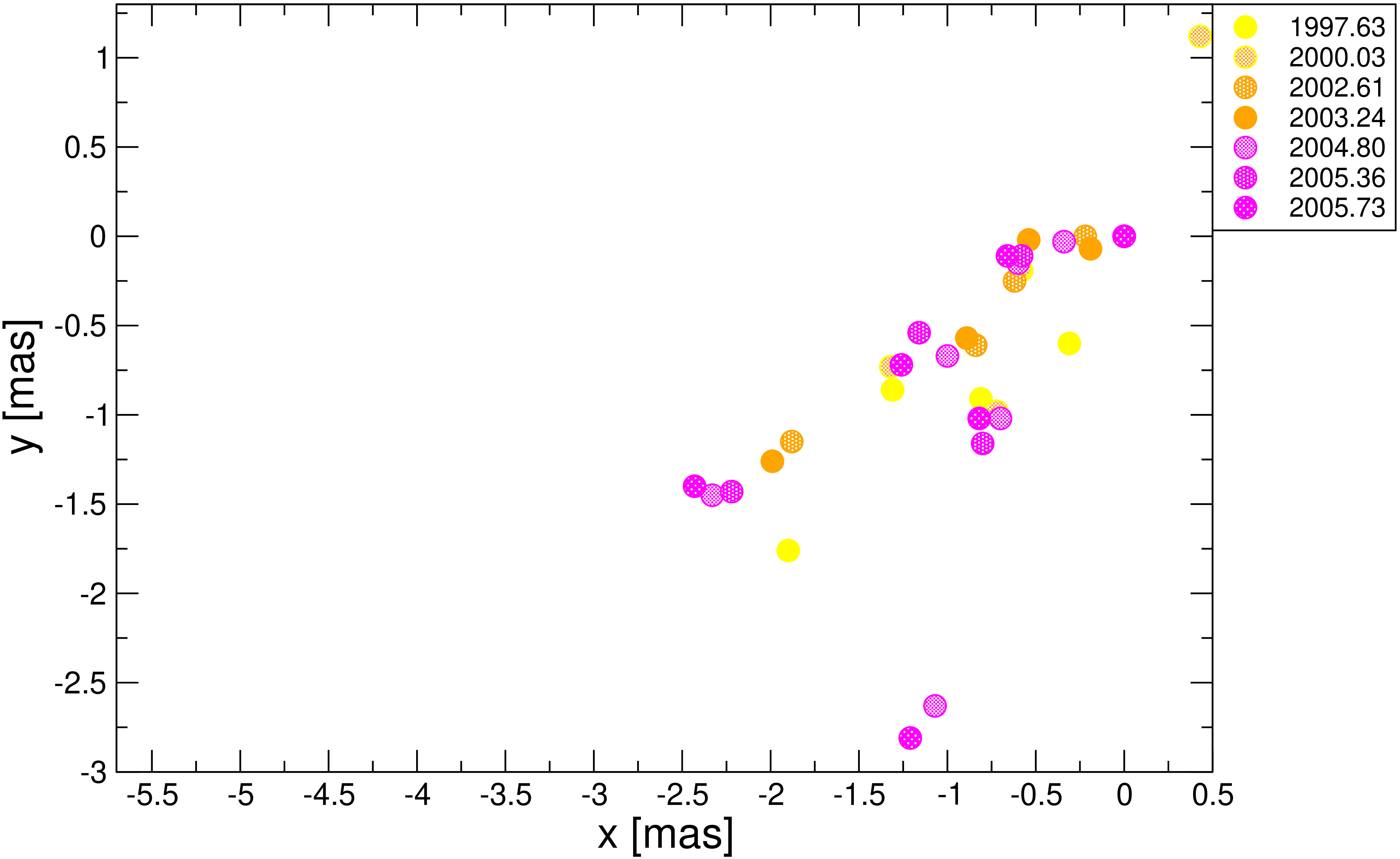}
\includegraphics[width=0.32\textwidth]{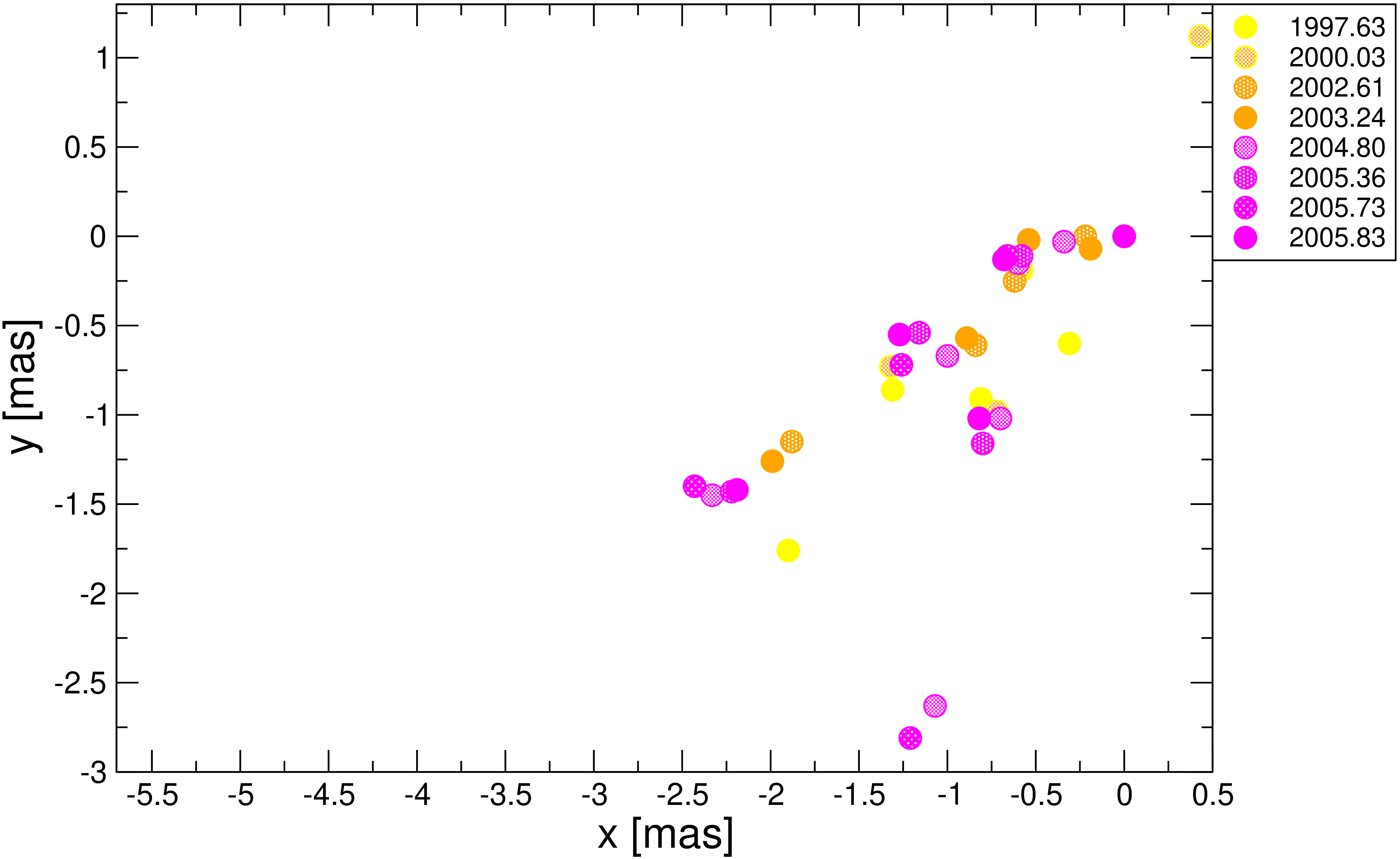}
\includegraphics[width=0.32\textwidth]{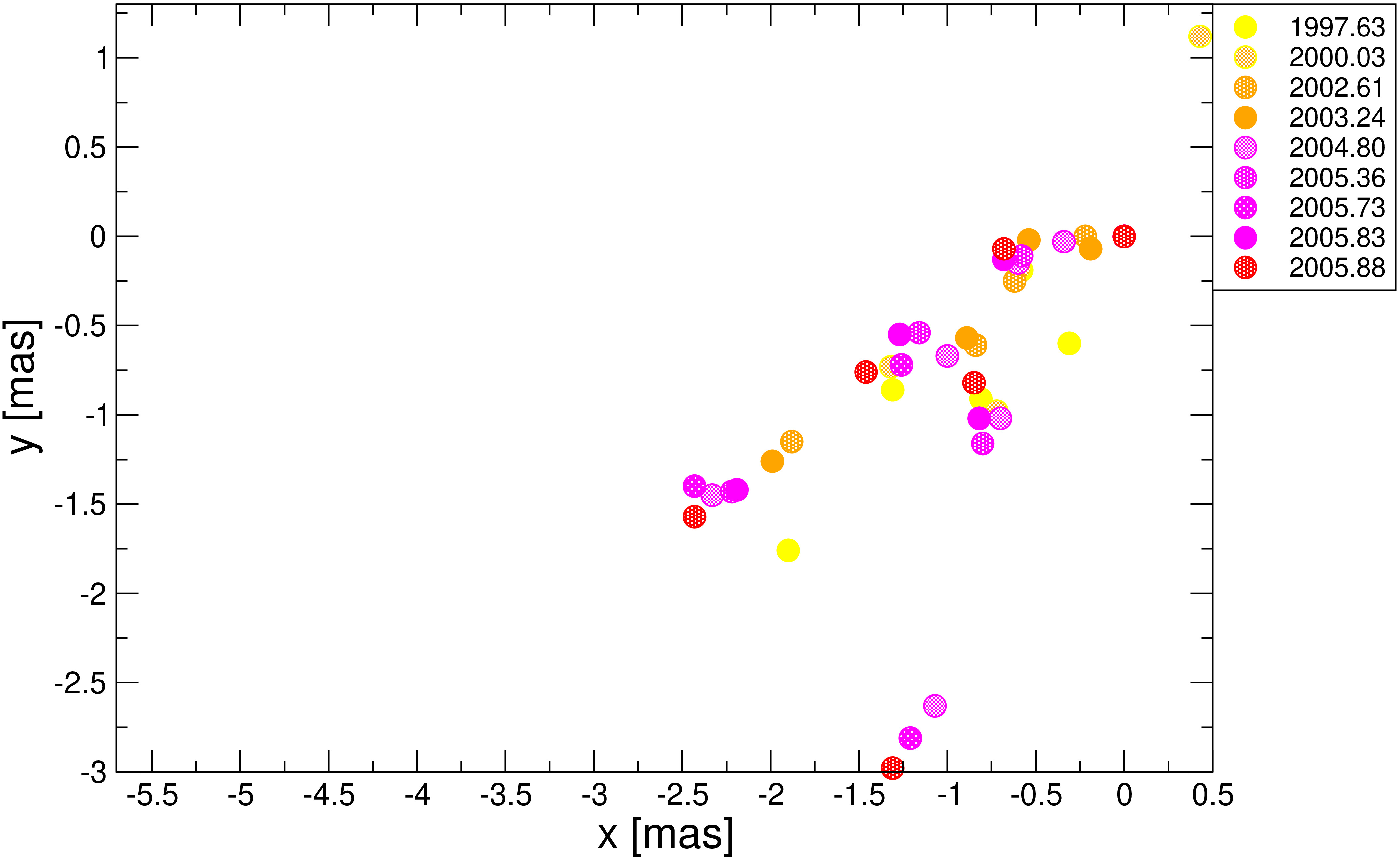}
\includegraphics[width=0.32\textwidth]{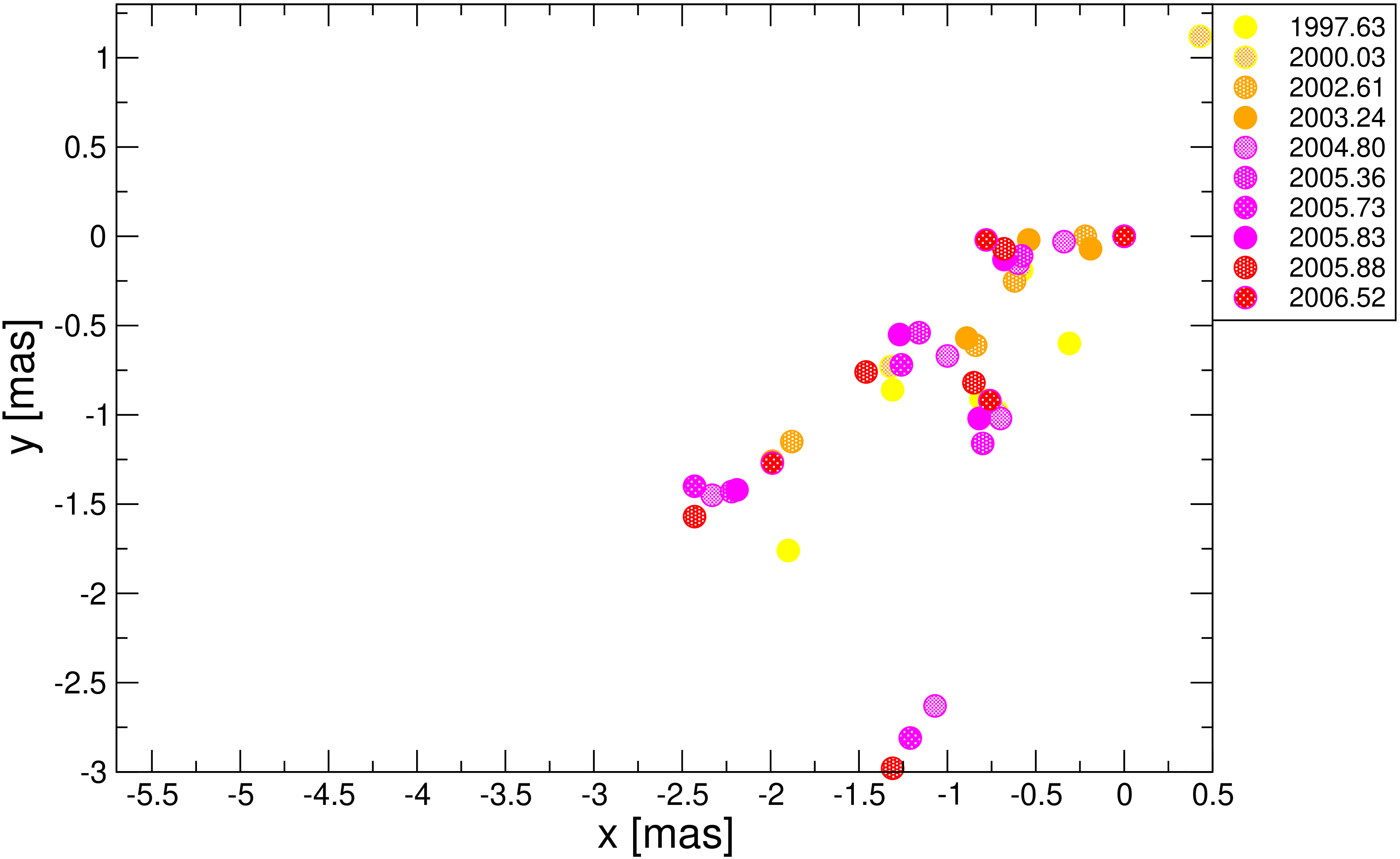}
\includegraphics[width=0.32\textwidth]{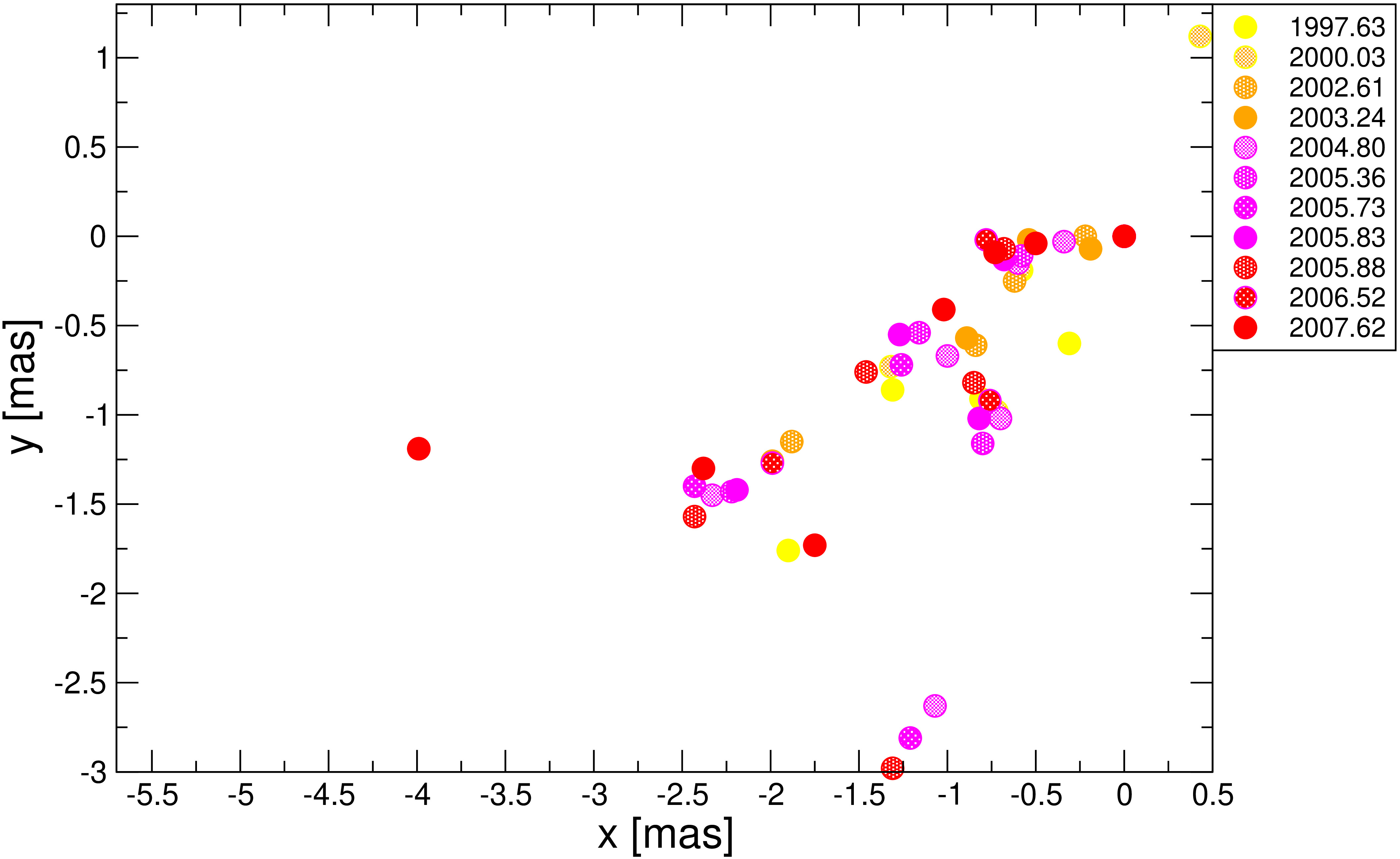}
\includegraphics[width=0.32\textwidth]{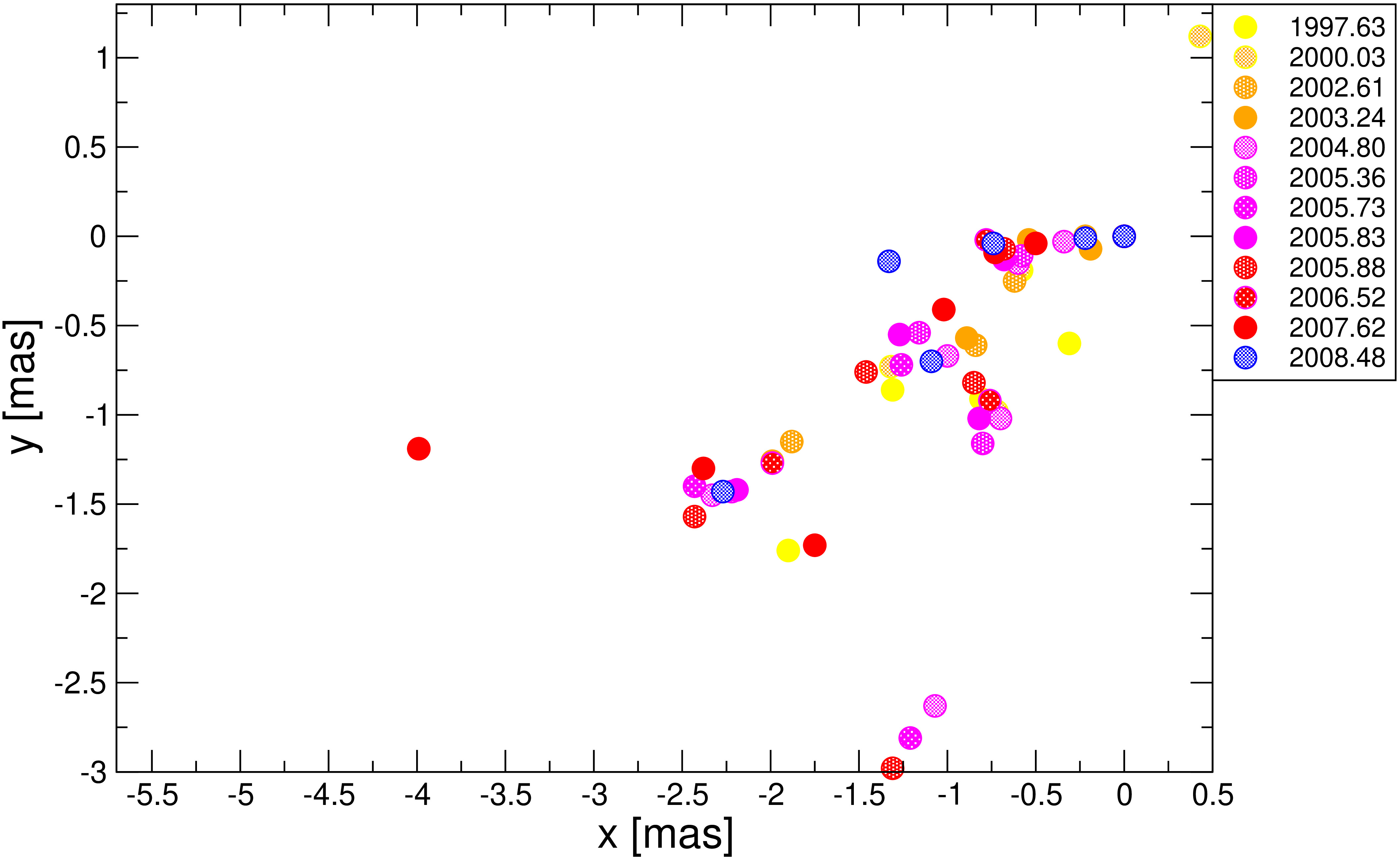}
\includegraphics[width=0.32\textwidth]{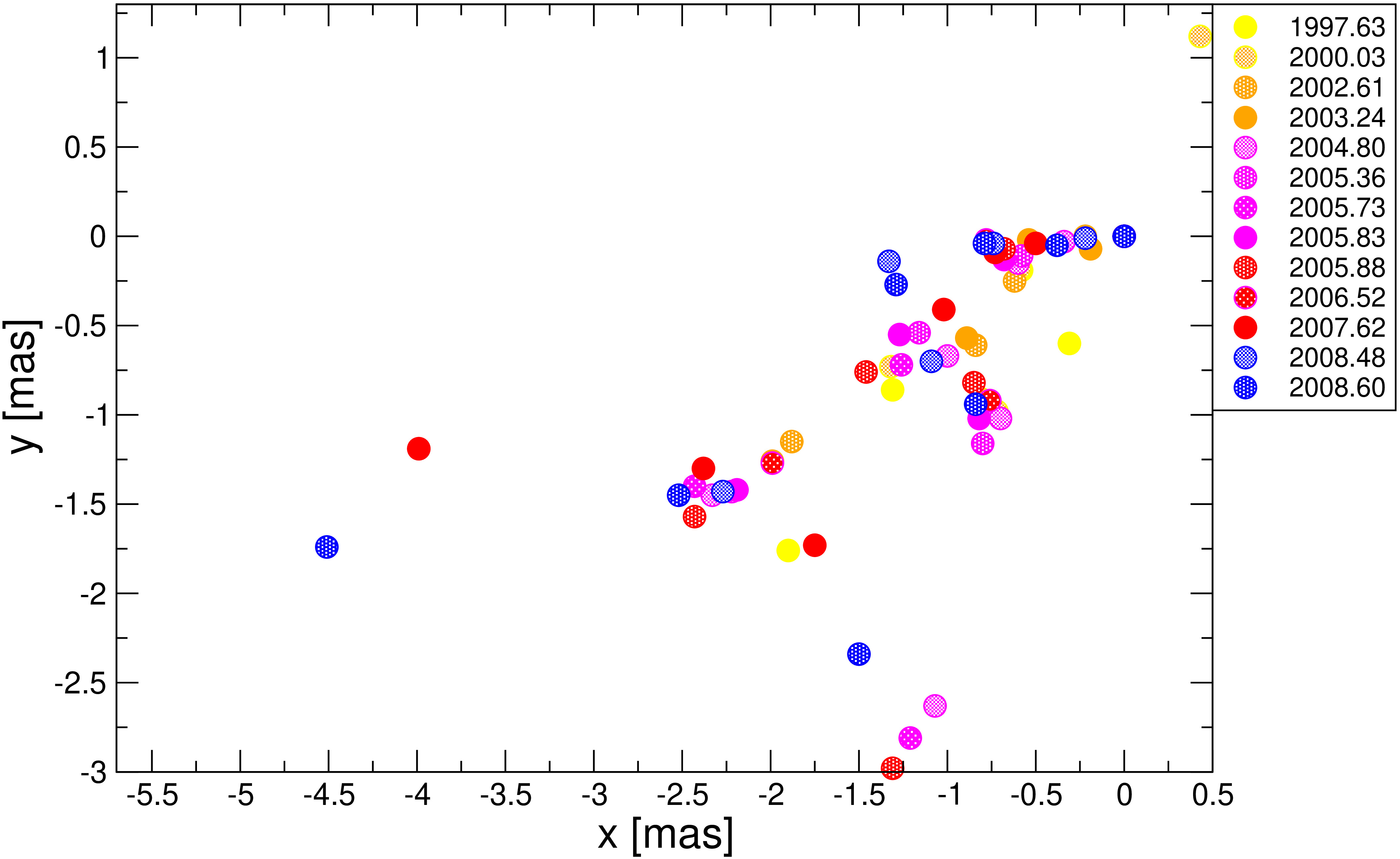}
\includegraphics[width=0.32\textwidth]{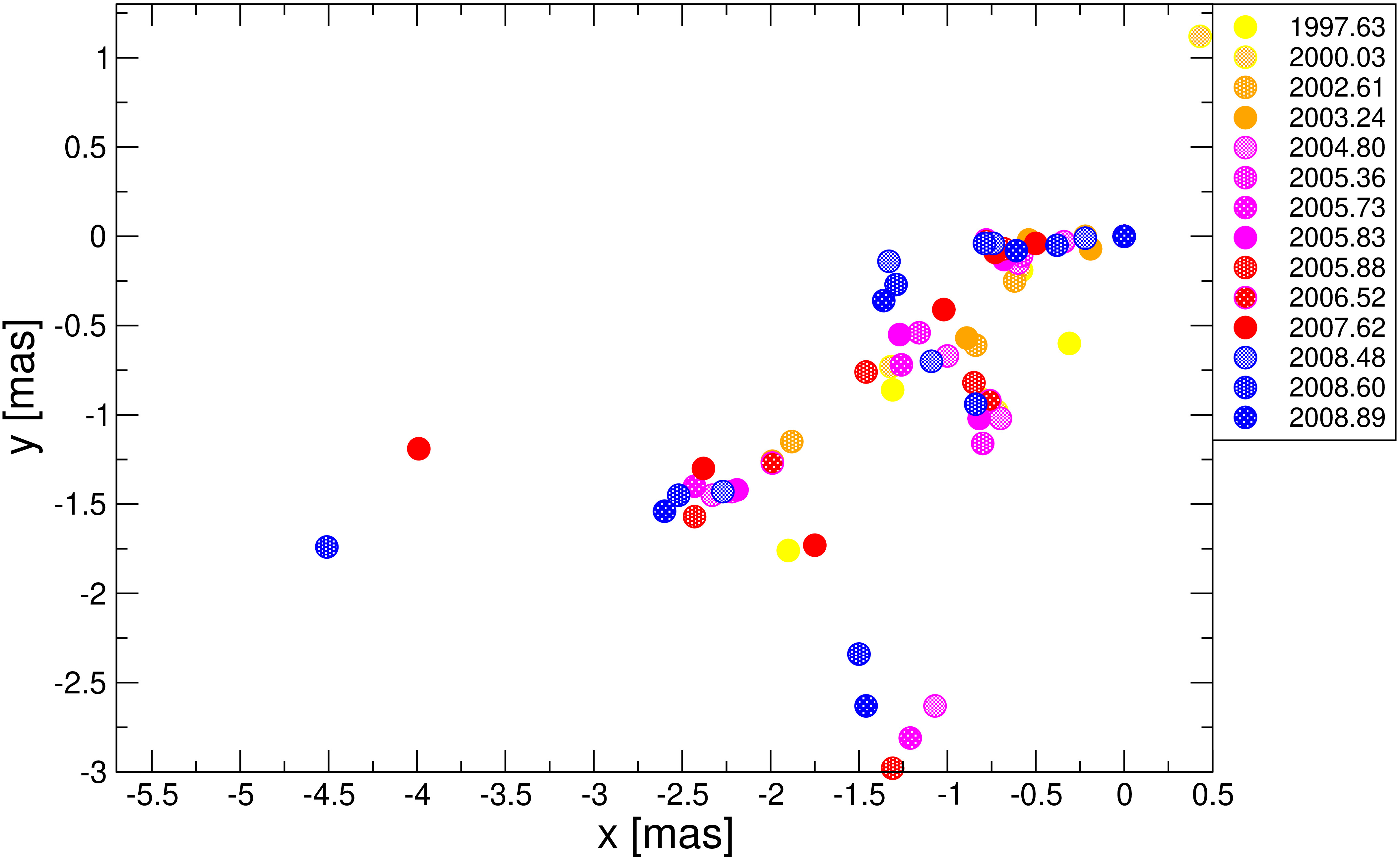}
\includegraphics[width=0.32\textwidth]{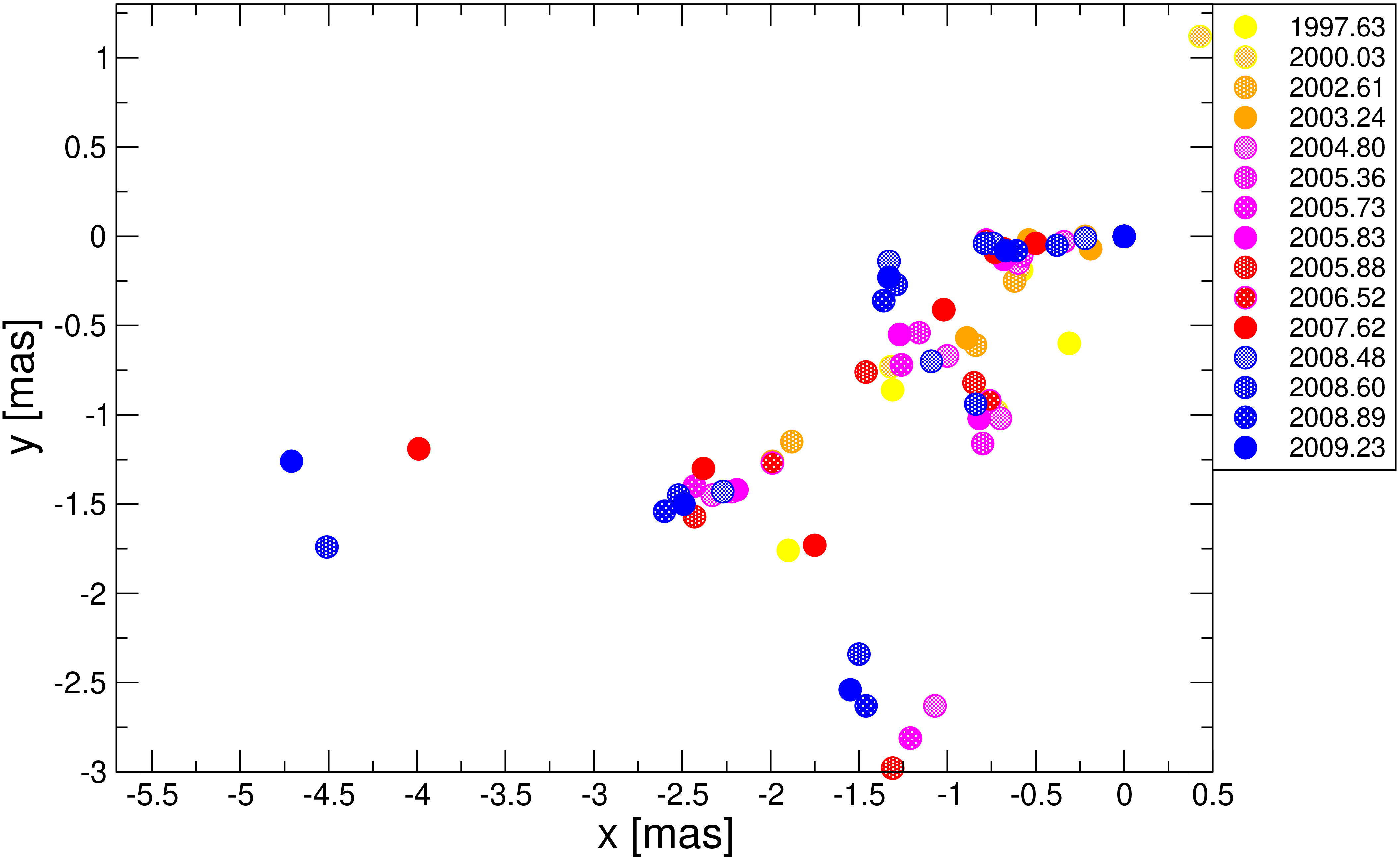}
\includegraphics[width=0.32\textwidth]{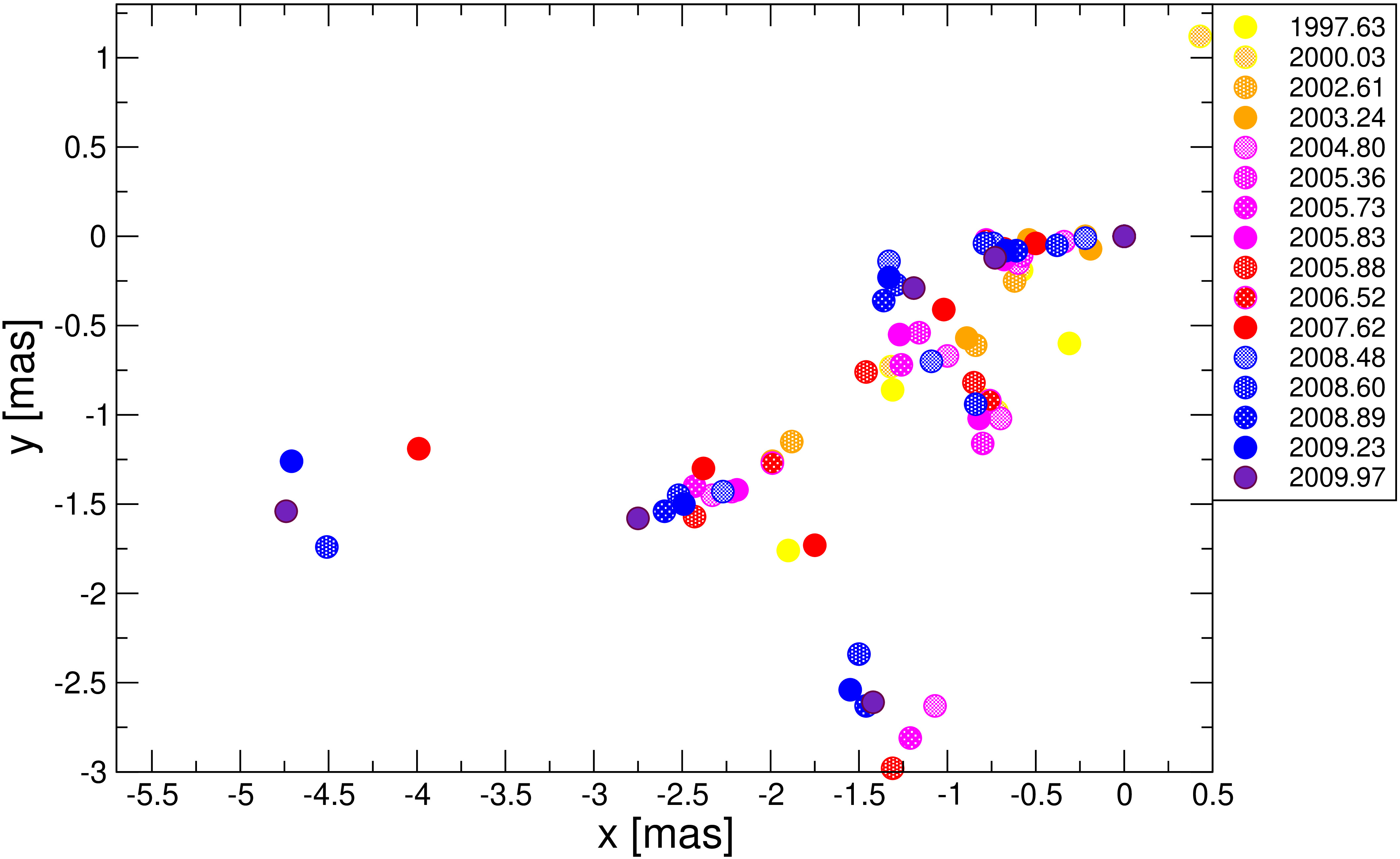}
\includegraphics[width=0.32\textwidth]{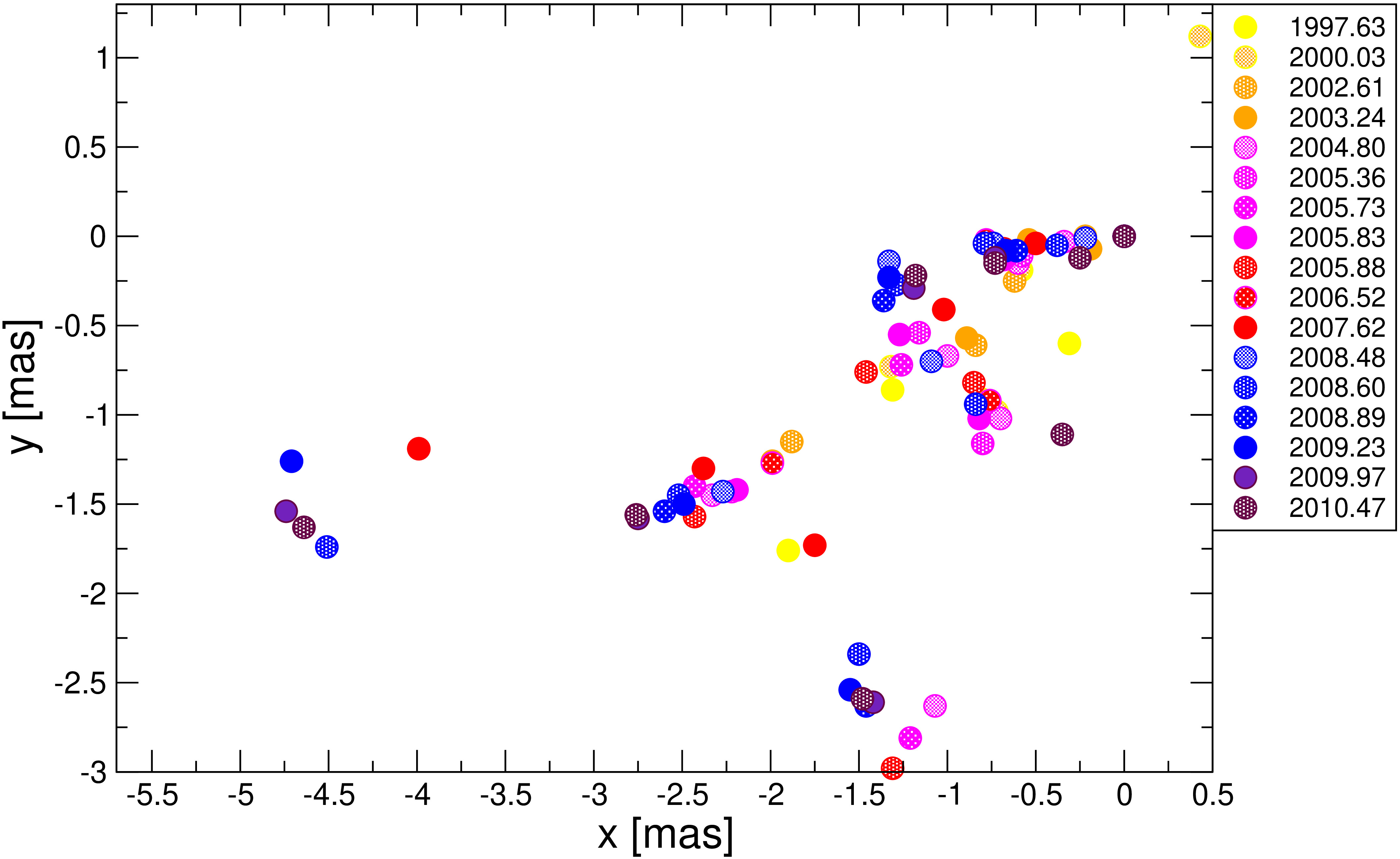}
\includegraphics[width=0.32\textwidth]{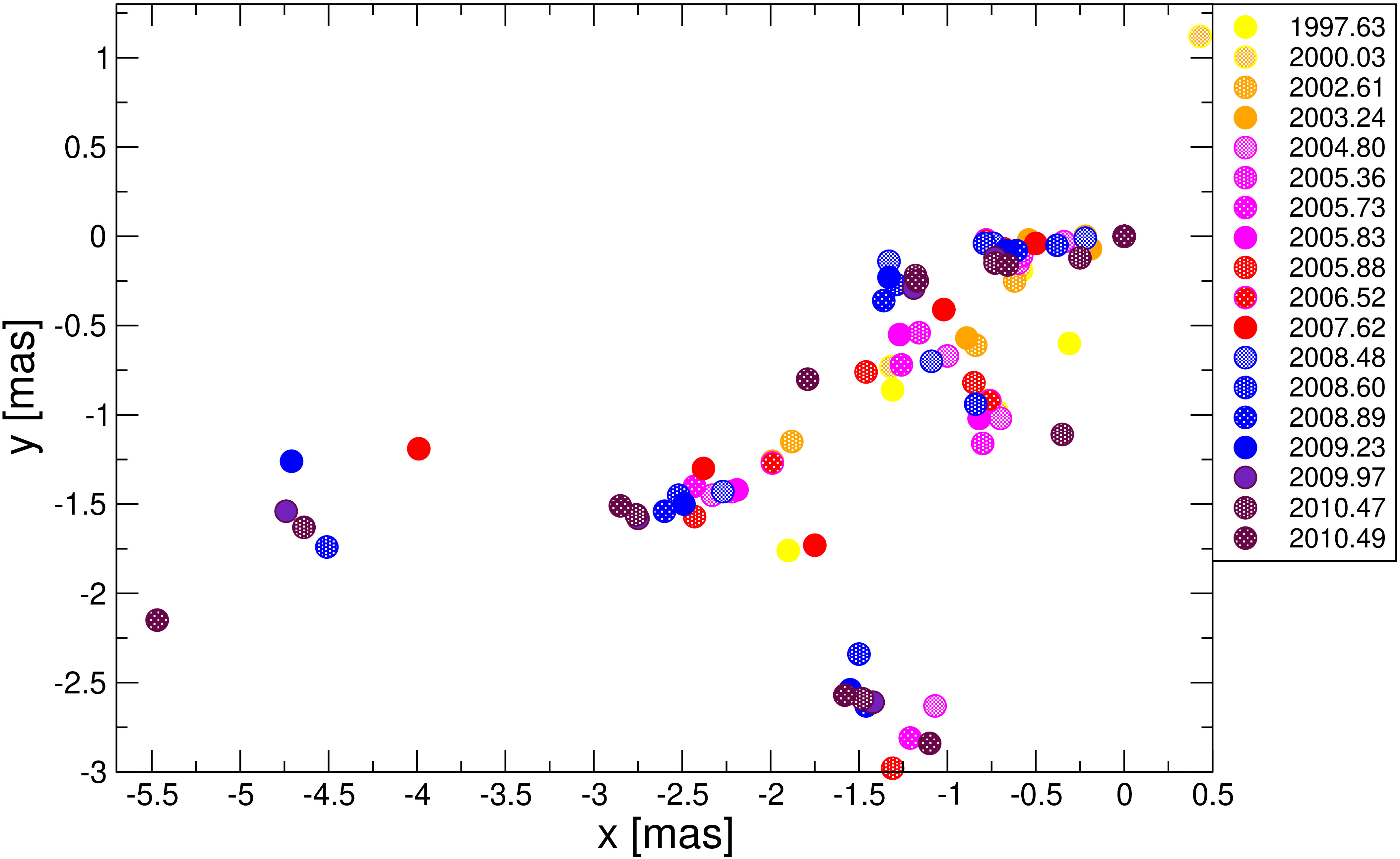}
\caption{The modelfit components in xy-coordinates as derived for the 15 GHz VLBA observations within the {\it difmap}-modelfit programme (1997.63 -- 2010.49).}
\label{xy_plot1}
\end{figure*}
\begin{figure*}
\centering
\includegraphics[width=0.32\textwidth]{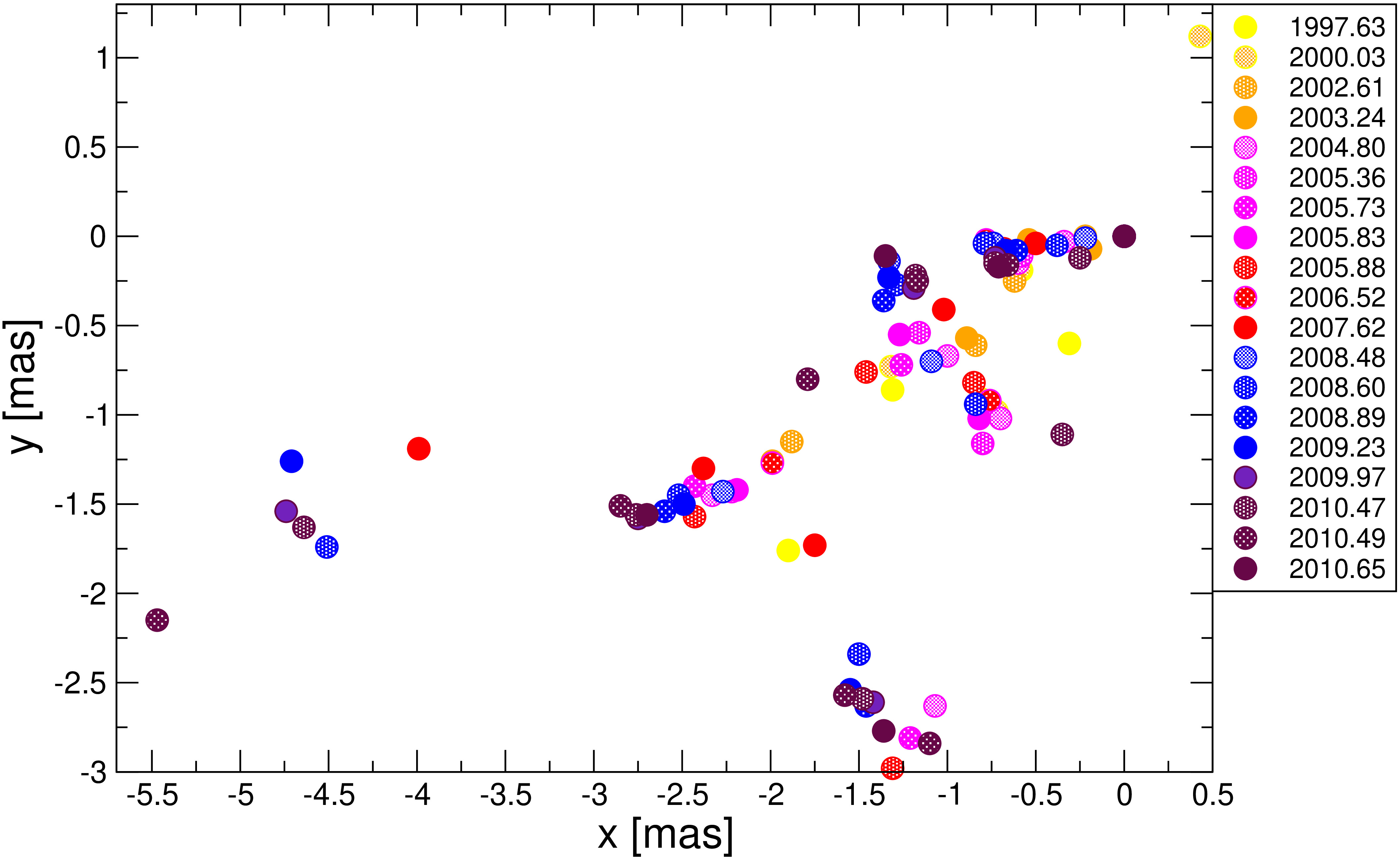}
\includegraphics[width=0.32\textwidth]{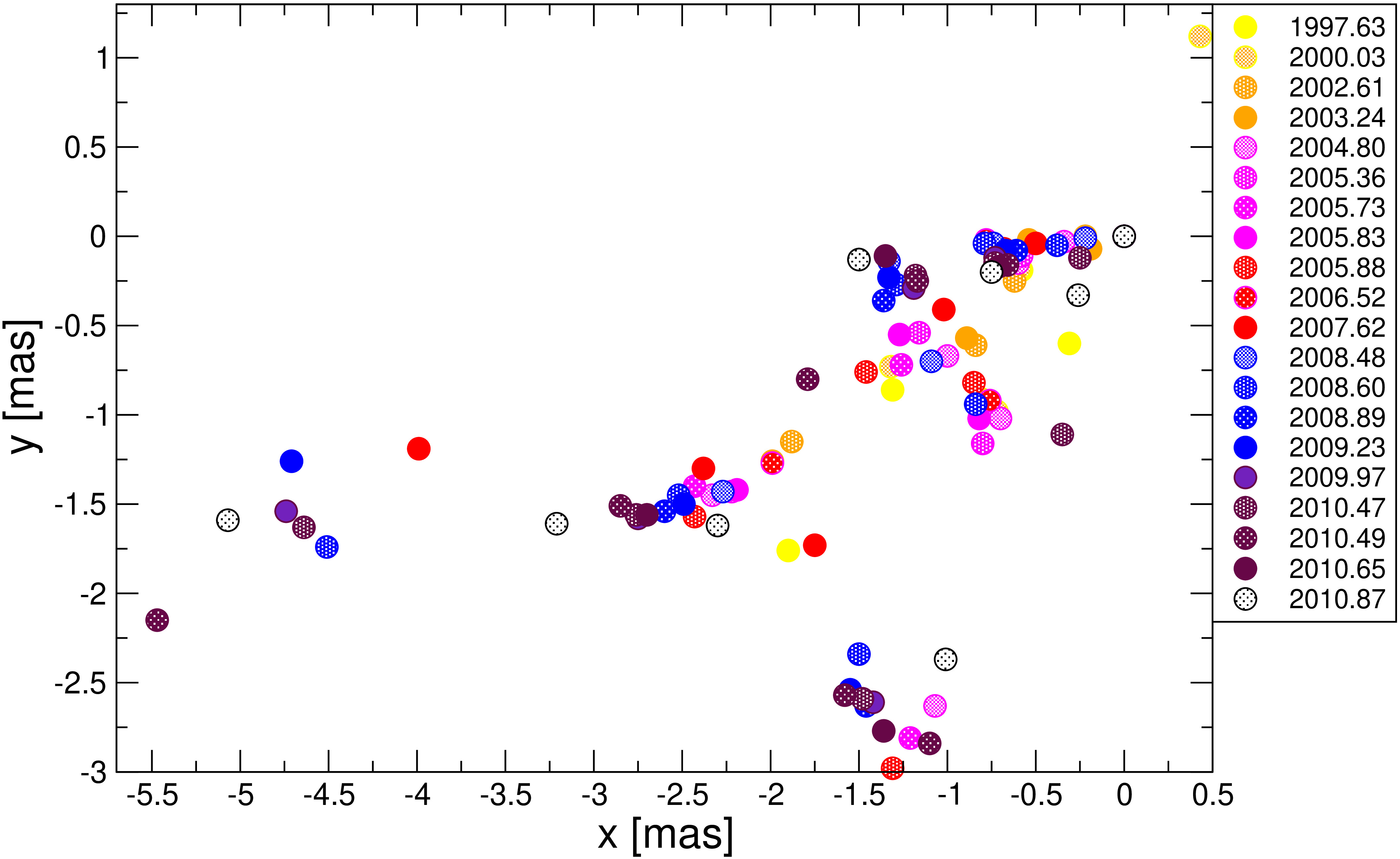}
\includegraphics[width=0.32\textwidth]{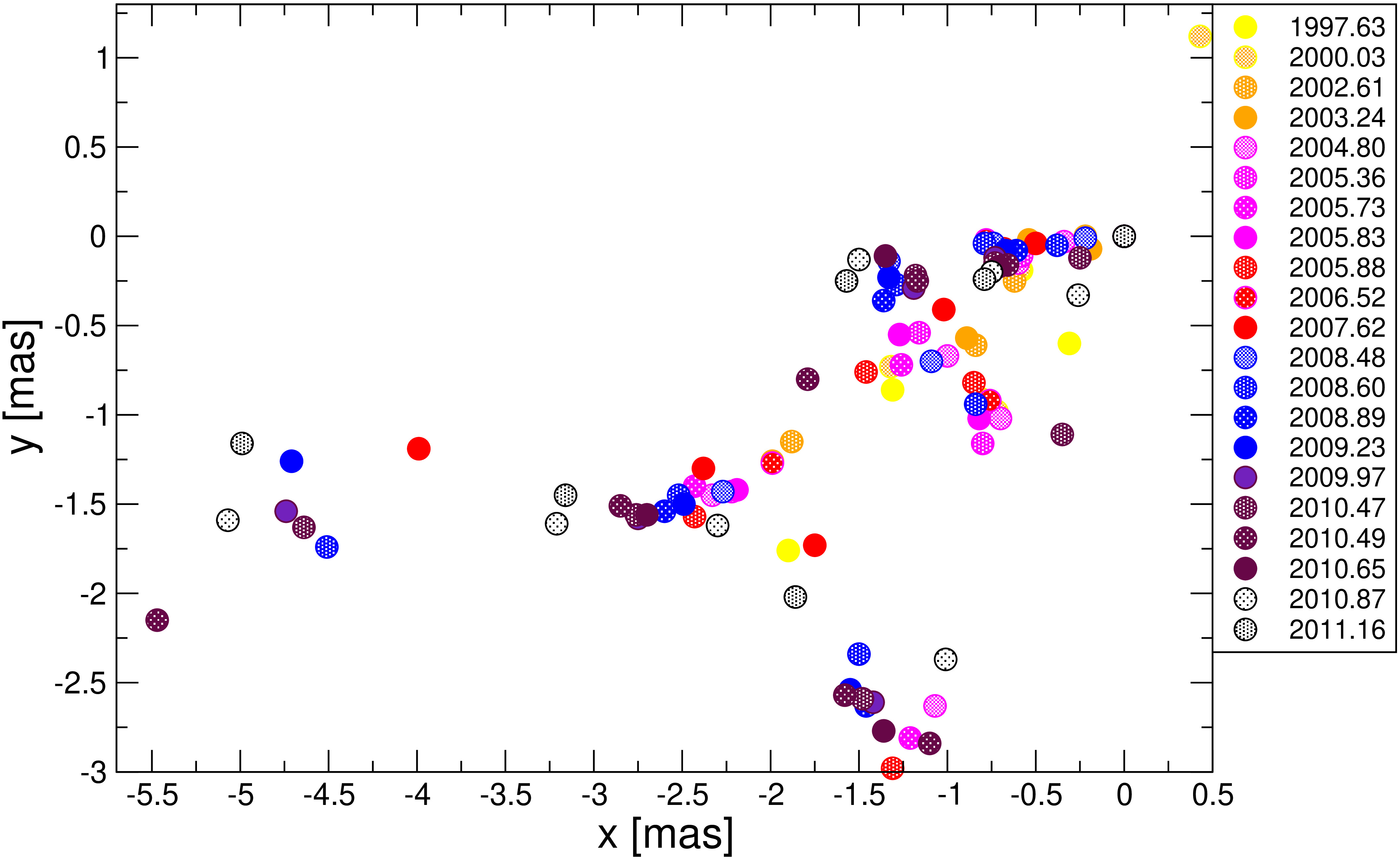}
\includegraphics[width=0.32\textwidth]{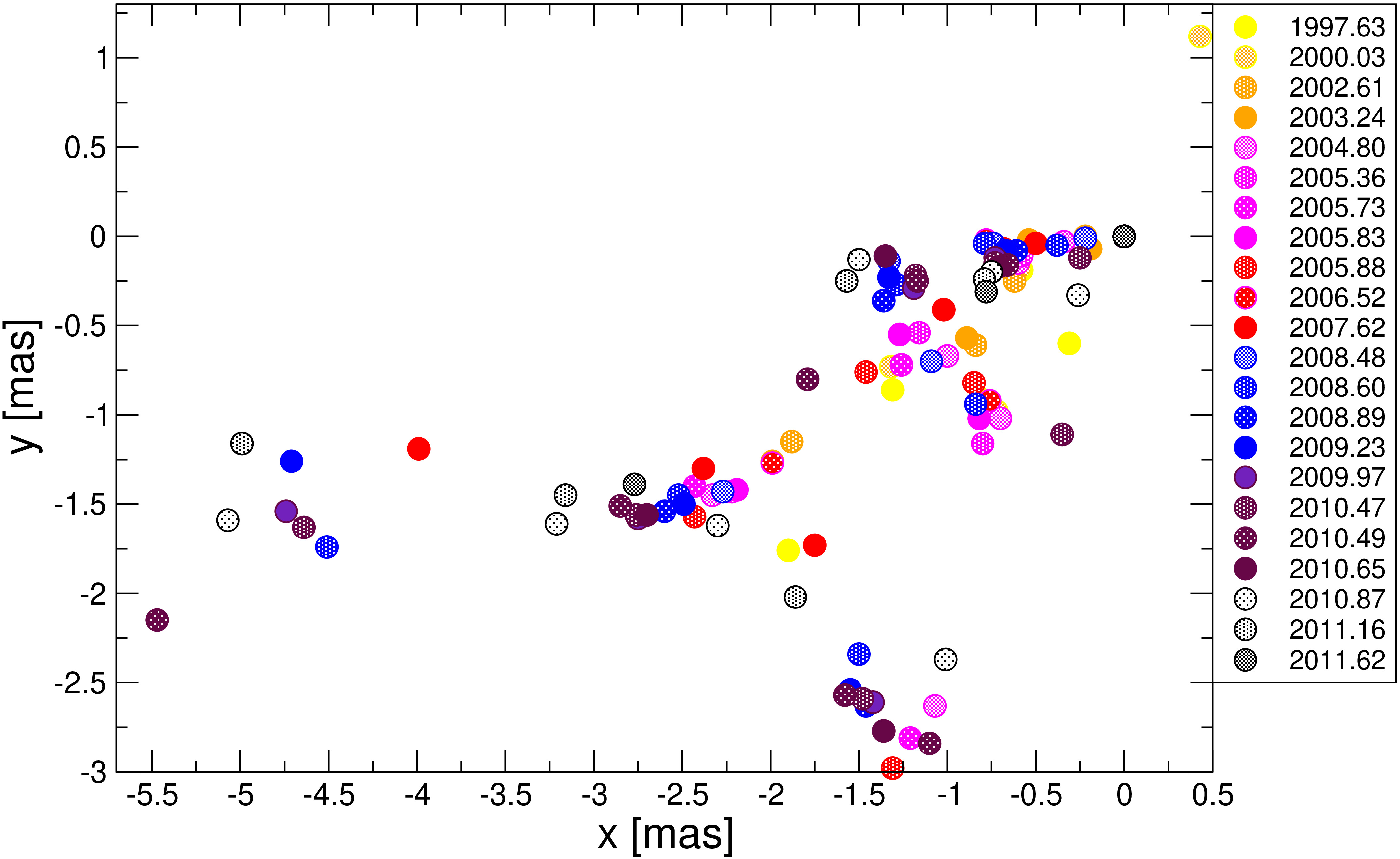}
\includegraphics[width=0.32\textwidth]{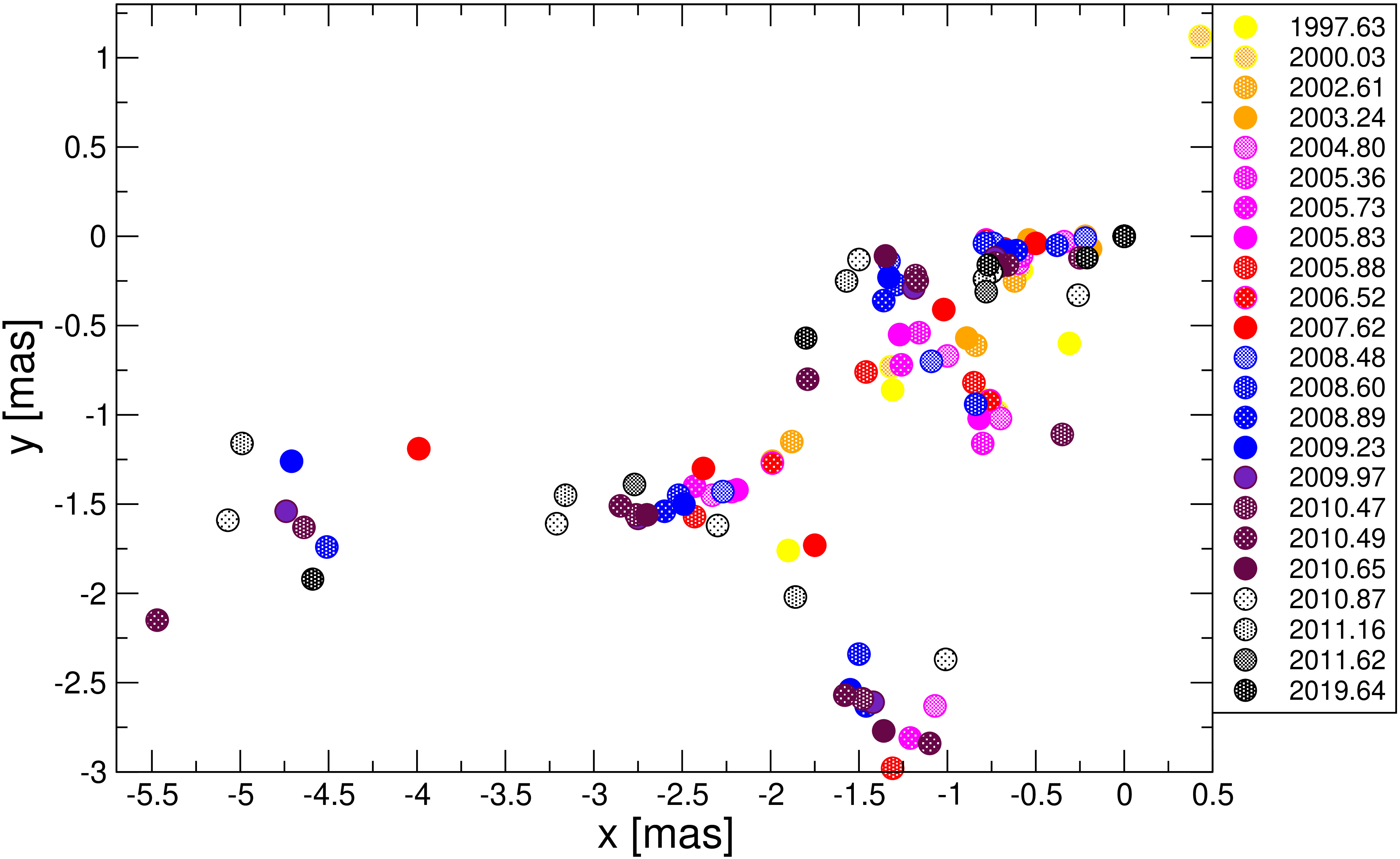}
\includegraphics[width=0.32\textwidth]{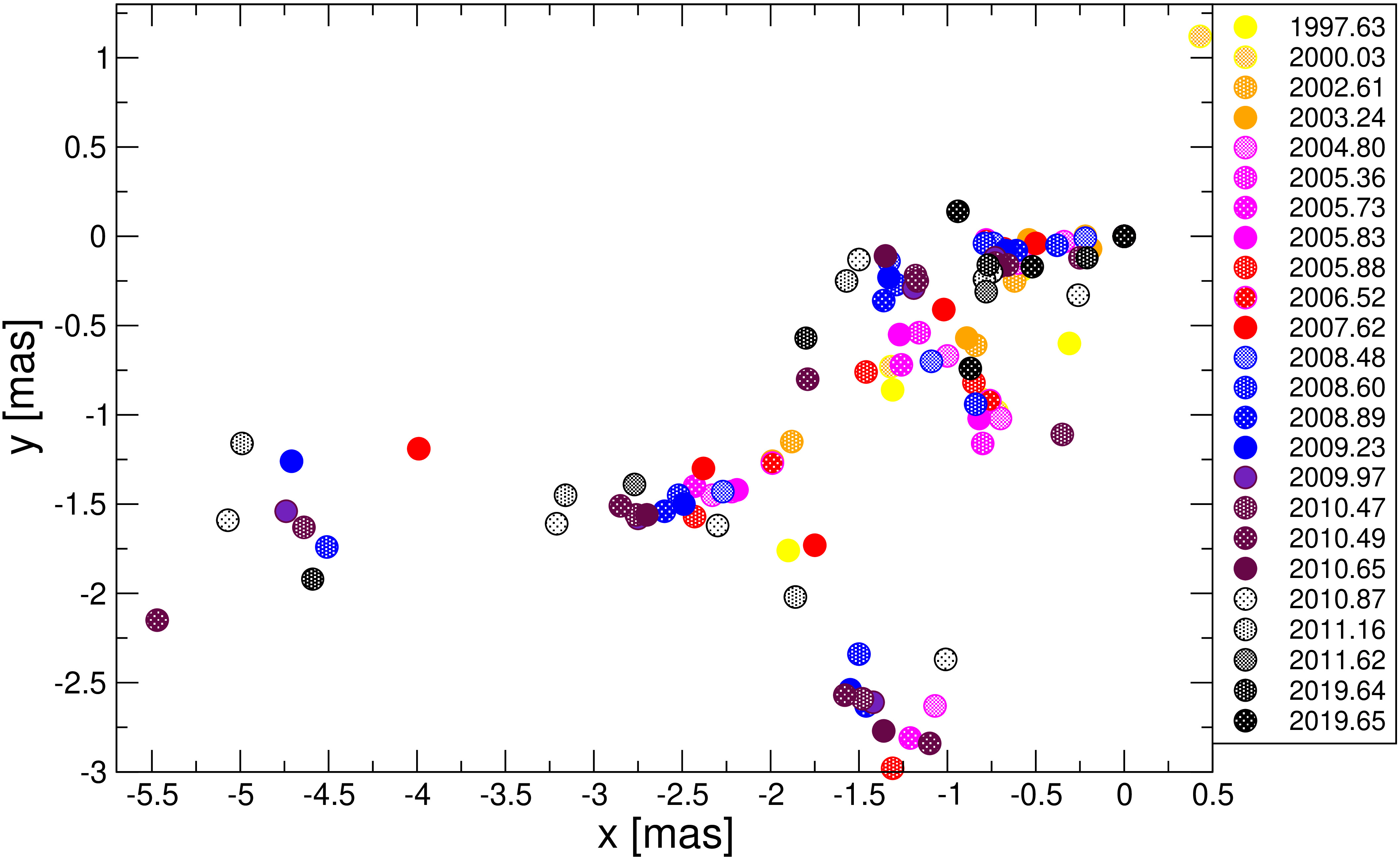}
\includegraphics[width=0.32\textwidth]{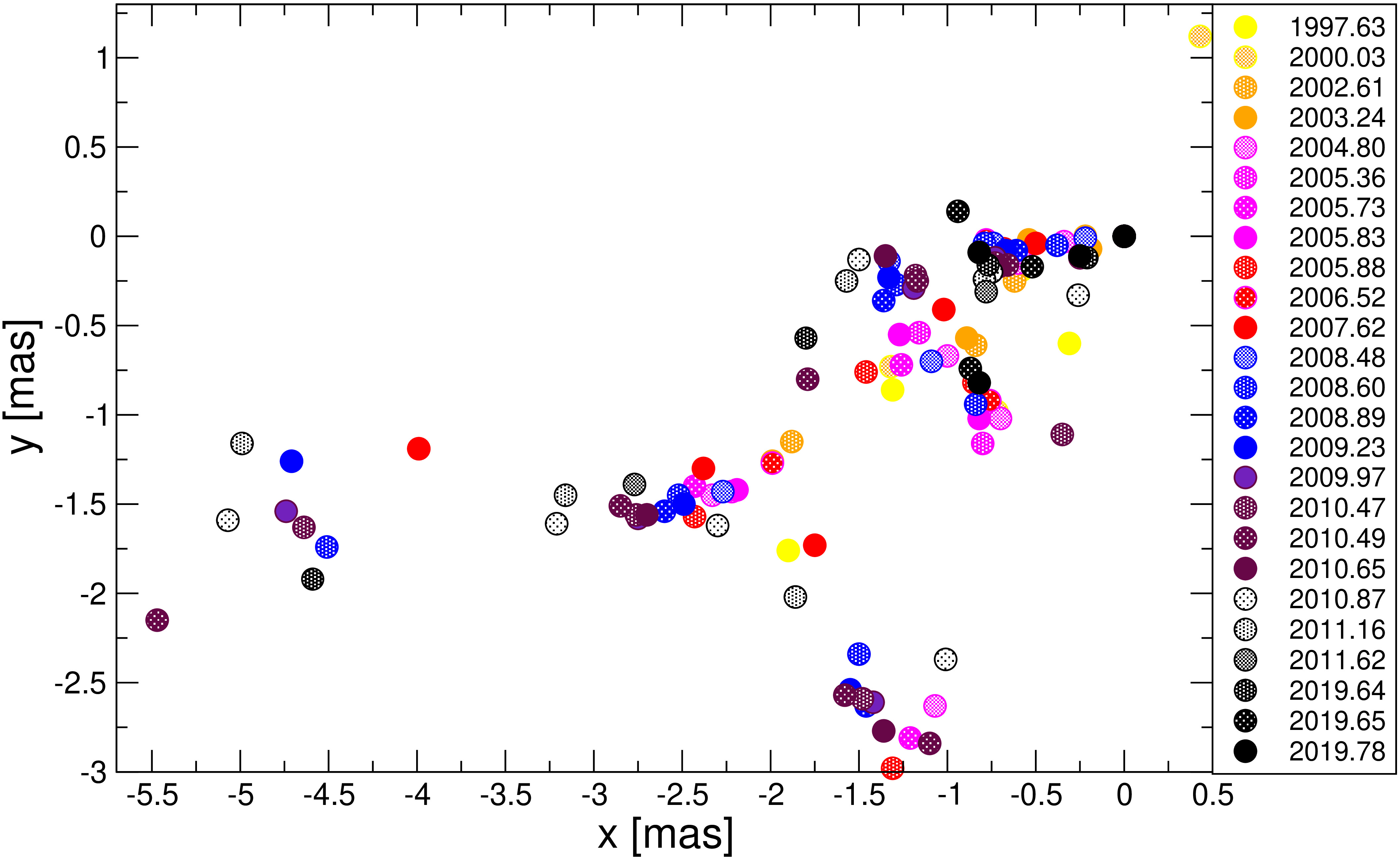}
\caption{The modelfit components in xy-coordinates as derived for the 15 GHz VLBA observations within the {\it difmap}-modelfit programme (2010.65 -- 2019.78).}
\label{xy_plot2}
\end{figure*}
\section{Polarisation information: Plots}
\begin{figure*}
  \begin{minipage}{8.3cm}
     \includegraphics[width=\linewidth]{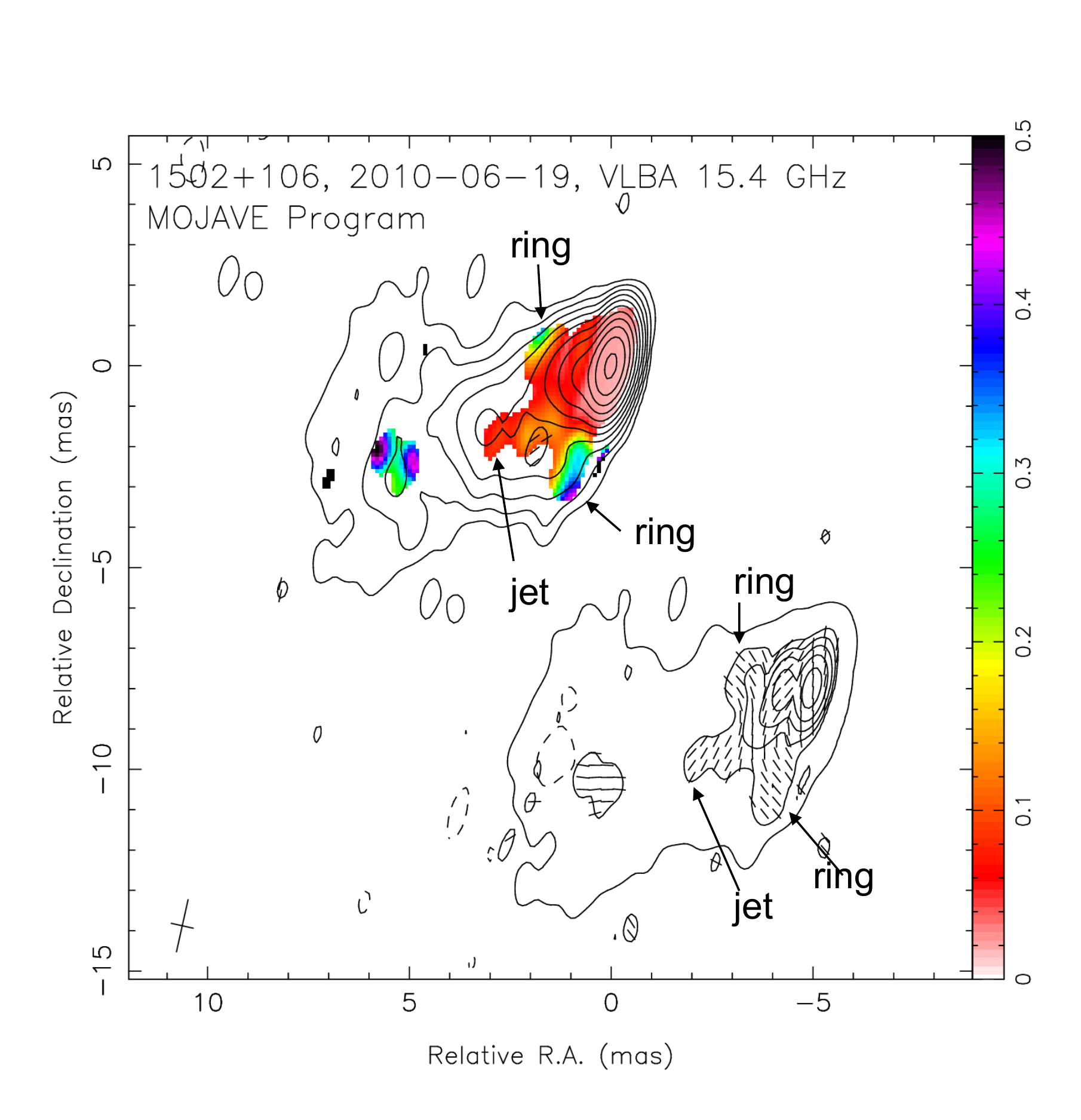}
     [a]
  \end{minipage}
   \begin{minipage}{8.3cm}
     \includegraphics[width=\linewidth]{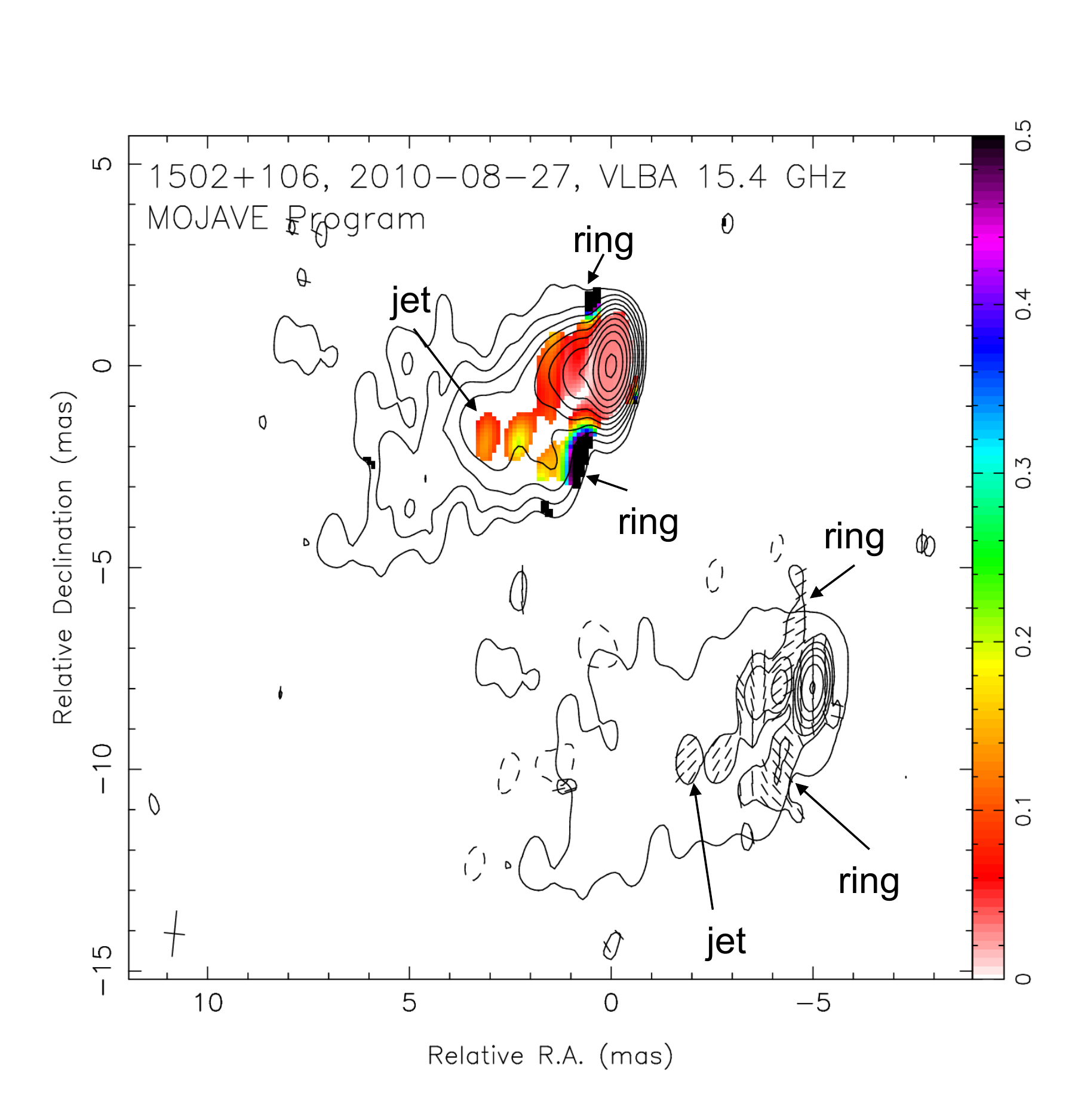}
     [b]
  \end{minipage}
   \begin{minipage}{8.3cm}
     \includegraphics[width=\linewidth]{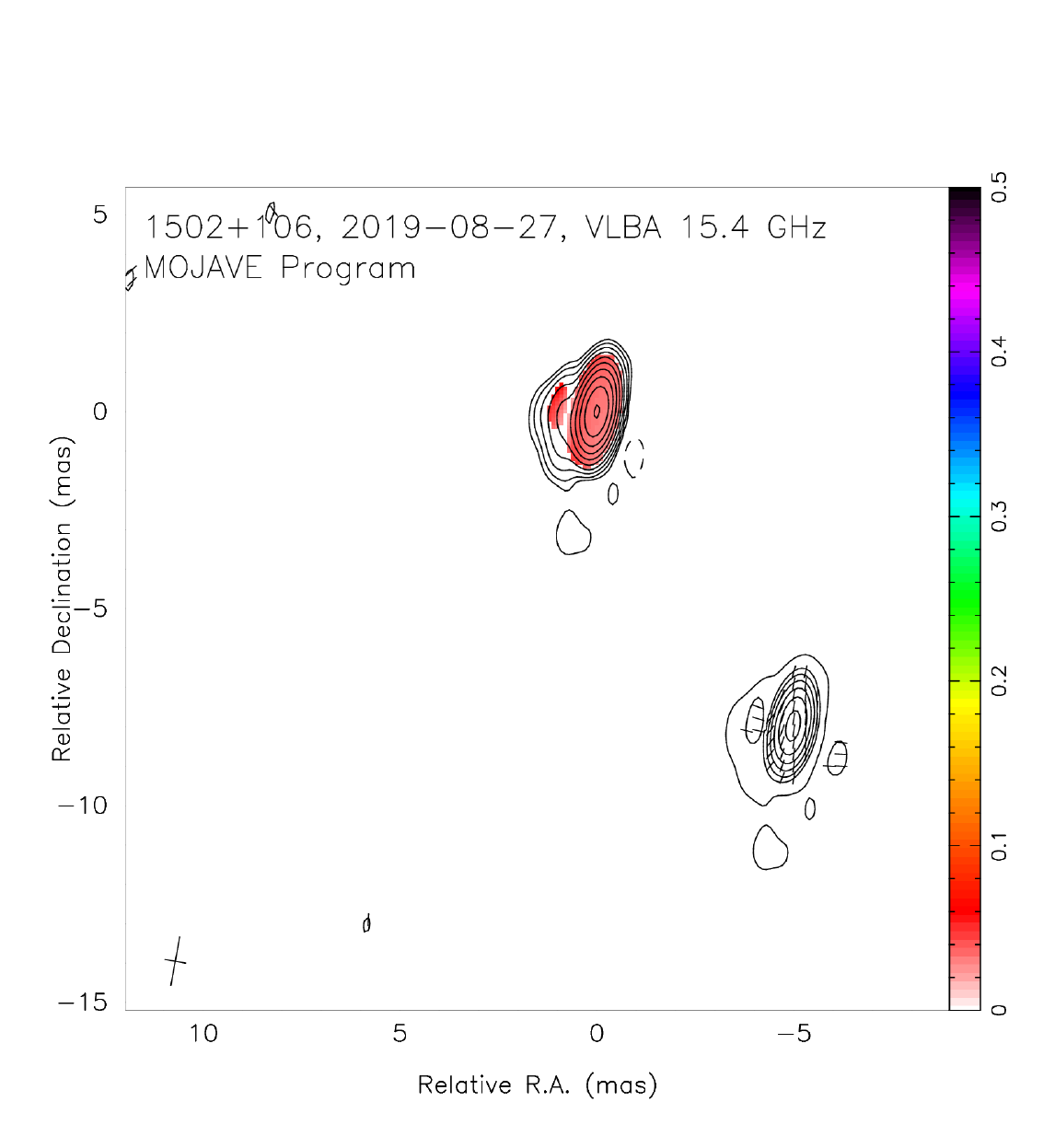}
     [c]
  \end{minipage}
   \begin{minipage}{8.3cm}
     \includegraphics[width=\linewidth]{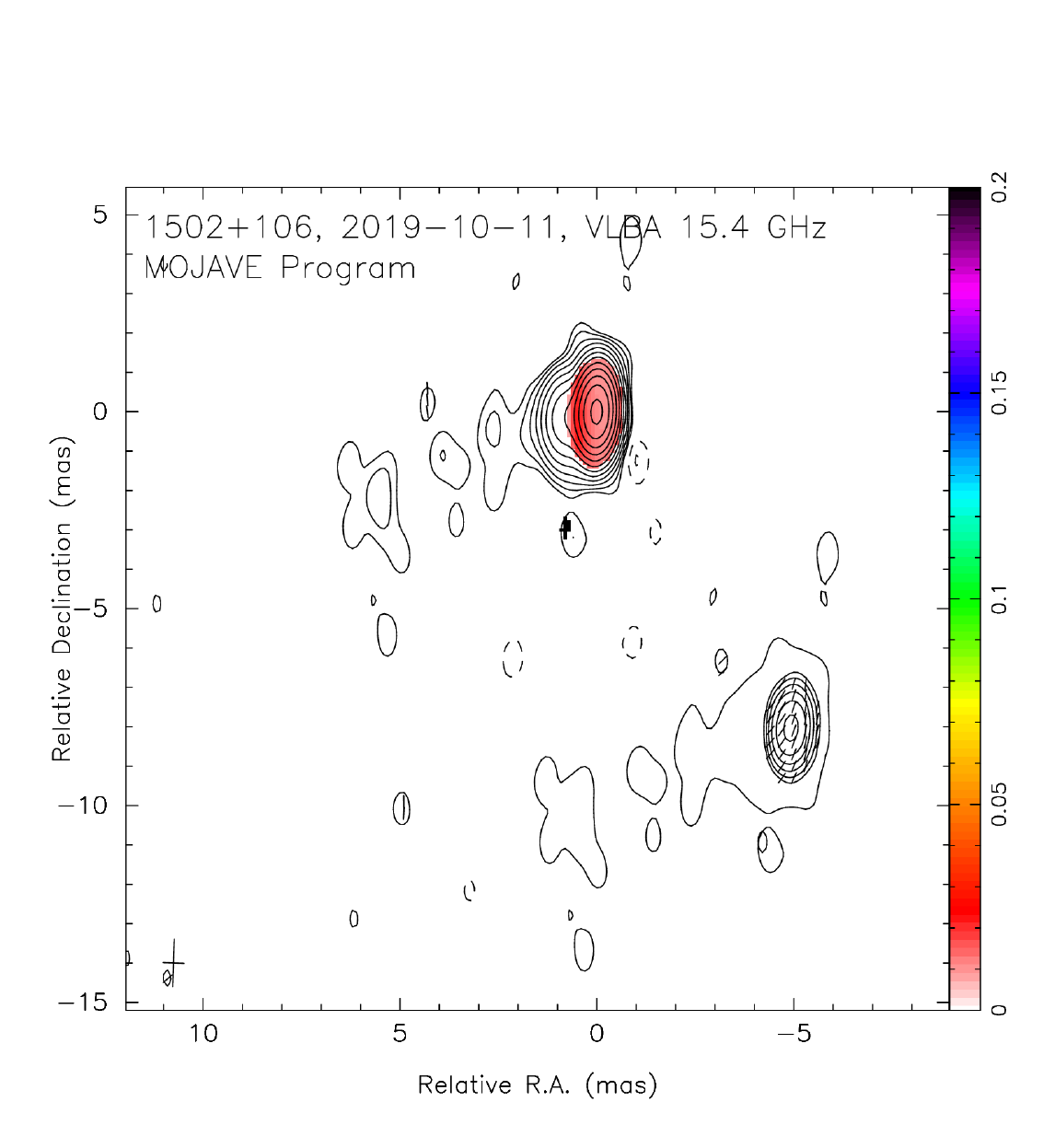}
     [d]
  \end{minipage}
\caption{MOJAVE polarisation images for four epochs ([a] 2010/06/19, [b] 2010/08/27, [c] 2019/08/27, and [d] 2019/10/11 adapted from the MOJAVE webpage). The peak intensity, bottom intensity contour level, peak polarisation, bottom contour polarisation level, and beam parameters for the four images are: 723 mJy, 0.6 mJy/beam, 14.4 mJy, 0.6 mJy/beam, 1.32$\times$0.56 mas at -12.6 deg [a]; 742, 0.60, 20.7, 0.61 mJy/beam, 1.13$\times$0.50 mas at -5.8 deg [b], 2313, 4.26, 72.4, 1.72 mJy/beam, 1.25$\times$0.53 mas at -10.4 deg [c]; 2489, 1, 27.5, 0.70 mJy/beam, 1.2$\times$0.54 at -1.9 deg [d]. We mark jet and ring for clarity. Each panel contains two contour maps of the radio source, the first consisting of total intensity contours in successive integer powers of two times the lowest contour level, with linear fractional polarisation overlaid according to the colour wedge. A single negative total intensity contour equal to the base contour level is also plotted. The second map includes the lowest positive total intensity contour from the first map, and linearly polarised intensity contours, also in increasing powers of two. The sticks indicate the electric polarisation vector directions, uncorrected for Faraday rotation. The FWHM dimensions and orientation of the elliptical Gaussian restoring beam are indicated by a cross in the lower left corner of the map (description according to \citet{Lister2018}). Clearly visible in [a] and [b] is the polarised emission in both directions perpendicular to the jet ridge line. In [c] and [d] the polarisation images shortly before and after the neutrino event are shown.}
\label{pola}
\end{figure*}
\section{Periodicity and correlation analysis of radio and $\gamma$-ray light curves: Plots}
\begin{figure}
\centering
\includegraphics[width=\columnwidth]{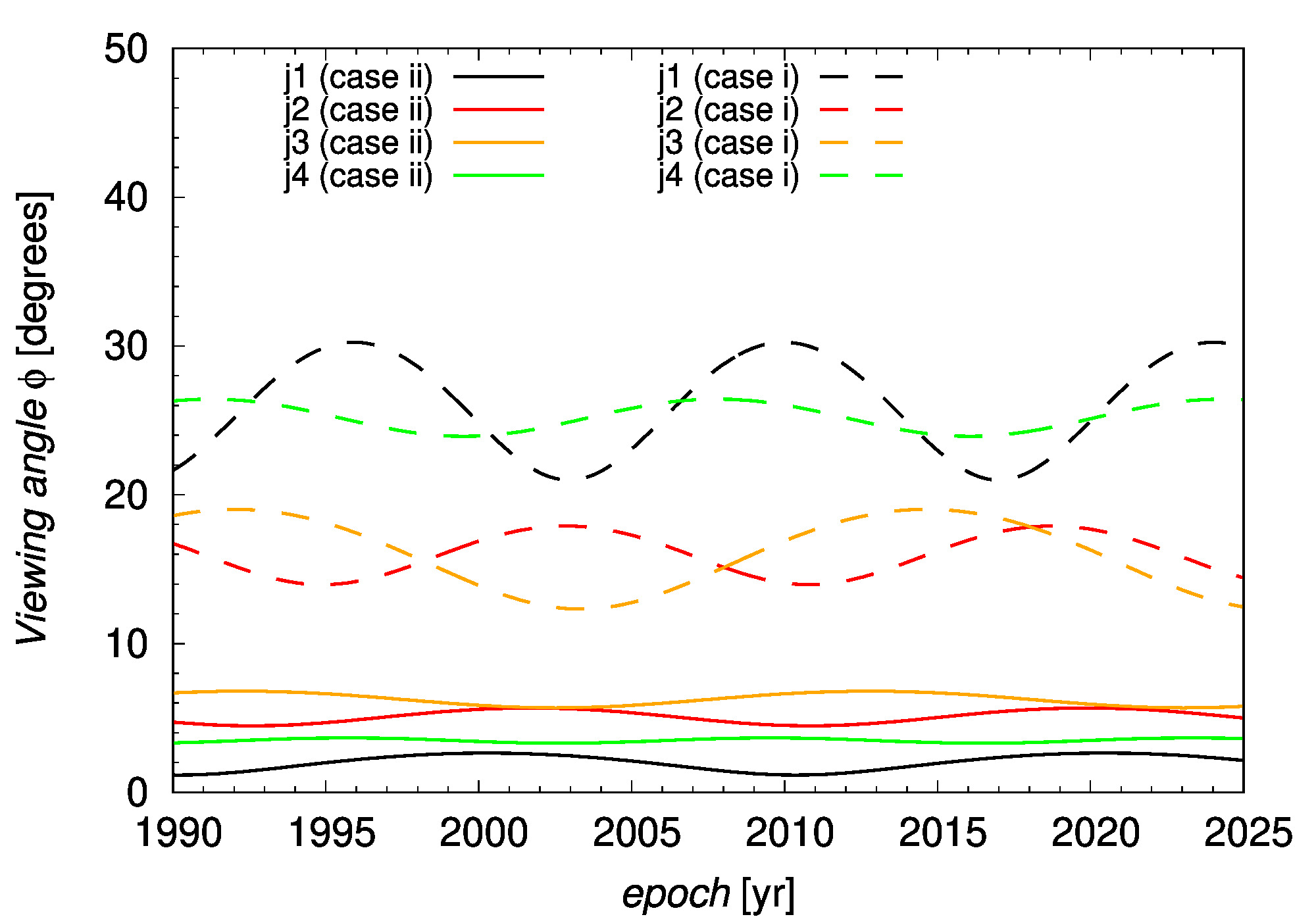}
\includegraphics[width=\columnwidth]{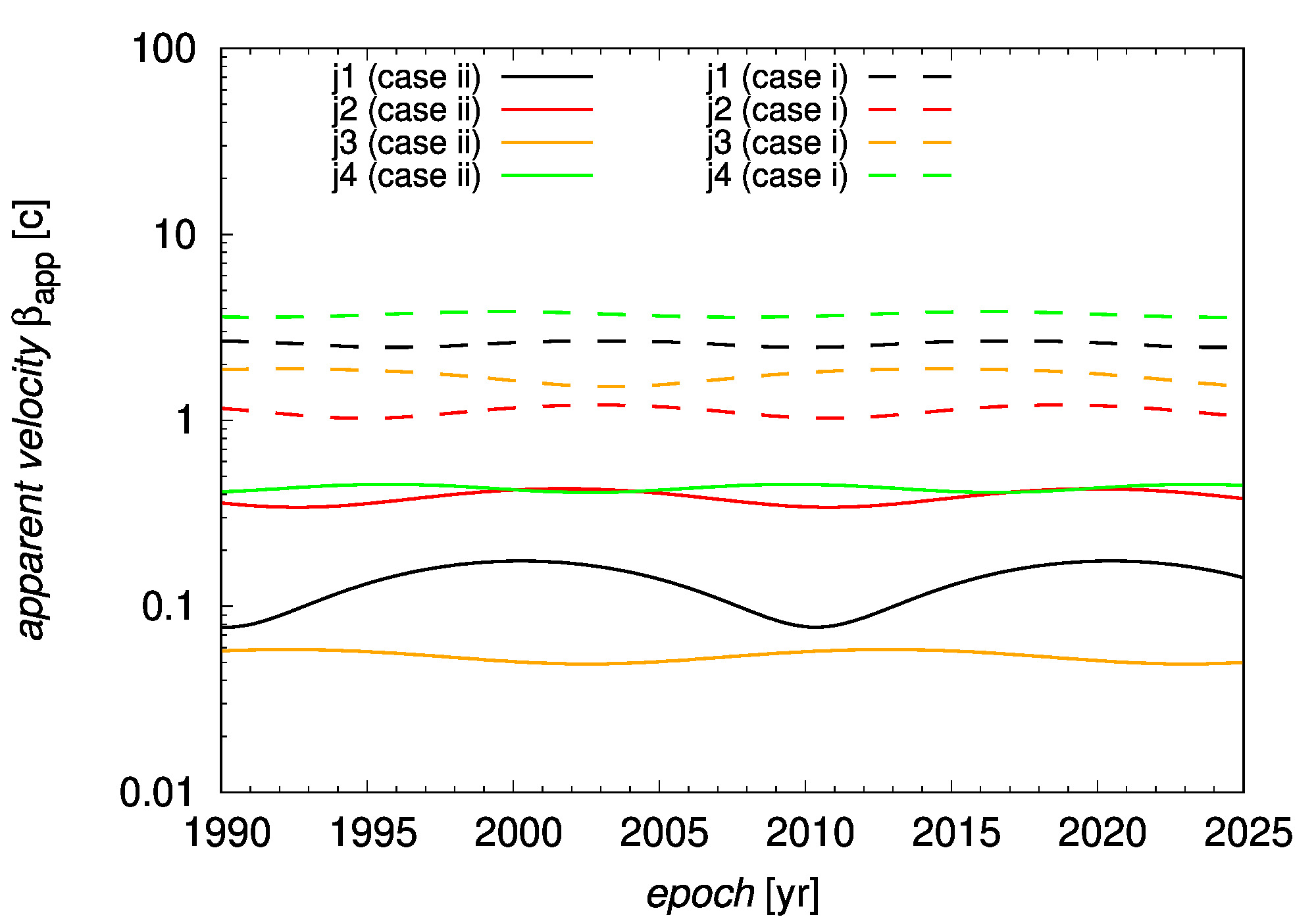}
\includegraphics[width=\columnwidth]{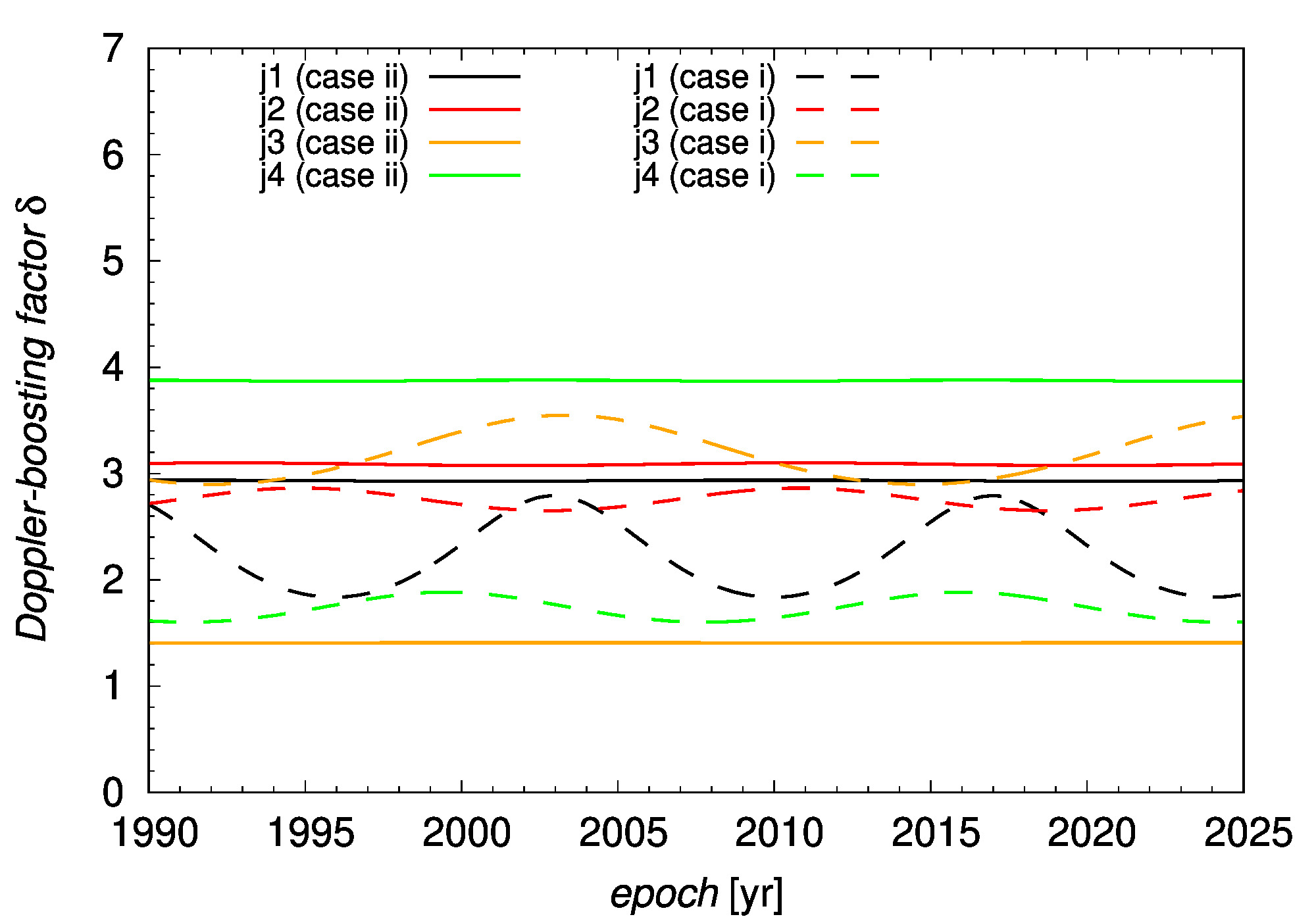}
\caption{Temporal evolution of the jet component viewing angle (top panel), apparent velocity (middle panel), and the Doppler-boosting factor based on the precession modelfitting to component flux densities and position angles. The scenario (i) fits (dashed lines) have on average a larger viewing angle (>10 degrees), while scenatio (ii) fits (solid lines) are with a smaller viewing angle (<10 degrees).}
\label{fig_precession_time}
\end{figure}
\begin{figure*}
\centering
\includegraphics[width=\columnwidth]{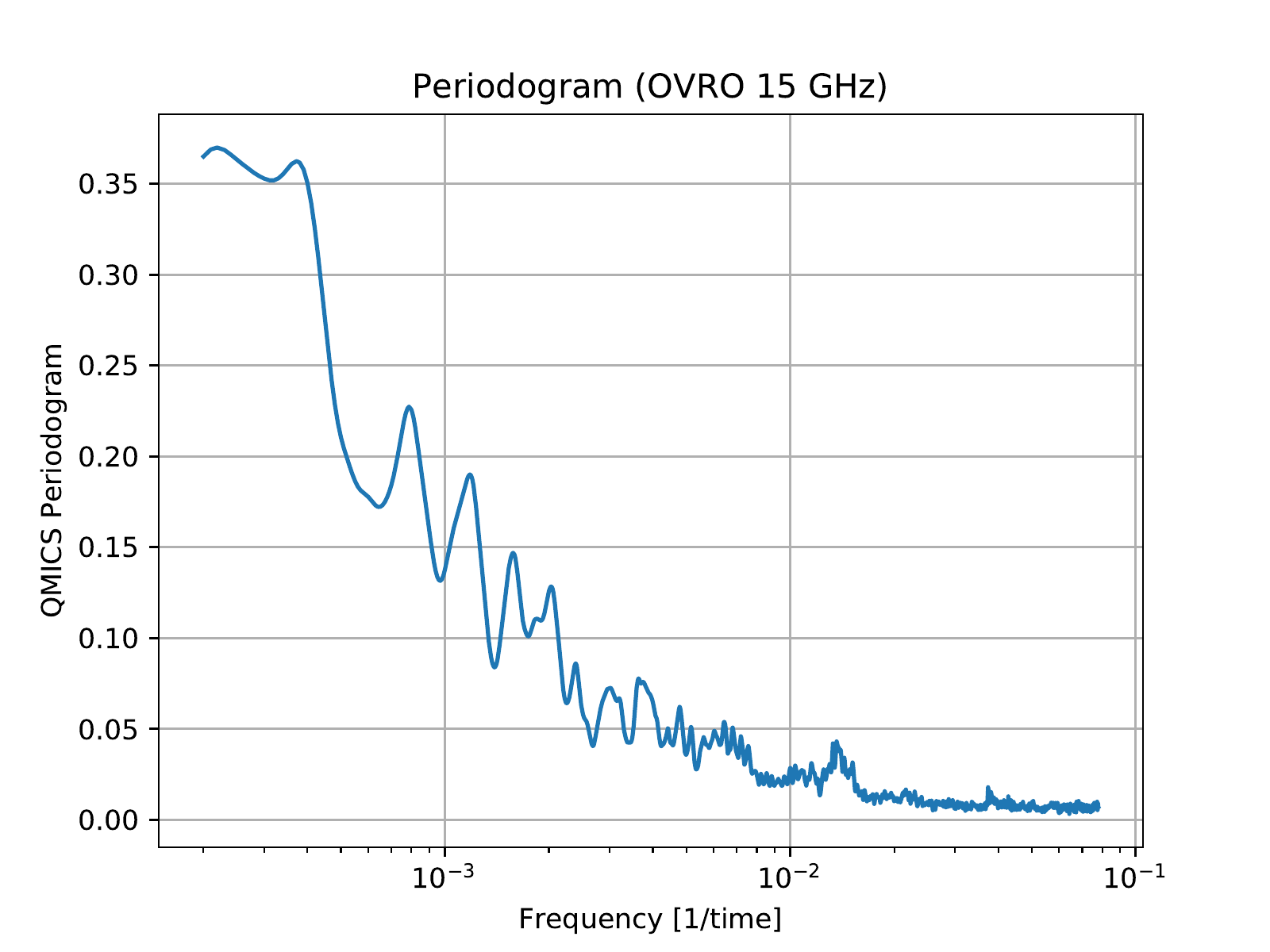}
\includegraphics[width=\columnwidth]{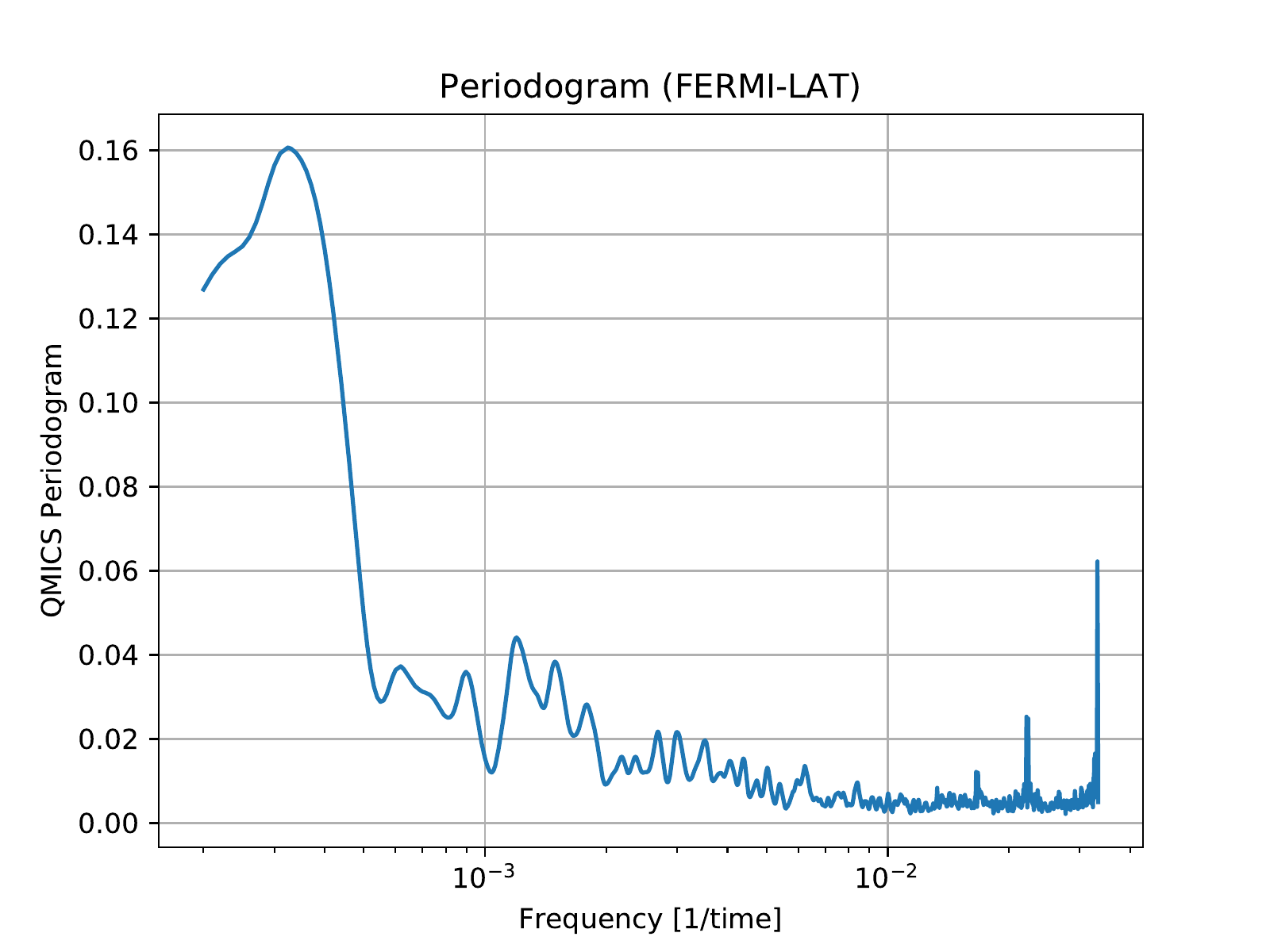}
\includegraphics[width=\columnwidth]{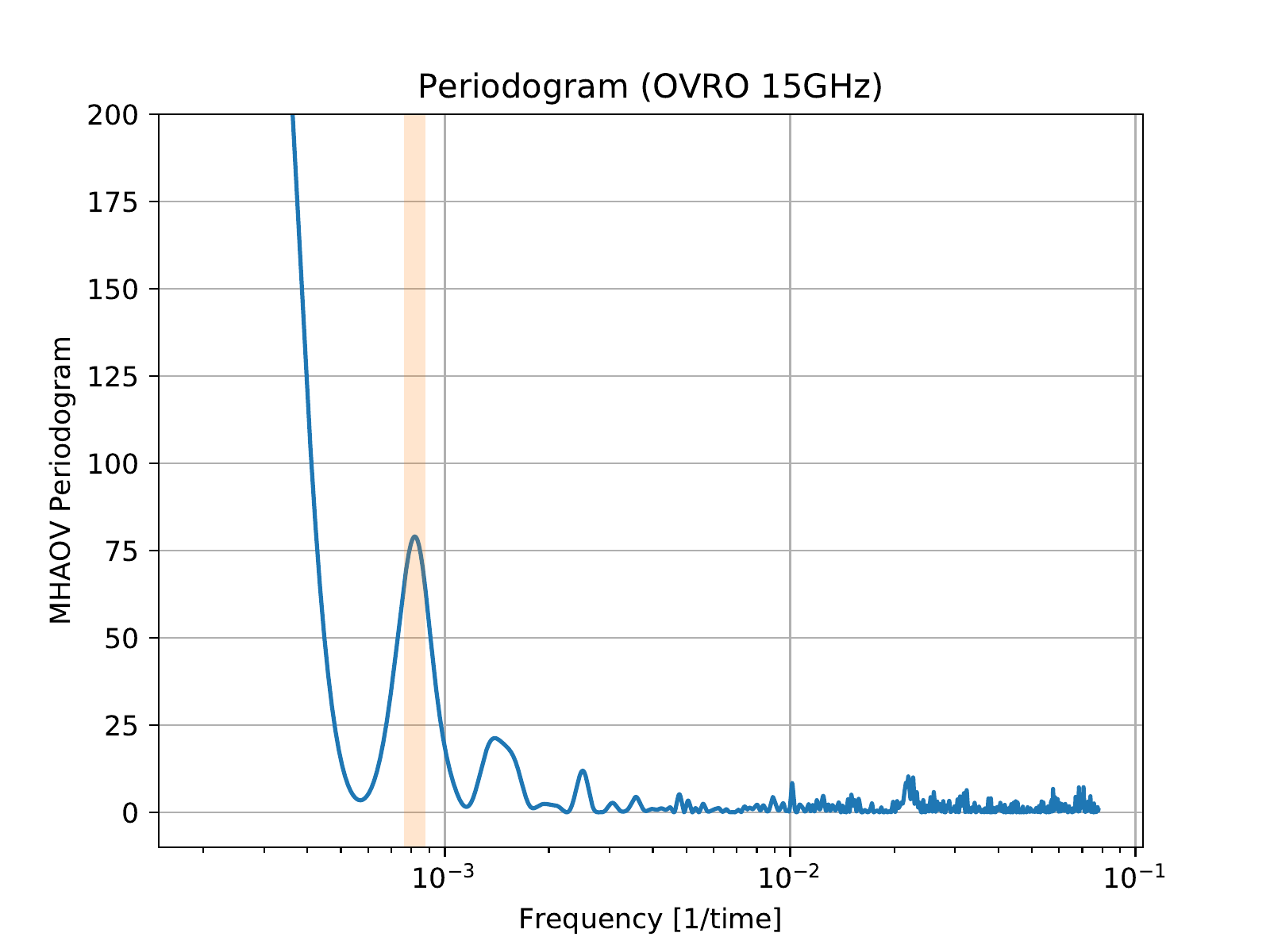}
\includegraphics[width=\columnwidth]{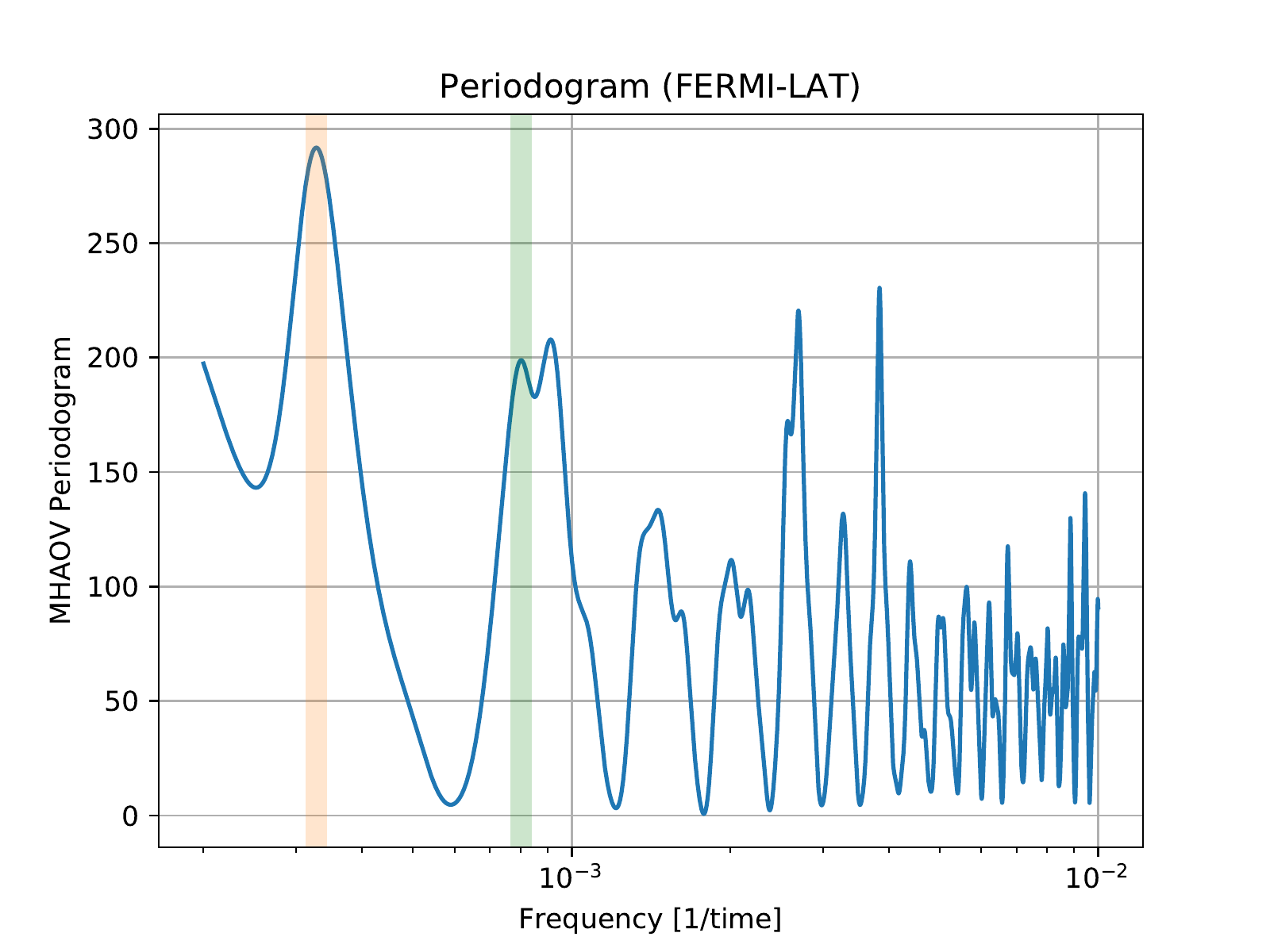}
\caption{Periodograms for the OVRO and \textit{Fermi}-LAT light curves. \textbf{Top left:} Quadratic Mutual Information Cauchy-Schwarz (QMICS) periodogram for the OVRO 15 GHz light curve. \textbf{Top right:} QMICS periodogram for the \textit{Fermi}-LAT light curve. \textbf{Bottom left:} Orthogonal multiharmonic analysis of variance (MHAOV) periodogram for the OVRO 15 GHz light curve. The red vertical line marks the most prominent peak. \textbf{Bottom right:} MHAOV periodogram for the \textit{Fermi}-LAT light curve. The red vertical line marks the most prominent peak. The green vertical line marks the prominent peak, which is consistent with the best peak of the OVRO light curve.}
\label{fig_QMI_periodograms}
\end{figure*}
\begin{figure*}
\centering
\includegraphics[width=\columnwidth]{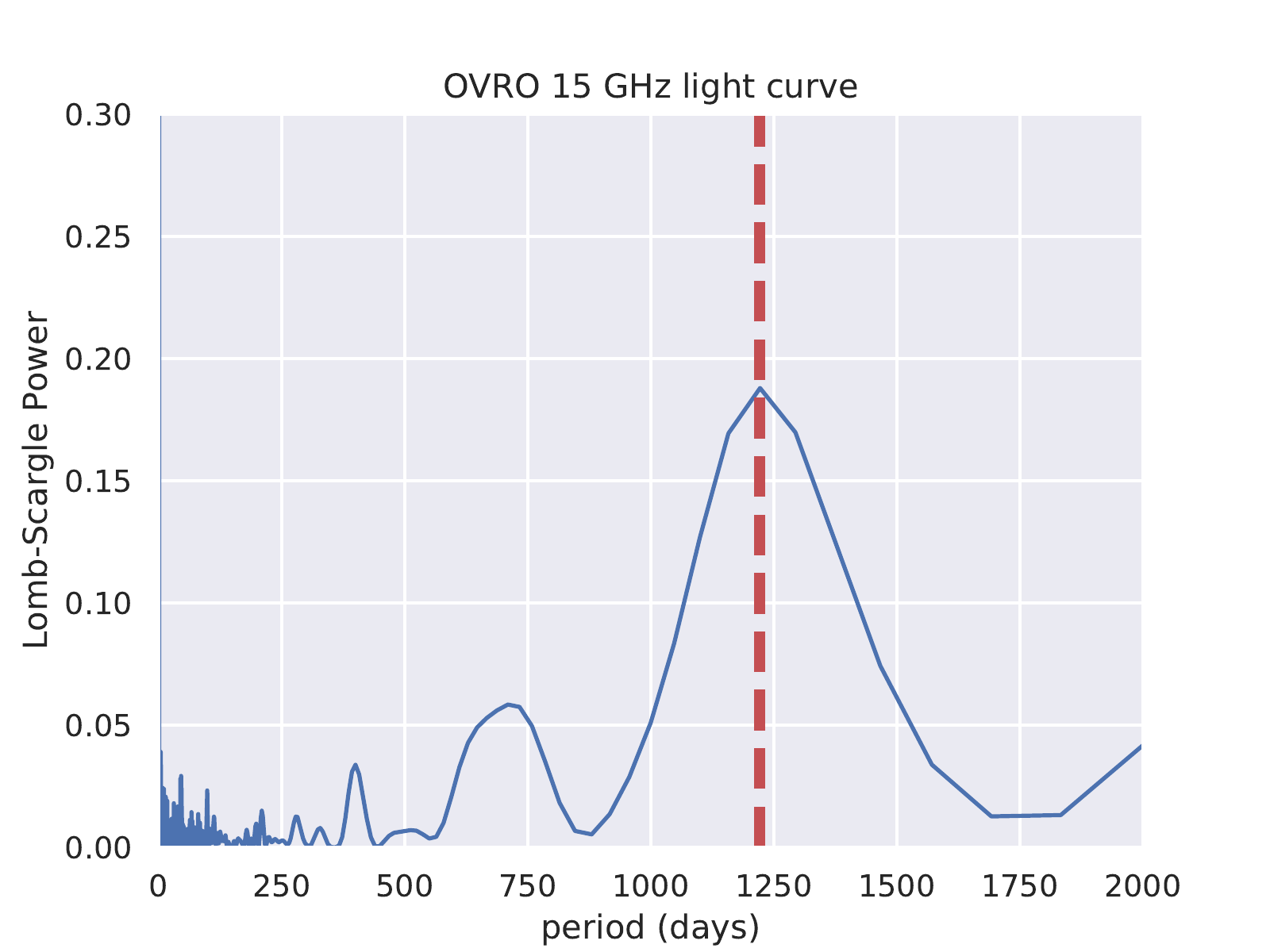}
\includegraphics[width=\columnwidth]{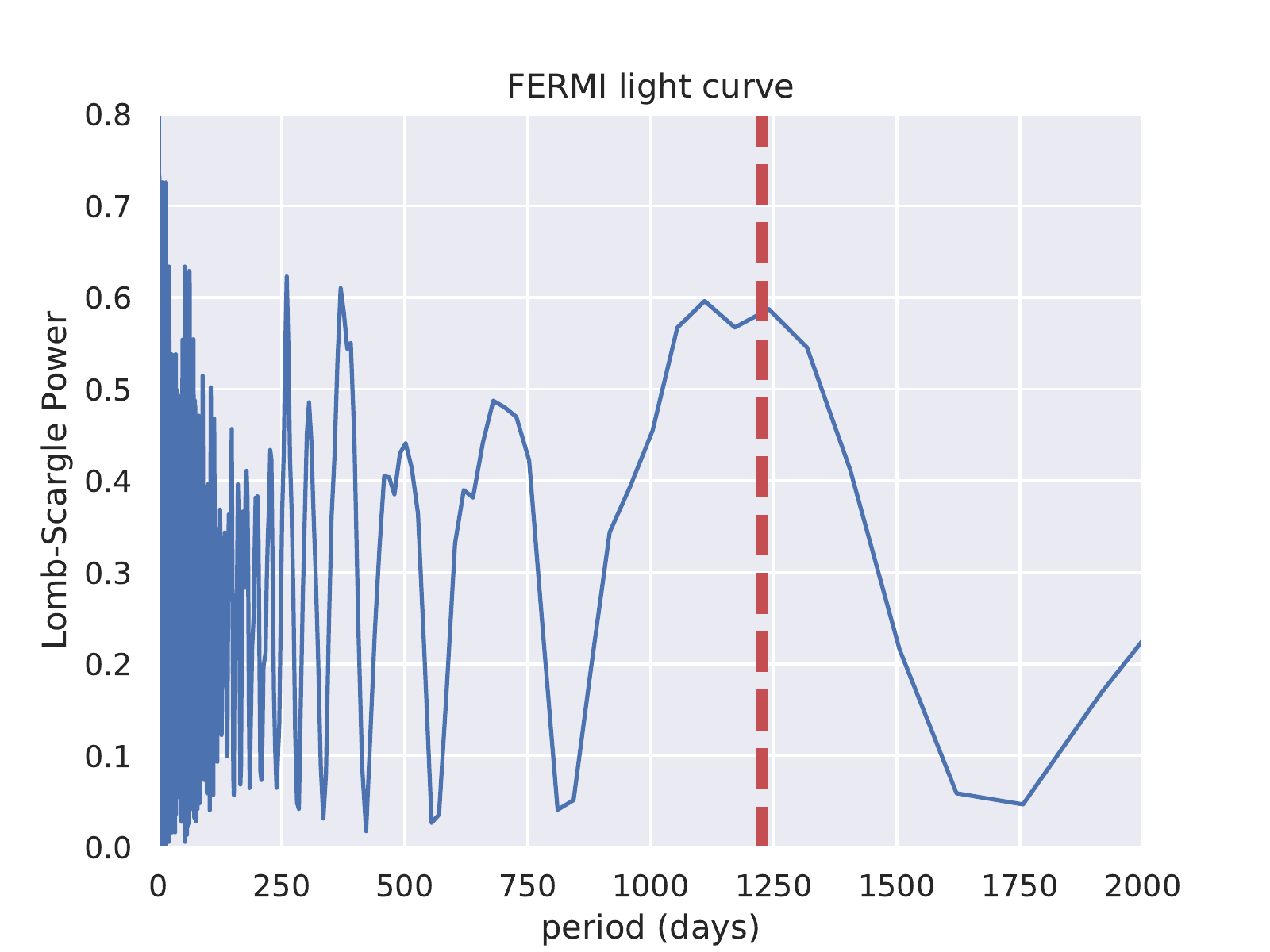}
\caption{The Lomb-Scargle periodograms for the OVRO 15 GHz light curve (left panel) and for the \textit{Fermi}-LAT $\gamma$-ray light curve (right panel). The dashed vertical lines mark the peak at $1220.6$ days and at $1225.9$ days for the left and the right panels, respectively.}
\label{fig_lomb_scargle}
\end{figure*}
\begin{figure*}
\centering
\includegraphics[width=\textwidth]{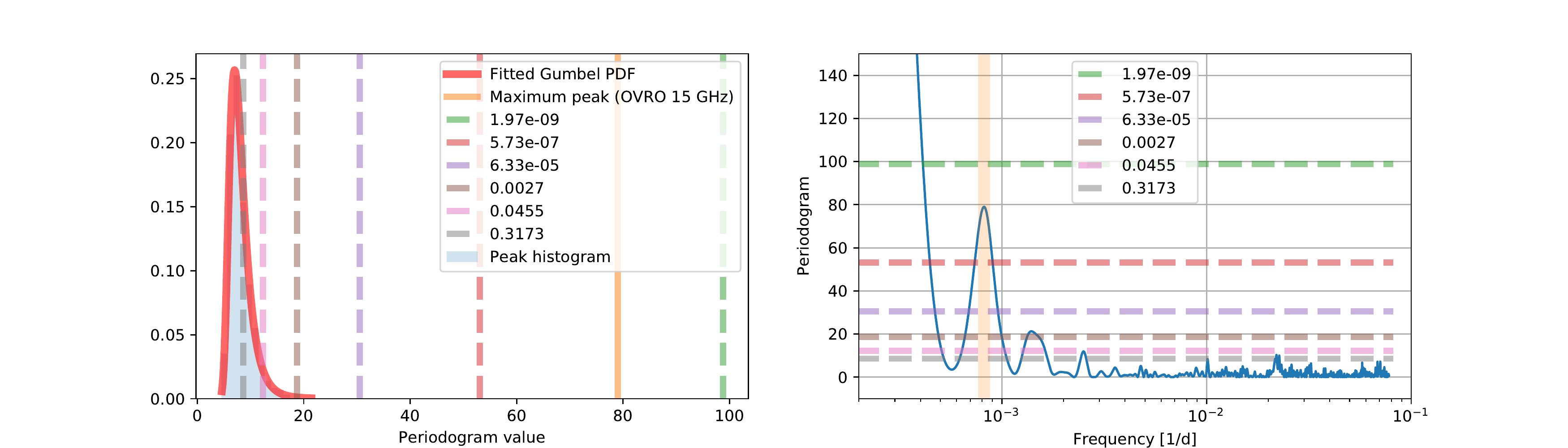}
\includegraphics[width=\textwidth]{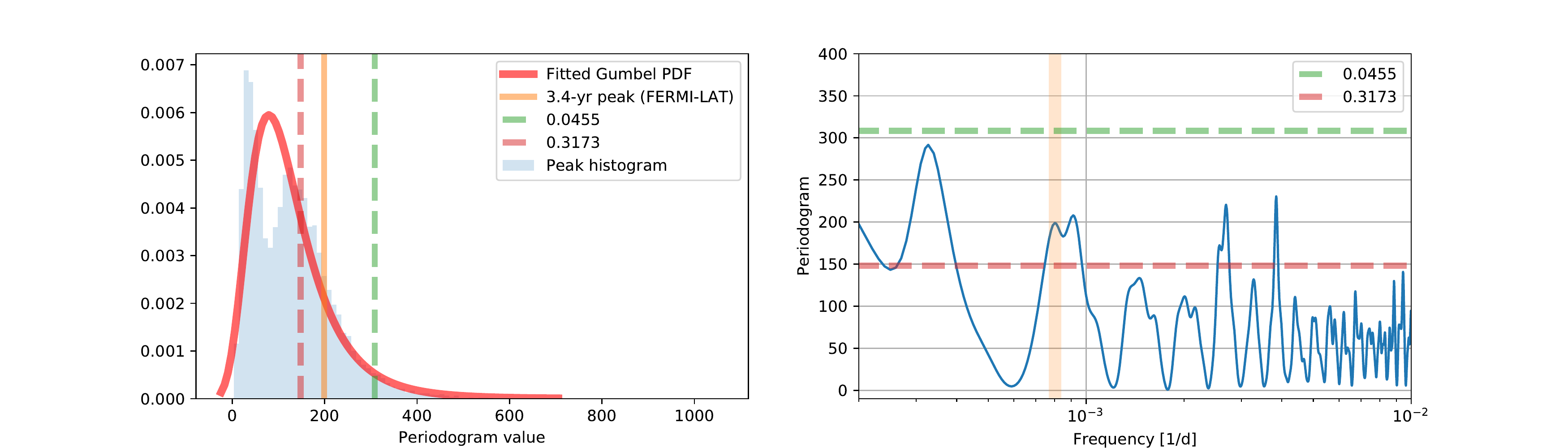}
\caption{Statistical significance of the period of $\sim 3.4$ years of the OVRO 15 GHz and the \textit{Fermi}-LAT light curves. \textbf{Top left panel:} A histogram of MHAOV periodogram peaks constructed from 200 surrogate light curves. From the fitted Gumbel PDF, we constructed confidence intervals to compare with the detected peak value in the OVRO 15 GHz light curve, which lies above the 5$\sigma$ level. \textbf{Top right panel:} The MHAOV periodogram constructed from the OVRO 15 GHz light curve with 6 confidence levels corresponding to 1$\sigma$ up to 6$\sigma$. The best peak lies above the 5$\sigma$ confidence level. \textbf{Bottom left panel:} A histogram of the MHAOV periodogram peaks constructed from 2000 surrogate light curves generated by bootstrap. We fitted the Gumbel PDF to infer the corresponding 1$\sigma$ and 2$\sigma$ confidence intervals. The peak at $\sim 3.4$ years is significant at the $1\sigma$ level. \textbf{Bottom right panel:} The MHAOV periodogram constructed from the \textit{Fermi}-LAT $\gamma$-ray light curve. In particular, we depict the peak at $f=0.000802\,{\rm d^{-1}}$, which corresponds to 1247 days or $3.41$ years, which is significant at the 1$\sigma$ level.}
\label{fig_histogram_confidence}
\end{figure*}
\begin{figure*}
\centering
\includegraphics[width=\textwidth]{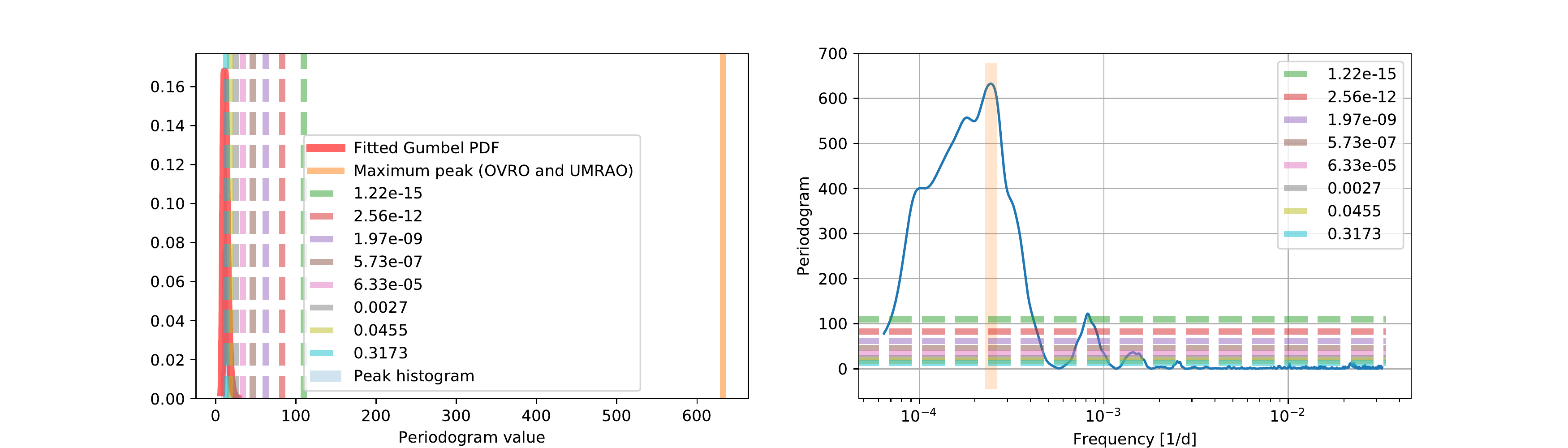}
\caption{Statistical significance of the longer period of $\sim 11.22$ years in the observer's frame of the combined OVRO 15 GHz and the UMRAO radio light curve. Left panel: The significance levels constructed from the Gumbel probability distribution function fitted to the histogram of bootstrapped best peaks. Right panel: The MHAOV periodogram of the whole radio light curve.}
\label{fig_radio_periodicity_all}
\end{figure*}
\begin{figure*}
    \centering
    \includegraphics[width=\columnwidth]{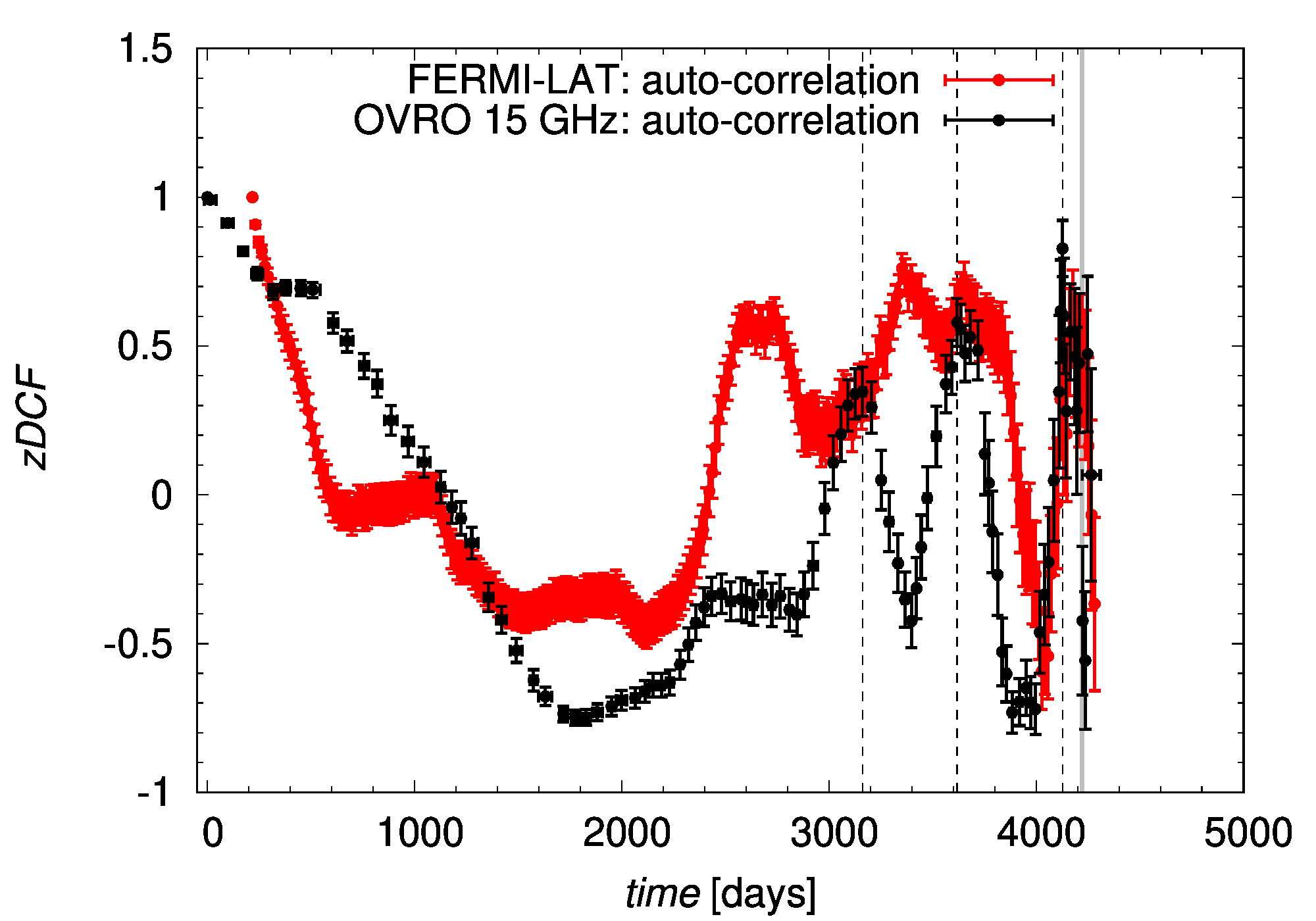}
    \includegraphics[width=\columnwidth]{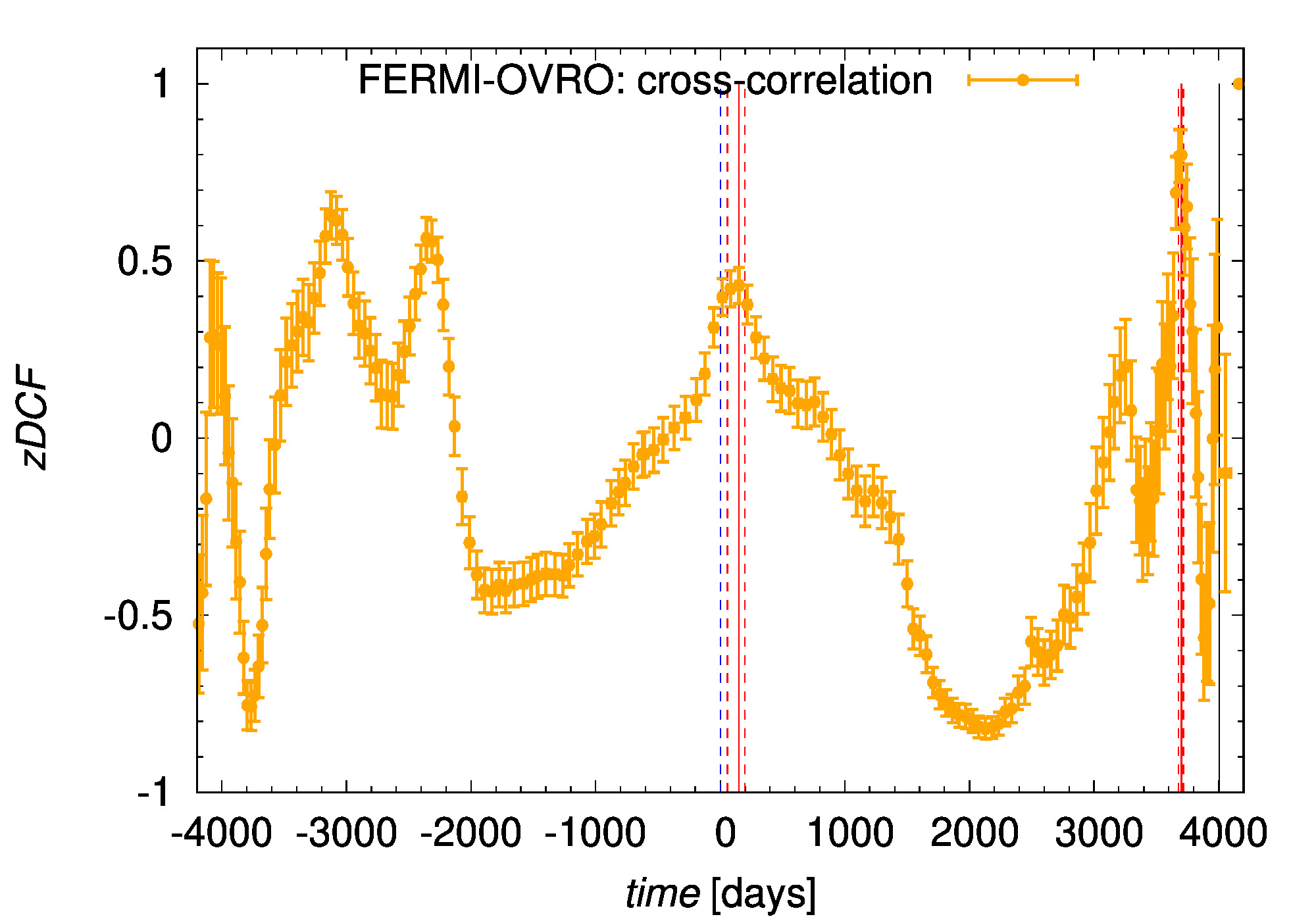}
    \caption{The auto-correlation of the radio OVRO 15 GHz light curve and the \textit{Fermi}-LAT light curve (left panel) and their cross-correlation (right panel) using $z$-transformed DCF. Left panel: The autocorrelation function of the radio OVRO (black points) and $\gamma$-ray \textit{Fermi}-LAT light curve (red points), with the \textit{Fermi}-LAT autocorrelation function shifted by $217$ days to correct for the different starting epoch of the light curve. Dashed vertical black lines depict the autocorrelation peaks of the radio light curve. The gray solid vertical line shows the neutrino emission epoch. Right panel: The cross-correlation indicates a time-lag of $151^{+46}_{-94}$ days between the $\gamma$-ray and the radio emission in the observer's frame (red vertical lines), with the radio emission lagging behind the $\gamma$-ray emission. A potentially longer time-delay of $3703^{+15}_{-25}$ days is also indicated by red vertical lines.}
    \label{fig_zDCF}
\end{figure*}
\begin{figure}
    \centering
    \includegraphics[width=\columnwidth]{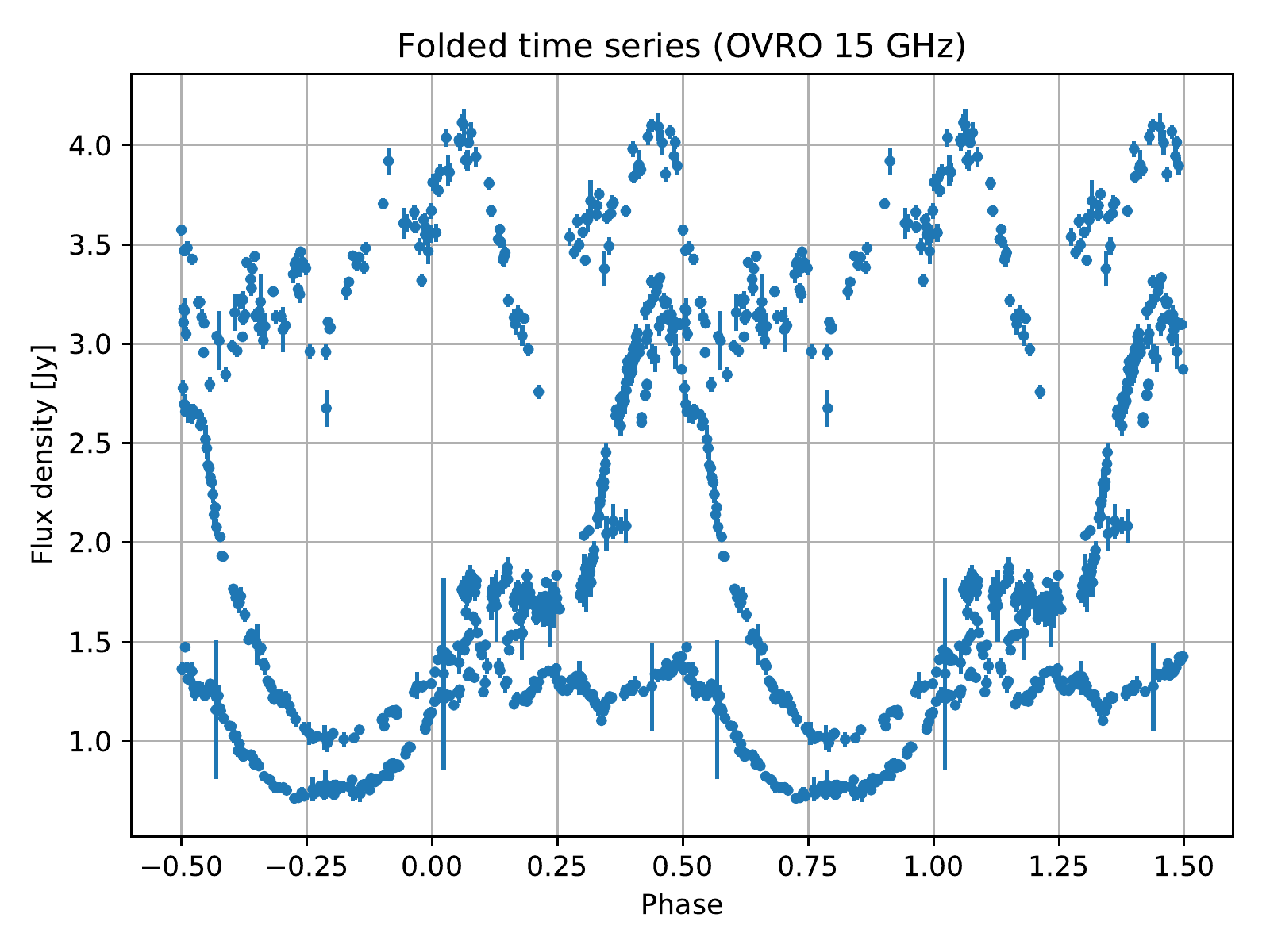}
    \includegraphics[width=\columnwidth]{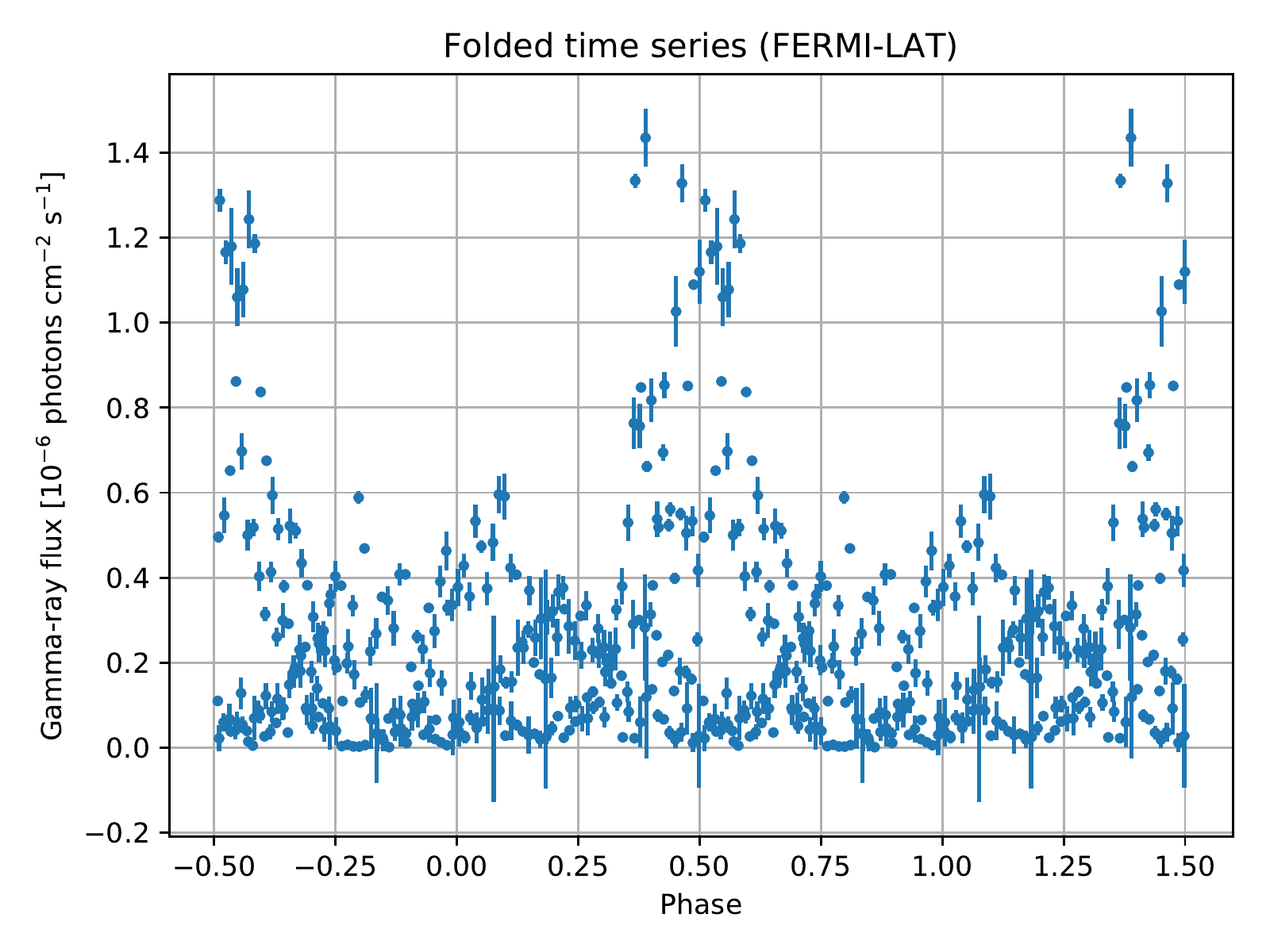}
    \caption{Folded light curves with the frequency of $f_{\rm best}\sim 8\times 10^{-4}\,{\rm d^{-1}}$, which corresponds to $\sim 3.4$ years. This period is significant in the radio light curve and it can also be found in the $\gamma$-ray light curve at the 1$\sigma$ level. In the top panel, we show the folded OVRO 15 GHz light curve, in the bottom panel, we display the folded \textit{Fermi}-LAT light curve.}
    \label{fig_folded_light_curves}
\end{figure}
\begin{figure}
    \centering
    \includegraphics[width=\columnwidth]{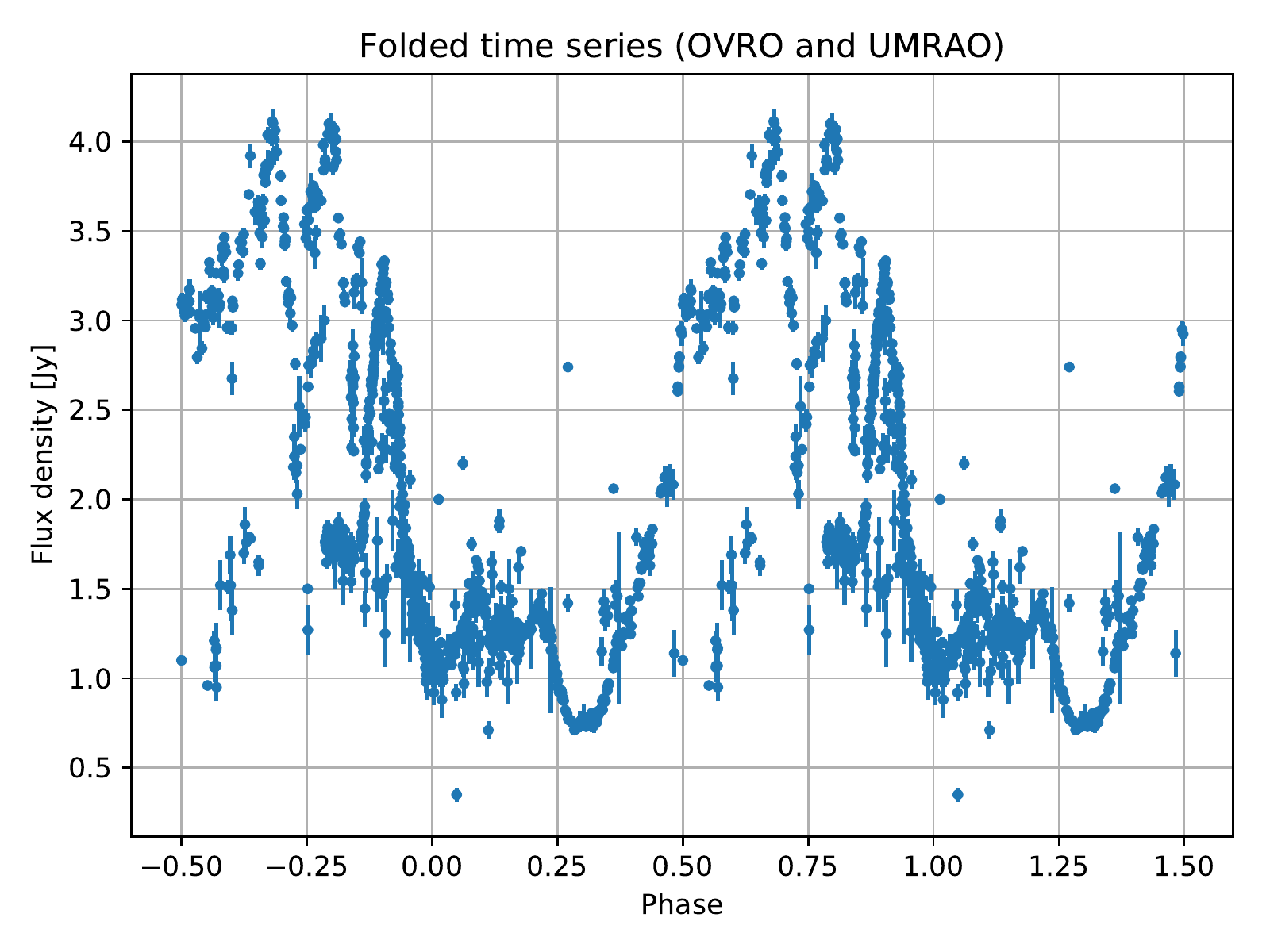}
    \caption{The radio light curve (OVRO and UMRAO dataset) folded with the most prominent frequency peak at $f_{\rm best}=2.4\times 10^{-4}\,{\rm d^{-1}}$, which corresponds to $11.22$ years in the observer's frame.}
    \label{fig_folded_radio_all}
\end{figure}
\section{Position angles of jet components as function of core distance}
\begin{figure}
\begin{minipage}{\linewidth}
  \includegraphics[width=\linewidth]{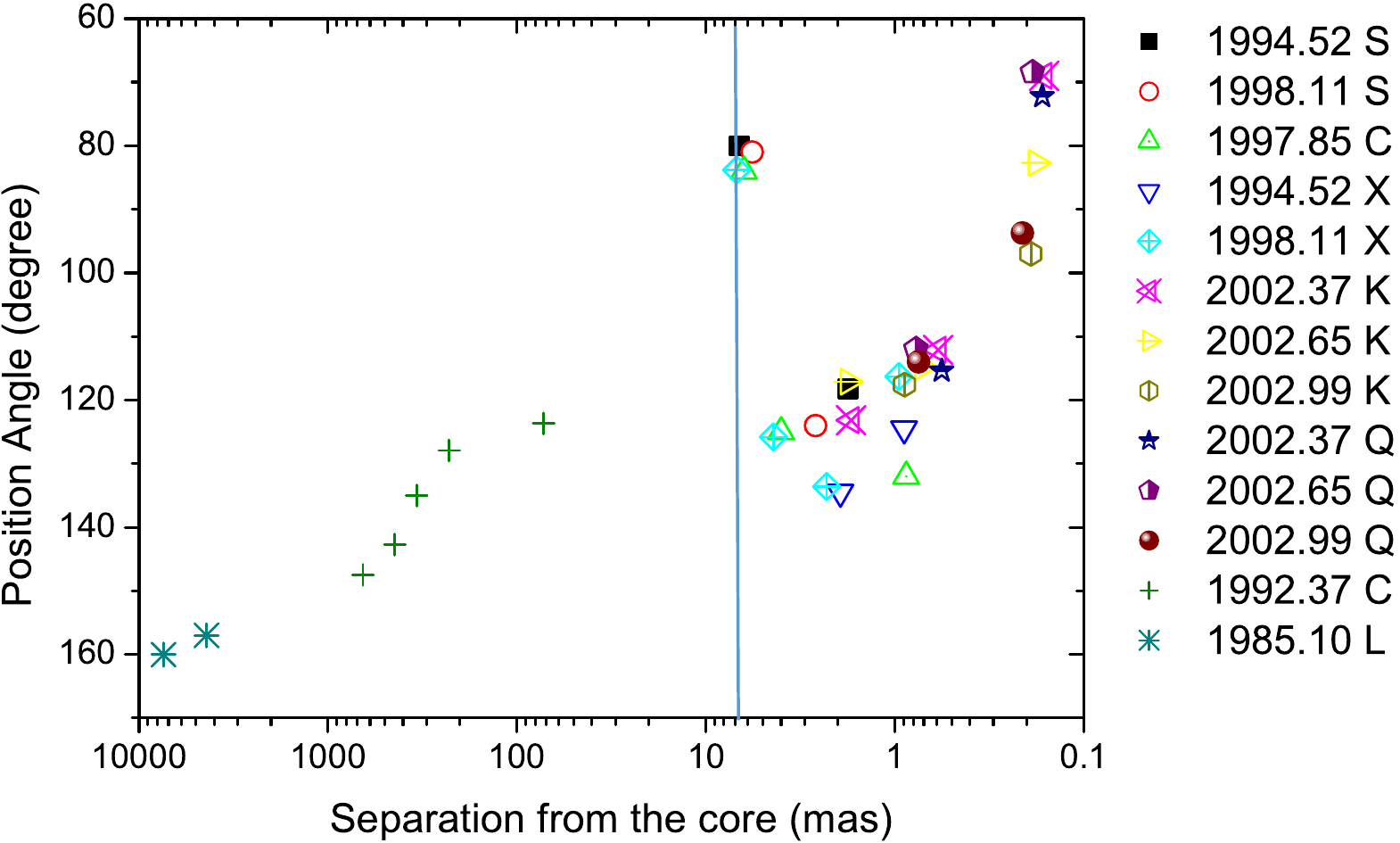}
  [a]
\end{minipage}
\begin{minipage}{7.5cm}
  \includegraphics[width=\linewidth]{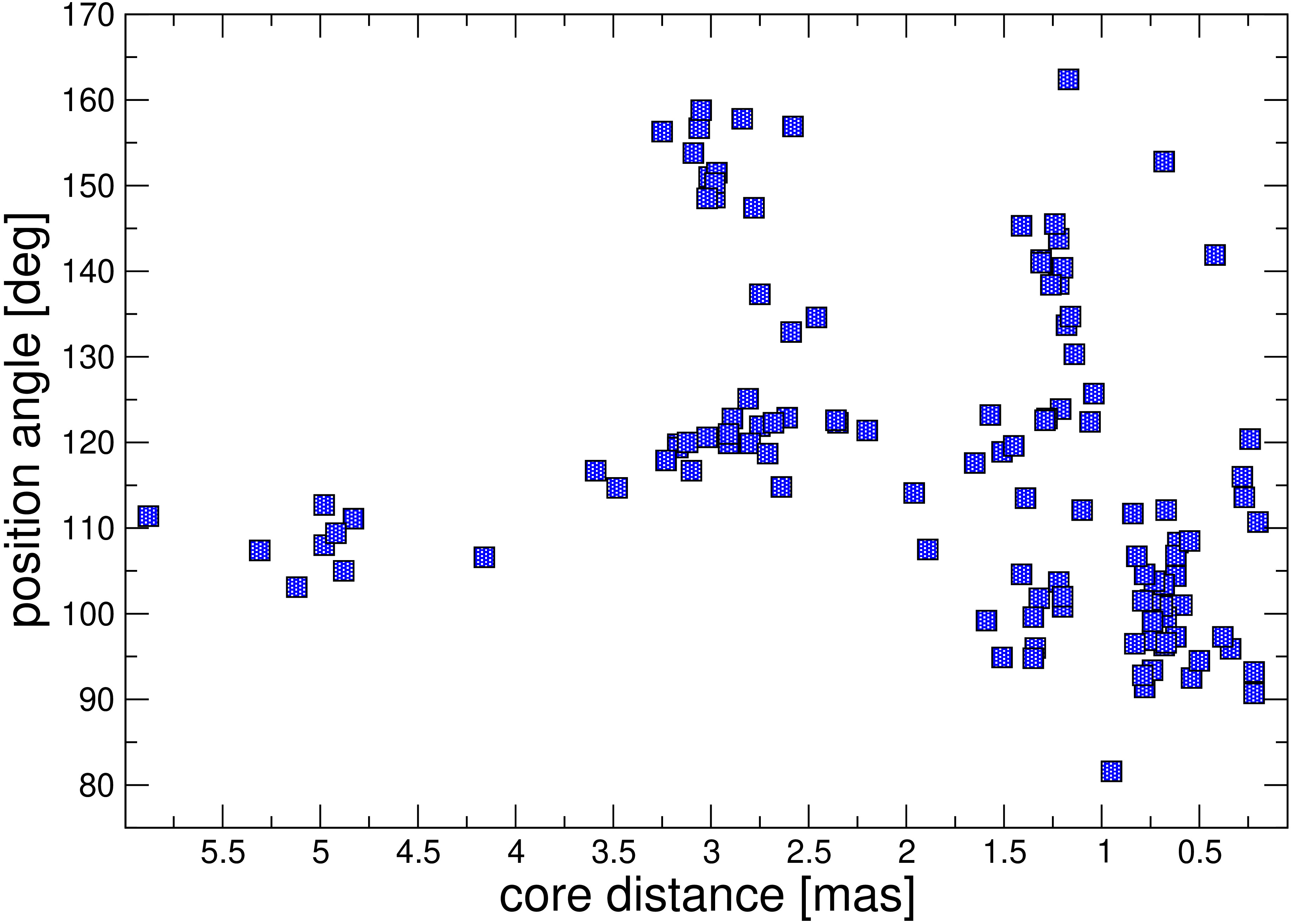}
  [b]
\end{minipage}
\caption{[a] The plot is taken from \citet{an} and shows the position angles of jet components at large separations from the core (based on multi waveband radio data). [b] We produced a similar plot with our VLBA data obtained at 15 GHz. Please note the different scales on the x-axis - the plot in [a] has a logarithmic scale. We mark (blue line) in [a] the portion of parameter space probed with the 15 GHz VLBA data for ease of comparison.}
\label{comparison}
\end{figure}

\bsp	
\label{lastpage}
\end{document}